\documentclass[11pt]{article}

\usepackage[letterpaper, margin=1in]{geometry}
\usepackage{amsmath}
\usepackage{amssymb}
\usepackage{mathrsfs}
\usepackage{graphicx}
\usepackage{booktabs}
\usepackage[authoryear, round]{natbib}
\usepackage{hyperref}
\hypersetup{colorlinks=true, linkcolor=blue, citecolor=blue, urlcolor=blue}
\usepackage{xcolor}
\usepackage{comment}
\usepackage{tikz}
\usepackage{pgfplots}
\pgfplotsset{compat=1.18}
\usetikzlibrary{plotmarks, arrows.meta}
\usepgfplotslibrary{patchplots}

\DeclareRobustCommand{\solidlegend}{%
    \tikz[baseline=-0.6ex]\draw[line width=0.5pt] (0,0) -- (0.25,0);%
}

\DeclareRobustCommand{\dashlegend}{%
    \tikz[baseline=-0.6ex]\draw[line width=0.5pt,dashed] (0,0) -- (0.25,0);%
}

\DeclareRobustCommand{\dottedlegend}{%
    \tikz[baseline=-0.6ex]\draw[line width=0.5pt,dotted] (0,0) -- (0.25,0);%
}

\DeclareRobustCommand{\circlegend}{%
    \tikz[baseline=-0.6ex]\fill (0.45,0) circle (0.11);%
}

\title{On the Rayleigh--Taylor instability of a diffuse interface}
\author{%
  Marildo Kola$^{1}$, Daniel Israel$^{2,3}$, Aaron Towne$^{1}$\\[6pt]
  \small $^{1}$Department of Mechanical Engineering, University of Michigan, Ann Arbor, MI, USA\\
  \small $^{2}$Los Alamos National Laboratory--Michigan SPARC, Ann Arbor, MI, USA\\
  \small $^{3}$Department of Aerospace Engineering, University of Michigan, Ann Arbor, MI, USA\\[4pt]
  \small\textit{Correspondence:} Marildo Kola, \texttt{marildo@umich.edu}%
}
\date{}

\newcommand{\difft}[2]{\frac{\mathrm{d} #1}{\mathrm{d} #2}}

\newcommand{\Atw}{\mathrm{A_t}}
\newcommand{\Sch}{\mathrm{Sc}}

\newcommand{\Frd}{\mathrm{Fr}}
\newcommand{\Pe}{\mathrm{Pe}}
\newcommand{\Amu}{\mathrm{A}_{\mu}}
\newcommand{\bnabla}{\boldsymbol{\nabla}}
\newcommand{\bcdot}{\boldsymbol{\cdot}}
\newcommand{\Rey}{\mathrm{Re}}

\DeclareMathOperator{\erf}{erf}
\DeclareMathAlphabet{\mathsfbi}{OT1}{phv}{b}{sl}
\newenvironment{appen}{\appendix}{}

\newif\ifloadfig
\loadfigtrue

\begin{document}
\maketitle

\begin{abstract}
We present a formulation for the linear stability of the miscible Rayleigh--Taylor instability, retaining viscosity, Fickian mass diffusion, and non-Boussinesq effects. Rather than adopting the ad hoc diffusive corrections of~\citet{duff-harlow-hirt-1962} or relying on the viscous eigenvalue problem of~\citet{chandrasekhar-1961}, the present formulation is derived directly from the incompressible variable-density governing equations. Both approaches are recovered as limiting cases, and the resulting framework can therefore be regarded as a formal extension of~\citet{morgan-likhachev-jacobs-2016}.
The governing equations admit a time-evolving self-similar diffusive base state in which an error-function density profile is accompanied by a nonzero vertical base velocity required by the non-solenoidal constraint. Linearization about this base state yields a reduced coupled formulation for the vertical velocity and density perturbations, suitable for both theoretical analysis and computation. Within the quasi-steady-state approximation, we show that the formulation reduces, in the infinite-Schmidt-number limit, to the classical viscous immiscible theory and recovers the Chandrasekhar eigenvalue problem. At finite Schmidt number, diffusion alters the growth-rate spectrum and its high-wavenumber structure. Beyond the spectral cutoff, we formally establish the scaling \(\omega_r \sim -\lambda_m k^2\), where \(\lambda_m\) is determined by the competition between the locally stratified kinematic viscosity and the reference diffusive scale. 
By isolating the effect of diffusion, we show that its stabilizing influence weakens with increasing density stratification. Buoyancy production becomes bimodal and shifts toward the lighter-fluid side, while pressure production grows to rival buoyancy and acts as an energy source on the heavier side of stratification. 
Comparison with the corresponding initial-value problem shows that the quasi-steady prediction remains accurate over much of the parameter space considered but can fail qualitatively when the base state evolves too rapidly. Together, these results clarify when classical diffusive corrections remain applicable and when diffusion and non-Boussinesq coupling must be retained.
\end{abstract}

\section{Introduction}%
The Rayleigh--Taylor instability (RTI) arises when two superposed fluids of distinct densities are subjected to an acceleration directed against their density stratification. First investigated by~\citet{lord_rayleigh-1879} and later extended by~\citet{taylor-1950}, it has since been recognized as a fundamental mechanism in a broad range of natural processes and engineering applications. In inertial confinement fusion, RTI represents a primary source of performance degradation~\citep{lindl-1995}; in astrophysics, it is a fundamental mechanism in core-collapse supernovae~\citep{muller-2020}; and in oceanography, it drives overturning at stratified interfaces~\citep{turner-1973}.
The RTI is commonly analyzed using linear stability theory (LST), in which infinitesimal perturbations to a prescribed base state are examined to determine whether they grow or decay in time, thereby classifying the flow as linearly stable or unstable. A comprehensive account of linear RTI theory and related developments is reported in the work of Zhou and co-workers~\citep{zhou-2017-a,zhou-2017-b,zhou-et-al-2021}. However, when diffusion is incorporated into RTI, linear stability analysis is not strictly applicable because of the time-dependent nature of the base state. Specifically, LST assumes the existence of a fixed point of the full nonlinear system about which linearization is performed, leading to a linear operator that is inherently time independent. This limitation was already recognized by~\citet{duff-harlow-hirt-1962}, who effectively circumvented it by treating the evolving base state as instantaneously frozen and subsequently validating the computed growth rates against experimental measurements. Although not stated explicitly, this assumption falls within the class of quasi-steady-state approximations described by~\citet{tan-homsy-1986}, in which the time scale associated with base-state evolution is much larger than that of perturbation growth, so that the diffusive base state may be treated as fixed.

In their seminal contribution,~\citet{duff-harlow-hirt-1962} identified two distinct mechanisms through which diffusion influences RTI evolution. The first, termed \emph{dynamic diffusion}, acts on the instability via a base state whose interface broadens over time. The second, termed \emph{static diffusion}, enters as a viscosity-like damping mechanism that preferentially attenuates short-wavelength perturbations. They analyzed the dynamic diffusion by extending the inviscid problem of \citet{chandrasekhar-1961}, arguing that a finite interface thickness enters as a correction to the inviscid growth rate. In this formulation, the growth rate is modified through a function \(\psi \ge 1\), yielding
\begin{equation}
   \omega_r = \sqrt{\frac{\Atw\,g\,k}{\psi(k,\delta,\Atw)}},
\label{eq:duff_growth_rate_correction}
\end{equation}
where
\begin{equation} 
   \Atw = \frac{\rho_2 - \rho_1}{\rho_2 + \rho_1}
\end{equation}
denotes the density Atwood number, \(\rho_1\) and \(\rho_2\) are the constant densities of the two fluids, \(\delta\) is the thickness of the diffusive layer, \(k\) is the magnitude of the in-plane wavenumber, and \(g\) is the gravitational acceleration. Within the same work, they incorporated viscous effects through an approximate treatment that omits the cross terms associated with the combined action of viscosity and dynamic diffusion, leading to
\begin{equation}
   \omega_r = \left( \frac{\Atw \, g \,k}{\psi(k,\delta,\Atw)} + \nu^2\,k^{4} \right)^{1/2} - \nu\,k^{2},
\label{eq:duff_quasi_viscous}
\end{equation}
where \(\nu\) represents the kinematic viscosity. This expression, originally proposed more than sixty years ago, was only recently re-examined by~\citet{morgan-likhachev-jacobs-2016}. To account for the complete coupling between viscous effects and dynamic diffusion, they solved the viscous eigenvalue problem of~\citet{chandrasekhar-1961} by prescribing a frozen density profile of thickness \(\delta\), rather than introducing a correction to the inviscid growth rate. This analysis revealed a large-wavenumber scaling \(\omega_r \propto k^{-2}\), which~\eqref{eq:duff_quasi_viscous} does not reproduce, and showed that, although the expression remains reasonably accurate for low-wavenumber modes, it does not accurately predict the growth rate associated with the most unstable wavenumber.

On the other hand, \citet{duff-harlow-hirt-1962} analyzed the static effect by observing that, if gravity were hypothetically removed at any stage of the instability, perturbations would no longer grow but would instead decay through diffusive spreading. Motivated by this physical argument, static diffusion was modeled as an additional viscosity-like contribution to the dispersion relation, ultimately leading to the well-known expression
\begin{equation}
  \omega_r = \left( \frac{\Atw \, g \,k}{\psi(k,\delta,\Atw)} + \nu^2\,k^{4} \right)^{1/2} - \left( \nu + \mathcal{D} \right) k^2,
\label{eq:duff_static_diffusion}
\end{equation}
where \(\mathcal{D}\) denotes the mass diffusivity. By contrast, diffusion-dominated configurations were excluded from the analysis of~\citet{morgan-likhachev-jacobs-2016}. Within linear stability analysis, the static effect of diffusion has received limited attention beyond the predictions of~\citet{duff-harlow-hirt-1962}. The influence of diffusion in miscible compressible flows was examined by~\citet{lafay-et-al-2007} and~\citet{gauthier-lecruer-2010}, who reported an attenuation of the growth rate and the emergence of a spectral cutoff beyond which the configuration becomes marginally stable. The origin of this marginal stability was not clarified, since the diffusion correction of~\citet{duff-harlow-hirt-1962} predicts negative growth rates beyond the cutoff.

Although the effect of mixing has been extensively investigated in nonlinear and turbulent regimes through direct numerical simulations and experiments~\citep{cook-dimotakis-2001,mueschke-schilling-2009,mueschke-et-al-2009,banerjee-mutnuri-et-al-2012}, its role in setting linear growth rates remains less clearly understood. In particular, existing reduced descriptions often rely on corrections to the inviscid dispersion relation, as discussed above, while the range of validity of such corrections remains unclear. This leaves open the question of how diffusion enters the linear stability problem when the governing transport equations are retained, and whether additional effects arise when viscosity and density stratification are treated simultaneously. It is within this research context that the present work is developed.

We examine the linear stability of the miscible Rayleigh--Taylor problem within the incompressible variable-density equations~\citep{sandoval-1995}, retaining both mass diffusion and non-Boussinesq effects. We first derive the laminar self-similar base-state solutions for incompressible variable-density flow. Building on this structure, we show that the linearized problem admits a reduced coupled density--vertical-velocity formulation, from which the complete initial-value problem is derived. To investigate the linear stability of the system, we then adopt the quasi-steady-state approximation. Within this hypothesis, we identify a large-wavenumber scaling of the dispersion relation, \(\omega_r \sim -\lambda_m k^2\), thereby placing the static-diffusion argument of~\citet{duff-harlow-hirt-1962} on firmer theoretical footing and showing that the Duff scaling is recovered only under suitable conditions; otherwise, the large-wavenumber damping depends on the local stratification of density and dynamic viscosity. We also show that diffusion cannot be characterized as a purely stabilizing correction because of the non-solenoidal nature of the flow.

Although the limitations of the quasi-steady-state approximation, particularly during the early
stages of instability evolution, are well known~\citep{tan-homsy-1986,riaz-et-al-2006,trevelyan-almarcha-wit-2011}, it remains the natural starting point for the present analysis. In related diffusive instability problems, approaches based on adjoint optimization and non-modal stability theory have successfully moved beyond this approximation~\citep{rapaka-et-al-2008,don-nils-amir-2013,prathama-pantano-2021}, but only after a thorough quasi-steady-state analysis had established a clear picture of the instability mechanism and identified where the approximation breaks down. To the authors' knowledge, no such foundation exists for the miscible RTI problem when diffusion and strong non-Boussinesq effects are both retained, and the present work represents a first attempt to build it. Rather than relying on ad hoc corrections, the theoretical framework developed here proceeds directly from the governing equations, and the quasi-steady-state approximation serves as its natural entry point, with its predictions subsequently assessed against those of the full initial-value problem.

The remainder of this paper is organized as follows. In \S\ref{sec:governing_equation}, we examine the governing equations, adopting the incompressible variable-density formulation~\citep{sandoval-1995}, whose linear-stability properties have received little attention despite extensive studies of the nonlinear and turbulent regimes~\citep{livescu-ristorcelli-2007,livescu-et-al-2010,wei-livescu-2012,livescu-2013,baltzer-livescu-2020}. In \S\ref{subsec:zeroth_order_self_similar}, we derive a laminar self-similar base state and discuss the quasi-steady-state approximation used to obtain a frozen, tractable linearized system. In \S\ref{subsec:first_order_equation}, we show that the linearized governing equations admit a density--vertical-velocity formulation. The numerical solver developed for both the initial-value problem and the corresponding quasi-steady eigenvalue problem is described in \S\ref{sec:numerical_implementation}. The main stability results are presented in \S\ref{sec:result_and_discussion}, beginning with the maximum growth rate \(\omega_r^{\max}\), the wavenumber \(k^{\max}\) at which this maximum occurs, and the spectral cutoff wavenumber \(k_c\). This characterization is followed by the derivation of a large-wavenumber scaling in \S\ref{subsec:kggkc} and by an inviscid diffusive-limit analysis that isolates the effect of diffusion on the unstable spectrum in \S\ref{subsec:budget_idl}. In \S\ref{sec:IVP}, we compare the quasi-steady-state predictions with the corresponding initial-value problem. Finally, conclusions are drawn in \S\ref{sec:conclusions}.

\section{Governing equations}%
\label{sec:governing_equation}
We consider the flow of two miscible fluids with different reference densities \(\rho_1\) and
\(\rho_2\), at low Mach number, so that acoustic waves are suppressed and the energy equation
decouples from the problem. The conservation equations for mass and momentum in
dimensional form read
\begin{subequations}
\begin{align}
    \label{eq:dimensional_continuity}
    \partial_t \rho + \bnabla\bcdot(\rho \boldsymbol{u}) &= 0, \\
    \rho\,\partial_t \boldsymbol{u} + \rho\,(\boldsymbol{u}\bcdot\bnabla)\boldsymbol{u} &= -\bnabla p + \bnabla\bcdot\mathsfbi{\tau} - \rho g \boldsymbol{e}_z,
\end{align}
\end{subequations}
where \(\rho(\boldsymbol{x},t)\) is the mixture density, \(\boldsymbol{u}(\boldsymbol{x},t)\) is the
velocity field, \(p(\boldsymbol{x},t)\) is the pressure, and \(\boldsymbol{e}_z\) points
opposite to gravity. The viscous stress tensor \(\mathsfbi{\tau}\) is given by
\begin{equation}
    \mathsfbi{\tau} =  2\,\mu\,\mathsfbi{S} - \frac{2}{3}\mu\,(\bnabla\bcdot\boldsymbol{u})\mathsfbi{I}, \qquad 2\,\mathsfbi{S} = \bnabla\boldsymbol{u} + (\bnabla\boldsymbol{u})^{\!\top},
\end{equation}
where \(\mu\) is the dynamic viscosity, \(\mathsfbi{S}\) is the strain-rate tensor, and
\(\mathsfbi{I}\) is the identity tensor. To describe the mixing of the two fluids, we introduce the
mass fractions \(Y_1(\boldsymbol{x},t)\) and \(Y_2(\boldsymbol{x},t)\), defined as the local mass
of each species divided by the local total mass, such that \(Y_1 + Y_2 = 1\). For a binary mixture in which each pure fluid is individually incompressible it can be assumed that~\citep{livescu-2020},
\begin{equation}
    \label{eq:specific_volume}
    \frac{1}{\rho} = \frac{Y_1}{\rho_1} + \frac{Y_2}{\rho_2}.
\end{equation}
The transport of each species is governed by Fick's law, with a binary diffusion coefficient \(\mathcal{D}\) assumed constant
throughout the domain and independent of mass fraction. For species \(\alpha\), this reads
\begin{equation}
    \label{eq:species}
    \partial_t (\rho Y_\alpha) + \bnabla\bcdot(\rho Y_\alpha \boldsymbol{u}) = \bnabla\bcdot\left(\rho \mathcal{D} \bnabla Y_\alpha\right), \qquad \alpha = 1,2.
\end{equation}
Substituting the mixing law~\eqref{eq:specific_volume} into the species equation~\eqref{eq:species} and making use of~\eqref{eq:dimensional_continuity}, the species transport equation can be
equivalently rewritten as an evolution equation for the density,
\begin{equation}
    \label{eq:dimensional_logrho}
    \partial_t \left(\log \rho\right) + \boldsymbol{u}\bcdot\bnabla \left(\log \rho\right) = \mathcal{D}\,\nabla^2 \left(\log \rho\right).
\end{equation}
Comparing~\eqref{eq:dimensional_logrho} with~\eqref{eq:dimensional_continuity}, it follows that the velocity field satisfies the divergence constraint
\begin{equation}
    \bnabla\bcdot\boldsymbol{u} = -\mathcal{D}\,\nabla^2 \log\rho,
\end{equation}
leading to a set of equations in which each of the two fluids is individually incompressible, yet the flow is not divergence-free~\citep{livescu-2013}.
To render the governing equations dimensionless, we introduce reference quantities for density \(\rho_r\), length \(L_r\), velocity \(U_r\), dynamic viscosity \(\mu_r\), gravitational acceleration \(g\), and mass diffusivity \(\mathcal{D}\). The Reynolds, Froude, and Schmidt numbers
follow as
\begin{equation}
    \Rey = \frac{\rho_r L_r U_r}{\mu_r}, \qquad \Frd = \frac{U_r}{\sqrt{g L_r}}, \qquad \Sch = \frac{\mu_r/\rho_r}{\mathcal{D}}.
\end{equation}
Since the present problem is not characterized by externally imposed length or velocity scales, it is convenient to adopt the classical viscous-gravitational reference scales introduced by~\citet{chandrasekhar-1961}. With this choice, the Reynolds and Froude numbers are set to unity, corresponding to
\begin{equation}
    \label{eq:chandrasekhar_scaling}
    U_r = (\nu_r g)^{1/3}, \qquad T_r = \left(\frac{\nu_r}{g^{2}}\right)^{1/3}, \qquad L_r = \left(\frac{\nu_r^{2}}{g}\right)^{1/3},
\end{equation}
where \(\nu_r=\mu_r/\rho_r\) denotes the reference kinematic viscosity. For the reference density and dynamic viscosity, we take
\begin{equation}
    \rho_r = \frac{\rho_1 + \rho_2}{2}, \qquad \mu_r = \frac{\mu_1 + \mu_2}{2},
\end{equation}
and introduce the density Atwood number and a corresponding viscosity Atwood number,
\begin{equation}
    \Atw = \frac{\rho_2 - \rho_1}{\rho_2 + \rho_1}, \qquad A_{\mu} = \frac{\mu_2 - \mu_1}{\mu_2 + \mu_1}.
\end{equation}
Referring to the nondimensional variables with the same notation as their dimensional counterparts, the governing system reads
\begin{subequations}
    \label{eq:IVD_NL}
    \begin{align}
        \partial_t \rho + \bnabla\bcdot(\rho \boldsymbol{u}) &= 0, \\
        \rho\,\partial_t \boldsymbol{u} + \rho\,(\boldsymbol{u}\bcdot\bnabla)\boldsymbol{u} &= -\,\bnabla p + \bnabla\bcdot\mathsfbi{\tau} - \rho\,\boldsymbol{e}_z, \\
        \bnabla \bcdot \boldsymbol{u} &= -\Sch^{-1}\,\nabla^2 \log\rho.
    \end{align}
\end{subequations}
Introducing the state vector \(\boldsymbol{Q}(\boldsymbol{x},t) = \left[ u, v, w, p, \rho \right]\), the nonlinear system~\eqref{eq:IVD_NL} can be written compactly as
\begin{equation}
    \mathscr{M}\,\partial_t \boldsymbol{Q} = \mathcal{N}\!\left( \boldsymbol{Q} \right),
\end{equation}
where \(\mathscr{M}\) is the identity matrix with the entry corresponding to \(p\) set to zero, since the pressure is not dynamically evolved but instead enforces the divergence constraint.
We consider the asymptotic expansion
\begin{equation}
    \boldsymbol{Q}(\boldsymbol{x},t)
    =
    \boldsymbol{Q}_0(\boldsymbol{x},t)
    +
    \epsilon \boldsymbol{Q}'(\boldsymbol{x},t)
    +
    O(\epsilon^2),
    \qquad 0<\epsilon\ll1 .
\end{equation}
Here, \(\boldsymbol{Q}_0(\boldsymbol{x},t)\) is a solution of the nonlinear equation, while \(\boldsymbol{Q}'(\boldsymbol{x},t)\) denotes the perturbation field. Equating terms at successive orders in \(\epsilon\) yields
\begin{subequations}
    \begin{align}
        \label{eq:compact_0th_equations}
        \mathscr{M}\,\partial_t \boldsymbol{Q}_0
        &=
        \mathcal{N}\!\left( \boldsymbol{Q}_0 \right),
        \\
        \label{eq:compact_1st_equations}
        \mathscr{M}\,\partial_t \boldsymbol{Q}'
        &=
        \left.
        \frac{\delta \mathcal{N}}{\delta \boldsymbol{Q}}
        \right|_{\boldsymbol{Q}_0}
        \boldsymbol{Q}' .
    \end{align}
\end{subequations}
The linearization of the equations follows the classical LST procedure, and the full operator is reported in Appendix~\ref{sec:appendix_reducedIVD}. Since viscosity and density are both determined by the local composition, a mixing model is required to close the system. In the present work, we adopt the linear mixing law of \citet{morgan-likhachev-jacobs-2016}
\begin{equation}
    \frac{\mu}{\rho} = \frac{1-A_{\mu}}{1-\Atw}Y_1 + \frac{1+A_{\mu}}{1+\Atw}Y_2.
    \label{eq:graham_mixing_law}
\end{equation}
The choice of mixing law is, to some extent, arbitrary. A detailed justification for the adopted model, together with a sensitivity analysis of the effect of the viscosity-law closure on the growth rates, is provided in Appendix~\ref{sec:appendix_viscosity_law}.

\subsection{Zeroth-order equations: self-similar evolution of the base state}%
\label{subsec:zeroth_order_self_similar}
In this work the base state is not a steady fixed point but a time-dependent solution of the nonlinear governing equations upon which a quasi-steady-state approximation is imposed. To clarify the precise meaning and implications of this approximation, it is necessary to examine the equations governing the base state itself. Expanding the compact form of~\eqref{eq:compact_0th_equations} yields the zeroth-order system
\begin{subequations}
\begin{align}
    \partial_t \rho_0 + \bnabla\bcdot(\rho_0 \boldsymbol{U}_0) &= 0, \label{eq:vd_zeroth_system_mass} \\
    \rho_0\,\partial_t \boldsymbol{U}_0 + \rho_0\,(\boldsymbol{U}_0\bcdot\bnabla)\boldsymbol{U}_0 &= -\,\bnabla p_0 + \bnabla\bcdot\mathsfbi{\tau}_0 - \rho_0\,\boldsymbol{e}_z, \\
    \bnabla\bcdot\boldsymbol{U}_0 &= -\Sch^{-1}\nabla^2 \log(\rho_0). \label{eq:vd_zeroth_system_divergence}
\end{align}
\end{subequations}
Since no intrinsic length or time scale is imposed on the system, the base-state evolution is assumed to depend only on the vertical coordinate and time~\citep{prathama-pantano-2021},
\begin{equation}
    \boldsymbol{U}_0 = \bigl(u_0(z,t),\, v_0(z,t),\, w_0(z,t)\bigr), \qquad \rho_0 = \rho_0(z,t).
\end{equation}
Equations~\eqref{eq:vd_zeroth_system_mass} and~\eqref{eq:vd_zeroth_system_divergence} then reduce to
\begin{subequations}
\label{eq:base_state_1D}
\begin{align}
    \partial_t \rho_0 + \partial_z (\rho_0 w_0) &= 0, \\
    \partial_z w_0 + \Sch^{-1}\,\partial_{zz} \log(\rho_0) &= 0.
\end{align}
\end{subequations}
Defining the similarity variable
\begin{equation}
    \eta = z \sqrt{\frac{\Sch}{t}} = \frac{2z}{\delta}, \qquad \delta(t) = 2\sqrt{\frac{t}{\Sch}},
\end{equation}
a self-similar solution of~\eqref{eq:base_state_1D} is obtained,
\begin{subequations}
\begin{align}
    \rho_{0}(\eta)
    &=
    1 + \Atw\,\erf\!\left(\frac{\eta}{2}\right),
    \label{eq:self_similar_base_state_density}
    \\
    w_0(\eta)
    &=
    -\,t^{-1/2}\,
    \frac{\Atw}{\sqrt{\pi\,\Sch}}\,
    \frac{e^{-\eta^{2}/4}}
    {1 + \Atw\,\erf\!\left(\eta/2\right)} .
    \label{eq:self_similar_base_state_velocity}
\end{align}
\end{subequations}
\begin{figure}
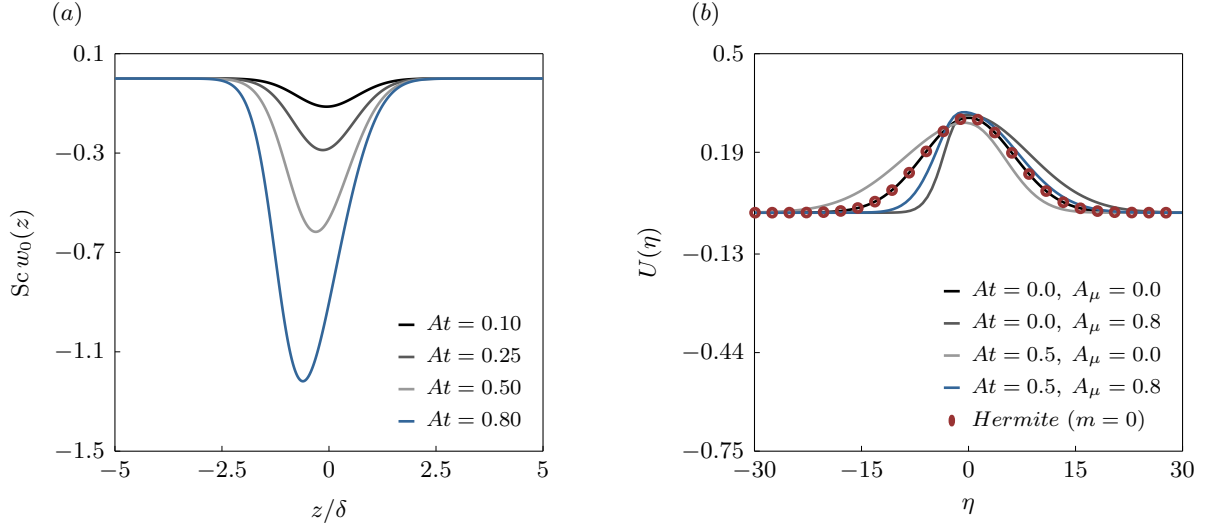

    \centering
    \begin{minipage}[t]{0.49\textwidth}
        \centering
        \vspace{0pt}
        \ifloadfig\input{figures/w0_profile}\fi
    \end{minipage}
    \hfill
    \begin{minipage}[t]{0.49\textwidth}
        \centering
        \vspace{0pt}
        \ifloadfig\input{figures/eigenvector_k1_comparison}\fi
    \end{minipage}
    \caption{Base-state velocity profiles from the self-similar derivation. (a) Scaled vertical velocity \(\Sch\,w_0\) for different Atwood numbers. (b) Horizontal velocity \(U(\eta)\) from the Sturm--Liouville eigenvalue problem at \(\Sch = 20\).}
    \label{fig:base_state_w0_ueta}
\end{figure}
It is worth emphasizing that the error-function form of the base-state density profile in the IVD framework is not imposed as an initial condition but instead arises directly from the governing equations. Although diffusive base states are common in miscible interfaces, the presence of a non-zero base-state vertical velocity is less frequently addressed in RTI analyses. This result is not surprising, since the divergence constraint directly requires the existence of a transverse velocity whenever density gradients are present \citep{bretonnet-et-al-2007}. Equation~\eqref{eq:self_similar_base_state_velocity} shows that this velocity vanishes in the Boussinesq limit, recovering the classical assumption of a quiescent base state. For large density contrasts, however, its magnitude becomes non-negligible and asymmetric with respect to the mid-plane, as illustrated in figure~\ref{fig:base_state_w0_ueta}(a).

The in-plane velocity components are not fixed by the vertical problem alone, and a self-similar solution must therefore be sought independently to complete the derivation of the admissible base state used in the stability analysis. We anticipate that the only solution consistent with far-field decay is the trivial one, \(u_0 = v_0 = 0\), as established by the following argument. Consider the base-state \(x\)-momentum equation, since the \(y\)-component is identical,
\begin{equation}
    \rho_0\,\partial_t u_0 + \rho_0 w_0\,\partial_z u_0 = \partial_z\left(\mu_0\,\partial_z u_0\right).
\end{equation}
Motivated by~\eqref{eq:self_similar_base_state_velocity}, we seek a self-similar solution of the form
\begin{equation}
    u_0(z,t) = t^{\alpha} U(\eta),
\end{equation}
which leads to the Sturm--Liouville eigenvalue problem
\begin{subequations}
\label{eq:sturm_liouville}
\begin{align}
    a(\eta)\,U_{\eta\eta} + b(\eta)\,U_{\eta} &= \alpha\,U, \\
    a(\eta) &= \Sch\,\frac{\mu_0(\eta)}{\rho_0(\eta)}, \\
    b(\eta) &= \frac{\eta}{2} + \frac{\rho_{0,\eta}}{\rho_0(\eta)} + \Sch\,\frac{\mu_{0,\eta}}{\rho_0(\eta)}, \\
    \lim_{|\eta|\to\infty} U &= \lim_{|\eta|\to\infty} U_{\eta} = 0.
\end{align}
\end{subequations}
For the case \(\alpha = 0\), a closed-form solution exists,
\begin{equation}
    U(\eta) = C_2 + C_1 \int^\eta \exp\!\left(-\int^r \frac{b(s)}{a(s)}\,\mathrm{d}s\right)\mathrm{d}r.
\end{equation}
For large \(|\eta|\) this behaves as
\begin{equation}
    U(\eta) \sim C_2 + C_1 \int^\eta \exp\!\left(-\frac{r^2}{4\,\Sch}\,\frac{1\pm\Atw}{1\pm\Amu}\right)\mathrm{d}r,
\end{equation}
which does not decay to zero at infinity, suggesting that the only admissible solution is the trivial one, \(C_1 = C_2 = 0\). For \(\alpha \neq 0\), the eigenvalues of the Sturm--Liouville problem~\eqref{eq:sturm_liouville} are negative and closed-form solutions are generally unavailable; numerical results are reported in figure~\ref{fig:base_state_w0_ueta}(b). In the limiting case \(\mathrm{A}_{\mu} = \Atw \to 0\), the problem reduces to
\begin{equation}
    \Sch\,U_{\eta\eta} + \frac{\eta}{2}\,U_{\eta} = \alpha U,
\end{equation}
where, introducing \(\eta = 2\sqrt{\Sch}\,x\) and writing
\begin{equation}
    U(\eta(x)) = \psi(x) = e^{-x^2} H(x),
\end{equation}
yields the Hermite-type problem
\begin{equation}
    H''(x) - 2x\,H'(x) - (4\alpha + 2)\,H(x) = 0.
\end{equation}
This equation admits polynomial eigenfunctions when
\begin{equation}
    \alpha = -\frac{m+1}{2}, \qquad m = 0,1,2,\dots,
\end{equation}
giving
\begin{equation}
    \psi_m(x) = e^{-x^2} H_m(x).
\end{equation}
These solutions possess a nontrivial Hermite--Gaussian vertical structure and are mathematically admissible within the IVD formulation; however, their inconsistency with the quiescent base state recovered in the standard Boussinesq RTI limit leads us to argue that the physically relevant base state is the one with identically zero in-plane velocity.

Although a detailed investigation of these solutions lies beyond the scope of the present work, they may prove relevant to related stability problems within the IVD framework. Notably, the fundamental mode coincides with the Gaussian vertical envelope commonly assumed in linear-stability models of interior stratified jets~\citep{harris-poulin-lamb-2022}, and the analysis reported here can therefore provide a useful starting point for density-stratified problems in which strong
non-Boussinesq effects play a role.

\subsection{First-order equations and reduction to the velocity--density formulation}%
\label{subsec:first_order_equation}
In the linearized equations, primes are omitted from perturbation variables for clarity and are
instead used exclusively to denote derivatives with respect to the vertical coordinate. The
linearized divergence constraint is
\begin{equation}
    \bnabla \bcdot \boldsymbol{u} = \mathscr{L}_\rho\,\rho, \qquad
    \mathscr{L}_\rho = -\Sch^{-1}\nabla^2\!\left(\frac{\star}{\rho_0}\right),
\end{equation}
where \(\star\) denotes the argument on which the operator acts. The linearized continuity
equation can be written in compact form as
\begin{subequations}
\label{eq:linearized_density}
\begin{align}
    \partial_t \rho &= \mathscr{A}_{\rho w}\,w + \mathscr{A}_{\rho\rho}\,\rho, \\
    \mathscr{A}_{\rho w} &= -\,\partial_z \rho_0, \\
    \mathscr{A}_{\rho\rho} &= -\left(w_0\,\partial_z + w_0'\right) - \rho_0\,\mathscr{L}_\rho.
\end{align}
\end{subequations}
Concerning the momentum equation, the key observation is that the linearized divergence constraint
depends only on the density perturbation. The horizontal velocity components can therefore be
eliminated, yielding a closed system for the vertical velocity and density perturbations. The
details of the reduction are given in Appendix~\ref{sec:appendix_reducedIVD}.

Before introducing the reduced momentum equation, two additional definitions are required. Since
the density field is determined by the composition through~\eqref{eq:specific_volume}, the viscosity perturbation follows directly from the density perturbation. From~\eqref{eq:graham_mixing_law} and~\eqref{eq:self_similar_base_state_density}, the
base-state viscosity is
\begin{equation}
    \mu_0 = 1 + A_{\mu}\,\erf\!\left(z/\delta\right),
\end{equation}
and the viscosity perturbation is linearly coupled to the density perturbation through
\begin{equation}
    \mu = C_{\mu\rho}\,\rho, \qquad
    C_{\mu\rho} = \left.\frac{\mathrm{d}\mu}{\mathrm{d}\rho}\right|_{\rho_0} = \frac{A_{\mu}}{\Atw}.
\end{equation}
We further define the time-differentiated operator
\begin{equation}
    \dot{\mathscr{L}}_\rho(\star)
    =
    \Sch^{-1}
    \nabla^2
    \left(
      \frac{\dot{\rho}_0}{\rho_0^{2}}\,\star
    \right),
\end{equation}
where overdots denote derivatives with respect to time, and
\(\nabla_{xy}^2 = \partial_{xx} + \partial_{yy}\) is the horizontal Laplacian. With these
definitions in hand, the reduced momentum equation reads
\begin{equation}
    \mathscr{B}_{ww}\,\partial_t w
    +
    \mathscr{B}_{w\rho}\,\partial_t \rho
    =
    \left(
      \mathscr{A}_{w_0 \rho}
      +
      \mathscr{A}_{w\rho}
    \right)\rho
    +
    \left(
      \mathscr{A}_{w_0 w}
      +
      \mathscr{A}_{ww}
    \right)w,
\end{equation}
where the operators are defined as
\begin{subequations}
\label{eq:A_B_operators}
\begin{align}
    \mathscr{B}_{ww} &= -\Big(\rho_0'\,\partial_z + \rho_0\,\nabla^2\Big), \\
    \mathscr{B}_{w\rho} &= \Big(\rho_0' + \rho_0\,\partial_z\Big)\mathscr{L}_{\rho}, \\
    \mathscr{A}_{ww} &= -\mu_0\,\nabla^4 - 2\,\mu_0'\,\partial_z\,\nabla^2 + \mu_0''\,\nabla^2 - 2\,\mu_0''\,\partial_{zz}, \\
    \mathscr{A}_{w_0 w} &= \Big(\rho_0\,w_0' + \rho_0\,w_0\,\partial_z\Big)\nabla^2 + w_0\,\rho_0'\,\partial_{zz}, \\
    \mathscr{A}_{w\rho} &= \Big(\mu_0\,\partial_z\,\nabla^2 + 2\,\mu_0'\,\nabla^2 + \mu_0''\,\partial_z\Big)\mathscr{L}_\rho - 2\,\partial_z\!\left(w_0'\,C_{\mu\rho}\,\nabla_{xy}^2\star\right) + \nabla_{xy}^2, \\
    \mathscr{A}_{w_0 \rho} &= -\Big((\rho_0 w_0)'\,\partial_z + \rho_0 w_0\,\partial_{zz}\Big)\mathscr{L}_{\rho} + \left(\dot{w}_0 + w_0 w_0'\right)\nabla_{xy}^2 - \Big(\rho_0' + \rho_0\,\partial_z\Big)\dot{\mathscr{L}}_{\rho}.
\end{align}
\end{subequations}
Together with~\eqref{eq:linearized_density}, these relations define the full initial-value problem
for a time-dependent base state. In compact form,
\begin{equation}
    \label{eq:linearized_continuous_system}
    \begin{pmatrix}
        \mathscr{B}_{ww} & \mathscr{B}_{w\rho} \\
        0 & 1
    \end{pmatrix}
    \begin{pmatrix}
        \partial_t\, w \\[4pt]
        \partial_t\,\rho
    \end{pmatrix}
    =
    \begin{pmatrix}
        \mathscr{A}_{w_0w}+\mathscr{A}_{ww} & \mathscr{A}_{w_0\rho}+\mathscr{A}_{w\rho} \\[4pt]
        \mathscr{A}_{\rho w} & \mathscr{A}_{\rho\rho}
    \end{pmatrix}
    \begin{pmatrix}
        w \\[4pt]
        \rho
    \end{pmatrix}.
\end{equation}
The boundary conditions enforce vanishing vertical velocity and its vertical gradient at the domain boundaries and require the density perturbation to vanish in the far field so that the base-state stratification remains undisturbed away from the diffusive layer,
\begin{equation}
    w(\pm L_z) = 0, \qquad \left.\frac{\partial w}{\partial z}\right|_{\pm L_z} = 0, \qquad \rho(\pm L_z) = 0.
\end{equation}
To perform a stability analysis, the quasi-steady-state approximation is introduced by freezing the base state at a prescribed diffusive layer width \(\delta_0\) and neglecting its temporal derivatives~\citep{junhao-et-al-2006}. Since the self-similar base state satisfies
\begin{equation}
    t_0 = \frac{\Sch \delta_0^2}{4},
\end{equation}
fixing \(\delta_0\) while varying \(\Sch\) implies that each case is evaluated at a different physical time \(t_0\). Comparisons at fixed \(\delta_0\) therefore correspond not to the same physical time, but to different instants at which the diffusive layer has the prescribed width. This choice follows earlier studies in which diffusive effects were characterized by fixing the layer width. For each \(\Sch\), the system is allowed to evolve until the layer reaches \(\delta_0\), although the time required to reach this state depends on \(\Sch\).

The validity of the quasi-steady-state approximation relies on a separation of time scales, whereby the characteristic time over which perturbations grow is assumed to be short compared with the diffusive time scale over which the base state evolves. Since the diffusive layer width is fixed, the temporal derivatives of the base-state quantities scale as
\begin{equation}
    \partial_t w_0 \sim \frac{\Atw}{\Sch^{2}}\,\delta_0^{-3}, \qquad \partial_t \rho_0 \sim \frac{\Atw}{\Sch}\,\delta_0^{-2}.
    \label{eq:estimation_qssa_failure}
\end{equation}
The quasi-steady-state approximation is therefore not uniformly valid across all Schmidt and Atwood numbers for a fixed \(\delta_0\). The magnitude of the neglected temporal derivatives increases with \(\Atw\) and decreases with \(\Sch\), so that for sufficiently small \(\Sch\) or sufficiently large \(\Atw\) the base state evolves on a time scale comparable to that of the perturbation growth and the frozen-profile assumption breaks down. In the present work, cases down to \(\Sch = 0.75\) and up to \(\Atw = 0.75\) are nevertheless considered, for which the separation of time scales may not be fully respected. Accordingly, results in that regime should be interpreted as indicative of the trends predicted by the quasi-steady-state approximation, rather than as outcomes supported by a clear asymptotic separation of scales.

Upon freezing the base state, the temporal derivatives of the base-state quantities are set to zero in~\eqref{eq:A_B_operators}. The resulting frozen operator system is then discretized and used to compute the growth rates. For the state vectors, we use bold italic symbols for vectors of continuous functions and bold roman symbols for spatially discretized vectors. The state vector associated with the reduced perturbation field is
\begin{equation}
    \boldsymbol{q}(\mathbf{x},t) =
    \begin{bmatrix}
        w(\mathbf{x},t) \\
        \rho(\mathbf{x},t)
    \end{bmatrix}.
\end{equation}
Assuming homogeneity in the in-plane directions \((x,y)\), the perturbation fields are decomposed into Fourier modes. The linearized operators in~\eqref{eq:A_B_operators} depend only on \(k^{2} = k_{x}^{2} + k_{y}^{2}\), where \(k_x\) and \(k_y\) are the wavenumbers in the \(x\) and \(y\) directions, respectively, and the state vector can therefore be expressed as
\begin{equation}
    \hat{\boldsymbol{q}}_{k}(z,t) =
    \begin{bmatrix}
        \hat{w}(z,k,t) \\
        \hat{\rho}(z,k,t)
    \end{bmatrix},
\end{equation}
where the subscript \(k\) denotes the Fourier coefficient at wavenumber \(k\). Evaluating the two components at the \(N_z\) vertical collocation points and stacking their values gives the spatially discretized state \(\hat{\mathbf{q}}_k(t)\in\mathbb{C}^{2N_z}\). This leads to the semi-discrete system
\begin{equation}
    \mathsfbi{B}_{k}\,\difft{\hat{\mathbf{q}}_{k}}{t} = \mathsfbi{A}_{k}\,\hat{\mathbf{q}}_{k},
\end{equation}
where the block matrices \(\mathsfbi{A}_{k}\) and \(\mathsfbi{B}_{k}\) follow from the discretization of~\eqref{eq:linearized_continuous_system}, and the growth rates are obtained by solving the generalized eigenvalue problem
\begin{equation}
    \mathsfbi{A}_{k}\,\hat{\mathbf{q}}_{k} = \omega(k)\,\mathsfbi{B}_{k}\,\hat{\mathbf{q}}_{k},
\end{equation}
where \(\omega(k)\) is the complex eigenvalue and its real part defines the temporal growth rate, \(\omega_r(k)\)

\section{Numerical implementation}%
\label{sec:numerical_implementation}
\begin{figure}
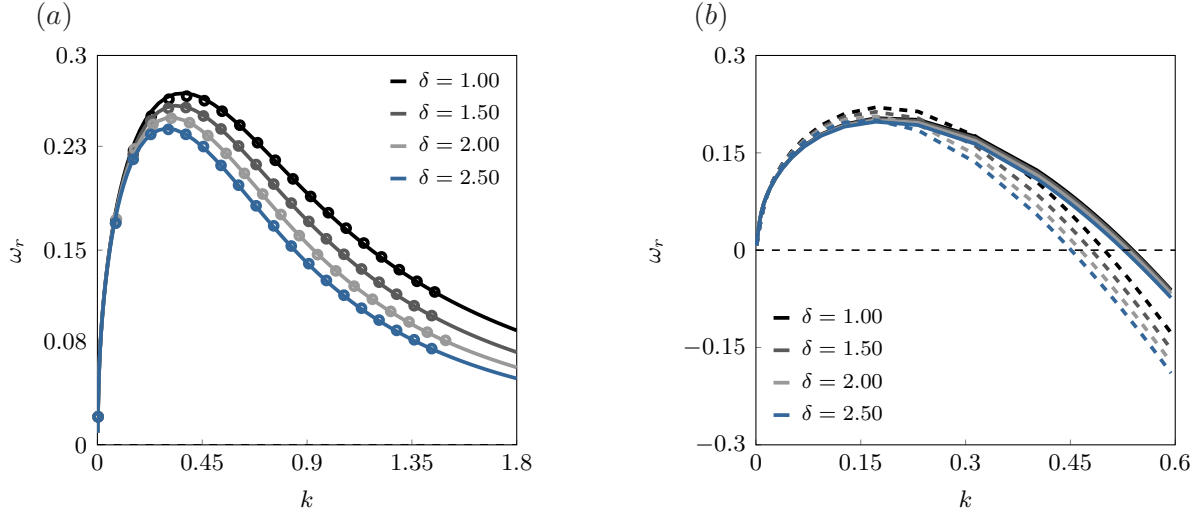

    \centering
    \begin{minipage}[t]{0.48\textwidth}
        \centering
        \vspace{0pt}
        \ifloadfig\input{figures/numerical_vs_duff_1}\fi
    \end{minipage}
    \hfill
    \begin{minipage}[t]{0.48\textwidth}
        \centering
        \vspace{0pt}
        \ifloadfig\input{figures/numerical_vs_duff_2}\fi
    \end{minipage}
    \caption{Growth-rate comparison for \(\Amu = 0.0\) and \(\Atw = 0.5\). (\solidlegend) denotes results based on the present analysis. (a) Diffusion-free limit, \(\Sch \to \infty\); (\circlegend) denotes the viscous growth rates reported by~\citet{morgan-likhachev-jacobs-2016}. (b) \(\Sch = 1\); (\dashlegend) denotes \citet{duff-harlow-hirt-1962} prediction.}
    \label{fig:duff_versus_prediction}
\end{figure}
Following the derivation of the governing equations, a solver employing a Chebyshev spectral method is developed using the MATLAB differentiation suite of~\cite{weideman-reddy-2000}. To resolve the sharp gradient of the initial profile, a coordinate transformation that concentrates collocation points within the diffusive layer is introduced. For a physical domain \([a,b]=[-L_{z},L_{z}]\), the mapping is
\begin{equation}
    z(\xi) = \frac{a+b}{2} + \frac{b-a}{2}\,\frac{\sinh\!\bigl(\beta\,\xi\bigr)}{\sinh(\beta)}, \quad \xi\in[-1,1],\; z\in[a,b],
\end{equation}
where the free parameter \(\beta\) controls the clustering, and for \(\beta\gg1\) most points fall near \(z=0\). The inverse mapping, required to construct the spectral matrices, follows from
\begin{equation}
    \frac{dz}{d\xi} = \frac{b-a}{2}\,\frac{\beta\,\cosh\!\bigl(\beta\,\xi\bigr)}{\sinh(\beta)}, \qquad
    \frac{d\xi}{dz} = \left(\frac{dz}{d\xi}\right)^{-1} = \frac{2}{b-a}\,\frac{\sinh(\beta)}{\beta\,\cosh\!\bigl(\beta\,\xi\bigr)}.
\end{equation}
Hence, the first-derivative matrix in physical space requires only the Chebyshev matrix \(D^{(\xi)}\),
\begin{equation}
    \left.\frac{df}{dz}\right|_{z_{i}} \approx \sum_{j=1}^{N} \underbrace{\Bigl(\tfrac{dz}{d\xi}\Bigr)^{-1}_{\xi_{i}}}_{w_{i}} D_{ij}^{(\xi)}\,f_{j}, \qquad
    D^{(z)} = \operatorname{diag}(w_{i})\,D^{(\xi)}.
\end{equation}
For the second derivative,
\begin{equation}
    \frac{d^{2}f}{dz^{2}} = \left(\frac{d\xi}{dz}\right)^{2}\frac{d^{2}f}{d\xi^{2}} + \frac{d^{2}\xi}{dz^{2}}\frac{df}{d\xi}, \qquad
    \frac{d^{2}\xi}{dz^{2}} = -\,\frac{d^{2}z}{d\xi^{2}}\left(\frac{dz}{d\xi}\right)^{-3},
\end{equation}
so that
\begin{equation}
    \bigl(D^{(z,2)}\mathbf{f}\bigr)_{i} \approx w_{i}^{2}\bigl(D^{(\xi,2)}\mathbf{f}\bigr)_{i} + u_{i}\bigl(D^{(\xi,1)}\mathbf{f}\bigr)_{i}, \qquad
    w_{i} = \left.\frac{d\xi}{dz}\right|_{z_{i}}, \; u_{i} = \left.\frac{d^{2}\xi}{dz^{2}}\right|_{z_{i}},
\end{equation}
which yields
\begin{equation}
    D^{(z,2)} = \operatorname{diag}\!\bigl(w_{i}^{2}\bigr)\,D^{(\xi,2)} + \operatorname{diag}\!\bigl(u_{i}\bigr)\,D^{(\xi,1)}.
\end{equation}
A detailed verification of the solver is reported in Appendix~\ref{sec:appendix_numerical_verification}. An additional remark is needed regarding the computation of growth rates in the limit \(k \to 0\), for which the unbounded-domain limit \(L_z \to \infty\) is formally required. In practice, the computational domain is taken sufficiently large, while the clustering parameter \(\beta\) associated with the nonlinear coordinate transformation is chosen so as to ensure that at least \(0.25N_z\) grid points lie within \(|z|\leq 4\delta_0\).

\section{Results and Discussion}%
\label{sec:result_and_discussion}
Before examining the instability mechanisms we note that the non-diffusive viscous theory of~\citet{chandrasekhar-1961} is recovered in the limit \(\Sch \to \infty\). In this limit mass diffusion vanishes, and the linearized IVD system reduces exactly to the equations of~\citet{chandrasekhar-1961}, as discussed in Appendix~\ref{sec:appendix_Sc_infinity_chandra_limit} and shown in figure~\ref{fig:duff_versus_prediction}(a). On this basis, we suggest that the formulation presented here may be viewed as an extension of Chandrasekhar's eigenvalue problem and therefore a consistent extension of the result of \citet{morgan-likhachev-jacobs-2016}. At finite Schmidt number, however, the effects of mass diffusion can no longer be neglected. At \(\Sch = 1\), the dispersion relation acquires a fundamentally different structure, as shown in figure~\ref{fig:duff_versus_prediction}(b). The \(k^{-2}\) decay predicted by~\citet{morgan-likhachev-jacobs-2016} is replaced by a cutoff wavenumber beyond which growth rates are negative. The Duff model reported in figure~\ref{fig:duff_versus_prediction}(b) overpredicts the maximum growth rate and, while it does predict a power-law decay at large wavenumbers, the associated prefactor depends sensitively on the region of parameter space considered. These discrepancies motivate a systematic analysis, which is carried out in the remainder of this section.

In \S\ref{subsec:growth_rate_and_spectral_cutoff_characterization}, we analyze the maximum growth rate \(\omega_r^{\max}\), which identifies the peak of each stability curve, while \(k^{\max}\) denotes the wavenumber at which this maximum occurs. The spectral cutoff wavenumber \(k_c\), defined as the wavenumber at which the growth rate changes sign, marks the boundary between linearly unstable and stable disturbances and traces the locus of marginal stability across parameter space. In \S\ref{subsec:kggkc}, we then examine the large-wavenumber regime \(k \gg k_c\), which characterizes the decay of the growth-rate curves beyond the cutoff. Finally, in \S\ref{subsec:budget_idl}, we isolate the role of diffusion through the inviscid diffusive limit and use the perturbation energy budget to interpret the trends observed in the spectral diagnostics.

\subsection{Growth rate and spectral cutoff characterization}
\label{subsec:growth_rate_and_spectral_cutoff_characterization}
\begin{figure}
    \centering
    \begin{minipage}[t]{0.32\textwidth}
        \centering
        \vspace{0pt}
        \ifloadfig
%
\definecolor{mycolor1}{rgb}{0.12941,0.12941,0.12941}%
\begin{tikzpicture}[trim axis left,trim axis right]

\begin{axis}[%
width=11.255in,
height=7.131in,
at={(1.888in,0.962in)},
scale only axis,
clip=true,
xmode=log,
xmin=0.01,
xmax=1.99526231496888,
xminorticks=true,
ymin=0,
ymax=0.6,
ytick={   0, 0.15,  0.3, 0.45,  0.6},
ylabel style={font=\color{mycolor1}},
ylabel={$\omega_r^{\max}$},
axis background/.style={fill=white},
title style={font=\bfseries\color{mycolor1}},
title={$A_{\mu} = -0.9$},
legend style={legend cell align=left, align=left},
clip=true,
clip mode=individual,
axis lines=box,
xtick pos=bottom,
ytick pos=left,
tick align=inside,
major tick length=2pt,
width=0.7\linewidth,
height=0.7\linewidth,
every axis/.append style={font=\fontsize{9}{9}\selectfont},xlabel style={font=\fontsize{9}{9}\selectfont},title style={font=\fontsize{9}{9}\selectfont},
legend style={font=\fontsize{8}{9}\selectfont,inner sep=1pt,row sep=1pt,column sep=2pt,nodes={scale=1.00},draw=none},legend image post style={xscale=0.35},
legend columns=1,
scaled x ticks=false,
tick label style={/pgf/number format/fixed,/pgf/number format/precision=2},
every x tick label/.append style={font=\fontsize{9}{9}\selectfont\color{black}},
every y tick label/.append style={font=\fontsize{9}{9}\selectfont\color{black}},
,,
colormap={sigmamap}{rgb(0.0000)=(0.0200,0.1880,0.3800); rgb(0.1000)=(0.0743,0.3456,0.5616); rgb(0.2000)=(0.1918,0.4980,0.7060); rgb(0.3000)=(0.3890,0.6708,0.8226); rgb(0.4000)=(0.7552,0.8682,0.9228); rgb(0.5000)=(0.9690,0.9689,0.9689); rgb(0.6000)=(0.9425,0.8044,0.7614); rgb(0.7000)=(0.8767,0.4910,0.4020); rgb(0.8000)=(0.7869,0.2866,0.2470); rgb(0.9000)=(0.6186,0.1239,0.1568); rgb(1.0000)=(0.4040,0.0000,0.1220)}
]
\addplot [color=black, line width=1.4pt]
  table[row sep=crcr]{%
0.01	0.0942586359836777\\
0.0125892541179417	0.0940758285784043\\
0.0158489319246111	0.0938481535298791\\
0.0199526231496888	0.0935698275672973\\
0.0251188643150958	0.0932330645449328\\
0.0316227766016838	0.0928204604373322\\
0.0398107170553497	0.0923273230514315\\
0.0501187233627272	0.0917368847137047\\
0.0630957344480193	0.0910339813173969\\
0.0794328234724281	0.0902067661145267\\
0.1	0.0892425617995714\\
0.125892541179417	0.0881280591911212\\
0.158489319246111	0.0868492718524562\\
0.199526231496888	0.0853962837427153\\
0.251188643150958	0.0837727181509143\\
0.316227766016838	0.0819568929869703\\
0.398107170553497	0.079966560942216\\
0.501187233627272	0.0777931937418598\\
0.630957344480193	0.0754381010055371\\
0.794328234724281	0.0729297322452033\\
1	0.0702720017811506\\
1.25892541179417	0.0674726197612465\\
1.58489319246111	0.0645512268171742\\
1.99526231496888	0.0615571862971885\\
};
\addlegendentry{$A_t=0.1$}

\addplot [color=black, dashed, line width=1.4pt, forget plot]
  table[row sep=crcr]{%
0.01	0.0962975153602883\\
0.0125892541179417	0.0962031015084586\\
0.0158489319246111	0.0960842415111648\\
0.0199526231496888	0.0959394819590804\\
0.0251188643150958	0.095760885859097\\
0.0316227766016838	0.0955360466903806\\
0.0398107170553497	0.0952529909473168\\
0.0501187233627272	0.0949069848322852\\
0.0630957344480193	0.0944822933758437\\
0.0794328234724281	0.0939566305637435\\
0.1	0.0933204145276264\\
0.125892541179417	0.0925517086165093\\
0.158489319246111	0.0916251171366777\\
0.199526231496888	0.0905223519488307\\
0.251188643150958	0.0892231317296694\\
0.316227766016838	0.0877051116860658\\
0.398107170553497	0.0859537112177226\\
0.501187233627272	0.0839731018331232\\
0.630957344480193	0.0817375953657291\\
0.794328234724281	0.079267175346622\\
1	0.0765829807780593\\
1.25892541179417	0.0736877940904068\\
1.58489319246111	0.0706082542813983\\
1.99526231496888	0.0674237450890013\\
};
\addplot [color=white!60!black, line width=1.4pt]
  table[row sep=crcr]{%
0.01	0.232926540589627\\
0.0125892541179417	0.232494049072786\\
0.0158489319246111	0.231958853245998\\
0.0199526231496888	0.231305918592346\\
0.0251188643150958	0.230505069296259\\
0.0316227766016838	0.229528933027649\\
0.0398107170553497	0.228352579466213\\
0.0501187233627272	0.226940126909442\\
0.0630957344480193	0.225261528349898\\
0.0794328234724281	0.223274715240933\\
0.1	0.220940165342061\\
0.125892541179417	0.218229860644617\\
0.158489319246111	0.215108999288527\\
0.199526231496888	0.211543192567253\\
0.251188643150958	0.207505248140603\\
0.316227766016838	0.202992165792675\\
0.398107170553497	0.197972490687439\\
0.501187233627272	0.192452504039498\\
0.630957344480193	0.186433936126914\\
0.794328234724281	0.179953794602879\\
1	0.17302402174557\\
1.25892541179417	0.165712871748712\\
1.58489319246111	0.158118103004402\\
1.99526231496888	0.150326911458491\\
};
\addlegendentry{$A_t=0.4$}

\addplot [color=white!60!black, dashed, line width=1.4pt, forget plot]
  table[row sep=crcr]{%
0.01	0.234109688134333\\
0.0125892541179417	0.233904645309307\\
0.0158489319246111	0.233646511686376\\
0.0199526231496888	0.233321540708828\\
0.0251188643150958	0.232912426487099\\
0.0316227766016838	0.232407162916309\\
0.0398107170553497	0.231781711037004\\
0.0501187233627272	0.230999258412079\\
0.0630957344480193	0.230043702635086\\
0.0794328234724281	0.228866608740174\\
0.1	0.227426760098122\\
0.125892541179417	0.225681083265141\\
0.158489319246111	0.22357845206686\\
0.199526231496888	0.221058224639927\\
0.251188643150958	0.218092303939079\\
0.316227766016838	0.214609898927795\\
0.398107170553497	0.210580393400892\\
0.501187233627272	0.205976589181567\\
0.630957344480193	0.20077672201325\\
0.794328234724281	0.195003951937102\\
1	0.188668555166456\\
1.25892541179417	0.18183866546972\\
1.58489319246111	0.174540665277468\\
1.99526231496888	0.166930232784276\\
};
\addplot [color=red!60!teal, line width=1.4pt]
  table[row sep=crcr]{%
0.01	0.356922968459693\\
0.0125892541179417	0.356205120378081\\
0.0158489319246111	0.355311593822146\\
0.0199526231496888	0.354209844988844\\
0.0251188643150958	0.352857886831417\\
0.0316227766016838	0.351202074754863\\
0.0398107170553497	0.349187916372775\\
0.0501187233627272	0.346758207506698\\
0.0630957344480193	0.343856005626556\\
0.0794328234724281	0.340417456788992\\
0.1	0.336390472192155\\
0.125892541179417	0.331734950231085\\
0.158489319246111	0.326439277207895\\
0.199526231496888	0.320512686367953\\
0.251188643150958	0.314028216064297\\
0.316227766016838	0.307096584883553\\
0.398107170553497	0.299922762139308\\
0.501187233627272	0.292813557296494\\
0.630957344480193	0.286207543216872\\
0.794328234724281	0.280707672387463\\
1	0.277194096096318\\
1.25892541179417	0.276738678147708\\
1.58489319246111	0.280841097961228\\
1.99526231496888	0.291468542134357\\
};
\addlegendentry{$A_t=0.75$}

\addplot [color=red!60!teal, dashed, line width=1.4pt, forget plot]
  table[row sep=crcr]{%
0.01	0.366193323603007\\
0.0125892541179417	0.365856500472884\\
0.0158489319246111	0.365441560551583\\
0.0199526231496888	0.36491918214029\\
0.0251188643150958	0.364261546683741\\
0.0316227766016838	0.36345107433306\\
0.0398107170553497	0.362437945669062\\
0.0501187233627272	0.361190240881419\\
0.0630957344480193	0.359645963767819\\
0.0794328234724281	0.357751269649458\\
0.1	0.355441084233261\\
0.125892541179417	0.35263677015736\\
0.158489319246111	0.349258917028929\\
0.199526231496888	0.345222651765689\\
0.251188643150958	0.340458317605567\\
0.316227766016838	0.334890312674737\\
0.398107170553497	0.328437427004685\\
0.501187233627272	0.321085215458011\\
0.630957344480193	0.312797106843565\\
0.794328234724281	0.303588214488961\\
1	0.293503519940607\\
1.25892541179417	0.282655225510767\\
1.58489319246111	0.27111532353595\\
1.99526231496888	0.259015977136474\\
};
\node[right, align=left, inner sep=0, font=\color{mycolor1}]
at (rel axis cs:-0.25,1.15) {$(a)$};
\end{axis}
\end{tikzpicture}%
\fi
    \end{minipage}
    \hfill
    \begin{minipage}[t]{0.32\textwidth}
        \centering
        \vspace{0pt}
        \ifloadfig
%
\definecolor{mycolor1}{rgb}{0.12941,0.12941,0.12941}%
\begin{tikzpicture}[trim axis left,trim axis right]

\begin{axis}[%
width=11.255in,
height=7.131in,
at={(1.888in,0.962in)},
scale only axis,
clip=true,
xmode=log,
xmin=0.01,
xmax=1.99526231496888,
xminorticks=true,
ymin=0,
ymax=0.6,
ytick={   0, 0.15,  0.3, 0.45,  0.6},
axis background/.style={fill=white},
title style={font=\bfseries\color{mycolor1}},
title={$A_{\mu} = 0.0$},
clip=true,
clip mode=individual,
axis lines=box,
xtick pos=bottom,
ytick pos=left,
tick align=inside,
major tick length=2pt,
width=0.7\linewidth,
height=0.7\linewidth,
every axis/.append style={font=\fontsize{9}{9}\selectfont},xlabel style={font=\fontsize{9}{9}\selectfont},title style={font=\fontsize{9}{9}\selectfont},
legend style={font=\fontsize{8}{9}\selectfont,inner sep=1pt,row sep=1pt,column sep=2pt,nodes={scale=1.00},draw=none},legend image post style={xscale=0.35},
legend columns=1,
scaled x ticks=false,
tick label style={/pgf/number format/fixed,/pgf/number format/precision=2},
every x tick label/.append style={font=\fontsize{9}{9}\selectfont\color{black}},
every y tick label/.append style={font=\fontsize{9}{9}\selectfont\color{black}},
,,
colormap={sigmamap}{rgb(0.0000)=(0.0200,0.1880,0.3800); rgb(0.1000)=(0.0743,0.3456,0.5616); rgb(0.2000)=(0.1918,0.4980,0.7060); rgb(0.3000)=(0.3890,0.6708,0.8226); rgb(0.4000)=(0.7552,0.8682,0.9228); rgb(0.5000)=(0.9690,0.9689,0.9689); rgb(0.6000)=(0.9425,0.8044,0.7614); rgb(0.7000)=(0.8767,0.4910,0.4020); rgb(0.8000)=(0.7869,0.2866,0.2470); rgb(0.9000)=(0.6186,0.1239,0.1568); rgb(1.0000)=(0.4040,0.0000,0.1220)}
]
\addplot [color=black, line width=1.4pt, forget plot]
  table[row sep=crcr]{%
0.01	0.093188093625324\\
0.0125892541179417	0.0930148120448383\\
0.0158489319246111	0.0928069373906492\\
0.0199526231496888	0.0925485791808597\\
0.0251188643150958	0.0922298087303794\\
0.0316227766016838	0.0918476638536247\\
0.0398107170553497	0.0913811473460477\\
0.0501187233627272	0.0908211835565532\\
0.0630957344480193	0.0901571378991911\\
0.0794328234724281	0.0893699932362037\\
0.1	0.0884460942789248\\
0.125892541179417	0.087370841446579\\
0.158489319246111	0.0861367231971655\\
0.199526231496888	0.0847319580141265\\
0.251188643150958	0.0831399276765019\\
0.316227766016838	0.0813724799348028\\
0.398107170553497	0.0794148141266089\\
0.501187233627272	0.077268008541997\\
0.630957344480193	0.0749578434134269\\
0.794328234724281	0.0724817332324935\\
1	0.069852251865404\\
1.25892541179417	0.0670854225311164\\
1.58489319246111	0.064216989754907\\
1.99526231496888	0.0612685999811111\\
};
\addplot [color=black, dashed, line width=1.4pt, forget plot]
  table[row sep=crcr]{%
0.01	0.0962975153602883\\
0.0125892541179417	0.0962031015084586\\
0.0158489319246111	0.0960842415111648\\
0.0199526231496888	0.0959394819590804\\
0.0251188643150958	0.095760885859097\\
0.0316227766016838	0.0955360466903806\\
0.0398107170553497	0.0952529909473168\\
0.0501187233627272	0.0949069848322852\\
0.0630957344480193	0.0944822933758437\\
0.0794328234724281	0.0939566305637435\\
0.1	0.0933204145276264\\
0.125892541179417	0.0925517086165093\\
0.158489319246111	0.0916251171366777\\
0.199526231496888	0.0905223519488307\\
0.251188643150958	0.0892231317296694\\
0.316227766016838	0.0877051116860658\\
0.398107170553497	0.0859537112177226\\
0.501187233627272	0.0839731018331232\\
0.630957344480193	0.0817375953657291\\
0.794328234724281	0.079267175346622\\
1	0.0765829807780593\\
1.25892541179417	0.0736877940904068\\
1.58489319246111	0.0706082542813983\\
1.99526231496888	0.0674237450890013\\
};
\addplot [color=white!60!black, line width=1.4pt, forget plot]
  table[row sep=crcr]{%
0.01	0.230788272292261\\
0.0125892541179417	0.230433522078495\\
0.0158489319246111	0.229988926098471\\
0.0199526231496888	0.22944583795685\\
0.0251188643150958	0.228768439370452\\
0.0316227766016838	0.227944443192755\\
0.0398107170553497	0.226933748172238\\
0.0501187233627272	0.225705220735626\\
0.0630957344480193	0.224222482022974\\
0.0794328234724281	0.222443215057965\\
0.1	0.220329672762682\\
0.125892541179417	0.217833662931069\\
0.158489319246111	0.214922274190182\\
0.199526231496888	0.211548740665257\\
0.251188643150958	0.207679938531821\\
0.316227766016838	0.203301586504884\\
0.398107170553497	0.198374546794827\\
0.501187233627272	0.192921474301483\\
0.630957344480193	0.186931665823048\\
0.794328234724281	0.1804371670716\\
1	0.173489083962391\\
1.25892541179417	0.16616994236346\\
1.58489319246111	0.158587660815083\\
1.99526231496888	0.150878127949441\\
};
\addplot [color=white!60!black, dashed, line width=1.4pt, forget plot]
  table[row sep=crcr]{%
0.01	0.234109688134333\\
0.0125892541179417	0.233904645309307\\
0.0158489319246111	0.233646511686376\\
0.0199526231496888	0.233321540708828\\
0.0251188643150958	0.232912426487099\\
0.0316227766016838	0.232407162916309\\
0.0398107170553497	0.231781711037004\\
0.0501187233627272	0.230999258412079\\
0.0630957344480193	0.230043702635086\\
0.0794328234724281	0.228866608740174\\
0.1	0.227426760098122\\
0.125892541179417	0.225681083265141\\
0.158489319246111	0.22357845206686\\
0.199526231496888	0.221058224639927\\
0.251188643150958	0.218092303939079\\
0.316227766016838	0.214609898927795\\
0.398107170553497	0.210580393400892\\
0.501187233627272	0.205976589181567\\
0.630957344480193	0.20077672201325\\
0.794328234724281	0.195003951937102\\
1	0.188668555166456\\
1.25892541179417	0.18183866546972\\
1.58489319246111	0.174540665277468\\
1.99526231496888	0.166930232784276\\
};
\addplot [color=red!60!teal, line width=1.4pt, forget plot]
  table[row sep=crcr]{%
0.01	0.35519323506708\\
0.0125892541179417	0.354699754890304\\
0.0158489319246111	0.354084605016238\\
0.0199526231496888	0.353313143856087\\
0.0251188643150958	0.352360181099463\\
0.0316227766016838	0.351178536786262\\
0.0398107170553497	0.34972060119611\\
0.0501187233627272	0.34793404039118\\
0.0630957344480193	0.345757133435115\\
0.0794328234724281	0.343129427210743\\
0.1	0.339983138723875\\
0.125892541179417	0.336269412272938\\
0.158489319246111	0.331943605330894\\
0.199526231496888	0.326984832639906\\
0.251188643150958	0.321413561324935\\
0.316227766016838	0.315312844623107\\
0.398107170553497	0.308854693806808\\
0.501187233627272	0.302322088193141\\
0.630957344480193	0.296147769357476\\
0.794328234724281	0.290960365499137\\
1	0.287637068964933\\
1.25892541179417	0.287396650131418\\
1.58489319246111	0.291783405253808\\
1.99526231496888	0.302831088733144\\
};
\addplot [color=red!60!teal, dashed, line width=1.4pt, forget plot]
  table[row sep=crcr]{%
0.01	0.366193323603007\\
0.0125892541179417	0.365856500472884\\
0.0158489319246111	0.365441560551583\\
0.0199526231496888	0.36491918214029\\
0.0251188643150958	0.364261546683741\\
0.0316227766016838	0.36345107433306\\
0.0398107170553497	0.362437945669062\\
0.0501187233627272	0.361190240881419\\
0.0630957344480193	0.359645963767819\\
0.0794328234724281	0.357751269649458\\
0.1	0.355441084233261\\
0.125892541179417	0.35263677015736\\
0.158489319246111	0.349258917028929\\
0.199526231496888	0.345222651765689\\
0.251188643150958	0.340458317605567\\
0.316227766016838	0.334890312674737\\
0.398107170553497	0.328437427004685\\
0.501187233627272	0.321085215458011\\
0.630957344480193	0.312797106843565\\
0.794328234724281	0.303588214488961\\
1	0.293503519940607\\
1.25892541179417	0.282655225510767\\
1.58489319246111	0.27111532353595\\
1.99526231496888	0.259015977136474\\
};
\node[right, align=left, inner sep=0, font=\color{mycolor1}]
at (rel axis cs:-0.25,1.15) {$(b)$};
\end{axis}
\end{tikzpicture}%
\fi
    \end{minipage}
    \hfill
    \begin{minipage}[t]{0.32\textwidth}
        \centering
        \vspace{0pt}
        \ifloadfig
%
\definecolor{mycolor1}{rgb}{0.12941,0.12941,0.12941}%
\begin{tikzpicture}[trim axis left,trim axis right]

\begin{axis}[%
width=11.255in,
height=7.131in,
at={(1.888in,0.962in)},
scale only axis,
clip=true,
xmode=log,
xmin=0.01,
xmax=1.99526231496888,
xminorticks=true,
ymin=0,
ymax=0.6,
ytick={   0, 0.15,  0.3, 0.45,  0.6},
axis background/.style={fill=white},
title style={font=\bfseries\color{mycolor1}},
title={$A_{\mu} = 0.9$},
clip=true,
clip mode=individual,
axis lines=box,
xtick pos=bottom,
ytick pos=left,
tick align=inside,
major tick length=2pt,
width=0.7\linewidth,
height=0.7\linewidth,
every axis/.append style={font=\fontsize{9}{9}\selectfont},xlabel style={font=\fontsize{9}{9}\selectfont},title style={font=\fontsize{9}{9}\selectfont},
legend style={font=\fontsize{8}{9}\selectfont,inner sep=1pt,row sep=1pt,column sep=2pt,nodes={scale=1.00},draw=none},legend image post style={xscale=0.35},
legend columns=1,
scaled x ticks=false,
tick label style={/pgf/number format/fixed,/pgf/number format/precision=2},
every x tick label/.append style={font=\fontsize{9}{9}\selectfont\color{black}},
every y tick label/.append style={font=\fontsize{9}{9}\selectfont\color{black}},
,,
colormap={sigmamap}{rgb(0.0000)=(0.0200,0.1880,0.3800); rgb(0.1000)=(0.0743,0.3456,0.5616); rgb(0.2000)=(0.1918,0.4980,0.7060); rgb(0.3000)=(0.3890,0.6708,0.8226); rgb(0.4000)=(0.7552,0.8682,0.9228); rgb(0.5000)=(0.9690,0.9689,0.9689); rgb(0.6000)=(0.9425,0.8044,0.7614); rgb(0.7000)=(0.8767,0.4910,0.4020); rgb(0.8000)=(0.7869,0.2866,0.2470); rgb(0.9000)=(0.6186,0.1239,0.1568); rgb(1.0000)=(0.4040,0.0000,0.1220)}
]
\addplot [color=black, line width=1.4pt, forget plot]
  table[row sep=crcr]{%
0.01	0.0941311931009808\\
0.0125892541179417	0.093963374157687\\
0.0158489319246111	0.0937528099107025\\
0.0199526231496888	0.0934889892863534\\
0.0251188643150958	0.0931680097142817\\
0.0316227766016838	0.0927727391268573\\
0.0398107170553497	0.09229403000533\\
0.0501187233627272	0.0917120828332615\\
0.0630957344480193	0.0910184547354298\\
0.0794328234724281	0.0901963482563398\\
0.1	0.0892282900175293\\
0.125892541179417	0.0881022830855456\\
0.158489319246111	0.0868106946536953\\
0.199526231496888	0.0853342447650544\\
0.251188643150958	0.0836737482544393\\
0.316227766016838	0.0818203246772269\\
0.398107170553497	0.0797867001391676\\
0.501187233627272	0.0775649140787725\\
0.630957344480193	0.0751656931032278\\
0.794328234724281	0.0726167444532431\\
1	0.0699276715534488\\
1.25892541179417	0.0671128882872677\\
1.58489319246111	0.0641895707158072\\
1.99526231496888	0.0611992851379526\\
};
\addplot [color=black, dashed, line width=1.4pt, forget plot]
  table[row sep=crcr]{%
0.01	0.0962975153602883\\
0.0125892541179417	0.0962031015084586\\
0.0158489319246111	0.0960842415111648\\
0.0199526231496888	0.0959394819590804\\
0.0251188643150958	0.095760885859097\\
0.0316227766016838	0.0955360466903806\\
0.0398107170553497	0.0952529909473168\\
0.0501187233627272	0.0949069848322852\\
0.0630957344480193	0.0944822933758437\\
0.0794328234724281	0.0939566305637435\\
0.1	0.0933204145276264\\
0.125892541179417	0.0925517086165093\\
0.158489319246111	0.0916251171366777\\
0.199526231496888	0.0905223519488307\\
0.251188643150958	0.0892231317296694\\
0.316227766016838	0.0877051116860658\\
0.398107170553497	0.0859537112177226\\
0.501187233627272	0.0839731018331232\\
0.630957344480193	0.0817375953657291\\
0.794328234724281	0.079267175346622\\
1	0.0765829807780593\\
1.25892541179417	0.0736877940904068\\
1.58489319246111	0.0706082542813983\\
1.99526231496888	0.0674237450890013\\
};
\addplot [color=white!60!black, line width=1.4pt, forget plot]
  table[row sep=crcr]{%
0.01	0.232709889543503\\
0.0125892541179417	0.23239727026561\\
0.0158489319246111	0.232001370116244\\
0.0199526231496888	0.231509223255748\\
0.0251188643150958	0.230890049068002\\
0.0316227766016838	0.230119752316016\\
0.0398107170553497	0.229159324065402\\
0.0501187233627272	0.227973407278312\\
0.0630957344480193	0.226520413515539\\
0.0794328234724281	0.224751852400967\\
0.1	0.22261169289092\\
0.125892541179417	0.220055790433847\\
0.158489319246111	0.217029059654231\\
0.199526231496888	0.213488999039021\\
0.251188643150958	0.209394138434278\\
0.316227766016838	0.204727546399439\\
0.398107170553497	0.199463706451952\\
0.501187233627272	0.193617077212126\\
0.630957344480193	0.187213806985187\\
0.794328234724281	0.180296174160189\\
1	0.172943497356142\\
1.25892541179417	0.165254640318777\\
1.58489319246111	0.157359242357626\\
1.99526231496888	0.149422901292408\\
};
\addplot [color=white!60!black, dashed, line width=1.4pt, forget plot]
  table[row sep=crcr]{%
0.01	0.234109688134333\\
0.0125892541179417	0.233904645309307\\
0.0158489319246111	0.233646511686376\\
0.0199526231496888	0.233321540708828\\
0.0251188643150958	0.232912426487099\\
0.0316227766016838	0.232407162916309\\
0.0398107170553497	0.231781711037004\\
0.0501187233627272	0.230999258412079\\
0.0630957344480193	0.230043702635086\\
0.0794328234724281	0.228866608740174\\
0.1	0.227426760098122\\
0.125892541179417	0.225681083265141\\
0.158489319246111	0.22357845206686\\
0.199526231496888	0.221058224639927\\
0.251188643150958	0.218092303939079\\
0.316227766016838	0.214609898927795\\
0.398107170553497	0.210580393400892\\
0.501187233627272	0.205976589181567\\
0.630957344480193	0.20077672201325\\
0.794328234724281	0.195003951937102\\
1	0.188668555166456\\
1.25892541179417	0.18183866546972\\
1.58489319246111	0.174540665277468\\
1.99526231496888	0.166930232784276\\
};
\addplot [color=red!60!teal, line width=1.4pt, forget plot]
  table[row sep=crcr]{%
0.01	0.359227939762784\\
0.0125892541179417	0.358933661477061\\
0.0158489319246111	0.358562566710888\\
0.0199526231496888	0.358086169522205\\
0.0251188643150958	0.357473894168415\\
0.0316227766016838	0.356700847324931\\
0.0398107170553497	0.355709055717636\\
0.0501187233627272	0.354450943780663\\
0.0630957344480193	0.352861660463128\\
0.0794328234724281	0.350863859429656\\
0.1	0.348380621663895\\
0.125892541179417	0.345316988954787\\
0.158489319246111	0.341599635745428\\
0.199526231496888	0.337166081879636\\
0.251188643150958	0.331984127087211\\
0.316227766016838	0.326095221913084\\
0.398107170553497	0.319633231827316\\
0.501187233627272	0.312857399287386\\
0.630957344480193	0.306223930716662\\
0.794328234724281	0.300422888955646\\
1	0.296425060376071\\
1.25892541179417	0.295579338207336\\
1.58489319246111	0.299627796965601\\
1.99526231496888	0.310736539808305\\
};
\addplot [color=red!60!teal, dashed, line width=1.4pt, forget plot]
  table[row sep=crcr]{%
0.01	0.366193323603007\\
0.0125892541179417	0.365856500472884\\
0.0158489319246111	0.365441560551583\\
0.0199526231496888	0.36491918214029\\
0.0251188643150958	0.364261546683741\\
0.0316227766016838	0.36345107433306\\
0.0398107170553497	0.362437945669062\\
0.0501187233627272	0.361190240881419\\
0.0630957344480193	0.359645963767819\\
0.0794328234724281	0.357751269649458\\
0.1	0.355441084233261\\
0.125892541179417	0.35263677015736\\
0.158489319246111	0.349258917028929\\
0.199526231496888	0.345222651765689\\
0.251188643150958	0.340458317605567\\
0.316227766016838	0.334890312674737\\
0.398107170553497	0.328437427004685\\
0.501187233627272	0.321085215458011\\
0.630957344480193	0.312797106843565\\
0.794328234724281	0.303588214488961\\
1	0.293503519940607\\
1.25892541179417	0.282655225510767\\
1.58489319246111	0.27111532353595\\
1.99526231496888	0.259015977136474\\
};
\node[right, align=left, inner sep=0, font=\color{mycolor1}]
at (rel axis cs:-0.25,1.15) {$(c)$};
\end{axis}
\end{tikzpicture}%
\fi
    \end{minipage}
    \\[0.75em]
    \begin{minipage}[t]{0.32\textwidth}
        \centering
        \vspace{0pt}
        \ifloadfig
%
\definecolor{mycolor1}{rgb}{0.12941,0.12941,0.12941}%
\begin{tikzpicture}[trim axis left,trim axis right]

\begin{axis}[%
width=11.255in,
height=7.131in,
at={(1.888in,0.962in)},
scale only axis,
clip=true,
xmode=log,
xmin=0.01,
xmax=1.99526231496888,
xminorticks=true,
xlabel style={font=\color{mycolor1}},
xlabel={$\mathrm{Sc}^{-1}$},
ymin=0,
ymax=0.6,
ytick={   0, 0.15,  0.3, 0.45,  0.6},
ylabel style={font=\color{mycolor1}},
ylabel={$k^{\max}$},
axis background/.style={fill=white},
clip=true,
clip mode=individual,
axis lines=box,
xtick pos=bottom,
ytick pos=left,
tick align=inside,
major tick length=2pt,
width=0.7\linewidth,
height=0.7\linewidth,
every axis/.append style={font=\fontsize{9}{9}\selectfont},xlabel style={font=\fontsize{9}{9}\selectfont},title style={font=\fontsize{9}{9}\selectfont},
legend style={font=\fontsize{8}{9}\selectfont,inner sep=1pt,row sep=1pt,column sep=2pt,nodes={scale=1.00},draw=none},legend image post style={xscale=0.35},
legend columns=1,
scaled x ticks=false,
tick label style={/pgf/number format/fixed,/pgf/number format/precision=2},
every x tick label/.append style={font=\fontsize{9}{9}\selectfont\color{black}},
every y tick label/.append style={font=\fontsize{9}{9}\selectfont\color{black}},
,,
colormap={sigmamap}{rgb(0.0000)=(0.0200,0.1880,0.3800); rgb(0.1000)=(0.0743,0.3456,0.5616); rgb(0.2000)=(0.1918,0.4980,0.7060); rgb(0.3000)=(0.3890,0.6708,0.8226); rgb(0.4000)=(0.7552,0.8682,0.9228); rgb(0.5000)=(0.9690,0.9689,0.9689); rgb(0.6000)=(0.9425,0.8044,0.7614); rgb(0.7000)=(0.8767,0.4910,0.4020); rgb(0.8000)=(0.7869,0.2866,0.2470); rgb(0.9000)=(0.6186,0.1239,0.1568); rgb(1.0000)=(0.4040,0.0000,0.1220)}
]
\addplot [color=black, line width=1.4pt, forget plot]
  table[row sep=crcr]{%
0.01	0.221105527638191\\
0.0125892541179417	0.221105527638191\\
0.0158489319246111	0.221105527638191\\
0.0199526231496888	0.21608040201005\\
0.0251188643150958	0.21608040201005\\
0.0316227766016838	0.21105527638191\\
0.0398107170553497	0.21105527638191\\
0.0501187233627272	0.206030150753769\\
0.0630957344480193	0.201005025125628\\
0.0794328234724281	0.195979899497487\\
0.1	0.190954773869347\\
0.125892541179417	0.185929648241206\\
0.158489319246111	0.180904522613065\\
0.199526231496888	0.170854271356784\\
0.251188643150958	0.165829145728643\\
0.316227766016838	0.155778894472362\\
0.398107170553497	0.150753768844221\\
0.501187233627272	0.14070351758794\\
0.630957344480193	0.130653266331658\\
0.794328234724281	0.125628140703518\\
1	0.115577889447236\\
1.25892541179417	0.105527638190955\\
1.58489319246111	0.100502512562814\\
1.99526231496888	0.0904522613065327\\
};
\addplot [color=black, dashed, line width=1.4pt, forget plot]
  table[row sep=crcr]{%
0.01	0.190954773869347\\
0.0125892541179417	0.190954773869347\\
0.0158489319246111	0.190954773869347\\
0.0199526231496888	0.185929648241206\\
0.0251188643150958	0.185929648241206\\
0.0316227766016838	0.185929648241206\\
0.0398107170553497	0.185929648241206\\
0.0501187233627272	0.180904522613065\\
0.0630957344480193	0.180904522613065\\
0.0794328234724281	0.175879396984925\\
0.1	0.175879396984925\\
0.125892541179417	0.170854271356784\\
0.158489319246111	0.165829145728643\\
0.199526231496888	0.160804020100503\\
0.251188643150958	0.155778894472362\\
0.316227766016838	0.150753768844221\\
0.398107170553497	0.14070351758794\\
0.501187233627272	0.135678391959799\\
0.630957344480193	0.125628140703518\\
0.794328234724281	0.120603015075377\\
1	0.110552763819095\\
1.25892541179417	0.100502512562814\\
1.58489319246111	0.0904522613065327\\
1.99526231496888	0.085427135678392\\
};
\addplot [color=white!60!black, line width=1.4pt, forget plot]
  table[row sep=crcr]{%
0.01	0.331658291457286\\
0.0125892541179417	0.331658291457286\\
0.0158489319246111	0.326633165829146\\
0.0199526231496888	0.326633165829146\\
0.0251188643150958	0.321608040201005\\
0.0316227766016838	0.316582914572864\\
0.0398107170553497	0.316582914572864\\
0.0501187233627272	0.306532663316583\\
0.0630957344480193	0.301507537688442\\
0.0794328234724281	0.296482412060302\\
0.1	0.28643216080402\\
0.125892541179417	0.281407035175879\\
0.158489319246111	0.271356783919598\\
0.199526231496888	0.261306532663317\\
0.251188643150958	0.251256281407035\\
0.316227766016838	0.236180904522613\\
0.398107170553497	0.226130653266332\\
0.501187233627272	0.21105527638191\\
0.630957344480193	0.201005025125628\\
0.794328234724281	0.185929648241206\\
1	0.170854271356784\\
1.25892541179417	0.160804020100503\\
1.58489319246111	0.14572864321608\\
1.99526231496888	0.130653266331658\\
};
\addplot [color=white!60!black, dashed, line width=1.4pt, forget plot]
  table[row sep=crcr]{%
0.01	0.281407035175879\\
0.0125892541179417	0.281407035175879\\
0.0158489319246111	0.281407035175879\\
0.0199526231496888	0.281407035175879\\
0.0251188643150958	0.281407035175879\\
0.0316227766016838	0.276381909547739\\
0.0398107170553497	0.276381909547739\\
0.0501187233627272	0.271356783919598\\
0.0630957344480193	0.271356783919598\\
0.0794328234724281	0.266331658291457\\
0.1	0.261306532663317\\
0.125892541179417	0.256281407035176\\
0.158489319246111	0.251256281407035\\
0.199526231496888	0.246231155778894\\
0.251188643150958	0.236180904522613\\
0.316227766016838	0.226130653266332\\
0.398107170553497	0.21608040201005\\
0.501187233627272	0.206030150753769\\
0.630957344480193	0.195979899497487\\
0.794328234724281	0.180904522613065\\
1	0.170854271356784\\
1.25892541179417	0.155778894472362\\
1.58489319246111	0.14070351758794\\
1.99526231496888	0.130653266331658\\
};
\addplot [color=red!60!teal, line width=1.4pt, forget plot]
  table[row sep=crcr]{%
0.01	0.381909547738693\\
0.0125892541179417	0.381909547738693\\
0.0158489319246111	0.376884422110553\\
0.0199526231496888	0.376884422110553\\
0.0251188643150958	0.371859296482412\\
0.0316227766016838	0.366834170854271\\
0.0398107170553497	0.361809045226131\\
0.0501187233627272	0.351758793969849\\
0.0630957344480193	0.346733668341709\\
0.0794328234724281	0.336683417085427\\
0.1	0.326633165829146\\
0.125892541179417	0.316582914572864\\
0.158489319246111	0.301507537688442\\
0.199526231496888	0.291457286432161\\
0.251188643150958	0.276381909547739\\
0.316227766016838	0.261306532663317\\
0.398107170553497	0.246231155778894\\
0.501187233627272	0.231155778894472\\
0.630957344480193	0.21608040201005\\
0.794328234724281	0.206030150753769\\
1	0.190954773869347\\
1.25892541179417	0.180904522613065\\
1.58489319246111	0.170854271356784\\
1.99526231496888	0.165829145728643\\
};
\addplot [color=red!60!teal, dashed, line width=1.4pt, forget plot]
  table[row sep=crcr]{%
0.01	0.361809045226131\\
0.0125892541179417	0.35678391959799\\
0.0158489319246111	0.35678391959799\\
0.0199526231496888	0.35678391959799\\
0.0251188643150958	0.35678391959799\\
0.0316227766016838	0.351758793969849\\
0.0398107170553497	0.351758793969849\\
0.0501187233627272	0.346733668341709\\
0.0630957344480193	0.341708542713568\\
0.0794328234724281	0.336683417085427\\
0.1	0.331658291457286\\
0.125892541179417	0.326633165829146\\
0.158489319246111	0.316582914572864\\
0.199526231496888	0.306532663316583\\
0.251188643150958	0.296482412060302\\
0.316227766016838	0.28643216080402\\
0.398107170553497	0.276381909547739\\
0.501187233627272	0.261306532663317\\
0.630957344480193	0.246231155778894\\
0.794328234724281	0.231155778894472\\
1	0.21105527638191\\
1.25892541179417	0.195979899497487\\
1.58489319246111	0.180904522613065\\
1.99526231496888	0.165829145728643\\
};
\node[right, align=left, inner sep=0, font=\color{mycolor1}]
at (rel axis cs:-0.25,1.15) {$(d)$};
\end{axis}
\end{tikzpicture}%
\fi
    \end{minipage}
    \hfill
    \begin{minipage}[t]{0.32\textwidth}
        \centering
        \vspace{0pt}
        \ifloadfig
%
\definecolor{mycolor1}{rgb}{0.12941,0.12941,0.12941}%
\begin{tikzpicture}[trim axis left,trim axis right]

\begin{axis}[%
width=11.255in,
height=7.131in,
at={(1.888in,0.962in)},
scale only axis,
clip=true,
xmode=log,
xmin=0.01,
xmax=1.99526231496888,
xminorticks=true,
xlabel style={font=\color{mycolor1}},
xlabel={$\mathrm{Sc}^{-1}$},
ymin=0,
ymax=0.6,
ytick={   0, 0.15,  0.3, 0.45,  0.6},
axis background/.style={fill=white},
clip=true,
clip mode=individual,
axis lines=box,
xtick pos=bottom,
ytick pos=left,
tick align=inside,
major tick length=2pt,
width=0.7\linewidth,
height=0.7\linewidth,
every axis/.append style={font=\fontsize{9}{9}\selectfont},xlabel style={font=\fontsize{9}{9}\selectfont},title style={font=\fontsize{9}{9}\selectfont},
legend style={font=\fontsize{8}{9}\selectfont,inner sep=1pt,row sep=1pt,column sep=2pt,nodes={scale=1.00},draw=none},legend image post style={xscale=0.35},
legend columns=1,
scaled x ticks=false,
tick label style={/pgf/number format/fixed,/pgf/number format/precision=2},
every x tick label/.append style={font=\fontsize{9}{9}\selectfont\color{black}},
every y tick label/.append style={font=\fontsize{9}{9}\selectfont\color{black}},
,,
colormap={sigmamap}{rgb(0.0000)=(0.0200,0.1880,0.3800); rgb(0.1000)=(0.0743,0.3456,0.5616); rgb(0.2000)=(0.1918,0.4980,0.7060); rgb(0.3000)=(0.3890,0.6708,0.8226); rgb(0.4000)=(0.7552,0.8682,0.9228); rgb(0.5000)=(0.9690,0.9689,0.9689); rgb(0.6000)=(0.9425,0.8044,0.7614); rgb(0.7000)=(0.8767,0.4910,0.4020); rgb(0.8000)=(0.7869,0.2866,0.2470); rgb(0.9000)=(0.6186,0.1239,0.1568); rgb(1.0000)=(0.4040,0.0000,0.1220)}
]
\addplot [color=black, line width=1.4pt, forget plot]
  table[row sep=crcr]{%
0.01	0.226130653266332\\
0.0125892541179417	0.226130653266332\\
0.0158489319246111	0.221105527638191\\
0.0199526231496888	0.221105527638191\\
0.0251188643150958	0.21608040201005\\
0.0316227766016838	0.21608040201005\\
0.0398107170553497	0.21105527638191\\
0.0501187233627272	0.21105527638191\\
0.0630957344480193	0.206030150753769\\
0.0794328234724281	0.201005025125628\\
0.1	0.195979899497487\\
0.125892541179417	0.190954773869347\\
0.158489319246111	0.180904522613065\\
0.199526231496888	0.175879396984925\\
0.251188643150958	0.165829145728643\\
0.316227766016838	0.160804020100503\\
0.398107170553497	0.150753768844221\\
0.501187233627272	0.14070351758794\\
0.630957344480193	0.135678391959799\\
0.794328234724281	0.125628140703518\\
1	0.115577889447236\\
1.25892541179417	0.105527638190955\\
1.58489319246111	0.100502512562814\\
1.99526231496888	0.0904522613065327\\
};
\addplot [color=black, dashed, line width=1.4pt, forget plot]
  table[row sep=crcr]{%
0.01	0.190954773869347\\
0.0125892541179417	0.190954773869347\\
0.0158489319246111	0.190954773869347\\
0.0199526231496888	0.185929648241206\\
0.0251188643150958	0.185929648241206\\
0.0316227766016838	0.185929648241206\\
0.0398107170553497	0.185929648241206\\
0.0501187233627272	0.180904522613065\\
0.0630957344480193	0.180904522613065\\
0.0794328234724281	0.175879396984925\\
0.1	0.175879396984925\\
0.125892541179417	0.170854271356784\\
0.158489319246111	0.165829145728643\\
0.199526231496888	0.160804020100503\\
0.251188643150958	0.155778894472362\\
0.316227766016838	0.150753768844221\\
0.398107170553497	0.14070351758794\\
0.501187233627272	0.135678391959799\\
0.630957344480193	0.125628140703518\\
0.794328234724281	0.120603015075377\\
1	0.110552763819095\\
1.25892541179417	0.100502512562814\\
1.58489319246111	0.0904522613065327\\
1.99526231496888	0.085427135678392\\
};
\addplot [color=white!60!black, line width=1.4pt, forget plot]
  table[row sep=crcr]{%
0.01	0.336683417085427\\
0.0125892541179417	0.336683417085427\\
0.0158489319246111	0.336683417085427\\
0.0199526231496888	0.331658291457286\\
0.0251188643150958	0.331658291457286\\
0.0316227766016838	0.326633165829146\\
0.0398107170553497	0.321608040201005\\
0.0501187233627272	0.316582914572864\\
0.0630957344480193	0.311557788944724\\
0.0794328234724281	0.306532663316583\\
0.1	0.296482412060302\\
0.125892541179417	0.291457286432161\\
0.158489319246111	0.281407035175879\\
0.199526231496888	0.271356783919598\\
0.251188643150958	0.256281407035176\\
0.316227766016838	0.246231155778894\\
0.398107170553497	0.236180904522613\\
0.501187233627272	0.221105527638191\\
0.630957344480193	0.206030150753769\\
0.794328234724281	0.190954773869347\\
1	0.175879396984925\\
1.25892541179417	0.165829145728643\\
1.58489319246111	0.150753768844221\\
1.99526231496888	0.135678391959799\\
};
\addplot [color=white!60!black, dashed, line width=1.4pt, forget plot]
  table[row sep=crcr]{%
0.01	0.281407035175879\\
0.0125892541179417	0.281407035175879\\
0.0158489319246111	0.281407035175879\\
0.0199526231496888	0.281407035175879\\
0.0251188643150958	0.281407035175879\\
0.0316227766016838	0.276381909547739\\
0.0398107170553497	0.276381909547739\\
0.0501187233627272	0.271356783919598\\
0.0630957344480193	0.271356783919598\\
0.0794328234724281	0.266331658291457\\
0.1	0.261306532663317\\
0.125892541179417	0.256281407035176\\
0.158489319246111	0.251256281407035\\
0.199526231496888	0.246231155778894\\
0.251188643150958	0.236180904522613\\
0.316227766016838	0.226130653266332\\
0.398107170553497	0.21608040201005\\
0.501187233627272	0.206030150753769\\
0.630957344480193	0.195979899497487\\
0.794328234724281	0.180904522613065\\
1	0.170854271356784\\
1.25892541179417	0.155778894472362\\
1.58489319246111	0.14070351758794\\
1.99526231496888	0.130653266331658\\
};
\addplot [color=red!60!teal, line width=1.4pt, forget plot]
  table[row sep=crcr]{%
0.01	0.396984924623116\\
0.0125892541179417	0.391959798994975\\
0.0158489319246111	0.391959798994975\\
0.0199526231496888	0.386934673366834\\
0.0251188643150958	0.386934673366834\\
0.0316227766016838	0.381909547738693\\
0.0398107170553497	0.376884422110553\\
0.0501187233627272	0.371859296482412\\
0.0630957344480193	0.366834170854271\\
0.0794328234724281	0.35678391959799\\
0.1	0.351758793969849\\
0.125892541179417	0.341708542713568\\
0.158489319246111	0.326633165829146\\
0.199526231496888	0.316582914572864\\
0.251188643150958	0.301507537688442\\
0.316227766016838	0.28643216080402\\
0.398107170553497	0.271356783919598\\
0.501187233627272	0.256281407035176\\
0.630957344480193	0.241206030150754\\
0.794328234724281	0.226130653266332\\
1	0.21608040201005\\
1.25892541179417	0.201005025125628\\
1.58489319246111	0.190954773869347\\
1.99526231496888	0.180904522613065\\
};
\addplot [color=red!60!teal, dashed, line width=1.4pt, forget plot]
  table[row sep=crcr]{%
0.01	0.361809045226131\\
0.0125892541179417	0.35678391959799\\
0.0158489319246111	0.35678391959799\\
0.0199526231496888	0.35678391959799\\
0.0251188643150958	0.35678391959799\\
0.0316227766016838	0.351758793969849\\
0.0398107170553497	0.351758793969849\\
0.0501187233627272	0.346733668341709\\
0.0630957344480193	0.341708542713568\\
0.0794328234724281	0.336683417085427\\
0.1	0.331658291457286\\
0.125892541179417	0.326633165829146\\
0.158489319246111	0.316582914572864\\
0.199526231496888	0.306532663316583\\
0.251188643150958	0.296482412060302\\
0.316227766016838	0.28643216080402\\
0.398107170553497	0.276381909547739\\
0.501187233627272	0.261306532663317\\
0.630957344480193	0.246231155778894\\
0.794328234724281	0.231155778894472\\
1	0.21105527638191\\
1.25892541179417	0.195979899497487\\
1.58489319246111	0.180904522613065\\
1.99526231496888	0.165829145728643\\
};
\node[right, align=left, inner sep=0, font=\color{mycolor1}]
at (rel axis cs:-0.25,1.15) {$(e)$};
\end{axis}
\end{tikzpicture}%
\fi
    \end{minipage}
    \hfill
    \begin{minipage}[t]{0.32\textwidth}
        \centering
        \vspace{0pt}
        \ifloadfig
%
\definecolor{mycolor1}{rgb}{0.12941,0.12941,0.12941}%
\begin{tikzpicture}[trim axis left,trim axis right]

\begin{axis}[%
width=11.255in,
height=7.131in,
at={(1.888in,0.962in)},
scale only axis,
clip=true,
xmode=log,
xmin=0.01,
xmax=1.99526231496888,
xminorticks=true,
xlabel style={font=\color{mycolor1}},
xlabel={$\mathrm{Sc}^{-1}$},
ymin=0,
ymax=0.6,
ytick={   0, 0.15,  0.3, 0.45,  0.6},
axis background/.style={fill=white},
clip=true,
clip mode=individual,
axis lines=box,
xtick pos=bottom,
ytick pos=left,
tick align=inside,
major tick length=2pt,
width=0.7\linewidth,
height=0.7\linewidth,
every axis/.append style={font=\fontsize{9}{9}\selectfont},xlabel style={font=\fontsize{9}{9}\selectfont},title style={font=\fontsize{9}{9}\selectfont},
legend style={font=\fontsize{8}{9}\selectfont,inner sep=1pt,row sep=1pt,column sep=2pt,nodes={scale=1.00},draw=none},legend image post style={xscale=0.35},
legend columns=1,
scaled x ticks=false,
tick label style={/pgf/number format/fixed,/pgf/number format/precision=2},
every x tick label/.append style={font=\fontsize{9}{9}\selectfont\color{black}},
every y tick label/.append style={font=\fontsize{9}{9}\selectfont\color{black}},
,,
colormap={sigmamap}{rgb(0.0000)=(0.0200,0.1880,0.3800); rgb(0.1000)=(0.0743,0.3456,0.5616); rgb(0.2000)=(0.1918,0.4980,0.7060); rgb(0.3000)=(0.3890,0.6708,0.8226); rgb(0.4000)=(0.7552,0.8682,0.9228); rgb(0.5000)=(0.9690,0.9689,0.9689); rgb(0.6000)=(0.9425,0.8044,0.7614); rgb(0.7000)=(0.8767,0.4910,0.4020); rgb(0.8000)=(0.7869,0.2866,0.2470); rgb(0.9000)=(0.6186,0.1239,0.1568); rgb(1.0000)=(0.4040,0.0000,0.1220)}
]
\addplot [color=black, line width=1.4pt, forget plot]
  table[row sep=crcr]{%
0.01	0.226130653266332\\
0.0125892541179417	0.221105527638191\\
0.0158489319246111	0.221105527638191\\
0.0199526231496888	0.221105527638191\\
0.0251188643150958	0.21608040201005\\
0.0316227766016838	0.21608040201005\\
0.0398107170553497	0.21105527638191\\
0.0501187233627272	0.206030150753769\\
0.0630957344480193	0.206030150753769\\
0.0794328234724281	0.201005025125628\\
0.1	0.195979899497487\\
0.125892541179417	0.185929648241206\\
0.158489319246111	0.180904522613065\\
0.199526231496888	0.175879396984925\\
0.251188643150958	0.165829145728643\\
0.316227766016838	0.160804020100503\\
0.398107170553497	0.150753768844221\\
0.501187233627272	0.14070351758794\\
0.630957344480193	0.130653266331658\\
0.794328234724281	0.125628140703518\\
1	0.115577889447236\\
1.25892541179417	0.105527638190955\\
1.58489319246111	0.0954773869346734\\
1.99526231496888	0.0904522613065327\\
};
\addplot [color=black, dashed, line width=1.4pt, forget plot]
  table[row sep=crcr]{%
0.01	0.190954773869347\\
0.0125892541179417	0.190954773869347\\
0.0158489319246111	0.190954773869347\\
0.0199526231496888	0.185929648241206\\
0.0251188643150958	0.185929648241206\\
0.0316227766016838	0.185929648241206\\
0.0398107170553497	0.185929648241206\\
0.0501187233627272	0.180904522613065\\
0.0630957344480193	0.180904522613065\\
0.0794328234724281	0.175879396984925\\
0.1	0.175879396984925\\
0.125892541179417	0.170854271356784\\
0.158489319246111	0.165829145728643\\
0.199526231496888	0.160804020100503\\
0.251188643150958	0.155778894472362\\
0.316227766016838	0.150753768844221\\
0.398107170553497	0.14070351758794\\
0.501187233627272	0.135678391959799\\
0.630957344480193	0.125628140703518\\
0.794328234724281	0.120603015075377\\
1	0.110552763819095\\
1.25892541179417	0.100502512562814\\
1.58489319246111	0.0904522613065327\\
1.99526231496888	0.085427135678392\\
};
\addplot [color=white!60!black, line width=1.4pt, forget plot]
  table[row sep=crcr]{%
0.01	0.341708542713568\\
0.0125892541179417	0.341708542713568\\
0.0158489319246111	0.341708542713568\\
0.0199526231496888	0.336683417085427\\
0.0251188643150958	0.336683417085427\\
0.0316227766016838	0.331658291457286\\
0.0398107170553497	0.326633165829146\\
0.0501187233627272	0.321608040201005\\
0.0630957344480193	0.316582914572864\\
0.0794328234724281	0.311557788944724\\
0.1	0.306532663316583\\
0.125892541179417	0.296482412060302\\
0.158489319246111	0.28643216080402\\
0.199526231496888	0.276381909547739\\
0.251188643150958	0.266331658291457\\
0.316227766016838	0.251256281407035\\
0.398107170553497	0.236180904522613\\
0.501187233627272	0.226130653266332\\
0.630957344480193	0.21105527638191\\
0.794328234724281	0.195979899497487\\
1	0.175879396984925\\
1.25892541179417	0.160804020100503\\
1.58489319246111	0.14572864321608\\
1.99526231496888	0.135678391959799\\
};
\addplot [color=white!60!black, dashed, line width=1.4pt, forget plot]
  table[row sep=crcr]{%
0.01	0.281407035175879\\
0.0125892541179417	0.281407035175879\\
0.0158489319246111	0.281407035175879\\
0.0199526231496888	0.281407035175879\\
0.0251188643150958	0.281407035175879\\
0.0316227766016838	0.276381909547739\\
0.0398107170553497	0.276381909547739\\
0.0501187233627272	0.271356783919598\\
0.0630957344480193	0.271356783919598\\
0.0794328234724281	0.266331658291457\\
0.1	0.261306532663317\\
0.125892541179417	0.256281407035176\\
0.158489319246111	0.251256281407035\\
0.199526231496888	0.246231155778894\\
0.251188643150958	0.236180904522613\\
0.316227766016838	0.226130653266332\\
0.398107170553497	0.21608040201005\\
0.501187233627272	0.206030150753769\\
0.630957344480193	0.195979899497487\\
0.794328234724281	0.180904522613065\\
1	0.170854271356784\\
1.25892541179417	0.155778894472362\\
1.58489319246111	0.14070351758794\\
1.99526231496888	0.130653266331658\\
};
\addplot [color=red!60!teal, line width=1.4pt, forget plot]
  table[row sep=crcr]{%
0.01	0.417085427135678\\
0.0125892541179417	0.412060301507538\\
0.0158489319246111	0.412060301507538\\
0.0199526231496888	0.412060301507538\\
0.0251188643150958	0.407035175879397\\
0.0316227766016838	0.407035175879397\\
0.0398107170553497	0.402010050251256\\
0.0501187233627272	0.402010050251256\\
0.0630957344480193	0.396984924623116\\
0.0794328234724281	0.386934673366834\\
0.1	0.381909547738693\\
0.125892541179417	0.371859296482412\\
0.158489319246111	0.361809045226131\\
0.199526231496888	0.351758793969849\\
0.251188643150958	0.336683417085427\\
0.316227766016838	0.321608040201005\\
0.398107170553497	0.306532663316583\\
0.501187233627272	0.291457286432161\\
0.630957344480193	0.271356783919598\\
0.794328234724281	0.256281407035176\\
1	0.236180904522613\\
1.25892541179417	0.221105527638191\\
1.58489319246111	0.21105527638191\\
1.99526231496888	0.201005025125628\\
};
\addplot [color=red!60!teal, dashed, line width=1.4pt, forget plot]
  table[row sep=crcr]{%
0.01	0.361809045226131\\
0.0125892541179417	0.35678391959799\\
0.0158489319246111	0.35678391959799\\
0.0199526231496888	0.35678391959799\\
0.0251188643150958	0.35678391959799\\
0.0316227766016838	0.351758793969849\\
0.0398107170553497	0.351758793969849\\
0.0501187233627272	0.346733668341709\\
0.0630957344480193	0.341708542713568\\
0.0794328234724281	0.336683417085427\\
0.1	0.331658291457286\\
0.125892541179417	0.326633165829146\\
0.158489319246111	0.316582914572864\\
0.199526231496888	0.306532663316583\\
0.251188643150958	0.296482412060302\\
0.316227766016838	0.28643216080402\\
0.398107170553497	0.276381909547739\\
0.501187233627272	0.261306532663317\\
0.630957344480193	0.246231155778894\\
0.794328234724281	0.231155778894472\\
1	0.21105527638191\\
1.25892541179417	0.195979899497487\\
1.58489319246111	0.180904522613065\\
1.99526231496888	0.165829145728643\\
};
\node[right, align=left, inner sep=0, font=\color{mycolor1}]
at (rel axis cs:-0.25,1.15) {$(f)$};
\end{axis}
\end{tikzpicture}%
\fi
    \end{minipage}
    \caption{Peak growth rate \(\omega_r^{\max}\) (top row a--c) and corresponding wavenumber \(k^{\max}\) (bottom row d--f) as functions of inverse Schmidt number. (\solidlegend) denotes the present quasi-steady-state approximation results, and (\dashlegend) denotes \citet{duff-harlow-hirt-1962} prediction~\eqref{eq:duff_static_diffusion}. Columns correspond to \(\Amu = -0.9\), \(0.0\), and \(0.9\), respectively.}
    \label{fig:wrmax_vs_Scinv}
\end{figure}
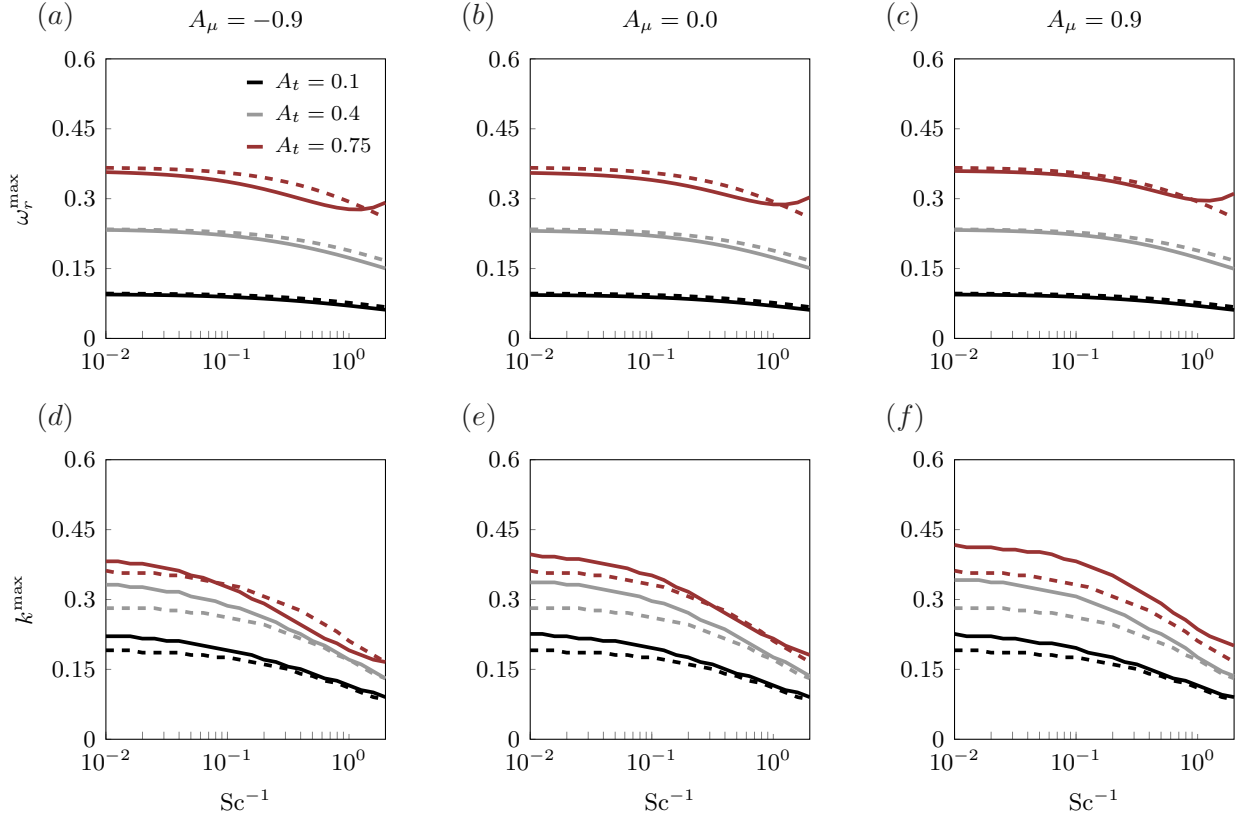
Figure~\ref{fig:wrmax_vs_Scinv} shows the peak growth rate \(\omega_r^{\max}\) and the corresponding wavenumber \(k^{\max}\) as functions of inverse Schmidt number. The peak growth rate is only weakly affected by the viscosity ratio across the parameter space examined, as can be seen from figures~\ref{fig:wrmax_vs_Scinv}(a,b,c), whose content is only slightly affected across different values of \(\Amu\). Fixing \(\Amu\), the dependence on diffusion, and hence on the Schmidt number, varies with the degree of density stratification. For \(\Atw\) between approximately \(0.1\) and \(0.5\), the peak growth rate decreases monotonically as \(\Sch^{-1}\) increases. For \(\Atw > 0.5\), by contrast, the growth rate first decreases down to \(\Sch \approx 1\) and then increases as \(\Sch^{-1}\) is increased further; a possible explanation of why the growth rate is destabilized as diffusion becomes more important is offered in \S\ref{subsec:budget_idl}. As \(\Sch^{-1}\) increases, the most amplified wavenumber \(k^{\max}\) generally shifts to smaller values, indicating that diffusion moves the dominant instability toward longer wavelengths. This trend is visible for all viscosity ratios, although the magnitude of the shift depends on both \(\Atw\) and \(\Amu\). Compared with the peak growth rate, \(k^{\max}\) is more visibly affected by viscosity stratification; \(\Amu > 0\) tends to maintain larger values of \(k^{\max}\), whereas \(\Amu < 0\) shifts the wavenumber toward smaller values. Thus, even when the maximum growth rate itself changes only weakly with \(\Amu\), the wavelength selected by the instability retains a clearer signature of viscosity stratification.

A comparison with the Duff prediction further clarifies these trends. Over most of the parameter space, the dashed curves follow the same qualitative dependence on \(\Sch^{-1}\) as the present quasi-steady-state approximation results, indicating that the static-diffusion model captures the leading variation of \(\omega_r^{\max}\) and \(k^{\max}\) with diffusion. Quantitatively, however, the Duff prediction is generally slightly larger than the present results for \(\omega_r^{\max}\), with the discrepancies remaining modest across different viscosity and density ratios. For \(k^{\max}\), the discrepancy is more sensitive to the viscosity ratio and becomes increasingly evident as diffusion effects become dominant. Thus, viscosity stratification does not strongly alter the overall trend of the peak growth rate with \(\Sch^{-1}\), but it introduces a more visible correction to the wavenumber selected by the instability.

\begin{figure}
    \centering
    \begin{minipage}[t]{0.32\textwidth}
        \centering
        \vspace{0pt}
        \ifloadfig
%
\definecolor{mycolor1}{rgb}{0.12941,0.12941,0.12941}%
\begin{tikzpicture}[trim axis left,trim axis right]

\begin{axis}[%
width=11.255in,
height=7.131in,
at={(1.888in,0.962in)},
scale only axis,
xmode=log,
xmin=0.01,
xmax=1.99526231496888,
xminorticks=true,
xlabel style={font=\color{mycolor1}},
xlabel={$\mathrm{Sc}^{-1}$},
ymin=0,
ymax=3,
ytick={   0, 0.75,  1.5, 2.25,    3},
ylabel style={font=\color{mycolor1}},
ylabel={$k_c$},
axis background/.style={fill=white},
title style={font=\bfseries\color{mycolor1}},
title={$A_{\mu} = -0.9$},
legend style={legend cell align=left, align=left},
clip=true,
clip mode=individual,
axis lines=box,
xtick pos=bottom,
ytick pos=left,
tick align=inside,
major tick length=2pt,
width=0.7\linewidth,
height=0.7\linewidth,
every axis/.append style={font=\fontsize{9}{9}\selectfont},xlabel style={font=\fontsize{9}{9}\selectfont},title style={font=\fontsize{9}{9}\selectfont},
legend style={font=\fontsize{8}{9}\selectfont,inner sep=1pt,row sep=1pt,column sep=2pt,nodes={scale=1.00},draw=none},legend image post style={xscale=0.35},
legend columns=1,
scaled x ticks=false,
tick label style={/pgf/number format/fixed,/pgf/number format/precision=2},
every x tick label/.append style={font=\fontsize{9}{9}\selectfont\color{black}},
every y tick label/.append style={font=\fontsize{9}{9}\selectfont\color{black}},
,,
colormap={sigmamap}{rgb(0.0000)=(0.0200,0.1880,0.3800); rgb(0.1000)=(0.0743,0.3456,0.5616); rgb(0.2000)=(0.1918,0.4980,0.7060); rgb(0.3000)=(0.3890,0.6708,0.8226); rgb(0.4000)=(0.7552,0.8682,0.9228); rgb(0.5000)=(0.9690,0.9689,0.9689); rgb(0.6000)=(0.9425,0.8044,0.7614); rgb(0.7000)=(0.8767,0.4910,0.4020); rgb(0.8000)=(0.7869,0.2866,0.2470); rgb(0.9000)=(0.6186,0.1239,0.1568); rgb(1.0000)=(0.4040,0.0000,0.1220)}
]
\addplot [color=black, line width=1.4pt]
  table[row sep=crcr]{%
0.01	1.50449411788983\\
0.0125892541179417	1.3962689297745\\
0.0158489319246111	1.29532760844303\\
0.0199526231496888	1.20128287231626\\
0.0251188643150958	1.1137407766894\\
0.0316227766016838	1.03232377701875\\
0.0398107170553497	0.956666247204815\\
0.0501187233627272	0.886412212628477\\
0.0630957344480193	0.821213924571229\\
0.0794328234724281	0.760738487532853\\
0.1	0.704666859541315\\
0.125892541179417	0.652694387243215\\
0.158489319246111	0.604530499943496\\
0.199526231496888	0.559901568312784\\
0.251188643150958	0.518548976009213\\
0.316227766016838	0.480226916281885\\
0.398107170553497	0.444707763715944\\
0.501187233627272	0.411774348197776\\
0.630957344480193	0.381225922305119\\
0.794328234724281	0.352874846803658\\
1	0.326527245597535\\
1.25892541179417	0.302039711083608\\
1.58489319246111	0.279247639404018\\
1.99526231496888	0.257975748082842\\
};
\addlegendentry{$A_t=0.1$}

\addplot [color=black, dashed, line width=1.4pt, forget plot]
  table[row sep=crcr]{%
0.01	1.32673259942863\\
0.0125892541179417	1.24254916525386\\
0.0158489319246111	1.1632260988395\\
0.0199526231496888	1.0884653553343\\
0.0251188643150958	1.01801979693477\\
0.0316227766016838	0.951627566985881\\
0.0398107170553497	0.889044312660551\\
0.0501187233627272	0.830037201680003\\
0.0630957344480193	0.774380333817176\\
0.0794328234724281	0.721858022977864\\
0.1	0.67226126382291\\
0.125892541179417	0.625389855259428\\
0.158489319246111	0.581051640174474\\
0.199526231496888	0.539063561430671\\
0.251188643150958	0.499252431971688\\
0.316227766016838	0.46146010595805\\
0.398107170553497	0.425539188280136\\
0.501187233627272	0.39137083747162\\
0.630957344480193	0.358844337188246\\
0.794328234724281	0.327896612501181\\
1	0.298473886336204\\
1.25892541179417	0.270572299356803\\
1.58489319246111	0.24418404723706\\
1.99526231496888	0.219359296393594\\
};
\addplot [color=white!60!black, line width=1.4pt]
  table[row sep=crcr]{%
0.01	2.29663676798059\\
0.0125892541179417	2.13458760846403\\
0.0158489319246111	1.98254600703386\\
0.0199526231496888	1.84007218673131\\
0.0251188643150958	1.70672331839165\\
0.0316227766016838	1.58206967071209\\
0.0398107170553497	1.46569340442208\\
0.0501187233627272	1.35717298054095\\
0.0630957344480193	1.25609412684011\\
0.0794328234724281	1.16204166659916\\
0.1	1.07460340612893\\
0.125892541179417	0.993368444797963\\
0.158489319246111	0.917929617544048\\
0.199526231496888	0.847884887595937\\
0.251188643150958	0.782838624709365\\
0.316227766016838	0.722403412787303\\
0.398107170553497	0.666201875206315\\
0.501187233627272	0.613862257619888\\
0.630957344480193	0.56504097097009\\
0.794328234724281	0.519390979565529\\
1	0.476591589126402\\
1.25892541179417	0.436387649084849\\
1.58489319246111	0.398581281892278\\
1.99526231496888	0.362999498011663\\
};
\addlegendentry{$A_t=0.4$}

\addplot [color=white!60!black, dashed, line width=1.4pt, forget plot]
  table[row sep=crcr]{%
0.01	1.97518431440134\\
0.0125892541179417	1.85300748153114\\
0.0158489319246111	1.73771494513307\\
0.0199526231496888	1.62892757732063\\
0.0251188643150958	1.52624564704255\\
0.0316227766016838	1.4293304694533\\
0.0398107170553497	1.33780792279772\\
0.0501187233627272	1.2513613666706\\
0.0630957344480193	1.16966693948619\\
0.0794328234724281	1.09241726894679\\
0.1	1.01931692418184\\
0.125892541179417	0.950083899969268\\
0.158489319246111	0.884443768969125\\
0.199526231496888	0.822134077673738\\
0.251188643150958	0.76291031289759\\
0.316227766016838	0.706543045942349\\
0.398107170553497	0.652830361133983\\
0.501187233627272	0.601590568991193\\
0.630957344480193	0.552689165258071\\
0.794328234724281	0.506012478991213\\
1	0.461515859980355\\
1.25892541179417	0.419175844784246\\
1.58489319246111	0.379029976873835\\
1.99526231496888	0.341143577919168\\
};
\addplot [color=red!60!teal, line width=1.4pt]
  table[row sep=crcr]{%
0.01	2.77215276761562\\
0.0125892541179417	2.57756069380292\\
0.0158489319246111	2.39420432814428\\
0.0199526231496888	2.2215437921035\\
0.0251188643150958	2.05907096709862\\
0.0316227766016838	1.90629604064494\\
0.0398107170553497	1.76274805750391\\
0.0501187233627272	1.62797187578129\\
0.0630957344480193	1.50153336714819\\
0.0794328234724281	1.38301091649301\\
0.1	1.27201640758989\\
0.125892541179417	1.16817728500461\\
0.158489319246111	1.07114824569396\\
0.199526231496888	0.980660494740156\\
0.251188643150958	0.896486078514631\\
0.316227766016838	0.81845085972969\\
0.398107170553497	0.746489904311865\\
0.501187233627272	0.680597690612494\\
0.630957344480193	0.620881772372808\\
0.794328234724281	0.567496943901398\\
1	0.520803323596725\\
1.25892541179417	0.480976387547947\\
1.58489319246111	0.448315215286372\\
1.99526231496888	0.423015112016037\\
};
\addlegendentry{$A_t=0.75$}

\addplot [color=red!60!teal, dashed, line width=1.4pt, forget plot]
  table[row sep=crcr]{%
0.01	2.46824181810177\\
0.0125892541179417	2.31654295644084\\
0.0158489319246111	2.17332043608431\\
0.0199526231496888	2.03807565688558\\
0.0251188643150958	1.91034173207392\\
0.0316227766016838	1.78967539374\\
0.0398107170553497	1.67566383721771\\
0.0501187233627272	1.56788139085858\\
0.0630957344480193	1.46594662075274\\
0.0794328234724281	1.36947899698772\\
0.1	1.27812613590967\\
0.125892541179417	1.19153342692267\\
0.158489319246111	1.10936745373315\\
0.199526231496888	1.03131102363565\\
0.251188643150958	0.95707108670961\\
0.316227766016838	0.886359138480226\\
0.398107170553497	0.818932246529735\\
0.501187233627272	0.754572691901745\\
0.630957344480193	0.693114341291851\\
0.794328234724281	0.634432427461626\\
1	0.578469602301791\\
1.25892541179417	0.525209171013812\\
1.58489319246111	0.474704536415614\\
1.99526231496888	0.427045615841519\\
};
\end{axis}
\end{tikzpicture}%
\fi
    \end{minipage}
    \hfill
    \begin{minipage}[t]{0.32\textwidth}
        \centering
        \vspace{0pt}
        \ifloadfig
%
\definecolor{mycolor1}{rgb}{0.12941,0.12941,0.12941}%
\begin{tikzpicture}[trim axis left,trim axis right]

\begin{axis}[%
width=11.255in,
height=7.131in,
at={(1.888in,0.962in)},
scale only axis,
xmode=log,
xmin=0.01,
xmax=1.99526231496888,
xminorticks=true,
xlabel style={font=\color{mycolor1}},
xlabel={$\mathrm{Sc}^{-1}$},
ymin=0,
ymax=3,
ytick={   0, 0.75,  1.5, 2.25,    3},
axis background/.style={fill=white},
title style={font=\bfseries\color{mycolor1}},
title={$A_{\mu} = 0.0$},
clip=true,
clip mode=individual,
axis lines=box,
xtick pos=bottom,
ytick pos=left,
tick align=inside,
major tick length=2pt,
width=0.7\linewidth,
height=0.7\linewidth,
every axis/.append style={font=\fontsize{9}{9}\selectfont},xlabel style={font=\fontsize{9}{9}\selectfont},title style={font=\fontsize{9}{9}\selectfont},
legend style={font=\fontsize{8}{9}\selectfont,inner sep=1pt,row sep=1pt,column sep=2pt,nodes={scale=1.00},draw=none},legend image post style={xscale=0.35},
legend columns=1,
scaled x ticks=false,
tick label style={/pgf/number format/fixed,/pgf/number format/precision=2},
every x tick label/.append style={font=\fontsize{9}{9}\selectfont\color{black}},
every y tick label/.append style={font=\fontsize{9}{9}\selectfont\color{black}},
,,
colormap={sigmamap}{rgb(0.0000)=(0.0200,0.1880,0.3800); rgb(0.1000)=(0.0743,0.3456,0.5616); rgb(0.2000)=(0.1918,0.4980,0.7060); rgb(0.3000)=(0.3890,0.6708,0.8226); rgb(0.4000)=(0.7552,0.8682,0.9228); rgb(0.5000)=(0.9690,0.9689,0.9689); rgb(0.6000)=(0.9425,0.8044,0.7614); rgb(0.7000)=(0.8767,0.4910,0.4020); rgb(0.8000)=(0.7869,0.2866,0.2470); rgb(0.9000)=(0.6186,0.1239,0.1568); rgb(1.0000)=(0.4040,0.0000,0.1220)}
]
\addplot [color=black, line width=1.4pt, forget plot]
  table[row sep=crcr]{%
0.01	1.44348639863925\\
0.0125892541179417	1.34579919550289\\
0.0158489319246111	1.25414388169548\\
0.0199526231496888	1.16819764202009\\
0.0251188643150958	1.08764924348904\\
0.0316227766016838	1.01223394334302\\
0.0398107170553497	0.941643357475115\\
0.0501187233627272	0.875631216713569\\
0.0630957344480193	0.813936020509056\\
0.0794328234724281	0.756312981651779\\
0.1	0.702527380584704\\
0.125892541179417	0.652353597247968\\
0.158489319246111	0.605574498817699\\
0.199526231496888	0.561985299310762\\
0.251188643150958	0.521383455666339\\
0.316227766016838	0.483580639864216\\
0.398107170553497	0.448393214865507\\
0.501187233627272	0.415645087358409\\
0.630957344480193	0.385168972586518\\
0.794328234724281	0.356814065186578\\
1	0.330396785138227\\
1.25892541179417	0.305795677454002\\
1.58489319246111	0.282846405848147\\
1.99526231496888	0.261427698679243\\
};
\addplot [color=black, dashed, line width=1.4pt, forget plot]
  table[row sep=crcr]{%
0.01	1.32673259942863\\
0.0125892541179417	1.24254916525386\\
0.0158489319246111	1.1632260988395\\
0.0199526231496888	1.0884653553343\\
0.0251188643150958	1.01801979693477\\
0.0316227766016838	0.951627566985881\\
0.0398107170553497	0.889044312660551\\
0.0501187233627272	0.830037201680003\\
0.0630957344480193	0.774380333817176\\
0.0794328234724281	0.721858022977864\\
0.1	0.67226126382291\\
0.125892541179417	0.625389855259428\\
0.158489319246111	0.581051640174474\\
0.199526231496888	0.539063561430671\\
0.251188643150958	0.499252431971688\\
0.316227766016838	0.46146010595805\\
0.398107170553497	0.425539188280136\\
0.501187233627272	0.39137083747162\\
0.630957344480193	0.358844337188246\\
0.794328234724281	0.327896612501181\\
1	0.298473886336204\\
1.25892541179417	0.270572299356803\\
1.58489319246111	0.24418404723706\\
1.99526231496888	0.219359296393594\\
};
\addplot [color=white!60!black, line width=1.4pt, forget plot]
  table[row sep=crcr]{%
0.01	2.17837371750019\\
0.0125892541179417	2.03656425144643\\
0.0158489319246111	1.90309372590101\\
0.0199526231496888	1.77750369016588\\
0.0251188643150958	1.65938436286646\\
0.0316227766016838	1.54833305805098\\
0.0398107170553497	1.44397632669965\\
0.0501187233627272	1.34594997645407\\
0.0630957344480193	1.25391002656488\\
0.0794328234724281	1.16752322925952\\
0.1	1.08647350763428\\
0.125892541179417	1.01044248787829\\
0.158489319246111	0.939150867195389\\
0.199526231496888	0.872278857235725\\
0.251188643150958	0.809547683565716\\
0.316227766016838	0.750665148495181\\
0.398107170553497	0.695350912948948\\
0.501187233627272	0.643324699872164\\
0.630957344480193	0.594294120450676\\
0.794328234724281	0.547999929696391\\
1	0.504165098242613\\
1.25892541179417	0.462574670549531\\
1.58489319246111	0.42302552485033\\
1.99526231496888	0.385423731531963\\
};
\addplot [color=white!60!black, dashed, line width=1.4pt, forget plot]
  table[row sep=crcr]{%
0.01	1.97518431440134\\
0.0125892541179417	1.85300748153114\\
0.0158489319246111	1.73771494513307\\
0.0199526231496888	1.62892757732063\\
0.0251188643150958	1.52624564704255\\
0.0316227766016838	1.4293304694533\\
0.0398107170553497	1.33780792279772\\
0.0501187233627272	1.2513613666706\\
0.0630957344480193	1.16966693948619\\
0.0794328234724281	1.09241726894679\\
0.1	1.01931692418184\\
0.125892541179417	0.950083899969268\\
0.158489319246111	0.884443768969125\\
0.199526231496888	0.822134077673738\\
0.251188643150958	0.76291031289759\\
0.316227766016838	0.706543045942349\\
0.398107170553497	0.652830361133983\\
0.501187233627272	0.601590568991193\\
0.630957344480193	0.552689165258071\\
0.794328234724281	0.506012478991213\\
1	0.461515859980355\\
1.25892541179417	0.419175844784246\\
1.58489319246111	0.379029976873835\\
1.99526231496888	0.341143577919168\\
};
\addplot [color=red!60!teal, line width=1.4pt, forget plot]
  table[row sep=crcr]{%
0.01	2.61813586192918\\
0.0125892541179417	2.44984527804521\\
0.0158489319246111	2.29112993923568\\
0.0199526231496888	2.14144818825026\\
0.0251188643150958	2.0003016761683\\
0.0316227766016838	1.86720956738951\\
0.0398107170553497	1.74169531975329\\
0.0501187233627272	1.62333145664029\\
0.0630957344480193	1.51169573074565\\
0.0794328234724281	1.4063974392362\\
0.1	1.30705886159367\\
0.125892541179417	1.21333077486437\\
0.158489319246111	1.12488879404362\\
0.199526231496888	1.04143652994154\\
0.251188643150958	0.962765157422263\\
0.316227766016838	0.888667740021644\\
0.398107170553497	0.81907580738216\\
0.501187233627272	0.754029353505024\\
0.630957344480193	0.693729026264503\\
0.794328234724281	0.638550834202137\\
1	0.589044191610459\\
1.25892541179417	0.54588599646929\\
1.58489319246111	0.509768350318709\\
1.99526231496888	0.481245245997523\\
};
\addplot [color=red!60!teal, dashed, line width=1.4pt, forget plot]
  table[row sep=crcr]{%
0.01	2.46824181810177\\
0.0125892541179417	2.31654295644084\\
0.0158489319246111	2.17332043608431\\
0.0199526231496888	2.03807565688558\\
0.0251188643150958	1.91034173207392\\
0.0316227766016838	1.78967539374\\
0.0398107170553497	1.67566383721771\\
0.0501187233627272	1.56788139085858\\
0.0630957344480193	1.46594662075274\\
0.0794328234724281	1.36947899698772\\
0.1	1.27812613590967\\
0.125892541179417	1.19153342692267\\
0.158489319246111	1.10936745373315\\
0.199526231496888	1.03131102363565\\
0.251188643150958	0.95707108670961\\
0.316227766016838	0.886359138480226\\
0.398107170553497	0.818932246529735\\
0.501187233627272	0.754572691901745\\
0.630957344480193	0.693114341291851\\
0.794328234724281	0.634432427461626\\
1	0.578469602301791\\
1.25892541179417	0.525209171013812\\
1.58489319246111	0.474704536415614\\
1.99526231496888	0.427045615841519\\
};
\end{axis}
\end{tikzpicture}%
\fi
    \end{minipage}
    \hfill
    \begin{minipage}[t]{0.32\textwidth}
        \centering
        \vspace{0pt}
        \ifloadfig
%
\definecolor{mycolor1}{rgb}{0.12941,0.12941,0.12941}%
\begin{tikzpicture}[trim axis left,trim axis right]

\begin{axis}[%
width=11.255in,
height=7.131in,
at={(1.888in,0.962in)},
scale only axis,
xmode=log,
xmin=0.01,
xmax=1.99526231496888,
xminorticks=true,
xlabel style={font=\color{mycolor1}},
xlabel={$\mathrm{Sc}^{-1}$},
ymin=0,
ymax=3,
ytick={   0, 0.75,  1.5, 2.25,    3},
axis background/.style={fill=white},
title style={font=\bfseries\color{mycolor1}},
title={$A_{\mu} = 0.9$},
clip=true,
clip mode=individual,
axis lines=box,
xtick pos=bottom,
ytick pos=left,
tick align=inside,
major tick length=2pt,
width=0.7\linewidth,
height=0.7\linewidth,
every axis/.append style={font=\fontsize{9}{9}\selectfont},xlabel style={font=\fontsize{9}{9}\selectfont},title style={font=\fontsize{9}{9}\selectfont},
legend style={font=\fontsize{8}{9}\selectfont,inner sep=1pt,row sep=1pt,column sep=2pt,nodes={scale=1.00},draw=none},legend image post style={xscale=0.35},
legend columns=1,
scaled x ticks=false,
tick label style={/pgf/number format/fixed,/pgf/number format/precision=2},
every x tick label/.append style={font=\fontsize{9}{9}\selectfont\color{black}},
every y tick label/.append style={font=\fontsize{9}{9}\selectfont\color{black}},
,,
colormap={sigmamap}{rgb(0.0000)=(0.0200,0.1880,0.3800); rgb(0.1000)=(0.0743,0.3456,0.5616); rgb(0.2000)=(0.1918,0.4980,0.7060); rgb(0.3000)=(0.3890,0.6708,0.8226); rgb(0.4000)=(0.7552,0.8682,0.9228); rgb(0.5000)=(0.9690,0.9689,0.9689); rgb(0.6000)=(0.9425,0.8044,0.7614); rgb(0.7000)=(0.8767,0.4910,0.4020); rgb(0.8000)=(0.7869,0.2866,0.2470); rgb(0.9000)=(0.6186,0.1239,0.1568); rgb(1.0000)=(0.4040,0.0000,0.1220)}
]
\addplot [color=black, line width=1.4pt, forget plot]
  table[row sep=crcr]{%
0.01	1.53252065803107\\
0.0125892541179417	1.42392987217753\\
0.0158489319246111	1.32248679636669\\
0.0199526231496888	1.22779599060623\\
0.0251188643150958	1.13949078685781\\
0.0316227766016838	1.05720344002104\\
0.0398107170553497	0.980577786895351\\
0.0501187233627272	0.90927713958549\\
0.0630957344480193	0.842971292759953\\
0.0794328234724281	0.781344666965305\\
0.1	0.724095762585151\\
0.125892541179417	0.670934417855421\\
0.158489319246111	0.621587972122948\\
0.199526231496888	0.575794600956073\\
0.251188643150958	0.533309512049994\\
0.316227766016838	0.493895901660792\\
0.398107170553497	0.457338541368075\\
0.501187233627272	0.423423236182125\\
0.630957344480193	0.391959769981313\\
0.794328234724281	0.362760322007755\\
1	0.33564915112181\\
1.25892541179417	0.310461356230237\\
1.58489319246111	0.287034949627221\\
1.99526231496888	0.2652226900649\\
};
\addplot [color=black, dashed, line width=1.4pt, forget plot]
  table[row sep=crcr]{%
0.01	1.32673259942863\\
0.0125892541179417	1.24254916525386\\
0.0158489319246111	1.1632260988395\\
0.0199526231496888	1.0884653553343\\
0.0251188643150958	1.01801979693477\\
0.0316227766016838	0.951627566985881\\
0.0398107170553497	0.889044312660551\\
0.0501187233627272	0.830037201680003\\
0.0630957344480193	0.774380333817176\\
0.0794328234724281	0.721858022977864\\
0.1	0.67226126382291\\
0.125892541179417	0.625389855259428\\
0.158489319246111	0.581051640174474\\
0.199526231496888	0.539063561430671\\
0.251188643150958	0.499252431971688\\
0.316227766016838	0.46146010595805\\
0.398107170553497	0.425539188280136\\
0.501187233627272	0.39137083747162\\
0.630957344480193	0.358844337188246\\
0.794328234724281	0.327896612501181\\
1	0.298473886336204\\
1.25892541179417	0.270572299356803\\
1.58489319246111	0.24418404723706\\
1.99526231496888	0.219359296393594\\
};
\addplot [color=white!60!black, line width=1.4pt, forget plot]
  table[row sep=crcr]{%
0.01	2.40948918205014\\
0.0125892541179417	2.24959578491967\\
0.0158489319246111	2.09925966496401\\
0.0199526231496888	1.95798286305659\\
0.0251188643150958	1.82529121342009\\
0.0316227766016838	1.70073237882307\\
0.0398107170553497	1.58386845629443\\
0.0501187233627272	1.4742828536466\\
0.0630957344480193	1.37156481345851\\
0.0794328234724281	1.2753243438139\\
0.1	1.18518008337866\\
0.125892541179417	1.10076460800202\\
0.158489319246111	1.02171018666689\\
0.199526231496888	0.947676899017111\\
0.251188643150958	0.878294790238715\\
0.316227766016838	0.81323401692539\\
0.398107170553497	0.752162314454904\\
0.501187233627272	0.694761948916436\\
0.630957344480193	0.640727275829658\\
0.794328234724281	0.589781277877005\\
1	0.541678872510115\\
1.25892541179417	0.496219447926657\\
1.58489319246111	0.453253493407215\\
1.99526231496888	0.412675156006183\\
};
\addplot [color=white!60!black, dashed, line width=1.4pt, forget plot]
  table[row sep=crcr]{%
0.01	1.97518431440134\\
0.0125892541179417	1.85300748153114\\
0.0158489319246111	1.73771494513307\\
0.0199526231496888	1.62892757732063\\
0.0251188643150958	1.52624564704255\\
0.0316227766016838	1.4293304694533\\
0.0398107170553497	1.33780792279772\\
0.0501187233627272	1.2513613666706\\
0.0630957344480193	1.16966693948619\\
0.0794328234724281	1.09241726894679\\
0.1	1.01931692418184\\
0.125892541179417	0.950083899969268\\
0.158489319246111	0.884443768969125\\
0.199526231496888	0.822134077673738\\
0.251188643150958	0.76291031289759\\
0.316227766016838	0.706543045942349\\
0.398107170553497	0.652830361133983\\
0.501187233627272	0.601590568991193\\
0.630957344480193	0.552689165258071\\
0.794328234724281	0.506012478991213\\
1	0.461515859980355\\
1.25892541179417	0.419175844784246\\
1.58489319246111	0.379029976873835\\
1.99526231496888	0.341143577919168\\
};
\addplot [color=red!60!teal, line width=1.4pt, forget plot]
  table[row sep=crcr]{%
0.01	3.00183524778189\\
0.0125892541179417	2.81420294030183\\
0.0158489319246111	2.63740075591988\\
0.0199526231496888	2.47081993107067\\
0.0251188643150958	2.3138776860204\\
0.0316227766016838	2.16601371250586\\
0.0398107170553497	2.0266908053733\\
0.0501187233627272	1.89539142003298\\
0.0630957344480193	1.77160520211544\\
0.0794328234724281	1.65483438683998\\
0.1	1.54461285950477\\
0.125892541179417	1.44044853914629\\
0.158489319246111	1.34188024074806\\
0.199526231496888	1.24848953270333\\
0.251188643150958	1.15985118884841\\
0.316227766016838	1.07561367808816\\
0.398107170553497	0.99560950846394\\
0.501187233627272	0.919750375982396\\
0.630957344480193	0.848307164251434\\
0.794328234724281	0.781877076879431\\
1	0.721373153448399\\
1.25892541179417	0.667866387819437\\
1.58489319246111	0.622320806744417\\
1.99526231496888	0.585371266463214\\
};
\addplot [color=red!60!teal, dashed, line width=1.4pt, forget plot]
  table[row sep=crcr]{%
0.01	2.46824181810177\\
0.0125892541179417	2.31654295644084\\
0.0158489319246111	2.17332043608431\\
0.0199526231496888	2.03807565688558\\
0.0251188643150958	1.91034173207392\\
0.0316227766016838	1.78967539374\\
0.0398107170553497	1.67566383721771\\
0.0501187233627272	1.56788139085858\\
0.0630957344480193	1.46594662075274\\
0.0794328234724281	1.36947899698772\\
0.1	1.27812613590967\\
0.125892541179417	1.19153342692267\\
0.158489319246111	1.10936745373315\\
0.199526231496888	1.03131102363565\\
0.251188643150958	0.95707108670961\\
0.316227766016838	0.886359138480226\\
0.398107170553497	0.818932246529735\\
0.501187233627272	0.754572691901745\\
0.630957344480193	0.693114341291851\\
0.794328234724281	0.634432427461626\\
1	0.578469602301791\\
1.25892541179417	0.525209171013812\\
1.58489319246111	0.474704536415614\\
1.99526231496888	0.427045615841519\\
};
\end{axis}
\end{tikzpicture}%
\fi
    \end{minipage}
    \caption{Cutoff wavenumber \(k_c\) as a function of inverse Schmidt number. (\solidlegend) denotes the present quasi-steady-state approximation results, and (\dashlegend) denotes \citet{duff-harlow-hirt-1962} prediction~\eqref{eq:duff_static_diffusion}.}
    \label{fig:kc_vs_Scinv}
\end{figure}
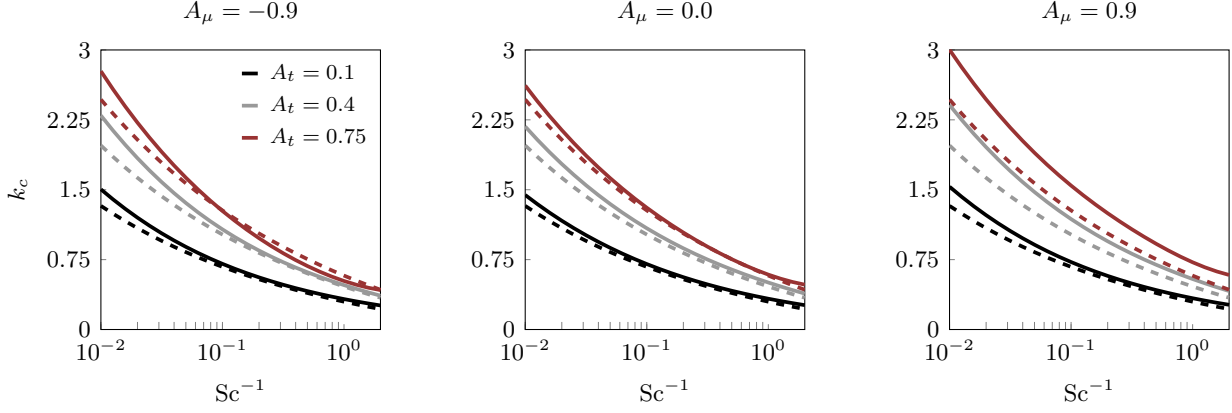

A complementary picture is provided by the cutoff wavenumber \(k_c\), defined by the neutral-stability condition \(\omega_r(k_c)=0\). This quantity represents the largest wavenumber for which perturbations remain unstable, and its variation across the parameter space is summarized in figure~\ref{fig:kc_vs_Scinv}. At fixed Atwood number, \(k_c\) decreases monotonically as \(\Sch^{-1}\) increases, so that stronger diffusion stabilizes progressively larger portions of the spectrum. The cutoff wavenumbers predicted by the present quasi-steady-state formulation lie systematically above those of the Duff model, indicating a broader unstable band. The difference in \(k_c\) is small at low \(\Atw\) but increases with density stratification, becoming particularly pronounced for positive viscosity ratio.

\subsection{Scaling law for \texorpdfstring{\( k \gg k_c \)}{k >> kc}}
\label{subsec:kggkc}
\begin{figure}
    \centering
    \begin{minipage}[b]{0.48\textwidth}
        \centering
        \vspace{0pt}
        \ifloadfig\input{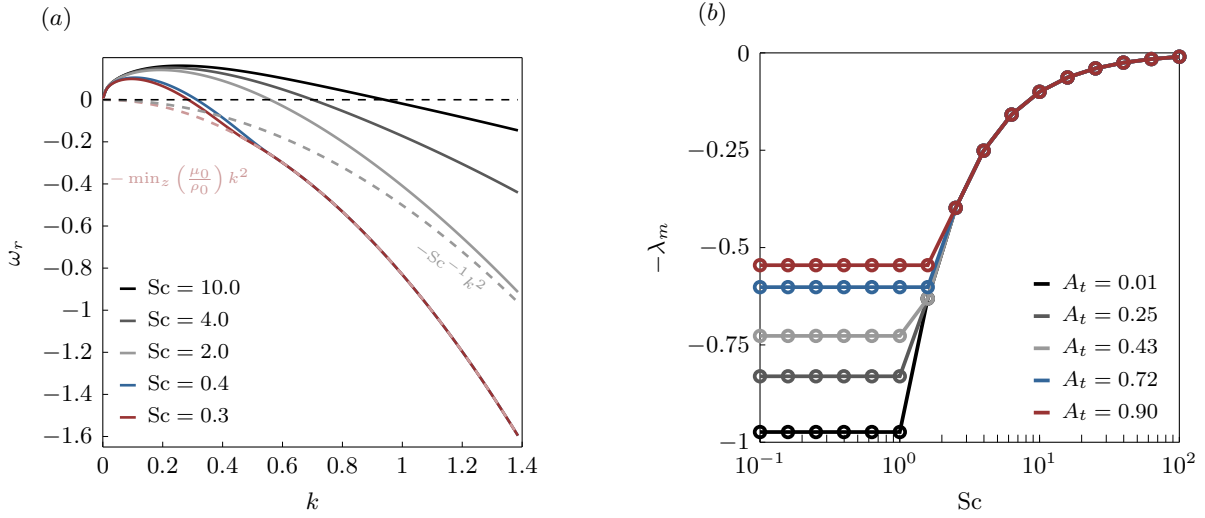}\fi
    \end{minipage}
    \hfill
    \begin{minipage}[b]{0.48\textwidth}
        \centering
        \vspace{0pt}
        \ifloadfig
%
\definecolor{mycolor1}{rgb}{0.20000,0.40000,0.60000}%
\begin{tikzpicture}

\begin{axis}[%
width=11.255in,
height=7.131in,
at={(1.888in,0.962in)},
scale only axis,
clip=true,
separate axis lines,
every outer x axis line/.append style={black},
every x tick label/.append style={font=\color{black}},
every x tick/.append style={black},
xmode=log,
xmin=0.1,
xmax=100,
xminorticks=true,
xlabel={$\mathrm{Sc}$},
every outer y axis line/.append style={black},
every y tick label/.append style={font=\color{black}},
every y tick/.append style={black},
ymin=-1,
ymax=0,
ytick={   -1, -0.75,  -0.5, -0.25,     0},
ylabel={$-\lambda_m$},
axis background/.style={fill=white},
legend style={at={(0.97,0.03)}, anchor=south east, legend cell align=left, align=left},
clip=true,
clip mode=individual,
axis lines=box,
xtick pos=bottom,
ytick pos=left,
tick align=inside,
major tick length=2pt,
width=0.7\linewidth,
height=0.65\linewidth,
every axis/.append style={font=\fontsize{9}{9}\selectfont},xlabel style={font=\fontsize{9}{9}\selectfont},title style={font=\fontsize{9}{9}\selectfont},
legend style={font=\fontsize{8}{9}\selectfont,inner sep=1pt,row sep=1pt,column sep=2pt,nodes={scale=1.00},draw=none},legend image post style={xscale=0.35},
legend columns=1,
scaled x ticks=false,
tick label style={/pgf/number format/fixed,/pgf/number format/precision=2},
every x tick label/.append style={font=\fontsize{9}{9}\selectfont\color{black}},
every y tick label/.append style={font=\fontsize{9}{9}\selectfont\color{black}},
,,
colormap={sigmamap}{rgb(0.0000)=(0.0200,0.1880,0.3800); rgb(0.1000)=(0.0743,0.3456,0.5616); rgb(0.2000)=(0.1918,0.4980,0.7060); rgb(0.3000)=(0.3890,0.6708,0.8226); rgb(0.4000)=(0.7552,0.8682,0.9228); rgb(0.5000)=(0.9690,0.9689,0.9689); rgb(0.6000)=(0.9425,0.8044,0.7614); rgb(0.7000)=(0.8767,0.4910,0.4020); rgb(0.8000)=(0.7869,0.2866,0.2470); rgb(0.9000)=(0.6186,0.1239,0.1568); rgb(1.0000)=(0.4040,0.0000,0.1220)}
]
\addplot [color=black, line width=1.4pt]
  table[row sep=crcr]{%
0.1	-0.973737373737374\\
0.158489319246111	-0.973737373737374\\
0.251188643150958	-0.973737373737374\\
0.398107170553497	-0.973737373737374\\
0.630957344480193	-0.973737373737374\\
1	-0.973737373737374\\
1.58489319246111	-0.630957344480193\\
2.51188643150958	-0.398107170553497\\
3.98107170553497	-0.251188643150958\\
6.30957344480193	-0.158489319246111\\
10	-0.1\\
15.8489319246111	-0.0630957344480193\\
25.1188643150958	-0.0398107170553497\\
39.8107170553497	-0.0251188643150958\\
63.0957344480193	-0.0158489319246111\\
100	-0.01\\
};
\addlegendentry{$A_t=0.01$}

\addplot [color=black, line width=1.4pt, only marks, mark size=2.0pt, mark=o, mark options={solid, black}, forget plot]
  table[row sep=crcr]{%
0.1	-0.973745389161803\\
0.158489319246111	-0.973745391491856\\
0.251188643150958	-0.973745392937564\\
0.398107170553497	-0.973745393834411\\
0.630957344480193	-0.973745394373424\\
1	-0.97374539373141\\
1.58489319246111	-0.630958324286094\\
2.51188643150958	-0.398107839224124\\
3.98107170553497	-0.251189062683724\\
6.30957344480193	-0.158489572254349\\
10	-0.100000145113947\\
15.8489319246111	-0.0630958100312529\\
25.1188643150958	-0.0398107473457575\\
39.8107170553497	-0.0251188639420104\\
63.0957344480193	-0.0158489086243755\\
100	-0.00999995569250334\\
};
\addplot [color=white!35!black, line width=1.4pt]
  table[row sep=crcr]{%
0.1	-0.830571886691609\\
0.158489319246111	-0.830571886691609\\
0.251188643150958	-0.830571886691609\\
0.398107170553497	-0.830571886691609\\
0.630957344480193	-0.830571886691609\\
1	-0.830571886691609\\
1.58489319246111	-0.630957344480193\\
2.51188643150958	-0.398107170553497\\
3.98107170553497	-0.251188643150958\\
6.30957344480193	-0.158489319246111\\
10	-0.1\\
15.8489319246111	-0.0630957344480193\\
25.1188643150958	-0.0398107170553497\\
39.8107170553497	-0.0251188643150958\\
63.0957344480193	-0.0158489319246111\\
100	-0.01\\
};
\addlegendentry{$A_t=0.25$}

\addplot [color=white!35!black, line width=1.4pt, only marks, mark size=2.0pt, mark=o, mark options={solid, white!35!black}, forget plot]
  table[row sep=crcr]{%
0.1	-0.830578816520805\\
0.158489319246111	-0.830578808540314\\
0.251188643150958	-0.830578802692611\\
0.398107170553497	-0.830578798632051\\
0.630957344480193	-0.830578795914141\\
1	-0.830578794143062\\
1.58489319246111	-0.630960724852934\\
2.51188643150958	-0.398107052359957\\
3.98107170553497	-0.251186451042511\\
6.30957344480193	-0.158485889904851\\
10	-0.0999954980848816\\
15.8489319246111	-0.0630899209115147\\
25.1188643150958	-0.0398030875590683\\
39.8107170553497	-0.0251087105656218\\
63.0957344480193	-0.0158354068621378\\
100	-0.00998222893480481\\
};
\addplot [color=white!60!black, line width=1.4pt]
  table[row sep=crcr]{%
0.1	-0.726847521047708\\
0.158489319246111	-0.726847521047708\\
0.251188643150958	-0.726847521047708\\
0.398107170553497	-0.726847521047708\\
0.630957344480193	-0.726847521047708\\
1	-0.726847521047708\\
1.58489319246111	-0.630957344480193\\
2.51188643150958	-0.398107170553497\\
3.98107170553497	-0.251188643150958\\
6.30957344480193	-0.158489319246111\\
10	-0.1\\
15.8489319246111	-0.0630957344480193\\
25.1188643150958	-0.0398107170553497\\
39.8107170553497	-0.0251188643150958\\
63.0957344480193	-0.0158489319246111\\
100	-0.01\\
};
\addlegendentry{$A_t=0.43$}

\addplot [color=white!60!black, line width=1.4pt, only marks, mark size=2.0pt, mark=o, mark options={solid, white!60!black}, forget plot]
  table[row sep=crcr]{%
0.1	-0.726853600717236\\
0.158489319246111	-0.726853593521454\\
0.251188643150958	-0.726853588075866\\
0.398107170553497	-0.726853584204801\\
0.630957344480193	-0.726853581572993\\
1	-0.726853579845425\\
1.58489319246111	-0.630962103558926\\
2.51188643150958	-0.398109315896435\\
3.98107170553497	-0.251187123278366\\
6.30957344480193	-0.158482725069204\\
10	-0.099989109718913\\
15.8489319246111	-0.0630806780961663\\
25.1188643150958	-0.0397907980548295\\
39.8107170553497	-0.0250928994805138\\
63.0957344480193	-0.0158155876177288\\
100	-0.0099581189630596\\
};
\addplot [color=mycolor1, line width=1.4pt]
  table[row sep=crcr]{%
0.1	-0.601626016260163\\
0.158489319246111	-0.601626016260163\\
0.251188643150958	-0.601626016260163\\
0.398107170553497	-0.601626016260163\\
0.630957344480193	-0.601626016260163\\
1	-0.601626016260163\\
1.58489319246111	-0.601626016260163\\
2.51188643150958	-0.398107170553497\\
3.98107170553497	-0.251188643150958\\
6.30957344480193	-0.158489319246111\\
10	-0.1\\
15.8489319246111	-0.0630957344480193\\
25.1188643150958	-0.0398107170553497\\
39.8107170553497	-0.0251188643150958\\
63.0957344480193	-0.0158489319246111\\
100	-0.01\\
};
\addlegendentry{$A_t=0.72$}

\addplot [color=mycolor1, line width=1.4pt, only marks, mark size=2.0pt, mark=o, mark options={solid, mycolor1}, forget plot]
  table[row sep=crcr]{%
0.1	-0.601631059360729\\
0.158489319246111	-0.601631053043741\\
0.251188643150958	-0.60163104805722\\
0.398107170553497	-0.601631044391008\\
0.630957344480193	-0.601631041835997\\
1	-0.601631040118206\\
1.58489319246111	-0.601631038397878\\
2.51188643150958	-0.398110217600035\\
3.98107170553497	-0.251190154923092\\
6.30957344480193	-0.158484400510257\\
10	-0.0999803385208192\\
15.8489319246111	-0.0630623462866434\\
25.1188643150958	-0.0397641620486538\\
39.8107170553497	-0.0250584638675556\\
63.0957344480193	-0.0157736307409226\\
100	-0.00990910467820801\\
};
\addplot [color=red!60!teal, line width=1.4pt]
  table[row sep=crcr]{%
0.1	-0.545263157894737\\
0.158489319246111	-0.545263157894737\\
0.251188643150958	-0.545263157894737\\
0.398107170553497	-0.545263157894737\\
0.630957344480193	-0.545263157894737\\
1	-0.545263157894737\\
1.58489319246111	-0.545263157894737\\
2.51188643150958	-0.398107170553497\\
3.98107170553497	-0.251188643150958\\
6.30957344480193	-0.158489319246111\\
10	-0.1\\
15.8489319246111	-0.0630957344480193\\
25.1188643150958	-0.0398107170553497\\
39.8107170553497	-0.0251188643150958\\
63.0957344480193	-0.0158489319246111\\
100	-0.01\\
};
\addlegendentry{$A_t=0.90$}

\addplot [color=red!60!teal, line width=1.4pt, only marks, mark size=2.0pt, mark=o, mark options={solid, red!60!teal}, forget plot]
  table[row sep=crcr]{%
0.1	-0.545267732361855\\
0.158489319246111	-0.545267726467689\\
0.251188643150958	-0.545267721717283\\
0.398107170553497	-0.545267718161649\\
0.630957344480193	-0.545267715649675\\
1	-0.545267713940934\\
1.58489319246111	-0.545267712700687\\
2.51188643150958	-0.398110345267729\\
3.98107170553497	-0.251190526859815\\
6.30957344480193	-0.158489815706571\\
10	-0.0999846627446276\\
15.8489319246111	-0.0630570374644955\\
25.1188643150958	-0.0397500370532821\\
39.8107170553497	-0.0250370485718064\\
63.0957344480193	-0.0157462353938479\\
100	-0.00987689109746054\\
};
\node[right, align=left, inner sep=0]
at (rel axis cs:-0.15,1.1) {$(b)$};
\end{axis}
\end{tikzpicture}%
\fi
    \end{minipage}
    \caption{Large-wavenumber asymptotic behavior of the growth-rate spectrum at \(\Amu = 0\). (a) Growth rates for different Schmidt numbers at \(\Atw = 0.25\). (b) Comparison across Atwood numbers between the numerically extracted leading-order coefficient and the theoretical prediction~\eqref{eq:theoretical_kgg1_lambdam}.}
    \label{fig:plot_kgg1}
\end{figure}

The asymptotic structure of the growth-rate spectrum at large wavenumbers is obtained by rescaling time according to \(\tau = k^{2}t\) in~\eqref{eq:linearized_continuous_system} and expanding the dependent variables in inverse powers of \(k^{2}\),
\begin{subequations}
\begin{align}
    \hat{w} &= \hat{w}^{(0)} + k^{-2}\hat{w}^{(1)} + O\!\left(k^{-4}\right), \\
    \hat{\rho} &= \hat{\rho}^{(0)} + k^{-2}\hat{\rho}^{(1)} + O\!\left(k^{-4}\right).
\end{align}
\end{subequations}
Letting \(k\to\infty\) yields the leading-order system
\begin{equation}
    \begin{pmatrix}
        \rho_0 & \Sch^{-1}\partial_z \\[4pt]
        0 & 1
    \end{pmatrix}
    \begin{pmatrix}
        \partial_\tau \hat{w}^{(0)}\\[4pt]
        \partial_\tau \hat{\rho}^{(0)}
    \end{pmatrix}
    =
    \begin{pmatrix}
        -\mu_0 & -\Sch^{-1}\left(\mu_0\partial_z + 2\mu_0'\right)\left(\star\,\rho_0^{-1}\right)\\[6pt]
        0 & -\Sch^{-1}
    \end{pmatrix}
    \begin{pmatrix}
        \hat{w}^{(0)}\\[4pt]
        \hat{\rho}^{(0)}
    \end{pmatrix}.
\end{equation}
Since both matrices are upper triangular, the eigenvalues of this generalized eigenvalue problem are given by the ratios of their diagonal entries,
\begin{equation}
    \lambda_1(z) = -\mu_0/\rho_0, \qquad \lambda_2 = -\Sch^{-1}.
\end{equation}
Defining
\begin{equation}
    \label{eq:theoretical_kgg1_lambdam}
    \lambda_m(\Atw,\mathrm{A}_{\mu},\Sch,\delta)
    =
    \min\left\{
        \min_{z}\!\left(\mu_0/\rho_0\right),
        \Sch^{-1}
    \right\},
\end{equation}
the asymptotic growth rate in the limit \(k \gg k_c\) is
\begin{equation}
    \omega_r \sim -\lambda_m(\Atw,\mathrm{A}_{\mu},\Sch,\delta)\,k^{2}.
\end{equation}
Figure~\ref{fig:plot_kgg1}(a) shows a reference case for varying Schmidt number at fixed \(\Atw = 0.25\) and \(\mathrm{A}_{\mu} = 0.0\). The figure highlights that, at large wavenumber, there are regimes in which all curves collapse onto the same branch, corresponding to \(\lambda_m = \min_{z}(\mu_0/\rho_0)\), and regimes in which the damping is instead controlled by the diffusive branch \(\lambda_m = \Sch^{-1}\). This result generalizes and proves from first principles the scaling proposed by~\citet{duff-harlow-hirt-1962}, who argued on heuristic grounds that mass diffusion acts as an effective additional viscosity at large wavenumber, leading to
\begin{equation}
    \omega_r \sim -\Sch^{-1} k^2.
\end{equation}
More generally, equation~\eqref{eq:theoretical_kgg1_lambdam} shows that the large-wavenumber behavior is set by the competition between the locally stratified kinematic viscosity, which varies across the diffusive layer, and the reference Schmidt number appearing in the nondimensional equations. The Duff scaling is recovered in two important cases. First, in the regime \(\Sch \gg 1\), the large-wavenumber behavior becomes independent of local stratification and depends only on the reference Schmidt number, yielding the \(\Sch^{-1}\) scaling. Second, when \(\Sch > 1\), the same scaling is recovered exactly by imposing \(\Atw = \mathrm{A}_{\mu}\), in which case the local Schmidt number is spatially uniform and coincides with the reference value. If instead \(\Atw = \mathrm{A}_{\mu}\) and the reference Schmidt number is less than one, \(\Sch < 1\), the large-wavenumber behavior becomes independent of the Schmidt number, with \(\lambda_m = 1\).

This particular choice has been widely adopted in direct numerical simulations of the IVD system; however, as noted by~\citet{livescu-2020}, it is not the only physically admissible one, and the effect of a spatially varying local Schmidt number has received little attention in the literature. The present analysis addresses this point from the LST perspective by deriving the large-wavenumber scaling directly from the governing equations and extending the Duff argument to the general case of non-uniform local Schmidt number. For moderate Schmidt numbers, the asymptotic behavior depends explicitly on both \(\Atw\) and the viscosity contrast \(\mathrm{A}_{\mu}\) through the stratified kinematic viscosity. The theoretical prediction is validated in figure~\ref{fig:plot_kgg1}(b), where the leading-order coefficient extracted from the numerically computed growth rates at large wavenumber is compared with~\eqref{eq:theoretical_kgg1_lambdam}. The agreement is excellent across all cases considered, confirming the asymptotic analysis.

\subsection{The effect of diffusion for \texorpdfstring{\(k < k_c\)}{k < kc}}
\label{subsec:budget_idl}
We now examine the effect of diffusion on the positive growth rates. To isolate this effect from the additional complexity introduced by viscosity stratification, we consider the inviscid diffusive limit (IDL), understood as \(\Rey \to \infty\) with \(\Sch \to 0\), such that the P\'{e}clet number
\begin{equation}
    \Pe = \Rey \, \Sch,
\end{equation}
remains finite. In this limit, viscous stresses vanish from the governing equations~\eqref{eq:linearized_continuous_system}, while mass diffusion remains active. The IDL is introduced here as an idealized mechanism-isolating limit, not as a quantitatively complete substitute for the full viscous problem. Care must therefore be taken when interpreting this limit, since referring to the Schmidt number is ill-posed in the absence of viscous effects. Moreover, the Chandrasekhar scaling introduced in~\eqref{eq:chandrasekhar_scaling} is no longer valid, since the Reynolds number is infinite in this regime. \(\Frd = 1\) is retained, and the characteristic length scale is set by the width of the diffusive layer \(\delta\), leading to the definition
\begin{equation}
    \mathrm{Pe}_\delta = \frac{\delta^{2}/\mathcal{D}}{\sqrt{\delta/g}},
\end{equation}
which measures the relative strength of advective to diffusive transport at the scale of the diffusive layer \(\delta\). Thus, \(\mathrm{Pe}_\delta \gg 1\) corresponds to weak diffusion and advection-dominated dynamics, whereas \(\mathrm{Pe}_\delta \ll 1\) indicates a strong diffusive regime. From an operational standpoint, the governing equations in this regime admit the same nondimensional form as~\eqref{eq:linearized_continuous_system} with \(\mu_0 = 0\), and the P\'{e}clet number replaces the Schmidt number in the corresponding linear operators.
\begin{figure}
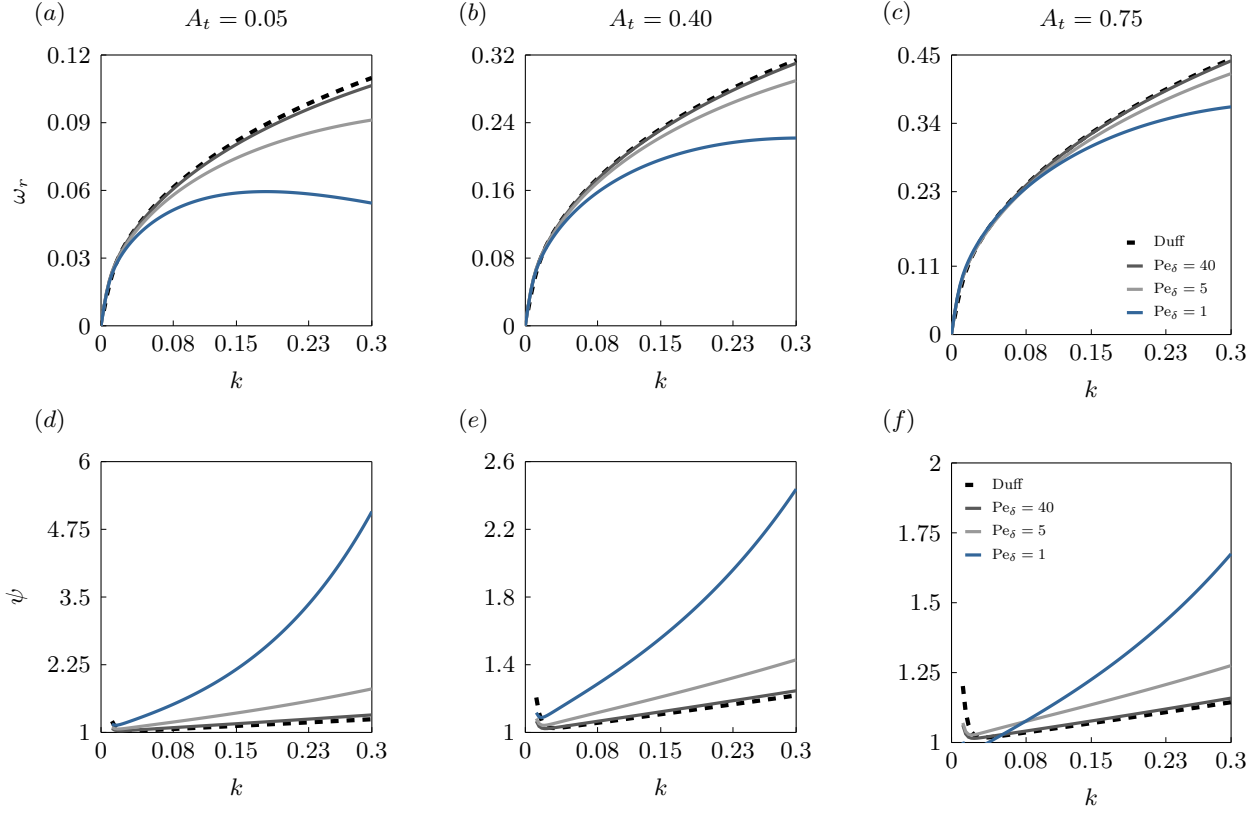

    \centering
    \begin{minipage}[t]{0.31\textwidth}
        \centering
        \vspace{0pt}
        \ifloadfig\input{figures/diffusion_effect_At0.05}\fi
    \end{minipage}
    \hfill
    \begin{minipage}[t]{0.31\textwidth}
        \centering
        \vspace{0pt}
        \ifloadfig\input{figures/diffusion_effect_At0.40}\fi
    \end{minipage}
    \hfill
    \begin{minipage}[t]{0.32\textwidth}
        \centering
        \vspace{0pt}
        \ifloadfig\input{figures/diffusion_effect_At0.75.tex}\fi
    \end{minipage}
    \begin{minipage}[t]{0.31\textwidth}
        \centering
        \vspace{0pt}
        \ifloadfig\input{figures/diffusion_effect_psi_At0.05}\fi
    \end{minipage}
    \hfill
    \begin{minipage}[t]{0.31\textwidth}
        \centering
        \vspace{0pt}
        \ifloadfig\input{figures/diffusion_effect_psi_At0.40}\fi
    \end{minipage}
    \hfill
    \begin{minipage}[t]{0.32\textwidth}
        \centering
        \vspace{0pt}
        \ifloadfig\input{figures/diffusion_effect_psi_At0.75}\fi
    \end{minipage}
    \caption{IDL growth rates and correction functions compared with the classical corrected inviscid prediction. (\dashlegend) denotes the prediction of~\citet{duff-harlow-hirt-1962}. Columns correspond to \(\Atw = 0.05\), \(0.40\), and \(0.75\), respectively.}
    \label{fig:IDL_growth_rates}
\end{figure}
To assess the role of diffusion, we compare the IDL growth rates with the classical corrected inviscid prediction~\eqref{eq:duff_growth_rate_correction}. Figure~\ref{fig:IDL_growth_rates} reports the corresponding growth rates and correction functions for three density ratios, namely \(\Atw = 0.05\), representative of the Boussinesq limit, \(\Atw = 0.40\), representative of an intermediate regime, and \(\Atw = 0.75\), representative of strong non-Boussinesq conditions. For each case, the diffusive strength is varied from a weakly diffusive regime at \(\mathrm{Pe}_\delta = 40\), through an intermediate regime at \(\mathrm{Pe}_\delta = 5\), to a high diffusive regime at \(\mathrm{Pe}_\delta = 1\). 

Figure~\ref{fig:IDL_growth_rates} shows that, in the weakly diffusive regime \((\Pe_{\delta}=40)\), the IDL growth rates and correction functions are nearly indistinguishable from the classical corrected inviscid prediction across all Atwood numbers considered. As the diffusive strength increases, however, diffusion suppresses the growth rates more strongly over the range of wavenumbers considered here. This is reflected in the correction functions, which progressively depart from the classical prediction and indicate an increasingly stabilizing influence. At fixed \(\Pe_{\delta}\), this stabilization weakens as \(\Atw\) increases, revealing a nontrivial coupling between diffusion and density contrast. This trend is particularly evident in figures~\ref{fig:IDL_growth_rates}(a,b,c), where the \(\Pe_{\delta}=1\) curve moves closer to the classical prediction as \(\Atw\) increases, especially for the smaller-wavenumber modes. 

We explain these observations by examining the perturbation kinetic-energy budget.
\begin{figure}
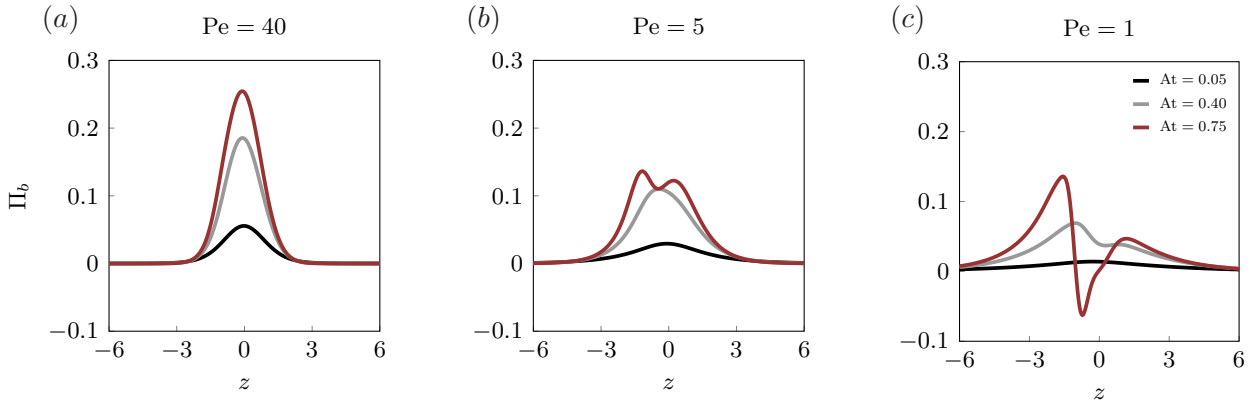

    \centering
    \begin{minipage}[t]{0.31\textwidth}
        \centering
        \vspace{0pt}
        \ifloadfig\input{figures/contribution_Pb_Pe40}\fi
    \end{minipage}
    \hfill
    \begin{minipage}[t]{0.31\textwidth}
        \centering
        \vspace{0pt}
        \ifloadfig\input{figures/contribution_Pb_Pe5}\fi
    \end{minipage}
    \hfill
    \begin{minipage}[t]{0.32\textwidth}
        \centering
        \vspace{0pt}
        \ifloadfig\input{figures/contribution_Pb_Pe1}\fi
    \end{minipage}
    \caption{Buoyancy production \(\Pi_b(z)\) at \(k=0.1\) in the IDL formulation, showing how its spatial structure changes with diffusive strength and density ratio.}
    \label{fig:contribution_buoyant}
\end{figure}
Since the perturbation fields are expressed as normal-mode eigenfunctions, we define the perturbation kinetic energy per unit mass as
\begin{equation}
    E_c = \frac{1}{2}\hat{u}_i^{*}\hat{u}_i,
\end{equation}
where \(\hat{u}_i\) denotes the eigenmode velocity component and the superscript \((\cdot)^{*}\) denotes its complex conjugate. The budget equation for the IDL system is obtained by multiplying the linearized momentum equation by \(\hat{u}_i^{*}\), summing this contracted form with its complex conjugate, and dividing by two. The result is then integrated over the domain, with boundary fluxes assumed to vanish. In what follows, all eigenfunctions are normalized such that the total perturbation kinetic energy satisfies
\begin{equation}
    \frac{1}{2}\int \rho_0 \, \hat{u}_i^{*}\hat{u}_i \,\mathrm{d}z = 1,
\end{equation}
so that the different production terms can be compared directly across density ratios and diffusive strengths \citep{kirthy-diwan-2020}. The budget equation for the linearized kinetic energy is
\begin{equation}
    \int \rho_0 \, \frac{\partial E_c}{\partial t}
    =
    - \int\underbrace{ \left( \rho_0 w_0' \hat{w}^{*}\hat{w} + \rho_0 w_0 \, \frac{\partial E_c}{\partial z} \right)}_{\text{energy transfer with the base state}}
    + \int\underbrace{ \operatorname{Re}\!\left(\hat{p} \, \bnabla \cdot \hat{\boldsymbol{u}}^{*}\right)}_{\Pi_p}
    - \int\underbrace{ \left( 1 + w_0 w_0' \right) \operatorname{Re}\!\left(\hat{\rho}\hat{w}^{*}\right)}_{-\Pi_b},
\end{equation}
where \(\Pi_b(z)\) denotes the buoyancy production and \(\Pi_p(z)\) the pressure production. The energy transfer with the base state is not discussed because it comprises only \(2\,\%\) of the total budget and is therefore not relevant to the instability mechanism. The buoyancy production quantifies the local rate of conversion from potential energy to perturbation kinetic energy. Where \(\Pi_b > 0\), the perturbation velocity and density fields are correlated such that light fluid moves upward and heavy fluid moves downward, thereby releasing gravitational potential energy into perturbation kinetic energy. Where \(\Pi_b < 0\), the perturbation works against gravity and locally consumes kinetic energy. The pressure production \(\Pi_p(z)\) represents the gain or loss of energy through work generated by the change in specific volume caused by mixing~\citep{livescu-ristorcelli-2007}.

\begin{figure}
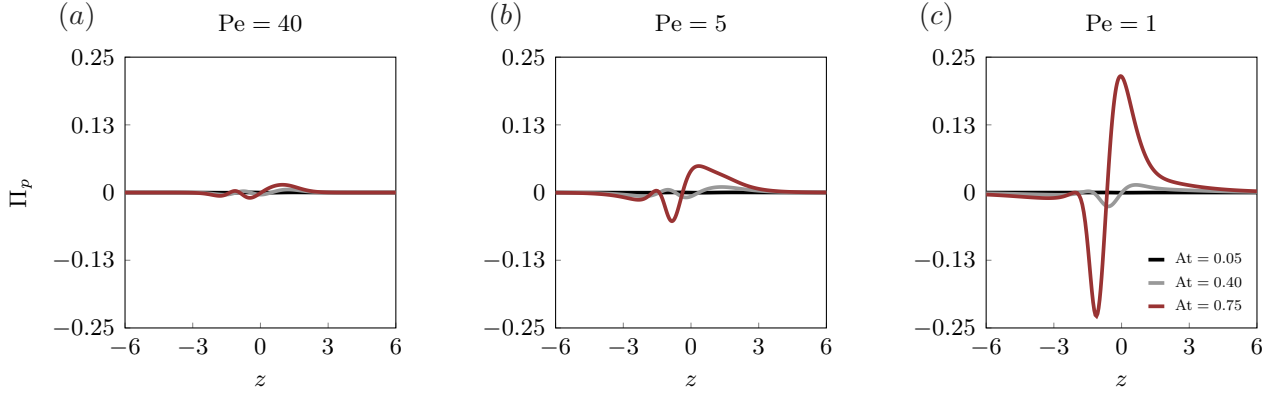

    \centering
    \begin{minipage}[t]{0.31\textwidth}
        \centering
        \vspace{0pt}
        \ifloadfig\input{figures/contribution_Pp_Pe40}\fi
    \end{minipage}
    \hfill
    \begin{minipage}[t]{0.31\textwidth}
        \centering
        \vspace{0pt}
        \ifloadfig\input{figures/contribution_Pp_Pe5}\fi
    \end{minipage}
    \hfill
    \begin{minipage}[t]{0.31\textwidth}
        \centering
        \vspace{0pt}
        \ifloadfig\input{figures/contribution_Pp_Pe1}\fi
    \end{minipage}
    \caption{Pressure production \(\Pi_p(z)\) at \(k=0.1\) in the IDL formulation, showing when pressure work becomes dynamically important.}
    \label{fig:contribution_pressure}
\end{figure}
Figure~\ref{fig:contribution_buoyant}(a) reports the buoyancy production for \(\Pe_{\delta} = 40\), showing that in this regime the profile is nearly symmetric and that the instability extracts most of its energy from the central region of the diffusive layer. As expected, increasing the Atwood number increases the magnitude of this contribution. Figure~\ref{fig:contribution_buoyant}(b) shows the case \(\Pe_{\delta} = 5\), where the buoyancy contribution spreads over a wider spatial region, consistent with the stabilizing effect of diffusion on the growth rates. For the larger Atwood numbers considered here, the production term still extracts most of the energy from the diffusive layer, although a mild asymmetry about \(z=0\) begins to appear. Figure~\ref{fig:contribution_buoyant}(c) shows the case \(\Pe_{\delta} = 1\), where, for the largest Atwood number, the buoyancy production develops a bimodal spatial distribution with a larger peak on the lighter side of the stratification. This suggests that when diffusion is important and the Atwood number is large, the lighter fluid, having lower inertia, experiences a larger acceleration for the same perturbation. Consequently, the light fluid responds more vigorously to the buoyancy forcing, generates larger velocity perturbations, and produces more perturbation kinetic energy. This is consistent with findings reported in the turbulent regime~\citep{livescu-ristorcelli-2008,baltzer-livescu-2020}, where the most intense turbulence migrates toward the lighter-fluid side as the density ratio increases. In the present linear-stability analysis, the same physics are manifested in the eigenmode structure, where the perturbation amplitude is larger on the light-fluid side, as is the buoyancy production. In the linear regime, this effect can be attributed directly to the effective buoyancy term \(b_e = 1 + w_0 w_0'\) appearing in the buoyancy production. The non-Boussinesq behavior is therefore caused by the presence of this additional buoyancy contribution, which is not sign-definite. In particular, it reduces the buoyancy forcing on the heavier side of the stratification while enhancing it on the lighter side, particularly in the highly diffusive and strongly stratified regime, where non-Boussinesq effects are expected to be most significant.

\begin{figure}
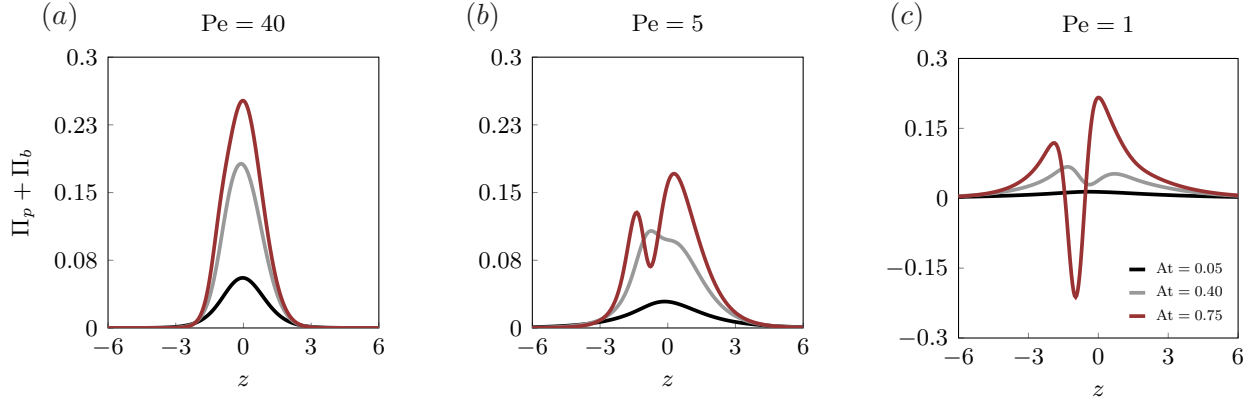

    \centering
    \begin{minipage}[t]{0.31\textwidth}
        \centering
        \vspace{0pt}
        \ifloadfig\input{figures/contribution_sum_Pe40}\fi
    \end{minipage}
    \hfill
    \begin{minipage}[t]{0.31\textwidth}
        \centering
        \vspace{0pt}
        \ifloadfig\input{figures/contribution_sum_Pe5}\fi
    \end{minipage}
    \hfill
    \begin{minipage}[t]{0.32\textwidth}
        \centering
        \vspace{0pt}
        \ifloadfig\input{figures/contribution_sum_Pe1}\fi
    \end{minipage}
    \caption{Net production \(\Pi_b(z) + \Pi_p(z)\) at \(k=0.1\) in the IDL formulation, showing how the overall energy input changes with diffusive strength.}
    \label{fig:contribution_sum_peclet}
\end{figure}

Although these observations support an enhanced stabilizing effect of diffusion relative to Duff prediction, they do not explain why this stabilization weakens as the Atwood number increases at fixed diffusive strength. To address this point, we turn our attention to the pressure production term. Figure~\ref{fig:contribution_pressure}(a) shows that, in the weakly diffusive regime, pressure production is much smaller than buoyancy production, suggesting that the instability is governed primarily by buoyancy. The same conclusion can be drawn from the net production term in figure~\ref{fig:contribution_sum_peclet}(a). Figure~\ref{fig:contribution_pressure}(b) shows that, in the moderately diffusive regime, pressure production, although still smaller than buoyancy production, already affects the overall energy balance. More specifically, it extracts energy from the lighter side of the fluid and injects it into the heavier portion of the stratification. Although this effect is not yet dominant, it is already visible in figure~\ref{fig:contribution_sum_peclet}(b), where two local maxima appear in the net production term for the largest Atwood number. Pressure production becomes important in the strongly diffusive regime, as shown in figure~\ref{fig:contribution_pressure}(c) and figure~\ref{fig:contribution_sum_peclet}(c), so that the overall production is largely affected by the pressure rather than just the classical buoyancy production. We therefore argue that this shift in the instability mechanism, from one dominated by the classical buoyancy-production term to one in which pressure production transfers energy from the lighter to the heavier part of the stratification, may explain why the stabilizing effect of diffusion weakens as the Atwood number increases.
The weakening of the stabilizing effect of diffusion with increasing Atwood number is most evident for large Atwood number and small P\'{e}clet number. This is precisely the strongly diffusive regime in which the quasi-steady-state approximation no longer predicts the growth rates accurately because the base state evolves too rapidly. The discussion in this section should therefore be interpreted as a qualitative description of the coupling introduced by mass diffusion, rather than as a definitive statement about the fully time-dependent problem. In particular, the budget analysis suggests that pressure production may play an increasingly important role in the highly diffusive regime and thereby may play a key role during the instability evolution.

\section{Initial-value problem comparison with the quasi-steady-state approximation}
\label{sec:IVP}
\begin{figure}
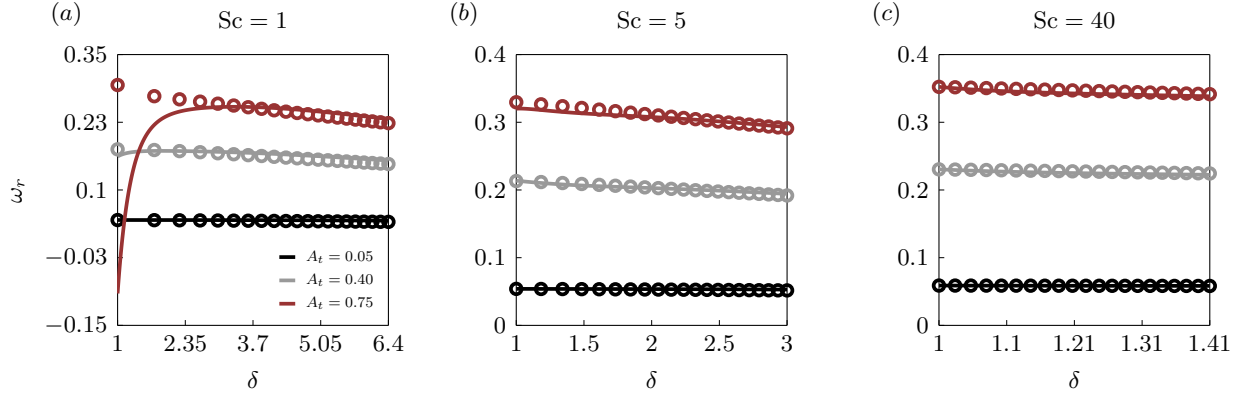

    \centering
    \begin{minipage}[t]{0.31\textwidth}
        \centering
        \vspace{0pt}
        \ifloadfig\input{figures/IVP1_growthrate_vsdelta_Sc1.tex}\fi
    \end{minipage}
    \hfill
    \begin{minipage}[t]{0.31\textwidth}
        \centering
        \vspace{0pt}
        \ifloadfig\input{figures/IVP1_growthrate_vsdelta_Sc5.tex}\fi
    \end{minipage}
    \hfill
    \begin{minipage}[t]{0.31\textwidth}
        \centering
        \vspace{0pt}
        \ifloadfig\input{figures/IVP1_growthrate_vsdelta_Sc40.tex}\fi
    \end{minipage}
    \caption{Comparison between the quasi-steady-state approximation growth rates and the growth rate \(\tilde{\omega}_r\) extracted from the initial-value problem at \(k^{\max}\). (\circlegend) denotes the quasi-steady-state approximation and (\solidlegend) the initial-value problem.}
    \label{fig:IVP_growthrate}
\end{figure}

The quasi-steady-state approximation provides a local-in-time prediction of the perturbation growth rate by treating the evolving base state as frozen at each instant. The purpose of the initial-value problem comparison is to assess this prediction once the full time dependence of the base state is retained, and to identify where the frozen-base approximation remains accurate and where it fails. To compare the evolution of the same disturbance in the two settings, the initial-value problem is initialized with the corresponding eigenvector obtained from the quasi-steady-state approximation. It must nevertheless be noted that other initial conditions may lead to different transient responses, since the early-time evolution of unstable diffusive systems is known to depend sensitively on the imposed disturbance~\citep{tan-homsy-1986,rapaka-et-al-2008,don-nils-amir-2013}. The instantaneous growth rate extracted from the initial-value problem is defined as
\begin{equation}
    \tilde{\omega}_r\!\left(k,t_0(\delta_0)\right)
    =
    \frac{1}{2}\difft{}{t}\log\!\left(\left\lVert \hat{\mathbf{q}}_k \right\rVert^2\right).
\end{equation}
The wavenumber chosen for the comparison reported in figure~\ref{fig:IVP_growthrate} is the most unstable wavenumber, \(k^{\max}\), predicted by the quasi-steady-state approximation at the initial thickness \(\delta_0\).
Discrepancies between the initial-value problem and quasi-steady-state approximation predictions at early times are to be expected, particularly in regimes where the base state evolves rapidly during the initial transient. Figure~\ref{fig:IVP_growthrate} shows, however, that the two approaches are in close agreement over most of the parameter space considered. The main exception arises for \(\Sch = 1\) at the largest Atwood number reported in figure~\ref{fig:IVP_growthrate}(a), precisely in the regime where the frozen-base approximation is expected to be least accurate based on the scaling analysis in~\eqref{eq:estimation_qssa_failure}. In this case, the initial-value problem predicts an initial decay before the predicted growth is recovered. As the diffusive layer broadens the initial-value problem growth rate approaches the quasi-steady-state approximation prediction evaluated along the evolving base state. This observation indicates that a configuration predicted to be unstable by the frozen-base problem may nevertheless be transiently stable during the early stage of the evolution when the base state changes rapidly. The comparison also shows that the breakdown of the quasi-steady-state approximation is not controlled by diffusion alone. Although a small Schmidt number promotes rapid changes of the base state, the discrepancy between the initial-value problem and the quasi-steady-state approximation becomes significant also when diffusion is accompanied by sufficiently large density stratification. This indicates that the density stratification plays an independent role in reducing the separation between the time scale of perturbation growth and that of base-state evolution. Thus, even when diffusion is only moderate, sufficiently strong stratification may still induce rapid variations of the base state and thereby compromise the validity of the quasi-steady-state approximation.

\section{Conclusions}%
\label{sec:conclusions}
This study presents a linear stability analysis of the miscible Rayleigh--Taylor instability in a configuration in which viscosity and mass diffusion are retained consistently within the incompressible variable-density (IVD) framework. Because compositional diffusion renders the velocity field non-solenoidal, the diffusive mixing layer is accompanied by a non-zero vertical base velocity. Rather than prescribing the base state a priori, a self-similar solution is derived directly from the governing equations, providing a physically consistent starting point for the stability analysis. The complete self-similar structure reveals features consistent with the asymmetric flow organization observed in direct numerical simulations of miscible variable-density Rayleigh--Taylor flows \citep{livescu-et-al-2010}. The existence of nontrivial self-similar solutions for the in-plane velocities, governed by a Sturm--Liouville eigenvalue problem with Hermite--Gaussian eigenfunctions, is also established, with detailed investigation left for future work.

Freezing this evolving base state within a quasi-steady approximation yields a tractable eigenvalue problem for the growth of perturbations. A scaling analysis of the neglected temporal derivatives shows that the quasi-steady-state approximation is not uniformly valid across the parameter space. The magnitude of the frozen terms increases with \(\Atw\) and decreases with \(\Sch\), so that the approximation is expected to deteriorate at large density contrasts and low Schmidt numbers.

The linearized IVD system is reduced to a coupled vertical-velocity--density formulation, and in the limit \(\Sch \to \infty\), the equations reduce exactly to those of~\citet{chandrasekhar-1961}, establishing the present framework as a consistent extension of the classical diffusion-free theory to account for mass diffusion. When mass diffusion is retained at finite Schmidt number, the structure of the linear spectrum changes qualitatively. An asymptotic analysis in the large-wavenumber limit establishes a \(k^2\) damping of the growth rate, generalizing the heuristic scaling of~\citet{duff-harlow-hirt-1962} and deriving it systematically from the governing equations. The damping coefficient is governed by a competition between the locally stratified kinematic viscosity and the reference Schmidt number, and reduces to the classical Duff scaling when the local Schmidt number is spatially uniform.

The analysis of the inviscid diffusive limit, introduced to isolate the relevant mechanism, reveals that the effect of mass diffusion on the instability is more nuanced than simple stabilization. At low to moderate density ratios, diffusion acts as a purely stabilizing mechanism. At large density ratios, however, a nontrivial coupling between diffusion and density stratification emerges. The buoyancy production develops a bimodal spatial distribution with a pronounced peak on the lighter side of the stratification, while the pressure production becomes comparable to the buoyancy production and acts as an energy source on the heavier side. Within the frozen-base framework, this coupling manifests itself as a weakening of diffusive stabilization rather than as a purely stabilizing correction. From the comparison with the initial-value problem, the strongly diffusive, strongly stratified regime should be interpreted as a regime in which diffusion--non-Boussinesq coupling becomes important, but also as one in which the quasi-steady-state approximation must be used with particular caution.

Capturing the effect of diffusion during the linear stages may improve the modeling of RTI. \citet{kola-israel-towne-2026} showed that, when the instability is analyzed within a statistical linear framework, the mean can exhibit a transient decay before the onset of unstable growth. The duration of this delay depends on the imposed initial spectrum and varies non-monotonically with the spectral content of the interface perturbation, suggesting that the diffusive contribution plays an important role in characterizing the subsequent evolution of the instability. Since the initial perturbation spectrum in Rayleigh--Taylor instability is naturally broadband and retains energy at higher wavenumbers~\citep{mueschke-andrews-schilling-2006}, the ability to characterize both the spectral cutoff and the large-wavenumber growth rates at large Atwood and Schmidt numbers may be relevant to the later stages of the instability, where these modes contribute to the self-similar evolution observed in the nonlinear and turbulent regimes~\citep{dimonte-et-al-2004,dimonte-2004}. Diffusion has also been shown to play an important role in miscible flows beyond the linear regime. In~\citet{roberts-jacobs-2016}, when small-wavelength perturbations are seeded into the flow, the growth parameter of the mixing width differs significantly between the miscible and immiscible cases at \(\Atw = 0.48\), despite being independent of the spectral content of the perturbation. This discrepancy was attributed to miscibility effects acting during the later stages of the instability. In that study, the representative mixture Schmidt number is \(\Sch \approx 1.4\times 10^{4}\), indicating that the miscible experiments of~\citet{roberts-jacobs-2016} correspond to a \(\Sch \to \infty\) regime. The present analysis may therefore provide a basis for designing dedicated experiments to investigate miscibility effects in regimes where the Schmidt number is of order unity, potentially offering additional insight into the growth parameter during the later stages of the instability evolution. Overall, deriving the effect of diffusion from first principles, together with its coupling to viscosity and density stratification, may improve predictive models of Rayleigh--Taylor instability in regimes where mass diffusion plays a significant role, such as plasma applications~\citep{vold-yin-albright-2021,keenan-sauppe-2023}.

\section*{Acknowledgments}
The authors gratefully acknowledge support provided by the U.S. Department of Energy through Los Alamos National Laboratory and the LANL Michigan SPARC. Los Alamos National Laboratory is operated by Triad National Security, LLC, for the National Nuclear Security Administration of the U.S. Department of Energy (Contract No. 89233218CNA000001). M.K. gratefully acknowledges Dr. Livescu for the discussion on the infinite Schmidt number limit.

\begin{appen}
\section{Incompressible variable-density reduced formulation}
\label{sec:appendix_reducedIVD}
In this appendix, we discuss the linearized equations reported in compact form in \eqref{eq:compact_1st_equations} and briefly outline the derivation of the reduced vertical-velocity--density formulation used in the linear stability analysis. The set of linearized equation, using Einstein summation notation is given by
\begin{subequations}
\begin{align}
    \frac{\partial \rho}{\partial t}
    + \frac{\partial (\rho_0 u_j)}{\partial x_j}
    + \frac{\partial (\rho u_{0j})}{\partial x_j}
    &=
    0,
    \\
    \rho \frac{\partial u_{0i}}{\partial t}
    + \rho_0 \frac{\partial u_i}{\partial t}
    + \rho u_{0j} \frac{\partial u_{0i}}{\partial x_j}
    + \rho_0 u_j \frac{\partial u_{0i}}{\partial x_j}
    + \rho_0 u_{0j} \frac{\partial u_i}{\partial x_j}
    &=
    -\frac{\partial p}{\partial x_i}
    - \rho \delta_{i3}
    + \frac{\partial \tau_{ij}^{(1)}}{\partial x_j},
    \label{eq:linearized_set_momentum}
\end{align}
\end{subequations}
where the linearized viscous stress \(\tau_{ij}^{(1)}\) is given by
\begin{subequations}
\begin{align}
    \tau_{ij}^{(1)} &= \tau_{ij}^{(1,0)} + \tau_{ij}^{(0,1)},
    \\
    \tau_{ij}^{(1,0)} &= \mu\left(2S_{0ij} - \frac{2}{3}\bnabla\bcdot\boldsymbol{u}_0\,\delta_{ij}\right),
    \\
    \tau_{ij}^{(0,1)} &= \mu_0\left(2S_{ij} - \frac{2}{3}\bnabla\bcdot\boldsymbol{u}\,\delta_{ij}\right).
\end{align}
\end{subequations}
with the linearized divergence constraint given by
\begin{equation}
    \bnabla\bcdot\boldsymbol{u}
    =
    -\Sch^{-1}\nabla^2\left(\frac{\rho}{\rho_0}\right).
\end{equation}
For the base state discussed in \S\ref{subsec:zeroth_order_self_similar}, \eqref{eq:linearized_set_momentum} can be written as
\begin{equation}\label{eq:linear_momentum_vertical_base}
  \rho \dot w_0\delta_{i3}
  + \rho_0 \frac{\partial u_i}{\partial t}
  + \rho w_0 w_0'\delta_{i3}
  + \rho_0 w_0' w \delta_{i3}
  + \rho_0 w_0 \frac{\partial u_i}{\partial z}
  =
  - \frac{\partial p}{\partial x_i}
  - \rho \delta_{i3}
  + \frac{\partial \tau_{ij}^{(1)}}{\partial x_j}.
\end{equation}
The vertical momentum equation follows by setting \(i=3\) in \eqref{eq:linear_momentum_vertical_base}. This gives
\begin{equation}\label{eq:vertical_momentum_step_5_1}
  \left(
    \dot w_0
    +
    w_0 w_0'
  \right)\rho
  +
  \rho_0 \frac{\partial w}{\partial t}
  +
  \left(
    \rho_0 w_0'
    +
    \rho_0 w_0 \frac{\partial}{\partial z}
  \right)w
  =
  -
  \frac{\partial p}{\partial z}
  -
  \rho
  +
  \frac{\partial \tau_{3j}^{(1)}}{\partial x_j}.
\end{equation}
We first define
\begin{subequations}\label{eq:vertical_momentum_coefficients}
\begin{gather}
  S_0
  =
  \dot w_0
  +
  w_0 w_0',
  \\
  C_0
  =
  \rho_0 w_0'
  +
  \rho_0 w_0 \frac{\partial}{\partial z}.
\end{gather}
\end{subequations}
Taking the Laplacian of \eqref{eq:vertical_momentum_step_5_1} and using \eqref{eq:vertical_momentum_coefficients} gives
\begin{equation}\label{eq:laplacian_vertical_momentum}
\begin{split}
  &
  \left(
    S_0 \nabla^2
    +
    2 S_0' \frac{\partial}{\partial z}
    +
    S_0''
  \right)\rho
  +
  \left(
    \rho_0''
    +
    2\rho_0' \frac{\partial}{\partial z}
    +
    \rho_0 \nabla^2
  \right)
  \frac{\partial w}{\partial t}
  +
  \left(
    C_0 \nabla^2
    +
    2 C_0' \frac{\partial}{\partial z}
    +
    C_0''
  \right)w
  \\
  &=
  -
  \frac{\partial}{\partial z}
  \left(
    \nabla^2 p
  \right)
  -
  \nabla^2 \rho
  +
  \nabla^2
  \left(
    \frac{\partial \tau_{3j}^{(1)}}{\partial x_j}
  \right).
\end{split}
\end{equation}
We next define the operators
\begin{subequations}\label{eq:poisson_operators}
\begin{gather}
  A_0
  =
  \left[
    \left(
      \dot w_0
    \right)'
    +
    \dot w_0
    \frac{\partial}{\partial z}
    +
    w_0 w_0'
    \frac{\partial}{\partial z}
    +
    \left(
      w_0 w_0'
    \right)'
  \right],
  \\
  B_0
  =
  \left[
    \left(
      \rho_0 w_0'
    \right)'
    +
    \rho_0 w_0'
    \frac{\partial}{\partial z}
    +
    \left(
      \rho_0 w_0
    \right)'
    \frac{\partial}{\partial z}
  \right].
\end{gather}
\end{subequations}
Taking the divergence of \eqref{eq:linear_momentum_vertical_base} and using \eqref{eq:poisson_operators} gives the pressure Poisson equation
\begin{equation}\label{eq:pressure_poisson_equation}
\begin{split}
  A_0 \rho
  +
  \rho_0 w_0
  \frac{\partial}{\partial z}
  \left(
    \bnabla\bcdot\boldsymbol{u}
  \right)
  +
  B_0 w
  +
  \rho_0'
  \frac{\partial w}{\partial t}
  +
  \rho_0
  \frac{\partial}{\partial t}
  \left(
    \bnabla\bcdot\boldsymbol{u}
  \right)
  =
  -
  \nabla^2 p
  -
  \frac{\partial \rho}{\partial z}
  +
  \frac{\partial^2 \tau_{ij}^{(1)}}{\partial x_i \partial x_j}.
\end{split}
\end{equation}
Taking the vertical derivative of \eqref{eq:pressure_poisson_equation} gives
\begin{equation}\label{eq:z_derivative_pressure_poisson_equation}
\begin{split}
  \left(
    \frac{\partial A_0}{\partial z}
    +
    A_0 \frac{\partial}{\partial z}
  \right)\rho
  +
  \frac{\partial}{\partial z}
  \left[
    \rho_0 w_0
    \frac{\partial}{\partial z}
    \left(
      \bnabla\bcdot\boldsymbol{u}
    \right)
  \right]
  +
  \left(
    \frac{\partial B_0}{\partial z}
    +
    B_0 \frac{\partial}{\partial z}
  \right)w
  +
  \frac{\partial}{\partial z}
  \left(
    \rho_0'
    \frac{\partial w}{\partial t}
  \right)
  \\
  \qquad
  +
  \frac{\partial}{\partial z}
  \left[
    \rho_0
    \frac{\partial}{\partial t}
    \left(
      \bnabla\bcdot\boldsymbol{u}
    \right)
  \right]
  =
  -
  \frac{\partial}{\partial z}
  \left(
    \nabla^2 p
  \right)
  -
  \frac{\partial^2 \rho}{\partial z^2}
  +
  \frac{\partial}{\partial z}
  \left(
    \frac{\partial^2 \tau_{ij}^{(1)}}{\partial x_i \partial x_j}
  \right).
\end{split}
\end{equation}
Subtracting \eqref{eq:laplacian_vertical_momentum} from \eqref{eq:z_derivative_pressure_poisson_equation} eliminates the pressure contribution. We define
\begin{subequations}\label{eq:w0_xy_operators}
\begin{gather}
  \mathscr{L}_{w_0\rho}
  =
  S_0 \nabla^2
  +
  2 S_0' \frac{\partial}{\partial z}
  +
  S_0''
  -
  \frac{\partial A_0}{\partial z}
  -
  A_0 \frac{\partial}{\partial z},
  \\
  \mathscr{L}_{w_0 w}
  =
  C_0''
  +
  2 C_0' \frac{\partial}{\partial z}
  +
  C_0 \nabla^2
  -
  \frac{\partial B_0}{\partial z}
  -
  B_0 \frac{\partial}{\partial z}.
\end{gather}
\end{subequations}
Then the reduced equation obtained from \eqref{eq:z_derivative_pressure_poisson_equation} minus \eqref{eq:laplacian_vertical_momentum} is
\begin{equation}\label{eq:reduced_vertical_momentum_poisson}
\begin{split}
  &
  -
  \left(
    \rho_0' \frac{\partial}{\partial z}
    +
    \rho_0 \nabla^2
  \right)
  \frac{\partial w}{\partial t}
  +
  \left(
    \rho_0'
    +
    \rho_0 \frac{\partial}{\partial z}
  \right)
  \frac{\partial}{\partial t}
  \left(
    \bnabla\bcdot\boldsymbol{u}
  \right)
  =
  -
  \frac{\partial}{\partial z}
  \left[
    \rho_0 w_0
    \frac{\partial}{\partial z}
    \left(
      \bnabla\bcdot\boldsymbol{u}
    \right)
  \right]
  \\
  &\quad
  +
  \mathscr{L}_{w_0\rho}\rho
  +
  \mathscr{L}_{w_0 w}w
  +
  \nabla_{xy}^2 \rho
  +
  \frac{\partial}{\partial z}
  \left(
    \frac{\partial^2 \tau_{ij}^{(1)}}{\partial x_i \partial x_j}
  \right)
  -
  \nabla^2
  \left(
    \frac{\partial \tau_{3j}^{(1)}}{\partial x_j}
  \right).
\end{split}
\end{equation}
The remaining stress contribution in \eqref{eq:reduced_vertical_momentum_poisson} can be shown to be
\begin{equation}\label{eq:stress_pressure_vertical_momentum_difference}
\begin{split}
  \frac{\partial}{\partial z}
  \left(
    \frac{\partial^2 \tau_{ij}^{(1)}}{\partial x_i \partial x_j}
  \right)
  -
  \nabla^2
  \left(
    \frac{\partial \tau_{3j}^{(1)}}{\partial x_j}
  \right)
  &=
  \mu_0
  \left(
    \frac{\partial}{\partial z}
    \left[
      \nabla^2
      \left(
        \bnabla\bcdot\boldsymbol{u}
      \right)
    \right]
    -
    \nabla^4 w
  \right)
  \\
  &\quad
  +
  \mu_0'
  \left(
    2 \nabla^2
    \left(
      \bnabla\bcdot\boldsymbol{u}
    \right)
    -
    2 \frac{\partial}{\partial z}
    \left(
      \nabla^2 w
    \right)
  \right)
  \\
  &\quad
  +
  \mu_0''
  \left(
    \nabla^2 w
    -
    2 \frac{\partial^2 w}{\partial z^2}
    +
    \frac{\partial}{\partial z}
    \left(
      \bnabla\bcdot\boldsymbol{u}
    \right)
  \right)
  \\
  &\quad
  -
  2 \frac{\partial}{\partial z}
  \left(
    w_0' \nabla_{xy}^2 \mu
  \right).
\end{split}
\end{equation}
Using the definitions in \eqref{eq:poisson_operators} and \eqref{eq:vertical_momentum_coefficients}, the operators in \eqref{eq:w0_xy_operators} reduce to
\begin{subequations}\label{eq:w0_operators}
\begin{gather}
  \mathscr{L}_{w_0\rho}
  =
  \left(
    \dot w_0
    +
    w_0 w_0'
  \right)
  \nabla_{xy}^2,
  \\
  \mathscr{L}_{w_0 w}
  =
  \rho_0' w_0
  \frac{\partial^2}{\partial z^2}
  +
  \left(
    \rho_0 w_0'
    +
    \rho_0 w_0 \frac{\partial}{\partial z}
  \right)
  \nabla^2.
\end{gather}
\end{subequations}
The last remaining term to rewrite is the time derivative of the velocity divergence. We use
\begin{subequations}\label{eq:divergence_time_derivative_identity}
\begin{gather}
  \frac{\partial}{\partial t}
  \left(
    \bnabla\bcdot\boldsymbol{u}
  \right)
  =
  \mathscr{L}_{\rho}
  \left(
    \frac{\partial \rho}{\partial t}
  \right)
  +
  \dot{\mathscr{L}}_{\rho}
  \left(
    \rho
  \right),
  \\
  \mathscr{L}_{\rho}\left(\star\right)
  =
  -
  \Sch^{-1}
  \nabla^2
  \left(
    \frac{\star}{\rho_0}
  \right),
  \\
  \dot{\mathscr{L}}_{\rho}\left(\star\right)
  =
  \Sch^{-1}
  \nabla^2
  \left(
    \frac{\dot{\rho}_0}{\rho_0^{2}}\,\star
  \right).
\end{gather}
\end{subequations}
Finally, using
\begin{equation}\label{eq:mu_density_linear_relation}
  \mu = C_{\mu\rho}\rho,
\end{equation}
together with \eqref{eq:stress_pressure_vertical_momentum_difference}, \eqref{eq:w0_operators}, and \eqref{eq:divergence_time_derivative_identity}, the rearrangement of \eqref{eq:reduced_vertical_momentum_poisson} gives \eqref{eq:A_B_operators}.

The derivation may be summarized as follows:
\begin{itemize}
  \item First, we take the Laplacian of \eqref{eq:vertical_momentum_step_5_1}, obtaining \eqref{eq:laplacian_vertical_momentum}.

  \item Second, we take the divergence of \eqref{eq:linear_momentum_vertical_base}, obtaining \eqref{eq:pressure_poisson_equation}. We then take the vertical derivative of \eqref{eq:pressure_poisson_equation}, obtaining \eqref{eq:z_derivative_pressure_poisson_equation}.

  \item Third, we subtract \eqref{eq:laplacian_vertical_momentum} from \eqref{eq:z_derivative_pressure_poisson_equation}, which eliminates the pressure term and gives \eqref{eq:reduced_vertical_momentum_poisson}.

  \item Finally, we use \eqref{eq:stress_pressure_vertical_momentum_difference}, \eqref{eq:w0_operators}, \eqref{eq:divergence_time_derivative_identity}, and \eqref{eq:mu_density_linear_relation} to rewrite \eqref{eq:reduced_vertical_momentum_poisson} in the final operator form \eqref{eq:A_B_operators}.
\end{itemize}

\section{Viscosity Mixing Law}%
\label{sec:appendix_viscosity_law}
In this appendix, we examine the sensitivity of the growth rates to the mixing law used to describe viscosity. For clarity, the discussion is presented in dimensional form. Here, \(\mu_1\) and \(\mu_2\) denote the dynamic viscosities of the two pure fluids, \(\rho_1\) and \(\rho_2\) their densities, and \(\mu\) and \(\rho\) the corresponding viscosity and density of the mixture. Let \(W_{\alpha}\), \(X_{\alpha}\), and \(Y_{\alpha}\) denote the molar mass, mole fraction, and mass fraction of species \(\alpha\), respectively. For a binary mixture, the mole fraction is defined by
\begin{equation}
    \label{eq:molar_fraction_definition}
    X_{\alpha} = \frac{Y_{\alpha}}{W_{\alpha}}
    \left(
        \sum_{\beta=1}^{2} \frac{Y_{\beta}}{W_{\beta}}
    \right)^{-1},
    \qquad \alpha = 1,2.
\end{equation}
and for the case we are considering
\begin{equation}
    \label{eq:amagat_law}
    \frac{1}{\rho} = \frac{Y_1}{\rho_1} + \frac{Y_2}{\rho_2}.
\end{equation}
The kinematic viscosity of the mixture is defined by \(\nu = \mu/\rho\), and the mass fractions satisfy
\begin{equation}
    \label{eq:Y_from_X}
    Y_1 = \frac{\rho_1 X_1}{\rho},
    \qquad
    Y_2 = \frac{\rho_2 X_2}{\rho}.
\end{equation}
Using~\eqref{eq:molar_fraction_definition}, the density law~\eqref{eq:amagat_law} can be expressed as
\begin{equation}
    \rho = \rho_1 X_1 + \rho_2 X_2.
\end{equation}
In particular, once the base-state density is specified by the self-similar diffusive profile~\eqref{eq:self_similar_base_state_density}, it follows that
\begin{equation}
    X_1 = \frac{1}{2}\left(1 - \erf\left(\frac{z}{\delta}\right)\right).
\end{equation}
This recovers the mole-fraction profile employed by~\citet{morgan-likhachev-jacobs-2016}, but here it arises directly from the governing equations rather than as an independent modeling prescription.
\begin{figure}
    \centering
    \begin{minipage}[t]{0.32\textwidth}
        \centering
        \ifloadfig
%
\definecolor{mycolor1}{rgb}{0.00000,0.45000,0.74000}%
\begin{tikzpicture}[trim axis left,trim axis right]

\begin{axis}[%
width=11.255in,
height=7.131in,
at={(1.888in,0.962in)},
scale only axis,
separate axis lines,
every outer x axis line/.append style={black},
every x tick label/.append style={font=\color{black}},
every x tick/.append style={black},
xmin=0,
xmax=2,
xlabel={$k$},
every outer y axis line/.append style={black},
every y tick label/.append style={font=\color{black}},
every y tick/.append style={black},
ymin=0,
ymax=0.3,
ylabel={$\omega_r$},
axis background/.style={fill=white},
title style={font=\bfseries},
title={$A_{\mu} = -0.90$},
legend style={legend cell align=left, align=left},
clip=true,
clip mode=individual,
axis lines=box,
xtick pos=bottom,
ytick pos=left,
tick align=inside,
major tick length=2pt,
width=0.7\linewidth,
height=0.65\linewidth,
every axis/.append style={font=\fontsize{9}{9}\selectfont},xlabel style={font=\fontsize{9}{9}\selectfont},title style={font=\fontsize{9}{9}\selectfont},
legend style={font=\fontsize{8}{9}\selectfont,inner sep=1pt,row sep=1pt,column sep=2pt,nodes={scale=1.00},draw=none},legend image post style={xscale=0.35},
legend columns=1,
scaled x ticks=false,
tick label style={/pgf/number format/fixed,/pgf/number format/precision=2},
every x tick label/.append style={font=\fontsize{9}{9}\selectfont\color{black}},
every y tick label/.append style={font=\fontsize{9}{9}\selectfont\color{black}},
,,
colormap={sigmamap}{rgb(0.0000)=(0.0200,0.1880,0.3800); rgb(0.1000)=(0.0743,0.3456,0.5616); rgb(0.2000)=(0.1918,0.4980,0.7060); rgb(0.3000)=(0.3890,0.6708,0.8226); rgb(0.4000)=(0.7552,0.8682,0.9228); rgb(0.5000)=(0.9690,0.9689,0.9689); rgb(0.6000)=(0.9425,0.8044,0.7614); rgb(0.7000)=(0.8767,0.4910,0.4020); rgb(0.8000)=(0.7869,0.2866,0.2470); rgb(0.9000)=(0.6186,0.1239,0.1568); rgb(1.0000)=(0.4040,0.0000,0.1220)}
]
\addplot [color=black, line width=1.0pt]
  table[row sep=crcr]{%
0.001	0.00701261889178508\\
0.00126087240768068	0.00893071582945259\\
0.00158979922845048	0.0113458961663005\\
0.00200453398090524	0.0143758503673875\\
0.00252746158678173	0.0181569614276944\\
0.00318680657624591	0.0228383960304555\\
0.00401815648060381	0.0285669467217202\\
0.00506638263613665	0.035457566574311\\
0.00638806207265721	0.0435480501977126\\
0.00805453120596493	0.0527485338310358\\
0.0101557361544042	0.0628191233518346\\
0.0128050874967733	0.0734229450480188\\
0.0161455815026184	0.0842657470495744\\
0.0203575182226111	0.0952437202997453\\
0.0256682330157469	0.106474738658859\\
0.0323643667634736	0.118177151812545\\
0.0408073370441215	0.13049872637852\\
0.0514528453098585	0.143418064570188\\
0.0648754729478629	0.156732457811551\\
0.0817996937751947	0.170062793231603\\
0.103138976837872	0.182830670836682\\
0.13004509005129	0.194218277033943\\
0.163970265800021	0.203144556569132\\
0.206745583827313	0.208302630800687\\
0.260679802057692	0.208316559461183\\
0.328683969654205	0.202060445588963\\
0.414428548183942	0.189090927891049\\
0.522541521360296	0.169985253576203\\
0.658858186150681	0.146320962809481\\
0.830736107491936	0.12026837854189\\
1.02040816326531	0.0969961733132146\\
1.12244897959184	0.0866030213436895\\
1.22448979591837	0.0774968042501197\\
1.3265306122449	0.0695260735287157\\
1.42857142857143	0.0625508663386185\\
1.53061224489796	0.0564442042306584\\
1.63265306122449	0.0510923417528904\\
1.73469387755102	0.0463944858421878\\
1.83673469387755	0.0422621846394841\\
1.93877551020408	0.0386184658329438\\
};
\addlegendentry{$\mu_g$}

\addplot [color=white!60!black, line width=1.0pt]
  table[row sep=crcr]{%
0.001	0.00719253574043249\\
0.00126087240768068	0.00913185935565453\\
0.00158979922845048	0.0115696554502736\\
0.00200453398090524	0.0146236574203191\\
0.00252746158678173	0.0184302380442533\\
0.00318680657624591	0.0231385549879256\\
0.00401815648060381	0.028895434973864\\
0.00506638263613665	0.0358161038178709\\
0.00638806207265721	0.0439391007243471\\
0.00805453120596493	0.053176097803891\\
0.0101557361544042	0.0632898463867165\\
0.0128050874967733	0.0739471106799277\\
0.0161455815026184	0.0848573958647341\\
0.0203575182226111	0.0959189403658989\\
0.0256682330157469	0.107248222967509\\
0.0323643667634736	0.119059560514533\\
0.0408073370441215	0.131500303699772\\
0.0514528453098585	0.144566056434578\\
0.0648754729478629	0.158112922589081\\
0.0817996937751947	0.171903922832785\\
0.103138976837872	0.185650995875211\\
0.13004509005129	0.199057355166015\\
0.163970265800021	0.211857806467695\\
0.206745583827313	0.223820223651911\\
0.260679802057692	0.234662609021246\\
0.328683969654205	0.243903028080425\\
0.414428548183942	0.250748600770025\\
0.522541521360296	0.254104108020628\\
0.658858186150681	0.252660577381455\\
0.830736107491936	0.245002190777772\\
1.02040816326531	0.231910067915691\\
1.12244897959184	0.223616513133382\\
1.22448979591837	0.21477253503475\\
1.3265306122449	0.205574791575123\\
1.42857142857143	0.19618727847568\\
1.53061224489796	0.186747659149366\\
1.63265306122449	0.177371256167264\\
1.73469387755102	0.1681536554919\\
1.83673469387755	0.159172553011016\\
1.93877551020408	0.150489258836154\\
};
\addlegendentry{$\mu_h$}

\addplot [color=mycolor1, dashed, line width=1.4pt, forget plot]
  table[row sep=crcr]{%
0.001	0.00691440978517865\\
0.0011492187010037	0.00800901486152831\\
0.00126087240768068	0.00882808420561014\\
0.0014490181504862	0.0102073551961804\\
0.00158979922845048	0.0112379671603847\\
0.00182702700417654	0.0129704098215086\\
0.00200453398090524	0.0142622233944057\\
0.00230364793765369	0.0164280296807092\\
0.00252746158678173	0.0180379355800905\\
0.00290460612159805	0.0207263949843451\\
0.00318680657624591	0.0227153959157676\\
0.00366233771390337	0.0260173034587817\\
0.00401815648060381	0.0284430700576134\\
0.00461774057106911	0.0324351741165621\\
0.00506638263613665	0.0353383391120558\\
0.00582238167188866	0.0400576891357577\\
0.00638806207265721	0.0434419202815761\\
0.00734128039707012	0.0488541772876411\\
0.00805453120596493	0.0526664134734893\\
0.00925641788971278	0.0586449327468994\\
0.0101557361544042	0.0627721139751994\\
0.0116711619111007	0.069117128777596\\
0.0128050874967733	0.0734181550022168\\
0.0147158460192806	0.0799327693313498\\
0.0161455815026184	0.0843024251189728\\
0.0185548042013884	0.0908905222567032\\
0.0203575182226111	0.0953128290877747\\
0.0233952406474482	0.102020218875866\\
0.0256682330157469	0.106562622189041\\
0.029498413403417	0.113519304090278\\
0.0323643667634736	0.118269913692396\\
0.0371937355307264	0.125582717728126\\
0.0408073370441215	0.130584297332856\\
0.0468965548692655	0.138264382225936\\
0.0514528453098585	0.143487810231695\\
0.0591305720499399	0.151439137265285\\
0.0648754729478629	0.156787495752287\\
0.0745561067481437	0.164816972775873\\
0.0817996937751947	0.1701290054103\\
0.0940057378228297	0.177942083526521\\
0.103138976837872	0.182982961446435\\
0.11852924098447	0.190162917981382\\
0.13004509005129	0.194607625221631\\
0.149450249460652	0.200589012030576\\
0.163970265800021	0.204004877168509\\
0.188437695865931	0.208050322858798\\
0.206745583827313	0.209885831800706\\
0.237595891284276	0.211081906872747\\
0.260679802057692	0.210678064560862\\
0.299578103498643	0.207988244804483\\
0.328683969654205	0.204650091234236\\
0.377729764646745	0.197104131555417\\
0.414428548183942	0.190300814353463\\
0.476269037802799	0.17743569803801\\
0.522541521360296	0.167194972312223\\
0.600514488398176	0.149758856011865\\
0.658858186150681	0.137137239572555\\
0.75717214883374	0.117555304476446\\
0.830736107491936	0.104603849586758\\
0.954697470328753	0.0861670027755723\\
1.02040816326531	0.0779843254384878\\
1.08163265306122	0.0712331940890663\\
1.12244897959184	0.0671543310054605\\
1.18367346938776	0.0616038736890199\\
1.22448979591837	0.0582464450703885\\
1.28571428571429	0.05366824864503\\
1.3265306122449	0.0508915283646499\\
1.38775510204082	0.0470934420185523\\
1.42857142857143	0.0447819917271014\\
1.48979591836735	0.0416090053232671\\
1.53061224489796	0.0396707929972464\\
1.59183673469388	0.0370002402886368\\
1.63265306122449	0.0353627848225483\\
1.69387755102041	0.033098290249629\\
1.73469387755102	0.0317046969474778\\
1.79591836734694	0.0297705791869807\\
1.83673469387755	0.0285761131407958\\
1.89795918367347	0.0269127477964376\\
1.93877551020408	0.0258820773083085\\
2	0.0244422384965295\\
};
\end{axis}
\end{tikzpicture}%
\fi
    \end{minipage}
    \begin{minipage}[t]{0.32\textwidth}
        \centering
        \ifloadfig
%
\definecolor{mycolor1}{rgb}{0.00000,0.45000,0.74000}%
\begin{tikzpicture}[trim axis left,trim axis right]

\begin{axis}[%
width=11.255in,
height=7.131in,
at={(1.888in,0.962in)},
scale only axis,
separate axis lines,
every outer x axis line/.append style={black},
every x tick label/.append style={font=\color{black}},
every x tick/.append style={black},
xmin=0,
xmax=2,
xlabel={$k$},
every outer y axis line/.append style={black},
every y tick label/.append style={font=\color{black}},
every y tick/.append style={black},
ymin=0,
ymax=0.25,
ylabel={$\omega_r$},
axis background/.style={fill=white},
title style={font=\bfseries},
title={$A_{\mu} = -0.50$},
legend style={legend cell align=left, align=left},
clip=true,
clip mode=individual,
axis lines=box,
xtick pos=bottom,
ytick pos=left,
tick align=inside,
major tick length=2pt,
width=0.7\linewidth,
height=0.65\linewidth,
every axis/.append style={font=\fontsize{9}{9}\selectfont},xlabel style={font=\fontsize{9}{9}\selectfont},title style={font=\fontsize{9}{9}\selectfont},
legend style={font=\fontsize{8}{9}\selectfont,inner sep=1pt,row sep=1pt,column sep=2pt,nodes={scale=1.00},draw=none},legend image post style={xscale=0.35},
legend columns=1,
scaled x ticks=false,
tick label style={/pgf/number format/fixed,/pgf/number format/precision=2},
every x tick label/.append style={font=\fontsize{9}{9}\selectfont\color{black}},
every y tick label/.append style={font=\fontsize{9}{9}\selectfont\color{black}},
,,
colormap={sigmamap}{rgb(0.0000)=(0.0200,0.1880,0.3800); rgb(0.1000)=(0.0743,0.3456,0.5616); rgb(0.2000)=(0.1918,0.4980,0.7060); rgb(0.3000)=(0.3890,0.6708,0.8226); rgb(0.4000)=(0.7552,0.8682,0.9228); rgb(0.5000)=(0.9690,0.9689,0.9689); rgb(0.6000)=(0.9425,0.8044,0.7614); rgb(0.7000)=(0.8767,0.4910,0.4020); rgb(0.8000)=(0.7869,0.2866,0.2470); rgb(0.9000)=(0.6186,0.1239,0.1568); rgb(1.0000)=(0.4040,0.0000,0.1220)}
]
\addplot [color=black, line width=1.0pt]
  table[row sep=crcr]{%
0.001	0.00692538802510537\\
0.00126087240768068	0.00884033120664907\\
0.00158979922845048	0.0112517493105919\\
0.00200453398090524	0.0142775517688799\\
0.00252746158678173	0.0180545457825412\\
0.00318680657624591	0.0227326156618203\\
0.00401815648060381	0.0284596452482577\\
0.00506638263613665	0.0353521728623102\\
0.00638806207265721	0.0434499088059428\\
0.00805453120596493	0.0526645843107681\\
0.0101557361544042	0.0627563604983773\\
0.0128050874967733	0.0733856676411022\\
0.0161455815026184	0.0842529886775008\\
0.0203575182226111	0.0952490346568986\\
0.0256682330157469	0.1064887730677\\
0.0323643667634736	0.118190790171408\\
0.0408073370441215	0.130504504338875\\
0.0514528453098585	0.143410572161766\\
0.0648754729478629	0.156710611031211\\
0.0817996937751947	0.170036094645842\\
0.103138976837872	0.182829333127778\\
0.13004509005129	0.194303185442504\\
0.163970265800021	0.203404476418371\\
0.206745583827313	0.20881188666175\\
0.260679802057692	0.209015261255599\\
0.328683969654205	0.202539961015791\\
0.414428548183942	0.188365248105596\\
0.522541521360296	0.166504588037498\\
0.658858186150681	0.138572173714692\\
0.830736107491936	0.107951710099737\\
1.02040816326531	0.0820160051639515\\
1.12244897959184	0.0711584235304426\\
1.22448979591837	0.0620847185718347\\
1.3265306122449	0.0544955644065098\\
1.42857142857143	0.0481262292856419\\
1.53061224489796	0.0427540418069229\\
1.63265306122449	0.0381966733133376\\
1.73469387755102	0.0343068496231797\\
1.83673469387755	0.030966290289278\\
1.93877551020408	0.0280800857306415\\
};
\addlegendentry{$\mu_g$}

\addplot [color=white!60!black, line width=1.0pt]
  table[row sep=crcr]{%
0.001	0.00697459881661048\\
0.00126087240768068	0.0088960400413729\\
0.00158979922845048	0.0113146495996273\\
0.00200453398090524	0.0143483850677399\\
0.00252746158678173	0.0181340495074154\\
0.00318680657624591	0.0228215236222856\\
0.00401815648060381	0.0285586999995713\\
0.00506638263613665	0.0354621857069769\\
0.00638806207265721	0.0435719060102183\\
0.00805453120596493	0.0528000548306447\\
0.0101557361544042	0.0629076378247046\\
0.0128050874967733	0.0735562446663991\\
0.0161455815026184	0.0844475501882279\\
0.0203575182226111	0.0954727670657747\\
0.0256682330157469	0.106745881183408\\
0.0323643667634736	0.118482816539896\\
0.0408073370441215	0.130829963157078\\
0.0514528453098585	0.143767938332018\\
0.0648754729478629	0.157107542354773\\
0.0817996937751947	0.170510028497631\\
0.103138976837872	0.183487862842876\\
0.13004509005129	0.195394574234892\\
0.163970265800021	0.205424969786464\\
0.206745583827313	0.212638034531234\\
0.260679802057692	0.216001767286015\\
0.328683969654205	0.214441348479612\\
0.414428548183942	0.206875727610512\\
0.522541521360296	0.19232076757657\\
0.658858186150681	0.170307192774408\\
0.830736107491936	0.141807322190198\\
1.02040816326531	0.113690258091147\\
1.12244897959184	0.100741178293694\\
1.22448979591837	0.0893705836477608\\
1.3265306122449	0.079479000807844\\
1.42857142857143	0.0709156843926778\\
1.53061224489796	0.0635141689123664\\
1.63265306122449	0.0571127256430587\\
1.73469387755102	0.0515643788811756\\
1.83673469387755	0.0467405603875836\\
1.93877551020408	0.0425312924886558\\
};
\addlegendentry{$\mu_h$}

\addplot [color=mycolor1, dashed, line width=1.4pt, forget plot]
  table[row sep=crcr]{%
0.001	0.00691440978517865\\
0.0011492187010037	0.00800901486152831\\
0.00126087240768068	0.00882808420561014\\
0.0014490181504862	0.0102073551961804\\
0.00158979922845048	0.0112379671603847\\
0.00182702700417654	0.0129704098215086\\
0.00200453398090524	0.0142622233944057\\
0.00230364793765369	0.0164280296807092\\
0.00252746158678173	0.0180379355800905\\
0.00290460612159805	0.0207263949843451\\
0.00318680657624591	0.0227153959157676\\
0.00366233771390337	0.0260173034587817\\
0.00401815648060381	0.0284430700576134\\
0.00461774057106911	0.0324351741165621\\
0.00506638263613665	0.0353383391120558\\
0.00582238167188866	0.0400576891357577\\
0.00638806207265721	0.0434419202815761\\
0.00734128039707012	0.0488541772876411\\
0.00805453120596493	0.0526664134734893\\
0.00925641788971278	0.0586449327468994\\
0.0101557361544042	0.0627721139751994\\
0.0116711619111007	0.069117128777596\\
0.0128050874967733	0.0734181550022168\\
0.0147158460192806	0.0799327693313498\\
0.0161455815026184	0.0843024251189728\\
0.0185548042013884	0.0908905222567032\\
0.0203575182226111	0.0953128290877747\\
0.0233952406474482	0.102020218875866\\
0.0256682330157469	0.106562622189041\\
0.029498413403417	0.113519304090278\\
0.0323643667634736	0.118269913692396\\
0.0371937355307264	0.125582717728126\\
0.0408073370441215	0.130584297332856\\
0.0468965548692655	0.138264382225936\\
0.0514528453098585	0.143487810231695\\
0.0591305720499399	0.151439137265285\\
0.0648754729478629	0.156787495752287\\
0.0745561067481437	0.164816972775873\\
0.0817996937751947	0.1701290054103\\
0.0940057378228297	0.177942083526521\\
0.103138976837872	0.182982961446435\\
0.11852924098447	0.190162917981382\\
0.13004509005129	0.194607625221631\\
0.149450249460652	0.200589012030576\\
0.163970265800021	0.204004877168509\\
0.188437695865931	0.208050322858798\\
0.206745583827313	0.209885831800706\\
0.237595891284276	0.211081906872747\\
0.260679802057692	0.210678064560862\\
0.299578103498643	0.207988244804483\\
0.328683969654205	0.204650091234236\\
0.377729764646745	0.197104131555417\\
0.414428548183942	0.190300814353463\\
0.476269037802799	0.17743569803801\\
0.522541521360296	0.167194972312223\\
0.600514488398176	0.149758856011865\\
0.658858186150681	0.137137239572555\\
0.75717214883374	0.117555304476446\\
0.830736107491936	0.104603849586758\\
0.954697470328753	0.0861670027755723\\
1.02040816326531	0.0779843254384878\\
1.08163265306122	0.0712331940890663\\
1.12244897959184	0.0671543310054605\\
1.18367346938776	0.0616038736890199\\
1.22448979591837	0.0582464450703885\\
1.28571428571429	0.05366824864503\\
1.3265306122449	0.0508915283646499\\
1.38775510204082	0.0470934420185523\\
1.42857142857143	0.0447819917271014\\
1.48979591836735	0.0416090053232671\\
1.53061224489796	0.0396707929972464\\
1.59183673469388	0.0370002402886368\\
1.63265306122449	0.0353627848225483\\
1.69387755102041	0.033098290249629\\
1.73469387755102	0.0317046969474778\\
1.79591836734694	0.0297705791869807\\
1.83673469387755	0.0285761131407958\\
1.89795918367347	0.0269127477964376\\
1.93877551020408	0.0258820773083085\\
2	0.0244422384965295\\
};
\end{axis}
\end{tikzpicture}%
\fi
    \end{minipage}
    \begin{minipage}[t]{0.32\textwidth}
        \centering
        \ifloadfig
%
\definecolor{mycolor1}{rgb}{0.00000,0.45000,0.74000}%
\begin{tikzpicture}[trim axis left,trim axis right]

\begin{axis}[%
width=11.255in,
height=7.131in,
at={(1.888in,0.962in)},
scale only axis,
separate axis lines,
every outer x axis line/.append style={black},
every x tick label/.append style={font=\color{black}},
every x tick/.append style={black},
xmin=0,
xmax=2,
xlabel={$k$},
every outer y axis line/.append style={black},
every y tick label/.append style={font=\color{black}},
every y tick/.append style={black},
ymin=0,
ymax=0.25,
ylabel={$\omega_r$},
axis background/.style={fill=white},
title style={font=\bfseries},
title={$A_{\mu} = -0.05$},
legend style={legend cell align=left, align=left},
clip=true,
clip mode=individual,
axis lines=box,
xtick pos=bottom,
ytick pos=left,
tick align=inside,
major tick length=2pt,
width=0.7\linewidth,
height=0.65\linewidth,
every axis/.append style={font=\fontsize{9}{9}\selectfont},xlabel style={font=\fontsize{9}{9}\selectfont},title style={font=\fontsize{9}{9}\selectfont},
legend style={font=\fontsize{8}{9}\selectfont,inner sep=1pt,row sep=1pt,column sep=2pt,nodes={scale=1.00},draw=none},legend image post style={xscale=0.35},
legend columns=1,
scaled x ticks=false,
tick label style={/pgf/number format/fixed,/pgf/number format/precision=2},
every x tick label/.append style={font=\fontsize{9}{9}\selectfont\color{black}},
every y tick label/.append style={font=\fontsize{9}{9}\selectfont\color{black}},
,,
colormap={sigmamap}{rgb(0.0000)=(0.0200,0.1880,0.3800); rgb(0.1000)=(0.0743,0.3456,0.5616); rgb(0.2000)=(0.1918,0.4980,0.7060); rgb(0.3000)=(0.3890,0.6708,0.8226); rgb(0.4000)=(0.7552,0.8682,0.9228); rgb(0.5000)=(0.9690,0.9689,0.9689); rgb(0.6000)=(0.9425,0.8044,0.7614); rgb(0.7000)=(0.8767,0.4910,0.4020); rgb(0.8000)=(0.7869,0.2866,0.2470); rgb(0.9000)=(0.6186,0.1239,0.1568); rgb(1.0000)=(0.4040,0.0000,0.1220)}
]
\addplot [color=black, line width=1.0pt]
  table[row sep=crcr]{%
0.001	0.00691245929977286\\
0.00126087240768068	0.00882617754303982\\
0.00158979922845048	0.0112361227130502\\
0.00200453398090524	0.0142604228833157\\
0.00252746158678173	0.0180361444111837\\
0.00318680657624591	0.022713552366474\\
0.00401815648060381	0.0284410877069624\\
0.00506638263613665	0.0353360528998079\\
0.00638806207265721	0.0434391105531887\\
0.00805453120596493	0.0526627894967124\\
0.0101557361544042	0.0627673551512581\\
0.0128050874967733	0.0734120194556452\\
0.0161455815026184	0.0842948179850988\\
0.0203575182226111	0.0953038245496432\\
0.0256682330157469	0.106552413461719\\
0.0323643667634736	0.11825875793583\\
0.0408073370441215	0.130572478482086\\
0.0514528453098585	0.143475494515869\\
0.0648754729478629	0.156774290950828\\
0.0817996937751947	0.170113056185958\\
0.103138976837872	0.18295946572368\\
0.13004509005129	0.194566888268356\\
0.163970265800021	0.203930648225718\\
0.206745583827313	0.209755902984828\\
0.260679802057692	0.210470654031652\\
0.328683969654205	0.204359442506121\\
0.414428548183942	0.189958625251257\\
0.522541521360296	0.166876500555818\\
0.658858186150681	0.136921636856204\\
0.830736107491936	0.104510091446463\\
1.02040816326531	0.0779629481030045\\
1.12244897959184	0.0671519584364876\\
1.22448979591837	0.0582549837135277\\
1.3265306122449	0.0509059849366698\\
1.42857142857143	0.0447993692173438\\
1.53061224489796	0.0396893261339868\\
1.63265306122449	0.035381453516258\\
1.73469387755102	0.0317229280158812\\
1.83673469387755	0.0285936002412808\\
1.93877551020408	0.0258986727050873\\
};
\addlegendentry{$\mu_g$}

\addplot [color=white!60!black, line width=1.0pt]
  table[row sep=crcr]{%
0.001	0.00691485572678654\\
0.00126087240768068	0.00882891249273605\\
0.00158979922845048	0.0112392384925237\\
0.00200453398090524	0.0142639646100638\\
0.00252746158678173	0.0180401554812489\\
0.00318680657624591	0.0227180821996053\\
0.00401815648060381	0.0284461832275534\\
0.00506638263613665	0.0353417609326923\\
0.00638806207265721	0.0434454915941407\\
0.00805453120596493	0.0526699318182965\\
0.0101557361544042	0.0627753907564885\\
0.0128050874967733	0.0734211406179281\\
0.0161455815026184	0.0843052867855661\\
0.0203575182226111	0.0953159344170849\\
0.0256682330157469	0.106566412877181\\
0.0323643667634736	0.118274756532466\\
0.0408073370441215	0.13059042522726\\
0.0514528453098585	0.143495330003509\\
0.0648754729478629	0.15679642116996\\
0.0817996937751947	0.170139410019558\\
0.103138976837872	0.182995576970011\\
0.13004509005129	0.194625551794063\\
0.163970265800021	0.204037587643353\\
0.206745583827313	0.209957121955149\\
0.260679802057692	0.210838892586754\\
0.328683969654205	0.204992215104451\\
0.414428548183942	0.190952982916607\\
0.522541521360296	0.168265592042633\\
0.658858186150681	0.138594305997601\\
0.830736107491936	0.106211238638919\\
1.02040816326531	0.0794728367725893\\
1.12244897959184	0.0685309391275541\\
1.22448979591837	0.0595046250375153\\
1.3265306122449	0.0520348415924865\\
1.42857142857143	0.0458187556104645\\
1.53061224489796	0.0406110657121461\\
1.63265306122449	0.0362167888018011\\
1.73469387755102	0.0324820644144956\\
1.83673469387755	0.0292855767250829\\
1.93877551020408	0.0265313905399028\\
};
\addlegendentry{$\mu_h$}

\addplot [color=mycolor1, dashed, line width=1.4pt, forget plot]
  table[row sep=crcr]{%
0.001	0.00691440978517865\\
0.0011492187010037	0.00800901486152831\\
0.00126087240768068	0.00882808420561014\\
0.0014490181504862	0.0102073551961804\\
0.00158979922845048	0.0112379671603847\\
0.00182702700417654	0.0129704098215086\\
0.00200453398090524	0.0142622233944057\\
0.00230364793765369	0.0164280296807092\\
0.00252746158678173	0.0180379355800905\\
0.00290460612159805	0.0207263949843451\\
0.00318680657624591	0.0227153959157676\\
0.00366233771390337	0.0260173034587817\\
0.00401815648060381	0.0284430700576134\\
0.00461774057106911	0.0324351741165621\\
0.00506638263613665	0.0353383391120558\\
0.00582238167188866	0.0400576891357577\\
0.00638806207265721	0.0434419202815761\\
0.00734128039707012	0.0488541772876411\\
0.00805453120596493	0.0526664134734893\\
0.00925641788971278	0.0586449327468994\\
0.0101557361544042	0.0627721139751994\\
0.0116711619111007	0.069117128777596\\
0.0128050874967733	0.0734181550022168\\
0.0147158460192806	0.0799327693313498\\
0.0161455815026184	0.0843024251189728\\
0.0185548042013884	0.0908905222567032\\
0.0203575182226111	0.0953128290877747\\
0.0233952406474482	0.102020218875866\\
0.0256682330157469	0.106562622189041\\
0.029498413403417	0.113519304090278\\
0.0323643667634736	0.118269913692396\\
0.0371937355307264	0.125582717728126\\
0.0408073370441215	0.130584297332856\\
0.0468965548692655	0.138264382225936\\
0.0514528453098585	0.143487810231695\\
0.0591305720499399	0.151439137265285\\
0.0648754729478629	0.156787495752287\\
0.0745561067481437	0.164816972775873\\
0.0817996937751947	0.1701290054103\\
0.0940057378228297	0.177942083526521\\
0.103138976837872	0.182982961446435\\
0.11852924098447	0.190162917981382\\
0.13004509005129	0.194607625221631\\
0.149450249460652	0.200589012030576\\
0.163970265800021	0.204004877168509\\
0.188437695865931	0.208050322858798\\
0.206745583827313	0.209885831800706\\
0.237595891284276	0.211081906872747\\
0.260679802057692	0.210678064560862\\
0.299578103498643	0.207988244804483\\
0.328683969654205	0.204650091234236\\
0.377729764646745	0.197104131555417\\
0.414428548183942	0.190300814353463\\
0.476269037802799	0.17743569803801\\
0.522541521360296	0.167194972312223\\
0.600514488398176	0.149758856011865\\
0.658858186150681	0.137137239572555\\
0.75717214883374	0.117555304476446\\
0.830736107491936	0.104603849586758\\
0.954697470328753	0.0861670027755723\\
1.02040816326531	0.0779843254384878\\
1.08163265306122	0.0712331940890663\\
1.12244897959184	0.0671543310054605\\
1.18367346938776	0.0616038736890199\\
1.22448979591837	0.0582464450703885\\
1.28571428571429	0.05366824864503\\
1.3265306122449	0.0508915283646499\\
1.38775510204082	0.0470934420185523\\
1.42857142857143	0.0447819917271014\\
1.48979591836735	0.0416090053232671\\
1.53061224489796	0.0396707929972464\\
1.59183673469388	0.0370002402886368\\
1.63265306122449	0.0353627848225483\\
1.69387755102041	0.033098290249629\\
1.73469387755102	0.0317046969474778\\
1.79591836734694	0.0297705791869807\\
1.83673469387755	0.0285761131407958\\
1.89795918367347	0.0269127477964376\\
1.93877551020408	0.0258820773083085\\
2	0.0244422384965295\\
};
\end{axis}
\end{tikzpicture}%
\fi
    \end{minipage}
    \caption{Sensitivity of the dispersion relation to the viscosity mixing law for \(\Atw = 0.5\), \(\delta = 5\), and \(\Sch \to \infty\) for negative viscosity ratios, \(\Amu < 0\). The \(\Amu = 0\) reference is denoted by the dashed blue line \((\dashlegend)\).}
    \label{fig:visc_law_comparison_neg}
\end{figure}
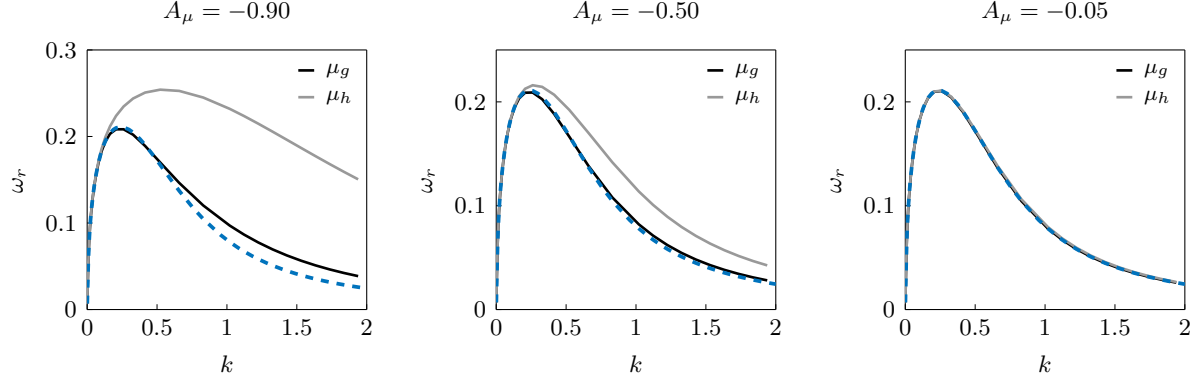
Two composition-dependent mixing laws for the dynamic viscosity are considered. The first is the linear law in mole fraction,
\begin{equation}
    \label{eq:mu_g}
    \mu_g = \mu_1 X_1 + \mu_2 X_2,
\end{equation}
as used by~\citet{morgan-likhachev-jacobs-2016} and equivalent to~\eqref{eq:graham_mixing_law} employed in the present work. Combining~\eqref{eq:Y_from_X} with~\eqref{eq:mu_g} gives in fact the corresponding kinematic viscosity,
\begin{equation}
    \label{eq:nu_g}
    \nu_g = \nu_1 Y_1 + \nu_2 Y_2.
\end{equation}
The second, following~\citet{baltzer-livescu-2020}, is the harmonic law; proceeding similarly,
\begin{equation}
    \label{eq:nu_h}
    \frac{1}{\mu_h} = \frac{Y_1}{\mu_1} + \frac{Y_2}{\mu_2}
    \quad\implies\quad
    \frac{1}{\nu_h} = \frac{X_1}{\nu_1} + \frac{X_2}{\nu_2}.
\end{equation}
To isolate the effect of viscosity stratification in the absence of mass diffusion, the limit \(\Sch \to \infty\) is considered.
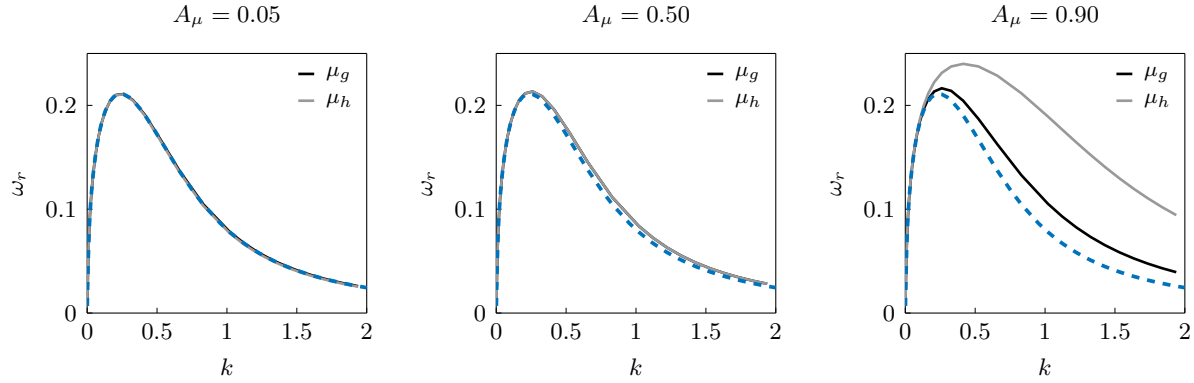
\begin{figure}
    \centering
    \begin{minipage}[t]{0.32\textwidth}
        \centering
        \ifloadfig
%
\definecolor{mycolor1}{rgb}{0.00000,0.45000,0.74000}%
\begin{tikzpicture}[trim axis left,trim axis right]

\begin{axis}[%
width=11.255in,
height=7.131in,
at={(1.888in,0.962in)},
scale only axis,
separate axis lines,
every outer x axis line/.append style={black},
every x tick label/.append style={font=\color{black}},
every x tick/.append style={black},
xmin=0,
xmax=2,
xlabel={$k$},
every outer y axis line/.append style={black},
every y tick label/.append style={font=\color{black}},
every y tick/.append style={black},
ymin=0,
ymax=0.25,
ylabel={$\omega_r$},
axis background/.style={fill=white},
title style={font=\bfseries},
title={$A_{\mu} = 0.05$},
legend style={legend cell align=left, align=left},
clip=true,
clip mode=individual,
axis lines=box,
xtick pos=bottom,
ytick pos=left,
tick align=inside,
major tick length=2pt,
width=0.7\linewidth,
height=0.65\linewidth,
every axis/.append style={font=\fontsize{9}{9}\selectfont},xlabel style={font=\fontsize{9}{9}\selectfont},title style={font=\fontsize{9}{9}\selectfont},
legend style={font=\fontsize{8}{9}\selectfont,inner sep=1pt,row sep=1pt,column sep=2pt,nodes={scale=1.00},draw=none},legend image post style={xscale=0.35},
legend columns=1,
scaled x ticks=false,
tick label style={/pgf/number format/fixed,/pgf/number format/precision=2},
every x tick label/.append style={font=\fontsize{9}{9}\selectfont\color{black}},
every y tick label/.append style={font=\fontsize{9}{9}\selectfont\color{black}},
,,
colormap={sigmamap}{rgb(0.0000)=(0.0200,0.1880,0.3800); rgb(0.1000)=(0.0743,0.3456,0.5616); rgb(0.2000)=(0.1918,0.4980,0.7060); rgb(0.3000)=(0.3890,0.6708,0.8226); rgb(0.4000)=(0.7552,0.8682,0.9228); rgb(0.5000)=(0.9690,0.9689,0.9689); rgb(0.6000)=(0.9425,0.8044,0.7614); rgb(0.7000)=(0.8767,0.4910,0.4020); rgb(0.8000)=(0.7869,0.2866,0.2470); rgb(0.9000)=(0.6186,0.1239,0.1568); rgb(1.0000)=(0.4040,0.0000,0.1220)}
]
\addplot [color=black, line width=1.0pt]
  table[row sep=crcr]{%
0.001	0.00691698167210052\\
0.00126087240768068	0.00883061768723799\\
0.00158979922845048	0.0112404744973955\\
0.00200453398090524	0.0142647100412068\\
0.00252746158678173	0.0180404344656826\\
0.00318680657624591	0.02271796975354\\
0.00401815648060381	0.0284458122924713\\
0.00506638263613665	0.0353413884449138\\
0.00638806207265721	0.0434454848505304\\
0.00805453120596493	0.0526707666868929\\
0.0101557361544042	0.0627775531114957\\
0.0128050874967733	0.0734249245354915\\
0.0161455815026184	0.0843106369660228\\
0.0203575182226111	0.0953224515768181\\
0.0256682330157469	0.106573508114171\\
0.0323643667634736	0.118281852954073\\
0.0408073370441215	0.130597045510914\\
0.0514528453098585	0.14350123780146\\
0.0648754729478629	0.156802033920128\\
0.0817996937751947	0.170146560518381\\
0.103138976837872	0.183008412484938\\
0.13004509005129	0.194650824334854\\
0.163970265800021	0.204082477564592\\
0.206745583827313	0.210021088825539\\
0.260679802057692	0.210895220483226\\
0.328683969654205	0.204959761895659\\
0.414428548183942	0.190678704675352\\
0.522541521360296	0.167572800917074\\
0.658858186150681	0.137435370017462\\
0.830736107491936	0.104790617704129\\
1.02040816326531	0.0780930587412806\\
1.12244897959184	0.0672378433477545\\
1.22448979591837	0.0583122700607214\\
1.3265306122449	0.0509447552841818\\
1.42857142857143	0.0448260398558839\\
1.53061224489796	0.0397079829041986\\
1.63265306122449	0.0353947203826978\\
1.73469387755102	0.0317325114201431\\
1.83673469387755	0.0286006261636588\\
1.93877551020408	0.0259038955237593\\
};
\addlegendentry{$\mu_g$}

\addplot [color=white!60!black, line width=1.0pt]
  table[row sep=crcr]{%
0.001	0.00691502614670306\\
0.00126087240768068	0.00882838981034593\\
0.00158979922845048	0.0112379279482769\\
0.00200453398090524	0.0142618073103968\\
0.00252746158678173	0.0180371450597729\\
0.00318680657624591	0.0227142450696891\\
0.00401815648060381	0.0284416172026866\\
0.00506638263613665	0.0353366838838215\\
0.00638806207265721	0.0434402196037713\\
0.00805453120596493	0.0526648652559621\\
0.0101557361544042	0.0627709063064018\\
0.0128050874967733	0.0734173725048628\\
0.0161455815026184	0.0843019597671274\\
0.0203575182226111	0.095312399967878\\
0.0256682330157469	0.106561873069492\\
0.0323643667634736	0.118268532715107\\
0.0408073370441215	0.130582062893653\\
0.0514528453098585	0.143484611124145\\
0.0648754729478629	0.156783378786565\\
0.0817996937751947	0.170124206330895\\
0.103138976837872	0.182977674409893\\
0.13004509005129	0.194600974163991\\
0.163970265800021	0.20399204685925\\
0.206745583827313	0.209851747670871\\
0.260679802057692	0.210586287594162\\
0.328683969654205	0.204429504132062\\
0.414428548183942	0.189845060470922\\
0.522541521360296	0.166407234957917\\
0.658858186150681	0.136032580163787\\
0.830736107491936	0.103367379301241\\
1.02040816326531	0.0768328071837919\\
1.12244897959184	0.0660877126176811\\
1.22448979591837	0.057270446281214\\
1.3265306122449	0.0500037835387407\\
1.42857142857143	0.0439763069355091\\
1.53061224489796	0.0389395444489248\\
1.63265306122449	0.0346981714785992\\
1.73469387755102	0.0310993418092868\\
1.83673469387755	0.0280233128514564\\
1.93877551020408	0.025375870147191\\
};
\addlegendentry{$\mu_h$}

\addplot [color=mycolor1, dashed, line width=1.4pt, forget plot]
  table[row sep=crcr]{%
0.001	0.00691440978517865\\
0.0011492187010037	0.00800901486152831\\
0.00126087240768068	0.00882808420561014\\
0.0014490181504862	0.0102073551961804\\
0.00158979922845048	0.0112379671603847\\
0.00182702700417654	0.0129704098215086\\
0.00200453398090524	0.0142622233944057\\
0.00230364793765369	0.0164280296807092\\
0.00252746158678173	0.0180379355800905\\
0.00290460612159805	0.0207263949843451\\
0.00318680657624591	0.0227153959157676\\
0.00366233771390337	0.0260173034587817\\
0.00401815648060381	0.0284430700576134\\
0.00461774057106911	0.0324351741165621\\
0.00506638263613665	0.0353383391120558\\
0.00582238167188866	0.0400576891357577\\
0.00638806207265721	0.0434419202815761\\
0.00734128039707012	0.0488541772876411\\
0.00805453120596493	0.0526664134734893\\
0.00925641788971278	0.0586449327468994\\
0.0101557361544042	0.0627721139751994\\
0.0116711619111007	0.069117128777596\\
0.0128050874967733	0.0734181550022168\\
0.0147158460192806	0.0799327693313498\\
0.0161455815026184	0.0843024251189728\\
0.0185548042013884	0.0908905222567032\\
0.0203575182226111	0.0953128290877747\\
0.0233952406474482	0.102020218875866\\
0.0256682330157469	0.106562622189041\\
0.029498413403417	0.113519304090278\\
0.0323643667634736	0.118269913692396\\
0.0371937355307264	0.125582717728126\\
0.0408073370441215	0.130584297332856\\
0.0468965548692655	0.138264382225936\\
0.0514528453098585	0.143487810231695\\
0.0591305720499399	0.151439137265285\\
0.0648754729478629	0.156787495752287\\
0.0745561067481437	0.164816972775873\\
0.0817996937751947	0.1701290054103\\
0.0940057378228297	0.177942083526521\\
0.103138976837872	0.182982961446435\\
0.11852924098447	0.190162917981382\\
0.13004509005129	0.194607625221631\\
0.149450249460652	0.200589012030576\\
0.163970265800021	0.204004877168509\\
0.188437695865931	0.208050322858798\\
0.206745583827313	0.209885831800706\\
0.237595891284276	0.211081906872747\\
0.260679802057692	0.210678064560862\\
0.299578103498643	0.207988244804483\\
0.328683969654205	0.204650091234236\\
0.377729764646745	0.197104131555417\\
0.414428548183942	0.190300814353463\\
0.476269037802799	0.17743569803801\\
0.522541521360296	0.167194972312223\\
0.600514488398176	0.149758856011865\\
0.658858186150681	0.137137239572555\\
0.75717214883374	0.117555304476446\\
0.830736107491936	0.104603849586758\\
0.954697470328753	0.0861670027755723\\
1.02040816326531	0.0779843254384878\\
1.08163265306122	0.0712331940890663\\
1.12244897959184	0.0671543310054605\\
1.18367346938776	0.0616038736890199\\
1.22448979591837	0.0582464450703885\\
1.28571428571429	0.05366824864503\\
1.3265306122449	0.0508915283646499\\
1.38775510204082	0.0470934420185523\\
1.42857142857143	0.0447819917271014\\
1.48979591836735	0.0416090053232671\\
1.53061224489796	0.0396707929972464\\
1.59183673469388	0.0370002402886368\\
1.63265306122449	0.0353627848225483\\
1.69387755102041	0.033098290249629\\
1.73469387755102	0.0317046969474778\\
1.79591836734694	0.0297705791869807\\
1.83673469387755	0.0285761131407958\\
1.89795918367347	0.0269127477964376\\
1.93877551020408	0.0258820773083085\\
2	0.0244422384965295\\
};
\end{axis}
\end{tikzpicture}%
\fi
    \end{minipage}
    \begin{minipage}[t]{0.32\textwidth}
        \centering
        \ifloadfig
%
\definecolor{mycolor1}{rgb}{0.00000,0.45000,0.74000}%
\begin{tikzpicture}[trim axis left,trim axis right]

\begin{axis}[%
width=11.255in,
height=7.131in,
at={(1.888in,0.962in)},
scale only axis,
separate axis lines,
every outer x axis line/.append style={black},
every x tick label/.append style={font=\color{black}},
every x tick/.append style={black},
xmin=0,
xmax=2,
xlabel={$k$},
every outer y axis line/.append style={black},
every y tick label/.append style={font=\color{black}},
every y tick/.append style={black},
ymin=0,
ymax=0.25,
ylabel={$\omega_r$},
axis background/.style={fill=white},
title style={font=\bfseries},
title={$A_{\mu} = 0.50$},
legend style={legend cell align=left, align=left},
clip=true,
clip mode=individual,
axis lines=box,
xtick pos=bottom,
ytick pos=left,
tick align=inside,
major tick length=2pt,
width=0.7\linewidth,
height=0.65\linewidth,
every axis/.append style={font=\fontsize{9}{9}\selectfont},xlabel style={font=\fontsize{9}{9}\selectfont},title style={font=\fontsize{9}{9}\selectfont},
legend style={font=\fontsize{8}{9}\selectfont,inner sep=1pt,row sep=1pt,column sep=2pt,nodes={scale=1.00},draw=none},legend image post style={xscale=0.35},
legend columns=1,
scaled x ticks=false,
tick label style={/pgf/number format/fixed,/pgf/number format/precision=2},
every x tick label/.append style={font=\fontsize{9}{9}\selectfont\color{black}},
every y tick label/.append style={font=\fontsize{9}{9}\selectfont\color{black}},
,,
colormap={sigmamap}{rgb(0.0000)=(0.0200,0.1880,0.3800); rgb(0.1000)=(0.0743,0.3456,0.5616); rgb(0.2000)=(0.1918,0.4980,0.7060); rgb(0.3000)=(0.3890,0.6708,0.8226); rgb(0.4000)=(0.7552,0.8682,0.9228); rgb(0.5000)=(0.9690,0.9689,0.9689); rgb(0.6000)=(0.9425,0.8044,0.7614); rgb(0.7000)=(0.8767,0.4910,0.4020); rgb(0.8000)=(0.7869,0.2866,0.2470); rgb(0.9000)=(0.6186,0.1239,0.1568); rgb(1.0000)=(0.4040,0.0000,0.1220)}
]
\addplot [color=black, line width=1.0pt]
  table[row sep=crcr]{%
0.001	0.00696921097661631\\
0.00126087240768068	0.00888363462172415\\
0.00158979922845048	0.0112944654407608\\
0.00200453398090524	0.014319920289638\\
0.00252746158678173	0.0180971755343923\\
0.00318680657624591	0.0227766193284808\\
0.00401815648060381	0.0285069253460641\\
0.00506638263613665	0.0354057595620901\\
0.00638806207265721	0.0435142817877908\\
0.00805453120596493	0.0527455894354881\\
0.0101557361544042	0.0628603423047125\\
0.0128050874967733	0.0735177618058294\\
0.0161455815026184	0.0844154495702377\\
0.0203575182226111	0.0954408412710943\\
0.0256682330157469	0.106706612489545\\
0.0323643667634736	0.118430082515931\\
0.0408073370441215	0.130760151465636\\
0.0514528453098585	0.143679862107972\\
0.0648754729478629	0.157002086240981\\
0.0817996937751947	0.170387771206964\\
0.103138976837872	0.183338569295194\\
0.13004509005129	0.195166388815459\\
0.163970265800021	0.204952948423805\\
0.206745583827313	0.211507152853407\\
0.260679802057692	0.213338519186146\\
0.328683969654205	0.208708655249976\\
0.414428548183942	0.195918498024872\\
0.522541521360296	0.174093691706079\\
0.658858186150681	0.144518128294618\\
0.830736107491936	0.111452218159546\\
1.02040816326531	0.0837474565066506\\
1.12244897959184	0.0723306009664185\\
1.22448979591837	0.0628822365106935\\
1.3265306122449	0.055043728466723\\
1.42857142857143	0.0485079099480314\\
1.53061224489796	0.0430235915315191\\
1.63265306122449	0.0383898049296419\\
1.73469387755102	0.0344472072662014\\
1.83673469387755	0.0310696974190794\\
1.93877551020408	0.028157264073598\\
};
\addlegendentry{$\mu_g$}

\addplot [color=white!60!black, line width=1.0pt]
  table[row sep=crcr]{%
0.001	0.00696921034229827\\
0.00126087240768068	0.00888363409510389\\
0.00158979922845048	0.0112944652582895\\
0.00200453398090524	0.014319919739432\\
0.00252746158678173	0.0180971759179179\\
0.00318680657624591	0.0227766203960559\\
0.00401815648060381	0.028506924534775\\
0.00506638263613665	0.0354057593769207\\
0.00638806207265721	0.043514281822822\\
0.00805453120596493	0.052745588711993\\
0.0101557361544042	0.0628603426858425\\
0.0128050874967733	0.073517761705578\\
0.0161455815026184	0.0844154488937342\\
0.0203575182226111	0.0954408409999973\\
0.0256682330157469	0.106706612176835\\
0.0323643667634736	0.118430082494903\\
0.0408073370441215	0.13076015125281\\
0.0514528453098585	0.143679862183664\\
0.0648754729478629	0.157002086334126\\
0.0817996937751947	0.170387771167622\\
0.103138976837872	0.183338569261652\\
0.13004509005129	0.195166388827614\\
0.163970265800021	0.204952948415034\\
0.206745583827313	0.211507152858154\\
0.260679802057692	0.21333851918609\\
0.328683969654205	0.20870865525262\\
0.414428548183942	0.195918498029205\\
0.522541521360296	0.174093691706698\\
0.658858186150681	0.144518128293644\\
0.830736107491936	0.111452218159878\\
1.02040816326531	0.0837474565066131\\
1.12244897959184	0.072330600966441\\
1.22448979591837	0.0628822365107204\\
1.3265306122449	0.0550437284667238\\
1.42857142857143	0.0485079099480271\\
1.53061224489796	0.0430235915315312\\
1.63265306122449	0.0383898049296461\\
1.73469387755102	0.0344472072661952\\
1.83673469387755	0.0310696974190763\\
1.93877551020408	0.0281572640735987\\
};
\addlegendentry{$\mu_h$}

\addplot [color=mycolor1, dashed, line width=1.4pt, forget plot]
  table[row sep=crcr]{%
0.001	0.00691440978517865\\
0.0011492187010037	0.00800901486152831\\
0.00126087240768068	0.00882808420561014\\
0.0014490181504862	0.0102073551961804\\
0.00158979922845048	0.0112379671603847\\
0.00182702700417654	0.0129704098215086\\
0.00200453398090524	0.0142622233944057\\
0.00230364793765369	0.0164280296807092\\
0.00252746158678173	0.0180379355800905\\
0.00290460612159805	0.0207263949843451\\
0.00318680657624591	0.0227153959157676\\
0.00366233771390337	0.0260173034587817\\
0.00401815648060381	0.0284430700576134\\
0.00461774057106911	0.0324351741165621\\
0.00506638263613665	0.0353383391120558\\
0.00582238167188866	0.0400576891357577\\
0.00638806207265721	0.0434419202815761\\
0.00734128039707012	0.0488541772876411\\
0.00805453120596493	0.0526664134734893\\
0.00925641788971278	0.0586449327468994\\
0.0101557361544042	0.0627721139751994\\
0.0116711619111007	0.069117128777596\\
0.0128050874967733	0.0734181550022168\\
0.0147158460192806	0.0799327693313498\\
0.0161455815026184	0.0843024251189728\\
0.0185548042013884	0.0908905222567032\\
0.0203575182226111	0.0953128290877747\\
0.0233952406474482	0.102020218875866\\
0.0256682330157469	0.106562622189041\\
0.029498413403417	0.113519304090278\\
0.0323643667634736	0.118269913692396\\
0.0371937355307264	0.125582717728126\\
0.0408073370441215	0.130584297332856\\
0.0468965548692655	0.138264382225936\\
0.0514528453098585	0.143487810231695\\
0.0591305720499399	0.151439137265285\\
0.0648754729478629	0.156787495752287\\
0.0745561067481437	0.164816972775873\\
0.0817996937751947	0.1701290054103\\
0.0940057378228297	0.177942083526521\\
0.103138976837872	0.182982961446435\\
0.11852924098447	0.190162917981382\\
0.13004509005129	0.194607625221631\\
0.149450249460652	0.200589012030576\\
0.163970265800021	0.204004877168509\\
0.188437695865931	0.208050322858798\\
0.206745583827313	0.209885831800706\\
0.237595891284276	0.211081906872747\\
0.260679802057692	0.210678064560862\\
0.299578103498643	0.207988244804483\\
0.328683969654205	0.204650091234236\\
0.377729764646745	0.197104131555417\\
0.414428548183942	0.190300814353463\\
0.476269037802799	0.17743569803801\\
0.522541521360296	0.167194972312223\\
0.600514488398176	0.149758856011865\\
0.658858186150681	0.137137239572555\\
0.75717214883374	0.117555304476446\\
0.830736107491936	0.104603849586758\\
0.954697470328753	0.0861670027755723\\
1.02040816326531	0.0779843254384878\\
1.08163265306122	0.0712331940890663\\
1.12244897959184	0.0671543310054605\\
1.18367346938776	0.0616038736890199\\
1.22448979591837	0.0582464450703885\\
1.28571428571429	0.05366824864503\\
1.3265306122449	0.0508915283646499\\
1.38775510204082	0.0470934420185523\\
1.42857142857143	0.0447819917271014\\
1.48979591836735	0.0416090053232671\\
1.53061224489796	0.0396707929972464\\
1.59183673469388	0.0370002402886368\\
1.63265306122449	0.0353627848225483\\
1.69387755102041	0.033098290249629\\
1.73469387755102	0.0317046969474778\\
1.79591836734694	0.0297705791869807\\
1.83673469387755	0.0285761131407958\\
1.89795918367347	0.0269127477964376\\
1.93877551020408	0.0258820773083085\\
2	0.0244422384965295\\
};
\end{axis}
\end{tikzpicture}%
\fi
    \end{minipage}
    \begin{minipage}[t]{0.32\textwidth}
        \centering
        \ifloadfig
%
\definecolor{mycolor1}{rgb}{0.00000,0.45000,0.74000}%
\begin{tikzpicture}[trim axis left,trim axis right]

\begin{axis}[%
width=11.255in,
height=7.131in,
at={(1.888in,0.962in)},
scale only axis,
separate axis lines,
every outer x axis line/.append style={black},
every x tick label/.append style={font=\color{black}},
every x tick/.append style={black},
xmin=0,
xmax=2,
xlabel={$k$},
every outer y axis line/.append style={black},
every y tick label/.append style={font=\color{black}},
every y tick/.append style={black},
ymin=0,
ymax=0.25,
ylabel={$\omega_r$},
axis background/.style={fill=white},
title style={font=\bfseries},
title={$A_{\mu} = 0.90$},
legend style={legend cell align=left, align=left},
clip=true,
clip mode=individual,
axis lines=box,
xtick pos=bottom,
ytick pos=left,
tick align=inside,
major tick length=2pt,
width=0.7\linewidth,
height=0.65\linewidth,
every axis/.append style={font=\fontsize{9}{9}\selectfont},xlabel style={font=\fontsize{9}{9}\selectfont},title style={font=\fontsize{9}{9}\selectfont},
legend style={font=\fontsize{8}{9}\selectfont,inner sep=1pt,row sep=1pt,column sep=2pt,nodes={scale=1.00},draw=none},legend image post style={xscale=0.35},
legend columns=1,
scaled x ticks=false,
tick label style={/pgf/number format/fixed,/pgf/number format/precision=2},
every x tick label/.append style={font=\fontsize{9}{9}\selectfont\color{black}},
every y tick label/.append style={font=\fontsize{9}{9}\selectfont\color{black}},
,,
colormap={sigmamap}{rgb(0.0000)=(0.0200,0.1880,0.3800); rgb(0.1000)=(0.0743,0.3456,0.5616); rgb(0.2000)=(0.1918,0.4980,0.7060); rgb(0.3000)=(0.3890,0.6708,0.8226); rgb(0.4000)=(0.7552,0.8682,0.9228); rgb(0.5000)=(0.9690,0.9689,0.9689); rgb(0.6000)=(0.9425,0.8044,0.7614); rgb(0.7000)=(0.8767,0.4910,0.4020); rgb(0.8000)=(0.7869,0.2866,0.2470); rgb(0.9000)=(0.6186,0.1239,0.1568); rgb(1.0000)=(0.4040,0.0000,0.1220)}
]
\addplot [color=black, line width=1.0pt]
  table[row sep=crcr]{%
0.001	0.00708512917933028\\
0.00126087240768068	0.00900372450287493\\
0.00158979922845048	0.0114190247811295\\
0.00200453398090524	0.0144491652066055\\
0.00252746158678173	0.0182311991370723\\
0.00318680657624591	0.0229153695480829\\
0.00401815648060381	0.0286501243257834\\
0.00506638263613665	0.0355528553638854\\
0.00638806207265721	0.0436644728597497\\
0.00805453120596493	0.0528980506965355\\
0.0101557361544042	0.0630148482033854\\
0.0128050874967733	0.0736756120288118\\
0.0161455815026184	0.0845802726169559\\
0.0203575182226111	0.0956184752823373\\
0.0256682330157469	0.106903735345398\\
0.0323643667634736	0.118652408436697\\
0.0408073370441215	0.131011842473613\\
0.0514528453098585	0.14396540803898\\
0.0648754729478629	0.157331089213233\\
0.0817996937751947	0.17078339690004\\
0.103138976837872	0.183851725975527\\
0.13004509005129	0.195898743702353\\
0.163970265800021	0.206092674764443\\
0.206745583827313	0.213382223220999\\
0.260679802057692	0.216491248270349\\
0.328683969654205	0.213980493847013\\
0.414428548183942	0.204471715606929\\
0.522541521360296	0.187145274748078\\
0.658858186150681	0.162472114592735\\
0.830736107491936	0.132757679843906\\
1.02040816326531	0.10525382810566\\
1.12244897959184	0.0930032075600035\\
1.22448979591837	0.0823934584104683\\
1.3265306122449	0.0732445176493792\\
1.42857142857143	0.0653655960703858\\
1.53061224489796	0.0585750353926153\\
1.63265306122449	0.0527095272856951\\
1.73469387755102	0.0476271228195358\\
1.83673469387755	0.0432069099200002\\
1.93877551020408	0.0393471146253238\\
};
\addlegendentry{$\mu_g$}

\addplot [color=white!60!black, line width=1.0pt]
  table[row sep=crcr]{%
0.001	0.00713801894605685\\
0.00126087240768068	0.00906501446622707\\
0.00158979922845048	0.0114896036732861\\
0.00200453398090524	0.0145298911318962\\
0.00252746158678173	0.0183228710620852\\
0.00318680657624591	0.0230187179230326\\
0.00401815648060381	0.0287658535932224\\
0.00506638263613665	0.0356817642310424\\
0.00638806207265721	0.0438076578080052\\
0.00805453120596493	0.053057301294645\\
0.0101557361544042	0.0631931664220148\\
0.0128050874967733	0.0738777692976807\\
0.0161455815026184	0.0848131932958501\\
0.0203575182226111	0.0958911993320212\\
0.0256682330157469	0.107227293629422\\
0.0323643667634736	0.119040900907663\\
0.0408073370441215	0.131487135116535\\
0.0514528453098585	0.144568614617903\\
0.0648754729478629	0.158144990707705\\
0.0817996937751947	0.171972003125301\\
0.103138976837872	0.185725380424511\\
0.13004509005129	0.199013232357711\\
0.163970265800021	0.211385809263196\\
0.206745583827313	0.222336490470399\\
0.260679802057692	0.231278377867654\\
0.328683969654205	0.237483636589053\\
0.414428548183942	0.239993332502658\\
0.522541521360296	0.237543301254246\\
0.658858186150681	0.228597127571924\\
0.830736107491936	0.211637500962305\\
1.02040816326531	0.189258286041089\\
1.12244897959184	0.176704792695394\\
1.22448979591837	0.164262405139757\\
1.3265306122449	0.152196087882291\\
1.42857142857143	0.140695665137008\\
1.53061224489796	0.129884627984622\\
1.63265306122449	0.119830940777284\\
1.73469387755102	0.110558796051224\\
1.83673469387755	0.102060056409297\\
1.93877551020408	0.0943044003059326\\
};
\addlegendentry{$\mu_h$}

\addplot [color=mycolor1, dashed, line width=1.4pt, forget plot]
  table[row sep=crcr]{%
0.001	0.00691440978517865\\
0.0011492187010037	0.00800901486152831\\
0.00126087240768068	0.00882808420561014\\
0.0014490181504862	0.0102073551961804\\
0.00158979922845048	0.0112379671603847\\
0.00182702700417654	0.0129704098215086\\
0.00200453398090524	0.0142622233944057\\
0.00230364793765369	0.0164280296807092\\
0.00252746158678173	0.0180379355800905\\
0.00290460612159805	0.0207263949843451\\
0.00318680657624591	0.0227153959157676\\
0.00366233771390337	0.0260173034587817\\
0.00401815648060381	0.0284430700576134\\
0.00461774057106911	0.0324351741165621\\
0.00506638263613665	0.0353383391120558\\
0.00582238167188866	0.0400576891357577\\
0.00638806207265721	0.0434419202815761\\
0.00734128039707012	0.0488541772876411\\
0.00805453120596493	0.0526664134734893\\
0.00925641788971278	0.0586449327468994\\
0.0101557361544042	0.0627721139751994\\
0.0116711619111007	0.069117128777596\\
0.0128050874967733	0.0734181550022168\\
0.0147158460192806	0.0799327693313498\\
0.0161455815026184	0.0843024251189728\\
0.0185548042013884	0.0908905222567032\\
0.0203575182226111	0.0953128290877747\\
0.0233952406474482	0.102020218875866\\
0.0256682330157469	0.106562622189041\\
0.029498413403417	0.113519304090278\\
0.0323643667634736	0.118269913692396\\
0.0371937355307264	0.125582717728126\\
0.0408073370441215	0.130584297332856\\
0.0468965548692655	0.138264382225936\\
0.0514528453098585	0.143487810231695\\
0.0591305720499399	0.151439137265285\\
0.0648754729478629	0.156787495752287\\
0.0745561067481437	0.164816972775873\\
0.0817996937751947	0.1701290054103\\
0.0940057378228297	0.177942083526521\\
0.103138976837872	0.182982961446435\\
0.11852924098447	0.190162917981382\\
0.13004509005129	0.194607625221631\\
0.149450249460652	0.200589012030576\\
0.163970265800021	0.204004877168509\\
0.188437695865931	0.208050322858798\\
0.206745583827313	0.209885831800706\\
0.237595891284276	0.211081906872747\\
0.260679802057692	0.210678064560862\\
0.299578103498643	0.207988244804483\\
0.328683969654205	0.204650091234236\\
0.377729764646745	0.197104131555417\\
0.414428548183942	0.190300814353463\\
0.476269037802799	0.17743569803801\\
0.522541521360296	0.167194972312223\\
0.600514488398176	0.149758856011865\\
0.658858186150681	0.137137239572555\\
0.75717214883374	0.117555304476446\\
0.830736107491936	0.104603849586758\\
0.954697470328753	0.0861670027755723\\
1.02040816326531	0.0779843254384878\\
1.08163265306122	0.0712331940890663\\
1.12244897959184	0.0671543310054605\\
1.18367346938776	0.0616038736890199\\
1.22448979591837	0.0582464450703885\\
1.28571428571429	0.05366824864503\\
1.3265306122449	0.0508915283646499\\
1.38775510204082	0.0470934420185523\\
1.42857142857143	0.0447819917271014\\
1.48979591836735	0.0416090053232671\\
1.53061224489796	0.0396707929972464\\
1.59183673469388	0.0370002402886368\\
1.63265306122449	0.0353627848225483\\
1.69387755102041	0.033098290249629\\
1.73469387755102	0.0317046969474778\\
1.79591836734694	0.0297705791869807\\
1.83673469387755	0.0285761131407958\\
1.89795918367347	0.0269127477964376\\
1.93877551020408	0.0258820773083085\\
2	0.0244422384965295\\
};
\end{axis}
\end{tikzpicture}%
\fi
    \end{minipage}
    \caption{Sensitivity of the dispersion relation to the viscosity mixing law for \(\Atw = 0.5\), \(\delta = 5\), and \(\Sch \to \infty\) for positive viscosity ratios, \(\Amu > 0\). The \(\Amu = 0\) reference is denoted by the dashed blue line \((\dashlegend)\).}
    \label{fig:visc_law_comparison_pos}
\end{figure}
Figures~\ref{fig:visc_law_comparison_neg} and~\ref{fig:visc_law_comparison_pos} show that the sensitivity of the dispersion relation to the closure depends on \(\Amu\). For \(\Amu < 0\), the harmonic law is consistently the most destabilizing and produces the largest departure from the reference case. For \(\Amu > 0\), the two models are nearly indistinguishable up to \(\Amu = 0.50\), whereas at \(\Amu = 0.90\) the harmonic law again yields the largest growth rates. The choice of mixing law is therefore not innocuous; at large viscosity ratios it can materially affect the predicted growth rates.

The two closures can be related in the reference case \(\nu_1 = \nu_2\). When the kinematic viscosities of the pure fluids coincide,~\eqref{eq:nu_g} and~\eqref{eq:nu_h} show that the harmonic and linear laws are identical, and both yield a constant kinematic viscosity throughout the domain,
\begin{equation}
    \label{eq:nu_const}
    \nu_g = \nu_h = \nu_1 = \nu_2.
\end{equation}
The harmonic law is often motivated by the observation of~\citet{livescu-2020} that several direct numerical simulations of IVD flows impose
\begin{equation}
    \frac{\mu}{\rho \mathcal{D}} = \Sch = \mathrm{constant},
\end{equation}
thereby enforcing a constant local Schmidt number; under this constraint \(\mu\) is proportional to \(\rho\), which leads to the harmonic closure. Equation~\eqref{eq:nu_const} shows, however, that the linear law recovers the same constant kinematic viscosity in the reference case; the harmonic closure is thus a sufficient, but not necessary, route to constant local kinematic viscosity.

At large viscosity ratios the harmonic closure differs substantially from the reference case. We argue that this behavior originates in the base-state viscosity, since the profile obtained from~\eqref{eq:nu_h} is not symmetric about the mid-plane \(z = 0\). This loss of symmetry may underlie the anomalously large growth rates produced by the harmonic law at large viscosity ratios. Aware of this sensitivity, in the present work we adopt the linear law~\eqref{eq:nu_g} since it coincides with the harmonic law in the reference case~\eqref{eq:nu_const} and is consistent with the observation that the growth rates are expected to be weakly dependent on the viscosity.
An in depth analysis of how different closures shapes the growth-rate spectrum is beyond the scope of the presente work and left for future investigations; see \citet{govindarajan-sahu-2014} for a related discussion.

\section{Linearized incompressible variable-density system in the limit of large Schmidt number}
\label{sec:appendix_Sc_infinity_chandra_limit}
In this appendix, we show that, in the limit \(\Sch \to \infty\), the incompressible variable-density formulation reduces to the classical non-diffusive Rayleigh--Taylor problem considered by~\citet{chandrasekhar-1961} and later employed by~\citet{morgan-likhachev-jacobs-2016}.

Taking~\eqref{eq:A_B_operators} in the limit \(\Sch \to \infty\), the velocity field becomes divergence-free and the base-state velocity \(w_0\), given by~\eqref{eq:self_similar_base_state_velocity}, vanishes identically. The governing equations reduce to
\begin{subequations}
\begin{align}
    \label{eq:appendix_chandraIVD_momentum}
    -\left(
        \rho_0' \partial_z
        +
        \rho_0 \nabla^2
     \right)
     \frac{\partial w}{\partial t}
    &=
    -\mu_0 \nabla^4 w
    -2\mu_0' \partial_z \nabla^2 w
    +\mu_0'' \nabla^2 w
    -2\mu_0'' \partial_z^2 w
    +\nabla_{xy}^2\rho,
    \\
    \label{eq:appendix_chandraIVD_density}
    \frac{\partial \rho}{\partial t}
    &=
    -\rho_0' w.
\end{align}
\end{subequations}
Since the in-plane directions \(x\) and \(y\) are homogeneous, we apply a Fourier transform in those directions. Let \(k_x\) and \(k_y\) denote the in-plane wavenumbers, and define \(k^2 = k_x^2 + k_y^2\). Using the identity
\begin{align}
    &-\mu_0 \nabla^4 \hat{w}
    -2\mu_0' \partial_z \nabla^2 \hat{w}
    +\mu_0'' \nabla^2 \hat{w}
    -2\mu_0'' \partial_z^2 \hat{w}
    =
    \notag\\
    &\qquad
    -\partial_z
    \left[
        \mu_0'
        \left(
            \partial_z^2 + k^2
        \right)
        +
        \mu_0
        \left(
            \partial_z^2 - k^2
        \right)\partial_z
    \right]\hat{w}
    +
    k^2
    \left[
        2\mu_0' \partial_z
        +
        \mu_0
        \left(
            \partial_z^2 - k^2
        \right)
    \right]\hat{w},
\end{align}
equation~\eqref{eq:appendix_chandraIVD_momentum} can be written as
\begin{align}
\label{eq:appendix_chandraIVD_momentum_fourier}
&\partial_z
\Bigg\{
    -\rho_0
    \partial_z
    \frac{\partial \hat{w}}{\partial t}
    +
    \mu_0'
    \left(
        \partial_z^2 + k^2
    \right)\hat{w}
    +
    \mu_0
    \left(
        \partial_z^2 - k^2
    \right)
    \partial_z \hat{w}
\Bigg\}
=
\notag\\
&\qquad
k^2
\Bigg\{
    -\rho_0
    \frac{\partial \hat{w}}{\partial t}
    +
    \mu_0
    \left(
        \partial_z^2 - k^2
    \right)\hat{w}
    -
    \hat{\rho}
    +
    2\mu_0' \partial_z \hat{w}
\Bigg\}.
\end{align}
Applying a Fourier transform in time, or equivalently adopting the normal-mode ansatz \(e^{nt}\) following~\citet{chandrasekhar-1961}, where \(n\) is the growth rate, reduces~\eqref{eq:appendix_chandraIVD_density} to
\begin{equation}
    \label{eq:appendix_density_fouriertime}
    \hat{\rho}
    =
    -\frac{\rho_0'}{n}\hat{w}.
\end{equation}
Substituting~\eqref{eq:appendix_density_fouriertime} into~\eqref{eq:appendix_chandraIVD_momentum_fourier} and dividing by \(n\) 
\begin{align}
    \label{eq:appendix_chandraIVD_final}
    &\partial_z
    \Bigg\{
        -\rho_0
        \partial_z \hat{w}
        +
        \frac{\mu_0'}{n}
        \left(
            \partial_z^2 + k^2
        \right)\hat{w}
        +
        \frac{\mu_0}{n}
        \left(
            \partial_z^2 - k^2
        \right)
        \partial_z \hat{w}
    \Bigg\}
    =
    \notag\\
    &\qquad
    k^2
    \Bigg\{
        -\rho_0 \hat{w}
        +
        \frac{\mu_0}{n}
        \left(
            \partial_z^2 - k^2
        \right)\hat{w}
        +
        \frac{\rho_0'}{n^2}\hat{w}
        +
        2\frac{\mu_0'}{n}
        \partial_z \hat{w}
    \Bigg\}.
\end{align}
Equation~\eqref{eq:appendix_chandraIVD_final} is precisely equation~(22) of~\citet{chandrasekhar-1961}, and is also the equation employed by~\citet{morgan-likhachev-jacobs-2016}. We argue that this result justifies interpreting the present set of equations as the direct extension of the Rayleigh--Taylor problem to gaseous interfaces, with diffusive effects, viscosity, and density stratification retained consistently.

A note is required regarding this limit. As \(\Sch\to\infty\) the interface thickness vanishes, \(\delta\to0\), so the initial density profile reduces to a sharp interface. This is consistent with the formulation of \citet{chandrasekhar-1961}, derived for two fluids of different density and viscosity, for which the appropriate interfacial boundary conditions can be recovered from the equations proposed here; we do not report them, and refer the interested reader to \citet{chandrasekhar-1961}.

\section{Numerical verification}
\label{sec:appendix_numerical_verification}

In this appendix, we document the numerical verification of the solver and organize the tests according to the numerical or physical limit being assessed. Section~\ref{subsec:numerical_convergence} first examines sensitivity to the domain half-height and grid resolution, isolating the discretization choices used in the computations. Section~\ref{subsec:chandrasekhar_vd} then verifies the non-diffusive variable-density formulation against the analytical sharp-interface results of~\citet{chandrasekhar-1961}. Section~\ref{subsec:morgan} compares the finite-thickness viscous predictions with the reference data of~\citet{morgan-likhachev-jacobs-2016}. Finally, Section~\ref{subsec:mikaelian} assesses the solver against the finite-layer semi-analytical results of~\citet{mikaelian-1996}, which include discontinuous density and viscosity.

\subsection{Numerical convergence}
\label{subsec:numerical_convergence}
\begin{figure}
    \centering
    \begin{minipage}[t]{0.32\textwidth}
        \centering\vspace{0pt}
        \ifloadfig
%
\begin{tikzpicture}

\begin{axis}[%
width=11.255in,
height=7.131in,
at={(1.888in,0.962in)},
scale only axis,
clip=true,
separate axis lines,
every outer x axis line/.append style={black},
every x tick label/.append style={font=\color{black}},
every x tick/.append style={black},
xmode=log,
xmin=128,
xmax=512,
xtick={128,256,394,512},
xticklabels={{128},{256},{394},{512}},
xminorticks=true,
xlabel={$L_z$},
every outer y axis line/.append style={black},
every y tick label/.append style={font=\color{black}},
every y tick/.append style={black},
ymin=0,
ymax=0.5,
ylabel={$\omega_r$},
axis background/.style={fill=white},
legend style={legend cell align=left, align=left},
clip=true,
clip mode=individual,
axis lines=box,
xtick pos=bottom,
ytick pos=left,
tick align=inside,
major tick length=2pt,
width=0.7\linewidth,
height=0.7\linewidth,
every axis/.append style={font=\fontsize{9}{9}\selectfont},xlabel style={font=\fontsize{9}{9}\selectfont},title style={font=\fontsize{9}{9}\selectfont},
legend style={font=\fontsize{8}{9}\selectfont,inner sep=1pt,row sep=1pt,column sep=2pt,nodes={scale=0.70},draw=none},legend image post style={xscale=0.35},
legend columns=1,
scaled x ticks=false,
tick label style={/pgf/number format/fixed,/pgf/number format/precision=2},
every x tick label/.append style={font=\fontsize{9}{9}\selectfont\color{black}},
every y tick label/.append style={font=\fontsize{9}{9}\selectfont\color{black}},
,,
colormap={sigmamap}{rgb(0.0000)=(0.0200,0.1880,0.3800); rgb(0.1000)=(0.0743,0.3456,0.5616); rgb(0.2000)=(0.1918,0.4980,0.7060); rgb(0.3000)=(0.3890,0.6708,0.8226); rgb(0.4000)=(0.7552,0.8682,0.9228); rgb(0.5000)=(0.9690,0.9689,0.9689); rgb(0.6000)=(0.9425,0.8044,0.7614); rgb(0.7000)=(0.8767,0.4910,0.4020); rgb(0.8000)=(0.7869,0.2866,0.2470); rgb(0.9000)=(0.6186,0.1239,0.1568); rgb(1.0000)=(0.4040,0.0000,0.1220)}
]
\addplot [color=black, line width=1.2pt, mark size=1.5pt, mark=o, mark options={solid, black}]
  table[row sep=crcr]{%
128	0.0153889793972656\\
256	0.0179069569993955\\
394	0.0181044671372273\\
512	0.0181167590937163\\
};
\addlegendentry{$A_t = 0.05$}

\addplot [color=black, dashed, line width=1.2pt, mark size=1.5pt, mark=o, mark options={solid, black}, forget plot]
  table[row sep=crcr]{%
128	0.0154122449079616\\
256	0.0179597722883003\\
394	0.0181575107257581\\
512	0.0181691558767306\\
};
\addplot [color=black, dotted, line width=1.2pt, mark size=1.5pt, mark=o, mark options={solid, black}, forget plot]
  table[row sep=crcr]{%
128	0.0155158254201716\\
256	0.017993195805832\\
394	0.0181812289696008\\
512	0.0181890669958651\\
};
\addplot [color=white!60!black, line width=1.2pt, mark size=1.5pt, mark=o, mark options={solid, white!60!black}]
  table[row sep=crcr]{%
128	0.0720014461552192\\
256	0.0791109756186219\\
394	0.0796760850873103\\
512	0.0797135141454948\\
};
\addlegendentry{$A_t = 0.40$}

\addplot [color=white!60!black, dashed, line width=1.2pt, mark size=1.5pt, mark=o, mark options={solid, white!60!black}, forget plot]
  table[row sep=crcr]{%
128	0.0721555681684248\\
256	0.0793096524368366\\
394	0.079877258674543\\
512	0.0799232925165549\\
};
\addplot [color=white!60!black, dotted, line width=1.2pt, mark size=1.5pt, mark=o, mark options={solid, white!60!black}, forget plot]
  table[row sep=crcr]{%
128	0.0725431931109221\\
256	0.0795771527574536\\
394	0.0801553204053335\\
512	0.0801753440671878\\
};
\addplot [color=red!60!teal, line width=1.2pt, mark size=1.5pt, mark=o, mark options={solid, red!60!teal}]
  table[row sep=crcr]{%
128	0.411919202509246\\
256	0.449580034861528\\
394	0.452643869715108\\
512	0.452842514435163\\
};
\addlegendentry{$A_t = 0.75$}

\addplot [color=red!60!teal, dashed, line width=1.2pt, mark size=1.5pt, mark=o, mark options={solid, red!60!teal}, forget plot]
  table[row sep=crcr]{%
128	0.412179887627474\\
256	0.450093744344902\\
394	0.453184220463465\\
512	0.453374721298282\\
};
\addplot [color=red!60!teal, dotted, line width=1.2pt, mark size=1.5pt, mark=o, mark options={solid, red!60!teal}, forget plot]
  table[row sep=crcr]{%
128	0.413186358139441\\
256	0.451002864101312\\
394	0.454087161901598\\
512	0.454287464873774\\
};
\node[right, align=left, inner sep=0]
at (rel axis cs:-0.25,1.15) {$(a)$};
\end{axis}
\end{tikzpicture}%
\fi
    \end{minipage}\hfill
    \begin{minipage}[t]{0.32\textwidth}
        \centering\vspace{0pt}
        \ifloadfig
%
\begin{tikzpicture}

\begin{axis}[%
width=11.255in,
height=7.131in,
at={(1.888in,0.962in)},
scale only axis,
clip=true,
separate axis lines,
every outer x axis line/.append style={black},
every x tick label/.append style={font=\color{black}},
every x tick/.append style={black},
xmode=log,
xmin=128,
xmax=512,
xtick={128,256,394,512},
xticklabels={{128},{256},{394},{512}},
xminorticks=true,
xlabel={$L_z$},
every outer y axis line/.append style={black},
every y tick label/.append style={font=\color{black}},
every y tick/.append style={black},
ymin=0.01,
ymax=0.09,
axis background/.style={fill=white},
clip=true,
clip mode=individual,
axis lines=box,
xtick pos=bottom,
ytick pos=left,
tick align=inside,
major tick length=2pt,
width=0.7\linewidth,
height=0.7\linewidth,
every axis/.append style={font=\fontsize{9}{9}\selectfont},xlabel style={font=\fontsize{9}{9}\selectfont},title style={font=\fontsize{9}{9}\selectfont},
legend style={font=\fontsize{8}{9}\selectfont,inner sep=1pt,row sep=1pt,column sep=2pt,nodes={scale=1.00},draw=none},legend image post style={xscale=0.35},
legend columns=1,
scaled x ticks=false,
tick label style={/pgf/number format/fixed,/pgf/number format/precision=2},
every x tick label/.append style={font=\fontsize{9}{9}\selectfont\color{black}},
every y tick label/.append style={font=\fontsize{9}{9}\selectfont\color{black}},
,,
colormap={sigmamap}{rgb(0.0000)=(0.0200,0.1880,0.3800); rgb(0.1000)=(0.0743,0.3456,0.5616); rgb(0.2000)=(0.1918,0.4980,0.7060); rgb(0.3000)=(0.3890,0.6708,0.8226); rgb(0.4000)=(0.7552,0.8682,0.9228); rgb(0.5000)=(0.9690,0.9689,0.9689); rgb(0.6000)=(0.9425,0.8044,0.7614); rgb(0.7000)=(0.8767,0.4910,0.4020); rgb(0.8000)=(0.7869,0.2866,0.2470); rgb(0.9000)=(0.6186,0.1239,0.1568); rgb(1.0000)=(0.4040,0.0000,0.1220)}
]
\addplot [color=black, line width=1.2pt, mark size=1.5pt, mark=o, mark options={solid, black}, forget plot]
  table[row sep=crcr]{%
128	0.0196439567233836\\
256	0.0214647547529239\\
394	0.021611332365737\\
512	0.0216208340459056\\
};
\addplot [color=black, dashed, line width=1.2pt, mark size=1.5pt, mark=o, mark options={solid, black}, forget plot]
  table[row sep=crcr]{%
128	0.0194199983019396\\
256	0.0213087455223332\\
394	0.0214608155727251\\
512	0.021470187319374\\
};
\addplot [color=black, dotted, line width=1.2pt, mark size=1.5pt, mark=o, mark options={solid, black}, forget plot]
  table[row sep=crcr]{%
128	0.019656550828726\\
256	0.0214770622312565\\
394	0.0216243966971243\\
512	0.0216330843207822\\
};
\addplot [color=white!60!black, line width=1.2pt, mark size=1.5pt, mark=o, mark options={solid, white!60!black}, forget plot]
  table[row sep=crcr]{%
128	0.0567396021543479\\
256	0.0615101810743804\\
394	0.0618964135209898\\
512	0.0619198787968389\\
};
\addplot [color=white!60!black, dashed, line width=1.2pt, mark size=1.5pt, mark=o, mark options={solid, white!60!black}, forget plot]
  table[row sep=crcr]{%
128	0.0565094818669938\\
256	0.0613858332659709\\
394	0.061780194548576\\
512	0.061804248673023\\
};
\addplot [color=white!60!black, dotted, line width=1.2pt, mark size=1.5pt, mark=o, mark options={solid, white!60!black}, forget plot]
  table[row sep=crcr]{%
128	0.0569328808131031\\
256	0.0617218920089409\\
394	0.062110490369209\\
512	0.0621354690331453\\
};
\addplot [color=red!60!teal, line width=1.2pt, mark size=1.5pt, mark=o, mark options={solid, red!60!teal}, forget plot]
  table[row sep=crcr]{%
128	0.0784638381384213\\
256	0.0848530942065096\\
394	0.0853709095211397\\
512	0.0854009979705437\\
};
\addplot [color=red!60!teal, dashed, line width=1.2pt, mark size=1.5pt, mark=o, mark options={solid, red!60!teal}, forget plot]
  table[row sep=crcr]{%
128	0.0783562037416023\\
256	0.084867131697303\\
394	0.0853942751987869\\
512	0.0854277114636466\\
};
\addplot [color=red!60!teal, dotted, line width=1.2pt, mark size=1.5pt, mark=o, mark options={solid, red!60!teal}, forget plot]
  table[row sep=crcr]{%
128	0.0788068491041872\\
256	0.0852543709366611\\
394	0.0857780113428995\\
512	0.085810613990898\\
};
\node[right, align=left, inner sep=0]
at (rel axis cs:-0.25,1.15) {$(b)$};
\end{axis}
\end{tikzpicture}%
\fi
    \end{minipage}\hfill
    \begin{minipage}[t]{0.32\textwidth}
        \centering\vspace{0pt}
        \ifloadfig
%
\begin{tikzpicture}

\begin{axis}[%
width=11.255in,
height=7.131in,
at={(1.888in,0.962in)},
scale only axis,
clip=true,
separate axis lines,
every outer x axis line/.append style={black},
every x tick label/.append style={font=\color{black}},
every x tick/.append style={black},
xmode=log,
xmin=128,
xmax=512,
xtick={128,256,394,512},
xticklabels={{128},{256},{394},{512}},
xminorticks=true,
xlabel={$L_z$},
every outer y axis line/.append style={black},
every y tick label/.append style={font=\color{black}},
every y tick/.append style={black},
ymin=0.01,
ymax=0.09,
axis background/.style={fill=white},
clip=true,
clip mode=individual,
axis lines=box,
xtick pos=bottom,
ytick pos=left,
tick align=inside,
major tick length=2pt,
width=0.7\linewidth,
height=0.7\linewidth,
every axis/.append style={font=\fontsize{9}{9}\selectfont},xlabel style={font=\fontsize{9}{9}\selectfont},title style={font=\fontsize{9}{9}\selectfont},
legend style={font=\fontsize{8}{9}\selectfont,inner sep=1pt,row sep=1pt,column sep=2pt,nodes={scale=1.00},draw=none},legend image post style={xscale=0.35},
legend columns=1,
scaled x ticks=false,
tick label style={/pgf/number format/fixed,/pgf/number format/precision=2},
every x tick label/.append style={font=\fontsize{9}{9}\selectfont\color{black}},
every y tick label/.append style={font=\fontsize{9}{9}\selectfont\color{black}},
,,
colormap={sigmamap}{rgb(0.0000)=(0.0200,0.1880,0.3800); rgb(0.1000)=(0.0743,0.3456,0.5616); rgb(0.2000)=(0.1918,0.4980,0.7060); rgb(0.3000)=(0.3890,0.6708,0.8226); rgb(0.4000)=(0.7552,0.8682,0.9228); rgb(0.5000)=(0.9690,0.9689,0.9689); rgb(0.6000)=(0.9425,0.8044,0.7614); rgb(0.7000)=(0.8767,0.4910,0.4020); rgb(0.8000)=(0.7869,0.2866,0.2470); rgb(0.9000)=(0.6186,0.1239,0.1568); rgb(1.0000)=(0.4040,0.0000,0.1220)}
]
\addplot [color=black, line width=1.2pt, mark size=1.5pt, mark=o, mark options={solid, black}, forget plot]
  table[row sep=crcr]{%
128	0.019783459728364\\
256	0.0215846202417624\\
394	0.0217303012413314\\
512	0.0217392569316356\\
};
\addplot [color=black, dashed, line width=1.2pt, mark size=1.5pt, mark=o, mark options={solid, black}, forget plot]
  table[row sep=crcr]{%
128	0.019529680739418\\
256	0.0214033113179657\\
394	0.0215545230139536\\
512	0.0215637291398573\\
};
\addplot [color=black, dotted, line width=1.2pt, mark size=1.5pt, mark=o, mark options={solid, black}, forget plot]
  table[row sep=crcr]{%
128	0.0197944255824842\\
256	0.0215962932053438\\
394	0.0217420804980316\\
512	0.0217509205483216\\
};
\addplot [color=white!60!black, line width=1.2pt, mark size=1.5pt, mark=o, mark options={solid, white!60!black}, forget plot]
  table[row sep=crcr]{%
128	0.0569735529244281\\
256	0.0617226593288172\\
394	0.0621075613356323\\
512	0.0621312641362954\\
};
\addplot [color=white!60!black, dashed, line width=1.2pt, mark size=1.5pt, mark=o, mark options={solid, white!60!black}, forget plot]
  table[row sep=crcr]{%
128	0.0566990587024551\\
256	0.0615621103176208\\
394	0.0619560535752779\\
512	0.061979990863544\\
};
\addplot [color=white!60!black, dotted, line width=1.2pt, mark size=1.5pt, mark=o, mark options={solid, white!60!black}, forget plot]
  table[row sep=crcr]{%
128	0.0571475461472267\\
256	0.0619210794636186\\
394	0.0623087306710372\\
512	0.0623321653232165\\
};
\addplot [color=red!60!teal, line width=1.2pt, mark size=1.5pt, mark=o, mark options={solid, red!60!teal}, forget plot]
  table[row sep=crcr]{%
128	0.0785344458669422\\
256	0.0848989313583755\\
394	0.0854145630696702\\
512	0.0854462594248644\\
};
\addplot [color=red!60!teal, dashed, line width=1.2pt, mark size=1.5pt, mark=o, mark options={solid, red!60!teal}, forget plot]
  table[row sep=crcr]{%
128	0.0783812231554221\\
256	0.0848763676999347\\
394	0.0854029988564309\\
512	0.0854352893046333\\
};
\addplot [color=red!60!teal, dotted, line width=1.2pt, mark size=1.5pt, mark=o, mark options={solid, red!60!teal}, forget plot]
  table[row sep=crcr]{%
128	0.0788607448988685\\
256	0.0852895553230683\\
394	0.0858122822003104\\
512	0.0858432547273326\\
};
\node[right, align=left, inner sep=0]
at (rel axis cs:-0.25,1.15) {$(c)$};
\end{axis}
\end{tikzpicture}%
\fi
    \end{minipage}\\[0.5em]
    \begin{minipage}[t]{0.32\textwidth}
        \centering\vspace{0pt}
        \ifloadfig
%
\begin{tikzpicture}

\begin{axis}[%
width=11.255in,
height=7.131in,
at={(1.888in,0.962in)},
scale only axis,
clip=true,
separate axis lines,
every outer x axis line/.append style={black},
every x tick label/.append style={font=\color{black}},
every x tick/.append style={black},
xmode=log,
xmin=128,
xmax=512,
xtick={128,256,394,512},
xticklabels={{128},{256},{394},{512}},
xminorticks=true,
xlabel={$L_z$},
every outer y axis line/.append style={black},
every y tick label/.append style={font=\color{black}},
every y tick/.append style={black},
ymin=-0.2,
ymax=1.6,
ylabel={$\omega_r$},
axis background/.style={fill=white},
clip=true,
clip mode=individual,
axis lines=box,
xtick pos=bottom,
ytick pos=left,
tick align=inside,
major tick length=2pt,
width=0.7\linewidth,
height=0.7\linewidth,
every axis/.append style={font=\fontsize{9}{9}\selectfont},xlabel style={font=\fontsize{9}{9}\selectfont},title style={font=\fontsize{9}{9}\selectfont},
legend style={font=\fontsize{8}{9}\selectfont,inner sep=1pt,row sep=1pt,column sep=2pt,nodes={scale=1.00},draw=none},legend image post style={xscale=0.35},
legend columns=1,
scaled x ticks=false,
tick label style={/pgf/number format/fixed,/pgf/number format/precision=2},
every x tick label/.append style={font=\fontsize{9}{9}\selectfont\color{black}},
every y tick label/.append style={font=\fontsize{9}{9}\selectfont\color{black}},
,,
colormap={sigmamap}{rgb(0.0000)=(0.0200,0.1880,0.3800); rgb(0.1000)=(0.0743,0.3456,0.5616); rgb(0.2000)=(0.1918,0.4980,0.7060); rgb(0.3000)=(0.3890,0.6708,0.8226); rgb(0.4000)=(0.7552,0.8682,0.9228); rgb(0.5000)=(0.9690,0.9689,0.9689); rgb(0.6000)=(0.9425,0.8044,0.7614); rgb(0.7000)=(0.8767,0.4910,0.4020); rgb(0.8000)=(0.7869,0.2866,0.2470); rgb(0.9000)=(0.6186,0.1239,0.1568); rgb(1.0000)=(0.4040,0.0000,0.1220)}
]
\addplot [color=black, line width=1.2pt, mark size=1.5pt, mark=o, mark options={solid, black}, forget plot]
  table[row sep=crcr]{%
128	-0.000993740340299722\\
256	-0.000963066342254914\\
394	-0.000957201092097769\\
512	-0.000955352533494663\\
};
\addplot [color=black, dashed, line width=1.2pt, mark size=1.5pt, mark=o, mark options={solid, black}, forget plot]
  table[row sep=crcr]{%
128	-0.00301358636765227\\
256	-0.00301356933547395\\
394	-0.00301359357637394\\
512	-0.00301359758853275\\
};
\addplot [color=black, dotted, line width=1.2pt, mark size=1.5pt, mark=o, mark options={solid, black}, forget plot]
  table[row sep=crcr]{%
128	-0.00106993805265659\\
256	-0.00105354446736899\\
394	-0.00105173875949664\\
512	-0.00105138742398111\\
};
\addplot [color=white!60!black, line width=1.2pt, mark size=1.5pt, mark=o, mark options={solid, white!60!black}, forget plot]
  table[row sep=crcr]{%
128	0.0983481682580654\\
256	0.0983481597381397\\
394	0.0983481521954307\\
512	0.0983481879141585\\
};
\addplot [color=white!60!black, dashed, line width=1.2pt, mark size=1.5pt, mark=o, mark options={solid, white!60!black}, forget plot]
  table[row sep=crcr]{%
128	0.101561488021553\\
256	0.101561472566478\\
394	0.101561497753094\\
512	0.101561466615898\\
};
\addplot [color=white!60!black, dotted, line width=1.2pt, mark size=1.5pt, mark=o, mark options={solid, white!60!black}, forget plot]
  table[row sep=crcr]{%
128	0.104171468304152\\
256	0.104171471836658\\
394	0.104171450251683\\
512	0.104171478865499\\
};
\addplot [color=red!60!teal, line width=1.2pt, mark size=1.5pt, mark=o, mark options={solid, red!60!teal}, forget plot]
  table[row sep=crcr]{%
128	1.47146919294046\\
256	1.47146926194063\\
394	1.47146920562613\\
512	1.47146913687379\\
};
\addplot [color=red!60!teal, dashed, line width=1.2pt, mark size=1.5pt, mark=o, mark options={solid, red!60!teal}, forget plot]
  table[row sep=crcr]{%
128	1.47611531434549\\
256	1.47611530493872\\
394	1.47611530950545\\
512	1.476115250815\\
};
\addplot [color=red!60!teal, dotted, line width=1.2pt, mark size=1.5pt, mark=o, mark options={solid, red!60!teal}, forget plot]
  table[row sep=crcr]{%
128	1.48563897911697\\
256	1.48563899807246\\
394	1.48563900547943\\
512	1.48563895027005\\
};
\node[right, align=left, inner sep=0]
at (rel axis cs:-0.25,1.15) {$(d)$};
\end{axis}
\end{tikzpicture}%
\fi
    \end{minipage}\hfill
    \begin{minipage}[t]{0.32\textwidth}
        \centering\vspace{0pt}
        \ifloadfig
%
\begin{tikzpicture}

\begin{axis}[%
width=11.255in,
height=7.131in,
at={(1.888in,0.962in)},
scale only axis,
clip=true,
separate axis lines,
every outer x axis line/.append style={black},
every x tick label/.append style={font=\color{black}},
every x tick/.append style={black},
xmode=log,
xmin=128,
xmax=512,
xtick={128,256,394,512},
xticklabels={{128},{256},{394},{512}},
xminorticks=true,
xlabel={$L_z$},
every outer y axis line/.append style={black},
every y tick label/.append style={font=\color{black}},
every y tick/.append style={black},
ymin=0.05,
ymax=0.25,
axis background/.style={fill=white},
clip=true,
clip mode=individual,
axis lines=box,
xtick pos=bottom,
ytick pos=left,
tick align=inside,
major tick length=2pt,
width=0.7\linewidth,
height=0.7\linewidth,
every axis/.append style={font=\fontsize{9}{9}\selectfont},xlabel style={font=\fontsize{9}{9}\selectfont},title style={font=\fontsize{9}{9}\selectfont},
legend style={font=\fontsize{8}{9}\selectfont,inner sep=1pt,row sep=1pt,column sep=2pt,nodes={scale=1.00},draw=none},legend image post style={xscale=0.35},
legend columns=1,
scaled x ticks=false,
tick label style={/pgf/number format/fixed,/pgf/number format/precision=2},
every x tick label/.append style={font=\fontsize{9}{9}\selectfont\color{black}},
every y tick label/.append style={font=\fontsize{9}{9}\selectfont\color{black}},
,,
colormap={sigmamap}{rgb(0.0000)=(0.0200,0.1880,0.3800); rgb(0.1000)=(0.0743,0.3456,0.5616); rgb(0.2000)=(0.1918,0.4980,0.7060); rgb(0.3000)=(0.3890,0.6708,0.8226); rgb(0.4000)=(0.7552,0.8682,0.9228); rgb(0.5000)=(0.9690,0.9689,0.9689); rgb(0.6000)=(0.9425,0.8044,0.7614); rgb(0.7000)=(0.8767,0.4910,0.4020); rgb(0.8000)=(0.7869,0.2866,0.2470); rgb(0.9000)=(0.6186,0.1239,0.1568); rgb(1.0000)=(0.4040,0.0000,0.1220)}
]
\addplot [color=black, line width=1.2pt, mark size=1.5pt, mark=o, mark options={solid, black}, forget plot]
  table[row sep=crcr]{%
128	0.0526429801940198\\
256	0.052642980944614\\
394	0.0526429791574219\\
512	0.0526429794710954\\
};
\addplot [color=black, dashed, line width=1.2pt, mark size=1.5pt, mark=o, mark options={solid, black}, forget plot]
  table[row sep=crcr]{%
128	0.0520798890662265\\
256	0.0520798898114293\\
394	0.0520798880012983\\
512	0.0520798889965739\\
};
\addplot [color=black, dotted, line width=1.2pt, mark size=1.5pt, mark=o, mark options={solid, black}, forget plot]
  table[row sep=crcr]{%
128	0.0525853772111606\\
256	0.0525853783028885\\
394	0.0525853787617508\\
512	0.0525853759705066\\
};
\addplot [color=white!60!black, line width=1.2pt, mark size=1.5pt, mark=o, mark options={solid, white!60!black}, forget plot]
  table[row sep=crcr]{%
128	0.172145893328032\\
256	0.17214589551027\\
394	0.172145892630669\\
512	0.172145890939879\\
};
\addplot [color=white!60!black, dashed, line width=1.2pt, mark size=1.5pt, mark=o, mark options={solid, white!60!black}, forget plot]
  table[row sep=crcr]{%
128	0.171022706338349\\
256	0.171022702890896\\
394	0.171022704190653\\
512	0.171022702083268\\
};
\addplot [color=white!60!black, dotted, line width=1.2pt, mark size=1.5pt, mark=o, mark options={solid, white!60!black}, forget plot]
  table[row sep=crcr]{%
128	0.17248431231321\\
256	0.172484312103778\\
394	0.172484313506082\\
512	0.172484310993431\\
};
\addplot [color=red!60!teal, line width=1.2pt, mark size=1.5pt, mark=o, mark options={solid, red!60!teal}, forget plot]
  table[row sep=crcr]{%
128	0.246163156785069\\
256	0.24616316271479\\
394	0.246163160433446\\
512	0.24616315822717\\
};
\addplot [color=red!60!teal, dashed, line width=1.2pt, mark size=1.5pt, mark=o, mark options={solid, red!60!teal}, forget plot]
  table[row sep=crcr]{%
128	0.245408637034539\\
256	0.245408634614974\\
394	0.245408635083651\\
512	0.245408633406366\\
};
\addplot [color=red!60!teal, dotted, line width=1.2pt, mark size=1.5pt, mark=o, mark options={solid, red!60!teal}, forget plot]
  table[row sep=crcr]{%
128	0.247305772618718\\
256	0.247305777754651\\
394	0.247305775476246\\
512	0.247305776780787\\
};
\node[right, align=left, inner sep=0]
at (rel axis cs:-0.25,1.15) {$(e)$};
\end{axis}
\end{tikzpicture}%
\fi
    \end{minipage}\hfill
    \begin{minipage}[t]{0.32\textwidth}
        \centering\vspace{0pt}
        \ifloadfig
%
\begin{tikzpicture}

\begin{axis}[%
width=11.255in,
height=7.131in,
at={(1.888in,0.962in)},
scale only axis,
clip=true,
separate axis lines,
every outer x axis line/.append style={black},
every x tick label/.append style={font=\color{black}},
every x tick/.append style={black},
xmode=log,
xmin=128,
xmax=512,
xtick={128,256,394,512},
xticklabels={{128},{256},{394},{512}},
xminorticks=true,
xlabel={$L_z$},
every outer y axis line/.append style={black},
every y tick label/.append style={font=\color{black}},
every y tick/.append style={black},
ymin=0.05,
ymax=0.25,
axis background/.style={fill=white},
clip=true,
clip mode=individual,
axis lines=box,
xtick pos=bottom,
ytick pos=left,
tick align=inside,
major tick length=2pt,
width=0.7\linewidth,
height=0.7\linewidth,
every axis/.append style={font=\fontsize{9}{9}\selectfont},xlabel style={font=\fontsize{9}{9}\selectfont},title style={font=\fontsize{9}{9}\selectfont},
legend style={font=\fontsize{8}{9}\selectfont,inner sep=1pt,row sep=1pt,column sep=2pt,nodes={scale=1.00},draw=none},legend image post style={xscale=0.35},
legend columns=1,
scaled x ticks=false,
tick label style={/pgf/number format/fixed,/pgf/number format/precision=2},
every x tick label/.append style={font=\fontsize{9}{9}\selectfont\color{black}},
every y tick label/.append style={font=\fontsize{9}{9}\selectfont\color{black}},
,,
colormap={sigmamap}{rgb(0.0000)=(0.0200,0.1880,0.3800); rgb(0.1000)=(0.0743,0.3456,0.5616); rgb(0.2000)=(0.1918,0.4980,0.7060); rgb(0.3000)=(0.3890,0.6708,0.8226); rgb(0.4000)=(0.7552,0.8682,0.9228); rgb(0.5000)=(0.9690,0.9689,0.9689); rgb(0.6000)=(0.9425,0.8044,0.7614); rgb(0.7000)=(0.8767,0.4910,0.4020); rgb(0.8000)=(0.7869,0.2866,0.2470); rgb(0.9000)=(0.6186,0.1239,0.1568); rgb(1.0000)=(0.4040,0.0000,0.1220)}
]
\addplot [color=black, line width=1.2pt, mark size=1.5pt, mark=o, mark options={solid, black}, forget plot]
  table[row sep=crcr]{%
128	0.0551657662861021\\
256	0.0551657671452841\\
394	0.0551657670740398\\
512	0.0551657662787053\\
};
\addplot [color=black, dashed, line width=1.2pt, mark size=1.5pt, mark=o, mark options={solid, black}, forget plot]
  table[row sep=crcr]{%
128	0.0544082204400931\\
256	0.0544082214671291\\
394	0.0544082219107181\\
512	0.0544082211526974\\
};
\addplot [color=black, dotted, line width=1.2pt, mark size=1.5pt, mark=o, mark options={solid, black}, forget plot]
  table[row sep=crcr]{%
128	0.0551352319432771\\
256	0.055135231204971\\
394	0.0551352322341404\\
512	0.0551352314177145\\
};
\addplot [color=white!60!black, line width=1.2pt, mark size=1.5pt, mark=o, mark options={solid, white!60!black}, forget plot]
  table[row sep=crcr]{%
128	0.175471928143913\\
256	0.175471928839631\\
394	0.175471928614474\\
512	0.17547192819531\\
};
\addplot [color=white!60!black, dashed, line width=1.2pt, mark size=1.5pt, mark=o, mark options={solid, white!60!black}, forget plot]
  table[row sep=crcr]{%
128	0.174205451609595\\
256	0.174205451477413\\
394	0.174205451144942\\
512	0.174205450486856\\
};
\addplot [color=white!60!black, dotted, line width=1.2pt, mark size=1.5pt, mark=o, mark options={solid, white!60!black}, forget plot]
  table[row sep=crcr]{%
128	0.175893008924764\\
256	0.17589300936865\\
394	0.175893008672956\\
512	0.175893008327821\\
};
\addplot [color=red!60!teal, line width=1.2pt, mark size=1.5pt, mark=o, mark options={solid, red!60!teal}, forget plot]
  table[row sep=crcr]{%
128	0.249057456773047\\
256	0.249057456595799\\
394	0.249057457261426\\
512	0.2490574589601\\
};
\addplot [color=red!60!teal, dashed, line width=1.2pt, mark size=1.5pt, mark=o, mark options={solid, red!60!teal}, forget plot]
  table[row sep=crcr]{%
128	0.248046962665486\\
256	0.248046962579005\\
394	0.248046962400925\\
512	0.248046960927033\\
};
\addplot [color=red!60!teal, dotted, line width=1.2pt, mark size=1.5pt, mark=o, mark options={solid, red!60!teal}, forget plot]
  table[row sep=crcr]{%
128	0.24994606976581\\
256	0.2499460699019\\
394	0.249946070292958\\
512	0.249946069178953\\
};
\node[right, align=left, inner sep=0]
at (rel axis cs:-0.25,1.15) {$(f)$};
\end{axis}
\end{tikzpicture}%
\fi
    \end{minipage}\\[0.5em]
    \begin{minipage}[t]{0.32\textwidth}
        \centering\vspace{0pt}
        \ifloadfig
%
\begin{tikzpicture}

\begin{axis}[%
width=11.255in,
height=7.131in,
at={(1.888in,0.962in)},
scale only axis,
clip=true,
separate axis lines,
every outer x axis line/.append style={black},
every x tick label/.append style={font=\color{black}},
every x tick/.append style={black},
xmode=log,
xmin=128,
xmax=512,
xtick={128,256,394,512},
xticklabels={{128},{256},{394},{512}},
xminorticks=true,
xlabel={$L_z$},
every outer y axis line/.append style={black},
every y tick label/.append style={font=\color{black}},
every y tick/.append style={black},
ymin=-1,
ymax=0,
ylabel={$\omega_r$},
axis background/.style={fill=white},
clip=true,
clip mode=individual,
axis lines=box,
xtick pos=bottom,
ytick pos=left,
tick align=inside,
major tick length=2pt,
width=0.7\linewidth,
height=0.7\linewidth,
every axis/.append style={font=\fontsize{9}{9}\selectfont},xlabel style={font=\fontsize{9}{9}\selectfont},title style={font=\fontsize{9}{9}\selectfont},
legend style={font=\fontsize{8}{9}\selectfont,inner sep=1pt,row sep=1pt,column sep=2pt,nodes={scale=1.00},draw=none},legend image post style={xscale=0.35},
legend columns=1,
scaled x ticks=false,
tick label style={/pgf/number format/fixed,/pgf/number format/precision=2},
every x tick label/.append style={font=\fontsize{9}{9}\selectfont\color{black}},
every y tick label/.append style={font=\fontsize{9}{9}\selectfont\color{black}},
,,
colormap={sigmamap}{rgb(0.0000)=(0.0200,0.1880,0.3800); rgb(0.1000)=(0.0743,0.3456,0.5616); rgb(0.2000)=(0.1918,0.4980,0.7060); rgb(0.3000)=(0.3890,0.6708,0.8226); rgb(0.4000)=(0.7552,0.8682,0.9228); rgb(0.5000)=(0.9690,0.9689,0.9689); rgb(0.6000)=(0.9425,0.8044,0.7614); rgb(0.7000)=(0.8767,0.4910,0.4020); rgb(0.8000)=(0.7869,0.2866,0.2470); rgb(0.9000)=(0.6186,0.1239,0.1568); rgb(1.0000)=(0.4040,0.0000,0.1220)}
]
\addplot [color=black, line width=1.2pt, mark size=1.5pt, mark=o, mark options={solid, black}, forget plot]
  table[row sep=crcr]{%
128	-0.0952982606773163\\
256	-0.0952527810696789\\
394	-0.095244243833798\\
512	-0.0952417234262464\\
};
\addplot [color=black, dashed, line width=1.2pt, mark size=1.5pt, mark=o, mark options={solid, black}, forget plot]
  table[row sep=crcr]{%
128	-0.95296321288197\\
256	-0.952525443979643\\
394	-0.952441794491463\\
512	-0.952416941891964\\
};
\addplot [color=black, dotted, line width=1.2pt, mark size=1.5pt, mark=o, mark options={solid, black}, forget plot]
  table[row sep=crcr]{%
128	-0.10532986633925\\
256	-0.105279414877403\\
394	-0.105269960555207\\
512	-0.105267171099137\\
};
\addplot [color=white!60!black, line width=1.2pt, mark size=1.5pt, mark=o, mark options={solid, white!60!black}, forget plot]
  table[row sep=crcr]{%
128	-0.0714737041396149\\
256	-0.0714395868337803\\
394	-0.0714331831543254\\
512	-0.0714312926960723\\
};
\addplot [color=white!60!black, dashed, line width=1.2pt, mark size=1.5pt, mark=o, mark options={solid, white!60!black}, forget plot]
  table[row sep=crcr]{%
128	-0.714731555359999\\
256	-0.714395203746993\\
394	-0.714331651221535\\
512	-0.714312845169765\\
};
\addplot [color=white!60!black, dotted, line width=1.2pt, mark size=1.5pt, mark=o, mark options={solid, white!60!black}, forget plot]
  table[row sep=crcr]{%
128	-0.166772918384776\\
256	-0.166692482512425\\
394	-0.166677457999391\\
512	-0.166673030172501\\
};
\addplot [color=red!60!teal, line width=1.2pt, mark size=1.5pt, mark=o, mark options={solid, red!60!teal}, forget plot]
  table[row sep=crcr]{%
128	-0.0571790116052954\\
256	-0.0571516752853596\\
394	-0.0571465480992065\\
512	-0.057145034871057\\
};
\addplot [color=red!60!teal, dashed, line width=1.2pt, mark size=1.5pt, mark=o, mark options={solid, red!60!teal}, forget plot]
  table[row sep=crcr]{%
128	-0.571786364240081\\
256	-0.571516299128528\\
394	-0.571465357956675\\
512	-0.571450292917447\\
};
\addplot [color=red!60!teal, dotted, line width=1.2pt, mark size=1.5pt, mark=o, mark options={solid, red!60!teal}, forget plot]
  table[row sep=crcr]{%
128	-0.400255712071431\\
256	-0.400062042724365\\
394	-0.400025922080265\\
512	-0.400015282773083\\
};
\node[right, align=left, inner sep=0]
at (rel axis cs:-0.25,1.15) {$(g)$};
\end{axis}
\end{tikzpicture}%
\fi
    \end{minipage}\hfill
    \begin{minipage}[t]{0.32\textwidth}
        \centering\vspace{0pt}
        \ifloadfig
%
\begin{tikzpicture}

\begin{axis}[%
width=11.255in,
height=7.131in,
at={(1.888in,0.962in)},
scale only axis,
clip=true,
separate axis lines,
every outer x axis line/.append style={black},
every x tick label/.append style={font=\color{black}},
every x tick/.append style={black},
xmode=log,
xmin=128,
xmax=512,
xtick={128,256,394,512},
xticklabels={{128},{256},{394},{512}},
xminorticks=true,
xlabel={$L_z$},
every outer y axis line/.append style={black},
every y tick label/.append style={font=\color{black}},
every y tick/.append style={black},
ymin=-0.2,
ymax=0.15,
axis background/.style={fill=white},
clip=true,
clip mode=individual,
axis lines=box,
xtick pos=bottom,
ytick pos=left,
tick align=inside,
major tick length=2pt,
width=0.7\linewidth,
height=0.7\linewidth,
every axis/.append style={font=\fontsize{9}{9}\selectfont},xlabel style={font=\fontsize{9}{9}\selectfont},title style={font=\fontsize{9}{9}\selectfont},
legend style={font=\fontsize{8}{9}\selectfont,inner sep=1pt,row sep=1pt,column sep=2pt,nodes={scale=1.00},draw=none},legend image post style={xscale=0.35},
legend columns=1,
scaled x ticks=false,
tick label style={/pgf/number format/fixed,/pgf/number format/precision=2},
every x tick label/.append style={font=\fontsize{9}{9}\selectfont\color{black}},
every y tick label/.append style={font=\fontsize{9}{9}\selectfont\color{black}},
,,
colormap={sigmamap}{rgb(0.0000)=(0.0200,0.1880,0.3800); rgb(0.1000)=(0.0743,0.3456,0.5616); rgb(0.2000)=(0.1918,0.4980,0.7060); rgb(0.3000)=(0.3890,0.6708,0.8226); rgb(0.4000)=(0.7552,0.8682,0.9228); rgb(0.5000)=(0.9690,0.9689,0.9689); rgb(0.6000)=(0.9425,0.8044,0.7614); rgb(0.7000)=(0.8767,0.4910,0.4020); rgb(0.8000)=(0.7869,0.2866,0.2470); rgb(0.9000)=(0.6186,0.1239,0.1568); rgb(1.0000)=(0.4040,0.0000,0.1220)}
]
\addplot [color=black, line width=1.2pt, mark size=1.5pt, mark=o, mark options={solid, black}, forget plot]
  table[row sep=crcr]{%
128	-0.0952982466641561\\
256	-0.0952527793812959\\
394	-0.0952442433764767\\
512	-0.0952417232189722\\
};
\addplot [color=black, dashed, line width=1.2pt, mark size=1.5pt, mark=o, mark options={solid, black}, forget plot]
  table[row sep=crcr]{%
128	-0.190458142669267\\
256	-0.190458142670646\\
394	-0.190458142673776\\
512	-0.19045814267581\\
};
\addplot [color=black, dotted, line width=1.2pt, mark size=1.5pt, mark=o, mark options={solid, black}, forget plot]
  table[row sep=crcr]{%
128	-0.105329598276117\\
256	-0.105279382579788\\
394	-0.105269951808352\\
512	-0.105267167134979\\
};
\addplot [color=white!60!black, line width=1.2pt, mark size=1.5pt, mark=o, mark options={solid, white!60!black}, forget plot]
  table[row sep=crcr]{%
128	-0.0679281743485471\\
256	-0.0679281743486295\\
394	-0.0679281743561332\\
512	-0.0679281743568297\\
};
\addplot [color=white!60!black, dashed, line width=1.2pt, mark size=1.5pt, mark=o, mark options={solid, white!60!black}, forget plot]
  table[row sep=crcr]{%
128	-0.069245139389848\\
256	-0.0692451393985892\\
394	-0.0692451393988058\\
512	-0.0692451394109186\\
};
\addplot [color=white!60!black, dotted, line width=1.2pt, mark size=1.5pt, mark=o, mark options={solid, white!60!black}, forget plot]
  table[row sep=crcr]{%
128	-0.0223082893654187\\
256	-0.0223082893794375\\
394	-0.0223082894065944\\
512	-0.0223082893636329\\
};
\addplot [color=red!60!teal, line width=1.2pt, mark size=1.5pt, mark=o, mark options={solid, red!60!teal}, forget plot]
  table[row sep=crcr]{%
128	0.00564379562323961\\
256	0.00564379567368133\\
394	0.00564379572054976\\
512	0.00564379559877124\\
};
\addplot [color=red!60!teal, dashed, line width=1.2pt, mark size=1.5pt, mark=o, mark options={solid, red!60!teal}, forget plot]
  table[row sep=crcr]{%
128	0.0393637048045937\\
256	0.039363704802539\\
394	0.0393637047464161\\
512	0.0393637047760262\\
};
\addplot [color=red!60!teal, dotted, line width=1.2pt, mark size=1.5pt, mark=o, mark options={solid, red!60!teal}, forget plot]
  table[row sep=crcr]{%
128	0.131974232597036\\
256	0.131974232679927\\
394	0.131974232756384\\
512	0.131974232513172\\
};
\node[right, align=left, inner sep=0]
at (rel axis cs:-0.25,1.15) {$(h)$};
\end{axis}
\end{tikzpicture}%
\fi
    \end{minipage}\hfill
    \begin{minipage}[t]{0.32\textwidth}
        \centering\vspace{0pt}
        \ifloadfig
%
\begin{tikzpicture}

\begin{axis}[%
width=11.255in,
height=7.131in,
at={(1.888in,0.962in)},
scale only axis,
clip=true,
separate axis lines,
every outer x axis line/.append style={black},
every x tick label/.append style={font=\color{black}},
every x tick/.append style={black},
xmode=log,
xmin=128,
xmax=512,
xtick={128,256,394,512},
xticklabels={{128},{256},{394},{512}},
xminorticks=true,
xlabel={$L_z$},
every outer y axis line/.append style={black},
every y tick label/.append style={font=\color{black}},
every y tick/.append style={black},
ymin=-0.05,
ymax=0.3,
axis background/.style={fill=white},
clip=true,
clip mode=individual,
axis lines=box,
xtick pos=bottom,
ytick pos=left,
tick align=inside,
major tick length=2pt,
width=0.7\linewidth,
height=0.7\linewidth,
every axis/.append style={font=\fontsize{9}{9}\selectfont},xlabel style={font=\fontsize{9}{9}\selectfont},title style={font=\fontsize{9}{9}\selectfont},
legend style={font=\fontsize{8}{9}\selectfont,inner sep=1pt,row sep=1pt,column sep=2pt,nodes={scale=1.00},draw=none},legend image post style={xscale=0.35},
legend columns=1,
scaled x ticks=false,
tick label style={/pgf/number format/fixed,/pgf/number format/precision=2},
every x tick label/.append style={font=\fontsize{9}{9}\selectfont\color{black}},
every y tick label/.append style={font=\fontsize{9}{9}\selectfont\color{black}},
,,
colormap={sigmamap}{rgb(0.0000)=(0.0200,0.1880,0.3800); rgb(0.1000)=(0.0743,0.3456,0.5616); rgb(0.2000)=(0.1918,0.4980,0.7060); rgb(0.3000)=(0.3890,0.6708,0.8226); rgb(0.4000)=(0.7552,0.8682,0.9228); rgb(0.5000)=(0.9690,0.9689,0.9689); rgb(0.6000)=(0.9425,0.8044,0.7614); rgb(0.7000)=(0.8767,0.4910,0.4020); rgb(0.8000)=(0.7869,0.2866,0.2470); rgb(0.9000)=(0.6186,0.1239,0.1568); rgb(1.0000)=(0.4040,0.0000,0.1220)}
]
\addplot [color=black, line width=1.2pt, mark size=1.5pt, mark=o, mark options={solid, black}, forget plot]
  table[row sep=crcr]{%
128	-0.00792568601868533\\
256	-0.00792568601903572\\
394	-0.0079256860188494\\
512	-0.00792568601850888\\
};
\addplot [color=black, dashed, line width=1.2pt, mark size=1.5pt, mark=o, mark options={solid, black}, forget plot]
  table[row sep=crcr]{%
128	-0.00941347291074579\\
256	-0.00941347290931994\\
394	-0.00941347290975314\\
512	-0.00941347291188175\\
};
\addplot [color=black, dotted, line width=1.2pt, mark size=1.5pt, mark=o, mark options={solid, black}, forget plot]
  table[row sep=crcr]{%
128	-0.00697642863090525\\
256	-0.00697642862933771\\
394	-0.0069764286295566\\
512	-0.0069764286322495\\
};
\addplot [color=white!60!black, line width=1.2pt, mark size=1.5pt, mark=o, mark options={solid, white!60!black}, forget plot]
  table[row sep=crcr]{%
128	0.117214372696531\\
256	0.117214372699866\\
394	0.117214372693807\\
512	0.117214372684362\\
};
\addplot [color=white!60!black, dashed, line width=1.2pt, mark size=1.5pt, mark=o, mark options={solid, white!60!black}, forget plot]
  table[row sep=crcr]{%
128	0.115201032656291\\
256	0.115201032669157\\
394	0.115201032667357\\
512	0.115201032641779\\
};
\addplot [color=white!60!black, dotted, line width=1.2pt, mark size=1.5pt, mark=o, mark options={solid, white!60!black}, forget plot]
  table[row sep=crcr]{%
128	0.130051336265891\\
256	0.130051336273315\\
394	0.130051336246591\\
512	0.130051336262111\\
};
\addplot [color=red!60!teal, line width=1.2pt, mark size=1.5pt, mark=o, mark options={solid, red!60!teal}, forget plot]
  table[row sep=crcr]{%
128	0.220280259136001\\
256	0.220280259131521\\
394	0.220280259141977\\
512	0.220280259129206\\
};
\addplot [color=red!60!teal, dashed, line width=1.2pt, mark size=1.5pt, mark=o, mark options={solid, red!60!teal}, forget plot]
  table[row sep=crcr]{%
128	0.224001034230161\\
256	0.224001034237988\\
394	0.22400103421568\\
512	0.224001034200623\\
};
\addplot [color=red!60!teal, dotted, line width=1.2pt, mark size=1.5pt, mark=o, mark options={solid, red!60!teal}, forget plot]
  table[row sep=crcr]{%
128	0.253697220175179\\
256	0.253697220163799\\
394	0.25369722019104\\
512	0.253697220130055\\
};
\node[right, align=left, inner sep=0]
at (rel axis cs:-0.25,1.15) {$(i)$};
\end{axis}
\end{tikzpicture}%
\fi
    \end{minipage}
    \caption{Sensitivity of the leading growth rate \(\omega_r\) to the domain half-height \(L_z\) at fixed \(N=256\), showing convergence with respect to domain size across representative parameters. Colors identify \(\Atw\), as indicated in (a), and line styles (\solidlegend, \dashlegend, \dottedlegend) identify \(\Amu \in \{-0.9, 0, 0.9\}\), respectively. (a) \(k=10^{-2}\), \(\Sch=0.05\). (b) \(k=10^{-2}\), \(\Sch=5\). (c) \(k=10^{-2}\), \(\Sch=40\). (d) \(k=10^{-1}\), \(\Sch=0.05\). (e) \(k=10^{-1}\), \(\Sch=5\). (f) \(k=10^{-1}\), \(\Sch=40\). (g) \(k=10^{0}\), \(\Sch=0.05\). (h) \(k=10^{0}\), \(\Sch=5\). (i) \(k=10^{0}\), \(\Sch=40\).}
    \label{fig:V16_wrLz}
\end{figure}
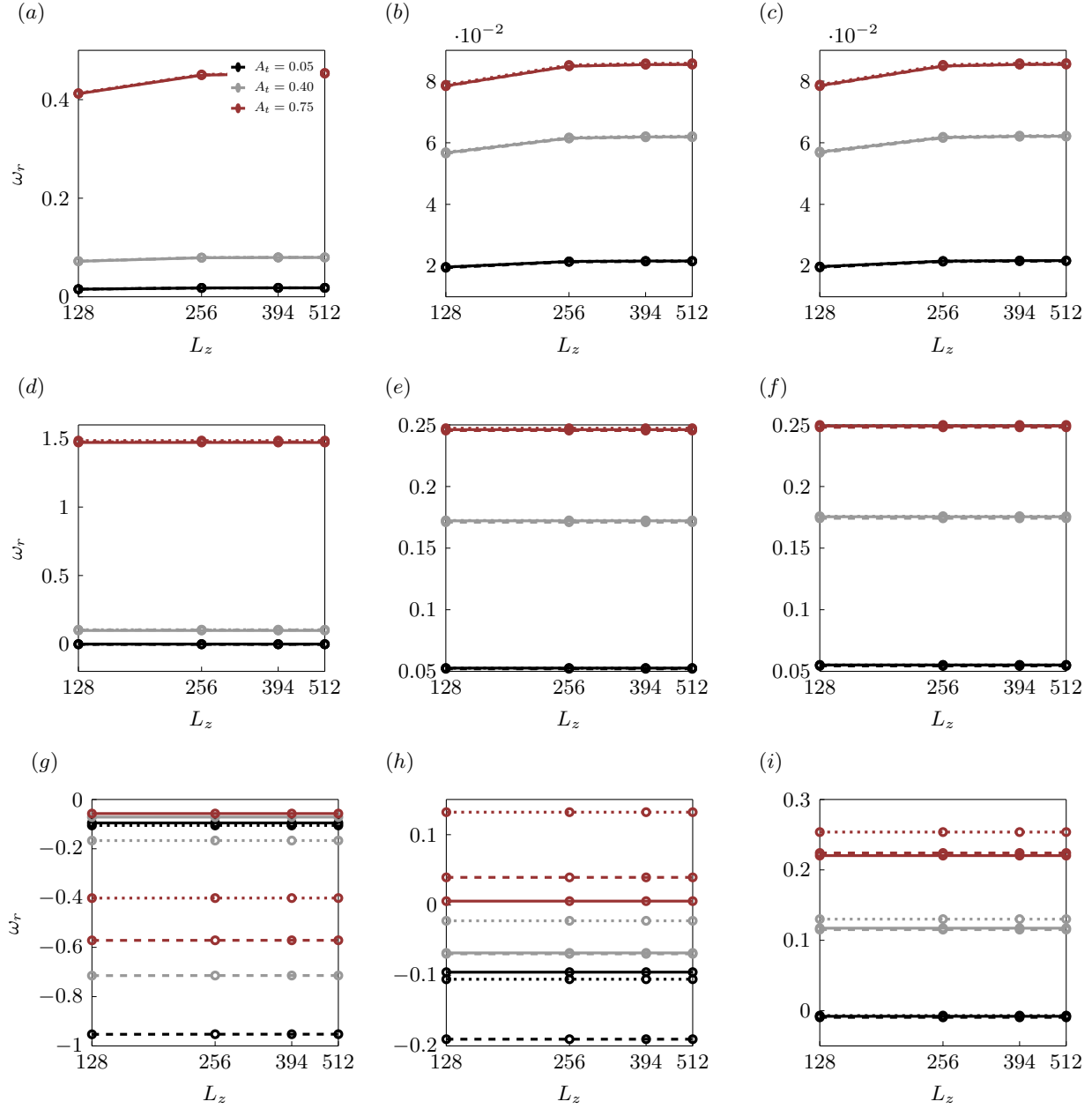
Figures~\ref{fig:V16_wrLz} and~\ref{fig:V16_wrN} examine the sensitivity of the leading growth rate \(\omega_r\) to the domain half-height \(L_z\) and the grid resolution \(N\) over the parameter space \(\Atw \in \{0.05, 0.4, 0.75\}\), \(\Amu \in \{-0.9, 0.0, 0.9\}\), and \(\Sch \in \{0.05, 5, 40\}\), at the representative wavenumbers \(k \in \{10^{-2},\,10^{-1},\,10^{0}\}\). In the domain study, \(L_z\) is varied from \(128\) to \(512\) at fixed \(N=256\). In the resolution study, \(N\) is varied from \(64\) to \(192\) at fixed \(L_z=128\). For each \((N,L_z)\) pair, the sinh-stretching parameter \(\beta\) is recomputed so as to maintain a fixed fraction of the Chebyshev nodes within the interfacial region.
\begin{figure}
    \centering
    \begin{minipage}[t]{0.32\textwidth}
        \centering\vspace{0pt}
        \ifloadfig
%
\begin{tikzpicture}

\begin{axis}[%
width=11.255in,
height=7.131in,
at={(1.888in,0.962in)},
scale only axis,
clip=true,
separate axis lines,
every outer x axis line/.append style={black},
every x tick label/.append style={font=\color{black}},
every x tick/.append style={black},
xmin=60,
xmax=260,
xtick={ 64,  96, 128, 192, 256},
xlabel={$N$},
every outer y axis line/.append style={black},
every y tick label/.append style={font=\color{black}},
every y tick/.append style={black},
ymin=0,
ymax=1.4,
ylabel={$\omega_r$},
axis background/.style={fill=white},
legend style={legend cell align=left, align=left},
clip=true,
clip mode=individual,
axis lines=box,
xtick pos=bottom,
ytick pos=left,
tick align=inside,
major tick length=2pt,
width=0.7\linewidth,
height=0.7\linewidth,
every axis/.append style={font=\fontsize{9}{9}\selectfont},xlabel style={font=\fontsize{9}{9}\selectfont},title style={font=\fontsize{9}{9}\selectfont},
legend style={font=\fontsize{8}{9}\selectfont,inner sep=1pt,row sep=1pt,column sep=2pt,nodes={scale=0.70},draw=none},legend image post style={xscale=0.35},
legend columns=1,
scaled x ticks=false,
tick label style={/pgf/number format/fixed,/pgf/number format/precision=2},
every x tick label/.append style={font=\fontsize{9}{9}\selectfont\color{black}},
every y tick label/.append style={font=\fontsize{9}{9}\selectfont\color{black}},
,,
colormap={sigmamap}{rgb(0.0000)=(0.0200,0.1880,0.3800); rgb(0.1000)=(0.0743,0.3456,0.5616); rgb(0.2000)=(0.1918,0.4980,0.7060); rgb(0.3000)=(0.3890,0.6708,0.8226); rgb(0.4000)=(0.7552,0.8682,0.9228); rgb(0.5000)=(0.9690,0.9689,0.9689); rgb(0.6000)=(0.9425,0.8044,0.7614); rgb(0.7000)=(0.8767,0.4910,0.4020); rgb(0.8000)=(0.7869,0.2866,0.2470); rgb(0.9000)=(0.6186,0.1239,0.1568); rgb(1.0000)=(0.4040,0.0000,0.1220)}
]
\addplot [color=black, line width=1.2pt, mark size=1.5pt, mark=o, mark options={solid, black}]
  table[row sep=crcr]{%
64	0.015175627386949\\
96	0.0153918395236571\\
128	0.0153889995498685\\
192	0.0153901674567938\\
256	0.0153889793972656\\
};
\addlegendentry{$A_t = 0.05$}

\addplot [color=black, dashed, line width=1.2pt, mark size=1.5pt, mark=o, mark options={solid, black}, forget plot]
  table[row sep=crcr]{%
64	0.015160021467809\\
96	0.0154153529163436\\
128	0.0154120926102264\\
192	0.015411751885999\\
256	0.0154122449079616\\
};
\addplot [color=black, dotted, line width=1.2pt, mark size=1.5pt, mark=o, mark options={solid, black}, forget plot]
  table[row sep=crcr]{%
64	0.0153353769083539\\
96	0.015518629806578\\
128	0.0155158277102104\\
192	0.0155142610854477\\
256	0.0155158254201716\\
};
\addplot [color=white!60!black, line width=1.2pt, mark size=1.5pt, mark=o, mark options={solid, white!60!black}]
  table[row sep=crcr]{%
64	0.0782269179656722\\
96	0.0719230103337248\\
128	0.0720038272867779\\
192	0.0720037615922001\\
256	0.0720014461552192\\
};
\addlegendentry{$A_t = 0.40$}

\addplot [color=white!60!black, dashed, line width=1.2pt, mark size=1.5pt, mark=o, mark options={solid, white!60!black}, forget plot]
  table[row sep=crcr]{%
64	0.0877620276264793\\
96	0.072101724438759\\
128	0.0721575578140409\\
192	0.0721519169415108\\
256	0.0721555681684248\\
};
\addplot [color=white!60!black, dotted, line width=1.2pt, mark size=1.5pt, mark=o, mark options={solid, white!60!black}, forget plot]
  table[row sep=crcr]{%
64	0.0922604830259305\\
96	0.0725335279496808\\
128	0.0725454010868035\\
192	0.0725460754416952\\
256	0.0725431931109221\\
};
\addplot [color=red!60!teal, line width=1.2pt, mark size=1.5pt, mark=o, mark options={solid, red!60!teal}]
  table[row sep=crcr]{%
64	0.297704487296635\\
96	0.403140330483603\\
128	0.41043911960375\\
192	0.411913605683142\\
256	0.411919202509246\\
};
\addlegendentry{$A_t = 0.75$}

\addplot [color=red!60!teal, dashed, line width=1.2pt, mark size=1.5pt, mark=o, mark options={solid, red!60!teal}, forget plot]
  table[row sep=crcr]{%
64	0.458643040011452\\
96	0.399529478038575\\
128	0.410831428284674\\
192	0.412174947311133\\
256	0.412179887627474\\
};
\addplot [color=red!60!teal, dotted, line width=1.2pt, mark size=1.5pt, mark=o, mark options={solid, red!60!teal}, forget plot]
  table[row sep=crcr]{%
64	1.24279577030075\\
96	0.394739605890123\\
128	0.412217828365355\\
192	0.413183140392196\\
256	0.413186358139441\\
};
\node[right, align=left, inner sep=0]
at (rel axis cs:-0.25,1.15) {$(a)$};
\end{axis}
\end{tikzpicture}%
\fi
    \end{minipage}\hfill
    \begin{minipage}[t]{0.32\textwidth}
        \centering\vspace{0pt}
        \ifloadfig
%
\begin{tikzpicture}

\begin{axis}[%
width=11.255in,
height=7.131in,
at={(1.888in,0.962in)},
scale only axis,
clip=true,
separate axis lines,
every outer x axis line/.append style={black},
every x tick label/.append style={font=\color{black}},
every x tick/.append style={black},
xmin=60,
xmax=260,
xtick={ 64,  96, 128, 192, 256},
xlabel={$N$},
every outer y axis line/.append style={black},
every y tick label/.append style={font=\color{black}},
every y tick/.append style={black},
ymin=0.01,
ymax=0.08,
axis background/.style={fill=white},
clip=true,
clip mode=individual,
axis lines=box,
xtick pos=bottom,
ytick pos=left,
tick align=inside,
major tick length=2pt,
width=0.7\linewidth,
height=0.7\linewidth,
every axis/.append style={font=\fontsize{9}{9}\selectfont},xlabel style={font=\fontsize{9}{9}\selectfont},title style={font=\fontsize{9}{9}\selectfont},
legend style={font=\fontsize{8}{9}\selectfont,inner sep=1pt,row sep=1pt,column sep=2pt,nodes={scale=1.00},draw=none},legend image post style={xscale=0.35},
legend columns=1,
scaled x ticks=false,
tick label style={/pgf/number format/fixed,/pgf/number format/precision=2},
every x tick label/.append style={font=\fontsize{9}{9}\selectfont\color{black}},
every y tick label/.append style={font=\fontsize{9}{9}\selectfont\color{black}},
,,
colormap={sigmamap}{rgb(0.0000)=(0.0200,0.1880,0.3800); rgb(0.1000)=(0.0743,0.3456,0.5616); rgb(0.2000)=(0.1918,0.4980,0.7060); rgb(0.3000)=(0.3890,0.6708,0.8226); rgb(0.4000)=(0.7552,0.8682,0.9228); rgb(0.5000)=(0.9690,0.9689,0.9689); rgb(0.6000)=(0.9425,0.8044,0.7614); rgb(0.7000)=(0.8767,0.4910,0.4020); rgb(0.8000)=(0.7869,0.2866,0.2470); rgb(0.9000)=(0.6186,0.1239,0.1568); rgb(1.0000)=(0.4040,0.0000,0.1220)}
]
\addplot [color=black, line width=1.2pt, mark size=1.5pt, mark=o, mark options={solid, black}, forget plot]
  table[row sep=crcr]{%
64	0.0196280142974325\\
96	0.0196441846240341\\
128	0.0196439674111534\\
192	0.0196440165476536\\
256	0.0196439567233836\\
};
\addplot [color=black, dashed, line width=1.2pt, mark size=1.5pt, mark=o, mark options={solid, black}, forget plot]
  table[row sep=crcr]{%
64	0.0194026894861962\\
96	0.0194202161975333\\
128	0.0194199802700521\\
192	0.0194199070663958\\
256	0.0194199983019396\\
};
\addplot [color=black, dotted, line width=1.2pt, mark size=1.5pt, mark=o, mark options={solid, black}, forget plot]
  table[row sep=crcr]{%
64	0.0196456206569742\\
96	0.0196567693750104\\
128	0.0196565707781555\\
192	0.0196565709361637\\
256	0.019656550828726\\
};
\addplot [color=white!60!black, line width=1.2pt, mark size=1.5pt, mark=o, mark options={solid, white!60!black}, forget plot]
  table[row sep=crcr]{%
64	0.0549854275576408\\
96	0.0567359835580714\\
128	0.0567397466976406\\
192	0.0567395259014492\\
256	0.0567396021543479\\
};
\addplot [color=white!60!black, dashed, line width=1.2pt, mark size=1.5pt, mark=o, mark options={solid, white!60!black}, forget plot]
  table[row sep=crcr]{%
64	0.0550610595307081\\
96	0.0565088085543453\\
128	0.0565097718821789\\
192	0.0565092707871196\\
256	0.0565094818669938\\
};
\addplot [color=white!60!black, dotted, line width=1.2pt, mark size=1.5pt, mark=o, mark options={solid, white!60!black}, forget plot]
  table[row sep=crcr]{%
64	0.0561844236656804\\
96	0.0569355106100715\\
128	0.0569331417503098\\
192	0.0569331037059805\\
256	0.0569328808131031\\
};
\addplot [color=red!60!teal, line width=1.2pt, mark size=1.5pt, mark=o, mark options={solid, red!60!teal}, forget plot]
  table[row sep=crcr]{%
64	0.0443407017924952\\
96	0.0768756756164607\\
128	0.0784509930909778\\
192	0.078463890146153\\
256	0.0784638381384213\\
};
\addplot [color=red!60!teal, dashed, line width=1.2pt, mark size=1.5pt, mark=o, mark options={solid, red!60!teal}, forget plot]
  table[row sep=crcr]{%
64	0.0402429720542245\\
96	0.0762321835317126\\
128	0.0783661323242585\\
192	0.0783557402781546\\
256	0.0783562037416023\\
};
\addplot [color=red!60!teal, dotted, line width=1.2pt, mark size=1.5pt, mark=o, mark options={solid, red!60!teal}, forget plot]
  table[row sep=crcr]{%
64	0.0418056066023027\\
96	0.076520407645959\\
128	0.0788374651051257\\
192	0.0788074472117122\\
256	0.0788068491041872\\
};
\node[right, align=left, inner sep=0]
at (rel axis cs:-0.25,1.15) {$(b)$};
\end{axis}
\end{tikzpicture}%
\fi
    \end{minipage}\hfill
    \begin{minipage}[t]{0.32\textwidth}
        \centering\vspace{0pt}
        \ifloadfig
%
\begin{tikzpicture}

\begin{axis}[%
width=11.255in,
height=7.131in,
at={(1.888in,0.962in)},
scale only axis,
clip=true,
separate axis lines,
every outer x axis line/.append style={black},
every x tick label/.append style={font=\color{black}},
every x tick/.append style={black},
xmin=60,
xmax=260,
xtick={ 64,  96, 128, 192, 256},
xlabel={$N$},
every outer y axis line/.append style={black},
every y tick label/.append style={font=\color{black}},
every y tick/.append style={black},
ymin=0.01,
ymax=0.08,
axis background/.style={fill=white},
clip=true,
clip mode=individual,
axis lines=box,
xtick pos=bottom,
ytick pos=left,
tick align=inside,
major tick length=2pt,
width=0.7\linewidth,
height=0.7\linewidth,
every axis/.append style={font=\fontsize{9}{9}\selectfont},xlabel style={font=\fontsize{9}{9}\selectfont},title style={font=\fontsize{9}{9}\selectfont},
legend style={font=\fontsize{8}{9}\selectfont,inner sep=1pt,row sep=1pt,column sep=2pt,nodes={scale=1.00},draw=none},legend image post style={xscale=0.35},
legend columns=1,
scaled x ticks=false,
tick label style={/pgf/number format/fixed,/pgf/number format/precision=2},
every x tick label/.append style={font=\fontsize{9}{9}\selectfont\color{black}},
every y tick label/.append style={font=\fontsize{9}{9}\selectfont\color{black}},
,,
colormap={sigmamap}{rgb(0.0000)=(0.0200,0.1880,0.3800); rgb(0.1000)=(0.0743,0.3456,0.5616); rgb(0.2000)=(0.1918,0.4980,0.7060); rgb(0.3000)=(0.3890,0.6708,0.8226); rgb(0.4000)=(0.7552,0.8682,0.9228); rgb(0.5000)=(0.9690,0.9689,0.9689); rgb(0.6000)=(0.9425,0.8044,0.7614); rgb(0.7000)=(0.8767,0.4910,0.4020); rgb(0.8000)=(0.7869,0.2866,0.2470); rgb(0.9000)=(0.6186,0.1239,0.1568); rgb(1.0000)=(0.4040,0.0000,0.1220)}
]
\addplot [color=black, line width=1.2pt, mark size=1.5pt, mark=o, mark options={solid, black}, forget plot]
  table[row sep=crcr]{%
64	0.0197794687276358\\
96	0.0197835246392758\\
128	0.0197834663783663\\
192	0.0197834646890771\\
256	0.019783459728364\\
};
\addplot [color=black, dashed, line width=1.2pt, mark size=1.5pt, mark=o, mark options={solid, black}, forget plot]
  table[row sep=crcr]{%
64	0.0195270053252691\\
96	0.0195297242756072\\
128	0.0195296725637299\\
192	0.019529643143256\\
256	0.019529680739418\\
};
\addplot [color=black, dotted, line width=1.2pt, mark size=1.5pt, mark=o, mark options={solid, black}, forget plot]
  table[row sep=crcr]{%
64	0.0197948458203032\\
96	0.0197944143673393\\
128	0.0197943903369736\\
192	0.0197944474146721\\
256	0.0197944255824842\\
};
\addplot [color=white!60!black, line width=1.2pt, mark size=1.5pt, mark=o, mark options={solid, white!60!black}, forget plot]
  table[row sep=crcr]{%
64	0.0570045264520412\\
96	0.0569746924397397\\
128	0.0569735449243181\\
192	0.0569734815761161\\
256	0.0569735529244281\\
};
\addplot [color=white!60!black, dashed, line width=1.2pt, mark size=1.5pt, mark=o, mark options={solid, white!60!black}, forget plot]
  table[row sep=crcr]{%
64	0.0567954690712955\\
96	0.0567013267451753\\
128	0.0566990585508207\\
192	0.056699035472971\\
256	0.0566990587024551\\
};
\addplot [color=white!60!black, dotted, line width=1.2pt, mark size=1.5pt, mark=o, mark options={solid, white!60!black}, forget plot]
  table[row sep=crcr]{%
64	0.0573105229508202\\
96	0.0571502195452484\\
128	0.0571475410911411\\
192	0.0571476044850767\\
256	0.0571475461472267\\
};
\addplot [color=red!60!teal, line width=1.2pt, mark size=1.5pt, mark=o, mark options={solid, red!60!teal}, forget plot]
  table[row sep=crcr]{%
64	0.0583145502767441\\
96	0.0776087324873886\\
128	0.0785651938820828\\
192	0.0785344518703883\\
256	0.0785344458669422\\
};
\addplot [color=red!60!teal, dashed, line width=1.2pt, mark size=1.5pt, mark=o, mark options={solid, red!60!teal}, forget plot]
  table[row sep=crcr]{%
64	0.0603680884987433\\
96	0.0773030910729599\\
128	0.0784261701912827\\
192	0.0783812211929558\\
256	0.0783812231554221\\
};
\addplot [color=red!60!teal, dotted, line width=1.2pt, mark size=1.5pt, mark=o, mark options={solid, red!60!teal}, forget plot]
  table[row sep=crcr]{%
64	0.0653544797396264\\
96	0.0778609126740445\\
128	0.0789064325120356\\
192	0.0788609365868329\\
256	0.0788607448988685\\
};
\node[right, align=left, inner sep=0]
at (rel axis cs:-0.25,1.15) {$(c)$};
\end{axis}
\end{tikzpicture}%
\fi
    \end{minipage}\\[0.5em]
    \begin{minipage}[t]{0.32\textwidth}
        \centering\vspace{0pt}
        \ifloadfig
%
\begin{tikzpicture}

\begin{axis}[%
width=11.255in,
height=7.131in,
at={(1.888in,0.962in)},
scale only axis,
clip=true,
separate axis lines,
every outer x axis line/.append style={black},
every x tick label/.append style={font=\color{black}},
every x tick/.append style={black},
xmin=60,
xmax=260,
xtick={ 64,  96, 128, 192, 256},
xlabel={$N$},
every outer y axis line/.append style={black},
every y tick label/.append style={font=\color{black}},
every y tick/.append style={black},
ymin=-0.2,
ymax=1.6,
ylabel={$\omega_r$},
axis background/.style={fill=white},
clip=true,
clip mode=individual,
axis lines=box,
xtick pos=bottom,
ytick pos=left,
tick align=inside,
major tick length=2pt,
width=0.7\linewidth,
height=0.7\linewidth,
every axis/.append style={font=\fontsize{9}{9}\selectfont},xlabel style={font=\fontsize{9}{9}\selectfont},title style={font=\fontsize{9}{9}\selectfont},
legend style={font=\fontsize{8}{9}\selectfont,inner sep=1pt,row sep=1pt,column sep=2pt,nodes={scale=1.00},draw=none},legend image post style={xscale=0.35},
legend columns=1,
scaled x ticks=false,
tick label style={/pgf/number format/fixed,/pgf/number format/precision=2},
every x tick label/.append style={font=\fontsize{9}{9}\selectfont\color{black}},
every y tick label/.append style={font=\fontsize{9}{9}\selectfont\color{black}},
,,
colormap={sigmamap}{rgb(0.0000)=(0.0200,0.1880,0.3800); rgb(0.1000)=(0.0743,0.3456,0.5616); rgb(0.2000)=(0.1918,0.4980,0.7060); rgb(0.3000)=(0.3890,0.6708,0.8226); rgb(0.4000)=(0.7552,0.8682,0.9228); rgb(0.5000)=(0.9690,0.9689,0.9689); rgb(0.6000)=(0.9425,0.8044,0.7614); rgb(0.7000)=(0.8767,0.4910,0.4020); rgb(0.8000)=(0.7869,0.2866,0.2470); rgb(0.9000)=(0.6186,0.1239,0.1568); rgb(1.0000)=(0.4040,0.0000,0.1220)}
]
\addplot [color=black, line width=1.2pt, mark size=1.5pt, mark=o, mark options={solid, black}, forget plot]
  table[row sep=crcr]{%
64	-0.000993761100114939\\
96	-0.000993739398262548\\
128	-0.000993739950969389\\
192	-0.000993741127721846\\
256	-0.000993740340299722\\
};
\addplot [color=black, dashed, line width=1.2pt, mark size=1.5pt, mark=o, mark options={solid, black}, forget plot]
  table[row sep=crcr]{%
64	-0.00301354305557083\\
96	-0.00301358113240564\\
128	-0.00301358148008734\\
192	-0.00301359296872225\\
256	-0.00301358636765227\\
};
\addplot [color=black, dotted, line width=1.2pt, mark size=1.5pt, mark=o, mark options={solid, black}, forget plot]
  table[row sep=crcr]{%
64	-0.00107002281983706\\
96	-0.00106993448183487\\
128	-0.00106993482082553\\
192	-0.00106993676027198\\
256	-0.00106993805265659\\
};
\addplot [color=white!60!black, line width=1.2pt, mark size=1.5pt, mark=o, mark options={solid, white!60!black}, forget plot]
  table[row sep=crcr]{%
64	0.0993504663763069\\
96	0.0983478963248349\\
128	0.0983481750926426\\
192	0.0983481085839242\\
256	0.0983481682580654\\
};
\addplot [color=white!60!black, dashed, line width=1.2pt, mark size=1.5pt, mark=o, mark options={solid, white!60!black}, forget plot]
  table[row sep=crcr]{%
64	0.10379489905161\\
96	0.101561514617896\\
128	0.101561498537618\\
192	0.101561520528346\\
256	0.101561488021553\\
};
\addplot [color=white!60!black, dotted, line width=1.2pt, mark size=1.5pt, mark=o, mark options={solid, white!60!black}, forget plot]
  table[row sep=crcr]{%
64	0.106006960142888\\
96	0.104171461417367\\
128	0.104171464594514\\
192	0.104171514686636\\
256	0.104171468304152\\
};
\addplot [color=red!60!teal, line width=1.2pt, mark size=1.5pt, mark=o, mark options={solid, red!60!teal}, forget plot]
  table[row sep=crcr]{%
64	1.27134620738182\\
96	1.47047066019017\\
128	1.4714719844176\\
192	1.47146913484269\\
256	1.47146919294046\\
};
\addplot [color=red!60!teal, dashed, line width=1.2pt, mark size=1.5pt, mark=o, mark options={solid, red!60!teal}, forget plot]
  table[row sep=crcr]{%
64	1.29769172627066\\
96	1.47525986432074\\
128	1.47611362437013\\
192	1.47611533782889\\
256	1.47611531434549\\
};
\addplot [color=red!60!teal, dotted, line width=1.2pt, mark size=1.5pt, mark=o, mark options={solid, red!60!teal}, forget plot]
  table[row sep=crcr]{%
64	1.36990485224449\\
96	1.48475646855977\\
128	1.48563606891222\\
192	1.48563896383182\\
256	1.48563897911697\\
};
\node[right, align=left, inner sep=0]
at (rel axis cs:-0.25,1.15) {$(d)$};
\end{axis}
\end{tikzpicture}%
\fi
    \end{minipage}\hfill
    \begin{minipage}[t]{0.32\textwidth}
        \centering\vspace{0pt}
        \ifloadfig
%
\begin{tikzpicture}

\begin{axis}[%
width=11.255in,
height=7.131in,
at={(1.888in,0.962in)},
scale only axis,
clip=true,
separate axis lines,
every outer x axis line/.append style={black},
every x tick label/.append style={font=\color{black}},
every x tick/.append style={black},
xmin=60,
xmax=260,
xtick={ 64,  96, 128, 192, 256},
xlabel={$N$},
every outer y axis line/.append style={black},
every y tick label/.append style={font=\color{black}},
every y tick/.append style={black},
ymin=0.05,
ymax=0.25,
axis background/.style={fill=white},
clip=true,
clip mode=individual,
axis lines=box,
xtick pos=bottom,
ytick pos=left,
tick align=inside,
major tick length=2pt,
width=0.7\linewidth,
height=0.7\linewidth,
every axis/.append style={font=\fontsize{9}{9}\selectfont},xlabel style={font=\fontsize{9}{9}\selectfont},title style={font=\fontsize{9}{9}\selectfont},
legend style={font=\fontsize{8}{9}\selectfont,inner sep=1pt,row sep=1pt,column sep=2pt,nodes={scale=1.00},draw=none},legend image post style={xscale=0.35},
legend columns=1,
scaled x ticks=false,
tick label style={/pgf/number format/fixed,/pgf/number format/precision=2},
every x tick label/.append style={font=\fontsize{9}{9}\selectfont\color{black}},
every y tick label/.append style={font=\fontsize{9}{9}\selectfont\color{black}},
,,
colormap={sigmamap}{rgb(0.0000)=(0.0200,0.1880,0.3800); rgb(0.1000)=(0.0743,0.3456,0.5616); rgb(0.2000)=(0.1918,0.4980,0.7060); rgb(0.3000)=(0.3890,0.6708,0.8226); rgb(0.4000)=(0.7552,0.8682,0.9228); rgb(0.5000)=(0.9690,0.9689,0.9689); rgb(0.6000)=(0.9425,0.8044,0.7614); rgb(0.7000)=(0.8767,0.4910,0.4020); rgb(0.8000)=(0.7869,0.2866,0.2470); rgb(0.9000)=(0.6186,0.1239,0.1568); rgb(1.0000)=(0.4040,0.0000,0.1220)}
]
\addplot [color=black, line width=1.2pt, mark size=1.5pt, mark=o, mark options={solid, black}, forget plot]
  table[row sep=crcr]{%
64	0.0526431388677941\\
96	0.0526429795630021\\
128	0.0526429797548603\\
192	0.0526429792917354\\
256	0.0526429801940198\\
};
\addplot [color=black, dashed, line width=1.2pt, mark size=1.5pt, mark=o, mark options={solid, black}, forget plot]
  table[row sep=crcr]{%
64	0.0520798917912442\\
96	0.0520798893716795\\
128	0.0520798893669788\\
192	0.052079891538816\\
256	0.0520798890662265\\
};
\addplot [color=black, dotted, line width=1.2pt, mark size=1.5pt, mark=o, mark options={solid, black}, forget plot]
  table[row sep=crcr]{%
64	0.0525857237135125\\
96	0.0525853772621551\\
128	0.0525853774297766\\
192	0.0525853792385804\\
256	0.0525853772111606\\
};
\addplot [color=white!60!black, line width=1.2pt, mark size=1.5pt, mark=o, mark options={solid, white!60!black}, forget plot]
  table[row sep=crcr]{%
64	0.172143201506431\\
96	0.172145894115205\\
128	0.172145892978588\\
192	0.172145898982499\\
256	0.172145893328032\\
};
\addplot [color=white!60!black, dashed, line width=1.2pt, mark size=1.5pt, mark=o, mark options={solid, white!60!black}, forget plot]
  table[row sep=crcr]{%
64	0.17102288165137\\
96	0.171022704496039\\
128	0.171022704473426\\
192	0.171022708621549\\
256	0.171022706338349\\
};
\addplot [color=white!60!black, dotted, line width=1.2pt, mark size=1.5pt, mark=o, mark options={solid, white!60!black}, forget plot]
  table[row sep=crcr]{%
64	0.172490058435502\\
96	0.172484309770518\\
128	0.172484311488093\\
192	0.172484310566561\\
256	0.17248431231321\\
};
\addplot [color=red!60!teal, line width=1.2pt, mark size=1.5pt, mark=o, mark options={solid, red!60!teal}, forget plot]
  table[row sep=crcr]{%
64	0.246953729174108\\
96	0.246162707221676\\
128	0.246163160662002\\
192	0.246163165347073\\
256	0.246163156785069\\
};
\addplot [color=red!60!teal, dashed, line width=1.2pt, mark size=1.5pt, mark=o, mark options={solid, red!60!teal}, forget plot]
  table[row sep=crcr]{%
64	0.245977416122871\\
96	0.245407818470729\\
128	0.245408634043551\\
192	0.245408632502495\\
256	0.245408637034539\\
};
\addplot [color=red!60!teal, dotted, line width=1.2pt, mark size=1.5pt, mark=o, mark options={solid, red!60!teal}, forget plot]
  table[row sep=crcr]{%
64	0.248086284297965\\
96	0.247304841697165\\
128	0.247305771402459\\
192	0.24730577501185\\
256	0.247305772618718\\
};
\node[right, align=left, inner sep=0]
at (rel axis cs:-0.25,1.15) {$(e)$};
\end{axis}
\end{tikzpicture}%
\fi
    \end{minipage}\hfill
    \begin{minipage}[t]{0.32\textwidth}
        \centering\vspace{0pt}
        \ifloadfig
%
\begin{tikzpicture}

\begin{axis}[%
width=11.255in,
height=7.131in,
at={(1.888in,0.962in)},
scale only axis,
clip=true,
separate axis lines,
every outer x axis line/.append style={black},
every x tick label/.append style={font=\color{black}},
every x tick/.append style={black},
xmin=60,
xmax=260,
xtick={ 64,  96, 128, 192, 256},
xlabel={$N$},
every outer y axis line/.append style={black},
every y tick label/.append style={font=\color{black}},
every y tick/.append style={black},
ymin=0.05,
ymax=0.250041387561402,
axis background/.style={fill=white},
clip=true,
clip mode=individual,
axis lines=box,
xtick pos=bottom,
ytick pos=left,
tick align=inside,
major tick length=2pt,
width=0.7\linewidth,
height=0.7\linewidth,
every axis/.append style={font=\fontsize{9}{9}\selectfont},xlabel style={font=\fontsize{9}{9}\selectfont},title style={font=\fontsize{9}{9}\selectfont},
legend style={font=\fontsize{8}{9}\selectfont,inner sep=1pt,row sep=1pt,column sep=2pt,nodes={scale=1.00},draw=none},legend image post style={xscale=0.35},
legend columns=1,
scaled x ticks=false,
tick label style={/pgf/number format/fixed,/pgf/number format/precision=2},
every x tick label/.append style={font=\fontsize{9}{9}\selectfont\color{black}},
every y tick label/.append style={font=\fontsize{9}{9}\selectfont\color{black}},
,,
colormap={sigmamap}{rgb(0.0000)=(0.0200,0.1880,0.3800); rgb(0.1000)=(0.0743,0.3456,0.5616); rgb(0.2000)=(0.1918,0.4980,0.7060); rgb(0.3000)=(0.3890,0.6708,0.8226); rgb(0.4000)=(0.7552,0.8682,0.9228); rgb(0.5000)=(0.9690,0.9689,0.9689); rgb(0.6000)=(0.9425,0.8044,0.7614); rgb(0.7000)=(0.8767,0.4910,0.4020); rgb(0.8000)=(0.7869,0.2866,0.2470); rgb(0.9000)=(0.6186,0.1239,0.1568); rgb(1.0000)=(0.4040,0.0000,0.1220)}
]
\addplot [color=black, line width=1.2pt, mark size=1.5pt, mark=o, mark options={solid, black}, forget plot]
  table[row sep=crcr]{%
64	0.0551661395571993\\
96	0.0551657665786296\\
128	0.0551657666634947\\
192	0.0551657667093517\\
256	0.0551657662861021\\
};
\addplot [color=black, dashed, line width=1.2pt, mark size=1.5pt, mark=o, mark options={solid, black}, forget plot]
  table[row sep=crcr]{%
64	0.0544082207367257\\
96	0.0544082206983774\\
128	0.0544082208540592\\
192	0.0544082206529434\\
256	0.0544082204400931\\
};
\addplot [color=black, dotted, line width=1.2pt, mark size=1.5pt, mark=o, mark options={solid, black}, forget plot]
  table[row sep=crcr]{%
64	0.0551357223293387\\
96	0.0551352316064079\\
128	0.0551352317804415\\
192	0.0551352310968435\\
256	0.0551352319432771\\
};
\addplot [color=white!60!black, line width=1.2pt, mark size=1.5pt, mark=o, mark options={solid, white!60!black}, forget plot]
  table[row sep=crcr]{%
64	0.175471476422374\\
96	0.17547192831869\\
128	0.175471928480403\\
192	0.17547192776773\\
256	0.175471928143913\\
};
\addplot [color=white!60!black, dashed, line width=1.2pt, mark size=1.5pt, mark=o, mark options={solid, white!60!black}, forget plot]
  table[row sep=crcr]{%
64	0.174205310154392\\
96	0.174205451538959\\
128	0.174205451683452\\
192	0.174205451850968\\
256	0.174205451609595\\
};
\addplot [color=white!60!black, dotted, line width=1.2pt, mark size=1.5pt, mark=o, mark options={solid, white!60!black}, forget plot]
  table[row sep=crcr]{%
64	0.175893947038991\\
96	0.175893008149834\\
128	0.175893008537562\\
192	0.175893009588341\\
256	0.175893008924764\\
};
\addplot [color=red!60!teal, line width=1.2pt, mark size=1.5pt, mark=o, mark options={solid, red!60!teal}, forget plot]
  table[row sep=crcr]{%
64	0.249059535004894\\
96	0.249057496036505\\
128	0.249057458109314\\
192	0.249057458206146\\
256	0.249057456773047\\
};
\addplot [color=red!60!teal, dashed, line width=1.2pt, mark size=1.5pt, mark=o, mark options={solid, red!60!teal}, forget plot]
  table[row sep=crcr]{%
64	0.248072846995502\\
96	0.248046954793252\\
128	0.248046963203167\\
192	0.248046961337035\\
256	0.248046962665486\\
};
\addplot [color=red!60!teal, dotted, line width=1.2pt, mark size=1.5pt, mark=o, mark options={solid, red!60!teal}, forget plot]
  table[row sep=crcr]{%
64	0.250041387561402\\
96	0.249946022251585\\
128	0.249946070443737\\
192	0.249946069133488\\
256	0.24994606976581\\
};
\node[right, align=left, inner sep=0]
at (rel axis cs:-0.25,1.15) {$(f)$};
\end{axis}
\end{tikzpicture}%
\fi
    \end{minipage}\\[0.5em]
    \begin{minipage}[t]{0.32\textwidth}
        \centering\vspace{0pt}
        \ifloadfig
%
\begin{tikzpicture}

\begin{axis}[%
width=11.255in,
height=7.131in,
at={(1.888in,0.962in)},
scale only axis,
clip=true,
separate axis lines,
every outer x axis line/.append style={black},
every x tick label/.append style={font=\color{black}},
every x tick/.append style={black},
xmin=60,
xmax=260,
xtick={ 64,  96, 128, 192, 256},
xlabel={$N$},
every outer y axis line/.append style={black},
every y tick label/.append style={font=\color{black}},
every y tick/.append style={black},
ymin=-1,
ymax=0,
ylabel={$\omega_r$},
axis background/.style={fill=white},
clip=true,
clip mode=individual,
axis lines=box,
xtick pos=bottom,
ytick pos=left,
tick align=inside,
major tick length=2pt,
width=0.7\linewidth,
height=0.7\linewidth,
every axis/.append style={font=\fontsize{9}{9}\selectfont},xlabel style={font=\fontsize{9}{9}\selectfont},title style={font=\fontsize{9}{9}\selectfont},
legend style={font=\fontsize{8}{9}\selectfont,inner sep=1pt,row sep=1pt,column sep=2pt,nodes={scale=1.00},draw=none},legend image post style={xscale=0.35},
legend columns=1,
scaled x ticks=false,
tick label style={/pgf/number format/fixed,/pgf/number format/precision=2},
every x tick label/.append style={font=\fontsize{9}{9}\selectfont\color{black}},
every y tick label/.append style={font=\fontsize{9}{9}\selectfont\color{black}},
,,
colormap={sigmamap}{rgb(0.0000)=(0.0200,0.1880,0.3800); rgb(0.1000)=(0.0743,0.3456,0.5616); rgb(0.2000)=(0.1918,0.4980,0.7060); rgb(0.3000)=(0.3890,0.6708,0.8226); rgb(0.4000)=(0.7552,0.8682,0.9228); rgb(0.5000)=(0.9690,0.9689,0.9689); rgb(0.6000)=(0.9425,0.8044,0.7614); rgb(0.7000)=(0.8767,0.4910,0.4020); rgb(0.8000)=(0.7869,0.2866,0.2470); rgb(0.9000)=(0.6186,0.1239,0.1568); rgb(1.0000)=(0.4040,0.0000,0.1220)}
]
\addplot [color=black, line width=1.2pt, mark size=1.5pt, mark=o, mark options={solid, black}, forget plot]
  table[row sep=crcr]{%
64	-0.0952983983829359\\
96	-0.0952982646972484\\
128	-0.0952982606997138\\
192	-0.0952982606772919\\
256	-0.0952982606773163\\
};
\addplot [color=black, dashed, line width=1.2pt, mark size=1.5pt, mark=o, mark options={solid, black}, forget plot]
  table[row sep=crcr]{%
64	-0.952964518941961\\
96	-0.952963251202302\\
128	-0.952963213095514\\
192	-0.952963212882\\
256	-0.95296321288197\\
};
\addplot [color=black, dotted, line width=1.2pt, mark size=1.5pt, mark=o, mark options={solid, black}, forget plot]
  table[row sep=crcr]{%
64	-0.105330018827117\\
96	-0.105329870810532\\
128	-0.105329866364215\\
192	-0.105329866339398\\
256	-0.10532986633925\\
};
\addplot [color=white!60!black, line width=1.2pt, mark size=1.5pt, mark=o, mark options={solid, white!60!black}, forget plot]
  table[row sep=crcr]{%
64	-0.0714737354158967\\
96	-0.0714737071023862\\
128	-0.0714737041562195\\
192	-0.0714737041395548\\
256	-0.0714737041396149\\
};
\addplot [color=white!60!black, dashed, line width=1.2pt, mark size=1.5pt, mark=o, mark options={solid, white!60!black}, forget plot]
  table[row sep=crcr]{%
64	-0.714732576305744\\
96	-0.714731584987193\\
128	-0.714731555525715\\
192	-0.714731555360003\\
256	-0.714731555359999\\
};
\addplot [color=white!60!black, dotted, line width=1.2pt, mark size=1.5pt, mark=o, mark options={solid, white!60!black}, forget plot]
  table[row sep=crcr]{%
64	-0.166773147686156\\
96	-0.166772925362327\\
128	-0.166772918420891\\
192	-0.166772918384189\\
256	-0.166772918384776\\
};
\addplot [color=red!60!teal, line width=1.2pt, mark size=1.5pt, mark=o, mark options={solid, red!60!teal}, forget plot]
  table[row sep=crcr]{%
64	-0.0570976250939065\\
96	-0.0571790070619081\\
128	-0.0571790113729609\\
192	-0.0571790116053394\\
256	-0.0571790116052954\\
};
\addplot [color=red!60!teal, dashed, line width=1.2pt, mark size=1.5pt, mark=o, mark options={solid, red!60!teal}, forget plot]
  table[row sep=crcr]{%
64	-0.571781851480997\\
96	-0.571786395657606\\
128	-0.571786364055038\\
192	-0.571786364239759\\
256	-0.571786364240081\\
};
\addplot [color=red!60!teal, dotted, line width=1.2pt, mark size=1.5pt, mark=o, mark options={solid, red!60!teal}, forget plot]
  table[row sep=crcr]{%
64	-0.400249602068264\\
96	-0.400255735287743\\
128	-0.400255711515608\\
192	-0.400255712077295\\
256	-0.400255712071431\\
};
\node[right, align=left, inner sep=0]
at (rel axis cs:-0.25,1.15) {$(g)$};
\end{axis}
\end{tikzpicture}%
\fi
    \end{minipage}\hfill
    \begin{minipage}[t]{0.32\textwidth}
        \centering\vspace{0pt}
        \ifloadfig
%
\begin{tikzpicture}

\begin{axis}[%
width=11.255in,
height=7.131in,
at={(1.888in,0.962in)},
scale only axis,
clip=true,
separate axis lines,
every outer x axis line/.append style={black},
every x tick label/.append style={font=\color{black}},
every x tick/.append style={black},
xmin=60,
xmax=260,
xtick={ 64,  96, 128, 192, 256},
xlabel={$N$},
every outer y axis line/.append style={black},
every y tick label/.append style={font=\color{black}},
every y tick/.append style={black},
ymin=-0.2,
ymax=0.15,
axis background/.style={fill=white},
clip=true,
clip mode=individual,
axis lines=box,
xtick pos=bottom,
ytick pos=left,
tick align=inside,
major tick length=2pt,
width=0.7\linewidth,
height=0.7\linewidth,
every axis/.append style={font=\fontsize{9}{9}\selectfont},xlabel style={font=\fontsize{9}{9}\selectfont},title style={font=\fontsize{9}{9}\selectfont},
legend style={font=\fontsize{8}{9}\selectfont,inner sep=1pt,row sep=1pt,column sep=2pt,nodes={scale=1.00},draw=none},legend image post style={xscale=0.35},
legend columns=1,
scaled x ticks=false,
tick label style={/pgf/number format/fixed,/pgf/number format/precision=2},
every x tick label/.append style={font=\fontsize{9}{9}\selectfont\color{black}},
every y tick label/.append style={font=\fontsize{9}{9}\selectfont\color{black}},
,,
colormap={sigmamap}{rgb(0.0000)=(0.0200,0.1880,0.3800); rgb(0.1000)=(0.0743,0.3456,0.5616); rgb(0.2000)=(0.1918,0.4980,0.7060); rgb(0.3000)=(0.3890,0.6708,0.8226); rgb(0.4000)=(0.7552,0.8682,0.9228); rgb(0.5000)=(0.9690,0.9689,0.9689); rgb(0.6000)=(0.9425,0.8044,0.7614); rgb(0.7000)=(0.8767,0.4910,0.4020); rgb(0.8000)=(0.7869,0.2866,0.2470); rgb(0.9000)=(0.6186,0.1239,0.1568); rgb(1.0000)=(0.4040,0.0000,0.1220)}
]
\addplot [color=black, line width=1.2pt, mark size=1.5pt, mark=o, mark options={solid, black}, forget plot]
  table[row sep=crcr]{%
64	-0.0952983842332047\\
96	-0.0952982506864624\\
128	-0.0952982466866296\\
192	-0.0952982466641546\\
256	-0.0952982466641561\\
};
\addplot [color=black, dashed, line width=1.2pt, mark size=1.5pt, mark=o, mark options={solid, black}, forget plot]
  table[row sep=crcr]{%
64	-0.190458142594693\\
96	-0.190458142671098\\
128	-0.190458142671199\\
192	-0.190458142668801\\
256	-0.190458142669267\\
};
\addplot [color=black, dotted, line width=1.2pt, mark size=1.5pt, mark=o, mark options={solid, black}, forget plot]
  table[row sep=crcr]{%
64	-0.105329750116102\\
96	-0.105329602717094\\
128	-0.105329598300934\\
192	-0.10532959827618\\
256	-0.105329598276117\\
};
\addplot [color=white!60!black, line width=1.2pt, mark size=1.5pt, mark=o, mark options={solid, white!60!black}, forget plot]
  table[row sep=crcr]{%
64	-0.0679286464951959\\
96	-0.0679281743339614\\
128	-0.0679281743517581\\
192	-0.0679281743310949\\
256	-0.0679281743485471\\
};
\addplot [color=white!60!black, dashed, line width=1.2pt, mark size=1.5pt, mark=o, mark options={solid, white!60!black}, forget plot]
  table[row sep=crcr]{%
64	-0.069245160075134\\
96	-0.0692451393885885\\
128	-0.0692451393854075\\
192	-0.0692451393760052\\
256	-0.069245139389848\\
};
\addplot [color=white!60!black, dotted, line width=1.2pt, mark size=1.5pt, mark=o, mark options={solid, white!60!black}, forget plot]
  table[row sep=crcr]{%
64	-0.022308042365215\\
96	-0.0223082897033318\\
128	-0.0223082893648711\\
192	-0.0223082892761897\\
256	-0.0223082893654187\\
};
\addplot [color=red!60!teal, line width=1.2pt, mark size=1.5pt, mark=o, mark options={solid, red!60!teal}, forget plot]
  table[row sep=crcr]{%
64	0.00564583914404336\\
96	0.00564380053790337\\
128	0.00564379566867459\\
192	0.00564379564412202\\
256	0.00564379562323961\\
};
\addplot [color=red!60!teal, dashed, line width=1.2pt, mark size=1.5pt, mark=o, mark options={solid, red!60!teal}, forget plot]
  table[row sep=crcr]{%
64	0.0393789338342475\\
96	0.0393637027834117\\
128	0.0393637047567001\\
192	0.0393637048632291\\
256	0.0393637048045937\\
};
\addplot [color=red!60!teal, dotted, line width=1.2pt, mark size=1.5pt, mark=o, mark options={solid, red!60!teal}, forget plot]
  table[row sep=crcr]{%
64	0.131967795446925\\
96	0.131974215452498\\
128	0.131974232534383\\
192	0.131974232671315\\
256	0.131974232597036\\
};
\node[right, align=left, inner sep=0]
at (rel axis cs:-0.25,1.15) {$(h)$};
\end{axis}
\end{tikzpicture}%
\fi
    \end{minipage}\hfill
    \begin{minipage}[t]{0.32\textwidth}
        \centering\vspace{0pt}
        \ifloadfig
%
\begin{tikzpicture}

\begin{axis}[%
width=11.255in,
height=7.131in,
at={(1.888in,0.962in)},
scale only axis,
clip=true,
separate axis lines,
every outer x axis line/.append style={black},
every x tick label/.append style={font=\color{black}},
every x tick/.append style={black},
xmin=60,
xmax=260,
xtick={ 64,  96, 128, 192, 256},
xlabel={$N$},
every outer y axis line/.append style={black},
every y tick label/.append style={font=\color{black}},
every y tick/.append style={black},
ymin=-0.05,
ymax=0.3,
axis background/.style={fill=white},
clip=true,
clip mode=individual,
axis lines=box,
xtick pos=bottom,
ytick pos=left,
tick align=inside,
major tick length=2pt,
width=0.7\linewidth,
height=0.7\linewidth,
every axis/.append style={font=\fontsize{9}{9}\selectfont},xlabel style={font=\fontsize{9}{9}\selectfont},title style={font=\fontsize{9}{9}\selectfont},
legend style={font=\fontsize{8}{9}\selectfont,inner sep=1pt,row sep=1pt,column sep=2pt,nodes={scale=1.00},draw=none},legend image post style={xscale=0.35},
legend columns=1,
scaled x ticks=false,
tick label style={/pgf/number format/fixed,/pgf/number format/precision=2},
every x tick label/.append style={font=\fontsize{9}{9}\selectfont\color{black}},
every y tick label/.append style={font=\fontsize{9}{9}\selectfont\color{black}},
,,
colormap={sigmamap}{rgb(0.0000)=(0.0200,0.1880,0.3800); rgb(0.1000)=(0.0743,0.3456,0.5616); rgb(0.2000)=(0.1918,0.4980,0.7060); rgb(0.3000)=(0.3890,0.6708,0.8226); rgb(0.4000)=(0.7552,0.8682,0.9228); rgb(0.5000)=(0.9690,0.9689,0.9689); rgb(0.6000)=(0.9425,0.8044,0.7614); rgb(0.7000)=(0.8767,0.4910,0.4020); rgb(0.8000)=(0.7869,0.2866,0.2470); rgb(0.9000)=(0.6186,0.1239,0.1568); rgb(1.0000)=(0.4040,0.0000,0.1220)}
]
\addplot [color=black, line width=1.2pt, mark size=1.5pt, mark=o, mark options={solid, black}, forget plot]
  table[row sep=crcr]{%
64	-0.00792566236976146\\
96	-0.00792568606431545\\
128	-0.00792568601830253\\
192	-0.00792568601891497\\
256	-0.00792568601868533\\
};
\addplot [color=black, dashed, line width=1.2pt, mark size=1.5pt, mark=o, mark options={solid, black}, forget plot]
  table[row sep=crcr]{%
64	-0.00941347284988948\\
96	-0.00941347291006718\\
128	-0.00941347290991271\\
192	-0.00941347290970737\\
256	-0.00941347291074579\\
};
\addplot [color=black, dotted, line width=1.2pt, mark size=1.5pt, mark=o, mark options={solid, black}, forget plot]
  table[row sep=crcr]{%
64	-0.0069763966934395\\
96	-0.00697642867852829\\
128	-0.0069764286295014\\
192	-0.00697642862853563\\
256	-0.00697642863090525\\
};
\addplot [color=white!60!black, line width=1.2pt, mark size=1.5pt, mark=o, mark options={solid, white!60!black}, forget plot]
  table[row sep=crcr]{%
64	0.117214354250601\\
96	0.11721437286601\\
128	0.117214372698788\\
192	0.117214372690895\\
256	0.117214372696531\\
};
\addplot [color=white!60!black, dashed, line width=1.2pt, mark size=1.5pt, mark=o, mark options={solid, white!60!black}, forget plot]
  table[row sep=crcr]{%
64	0.115201027728144\\
96	0.115201032663785\\
128	0.115201032665002\\
192	0.115201032658196\\
256	0.115201032656291\\
};
\addplot [color=white!60!black, dotted, line width=1.2pt, mark size=1.5pt, mark=o, mark options={solid, white!60!black}, forget plot]
  table[row sep=crcr]{%
64	0.130051378102756\\
96	0.130051336342246\\
128	0.130051336263221\\
192	0.130051336259834\\
256	0.130051336265891\\
};
\addplot [color=red!60!teal, line width=1.2pt, mark size=1.5pt, mark=o, mark options={solid, red!60!teal}, forget plot]
  table[row sep=crcr]{%
64	0.220279754122664\\
96	0.220280257995508\\
128	0.220280259137585\\
192	0.220280259149969\\
256	0.220280259136001\\
};
\addplot [color=red!60!teal, dashed, line width=1.2pt, mark size=1.5pt, mark=o, mark options={solid, red!60!teal}, forget plot]
  table[row sep=crcr]{%
64	0.224001540833978\\
96	0.224001035040474\\
128	0.224001034229596\\
192	0.224001034240017\\
256	0.224001034230161\\
};
\addplot [color=red!60!teal, dotted, line width=1.2pt, mark size=1.5pt, mark=o, mark options={solid, red!60!teal}, forget plot]
  table[row sep=crcr]{%
64	0.253693751107552\\
96	0.253697223035508\\
128	0.253697220159432\\
192	0.253697220121145\\
256	0.253697220175179\\
};
\node[right, align=left, inner sep=0]
at (rel axis cs:-0.25,1.15) {$(i)$};
\end{axis}
\end{tikzpicture}%
\fi
    \end{minipage}
    \caption{Sensitivity of the leading growth rate \(\omega_r\) to the grid resolution \(N\) at fixed \(L_z=128\), showing convergence with respect to discretization across representative parameters. Colors identify \(\Atw\), as indicated in (a), and line styles (\solidlegend, \dashlegend, \dottedlegend) identify \(\Amu \in \{-0.9, 0, 0.9\}\), respectively. (a) \(k=10^{-2}\), \(\Sch=0.05\). (b) \(k=10^{-2}\), \(\Sch=5\). (c) \(k=10^{-2}\), \(\Sch=40\). (d) \(k=10^{-1}\), \(\Sch=0.05\). (e) \(k=10^{-1}\), \(\Sch=5\). (f) \(k=10^{-1}\), \(\Sch=40\). (g) \(k=10^{0}\), \(\Sch=0.05\). (h) \(k=10^{0}\), \(\Sch=5\). (i) \(k=10^{0}\), \(\Sch=40\).}
    \label{fig:V16_wrN}
\end{figure}
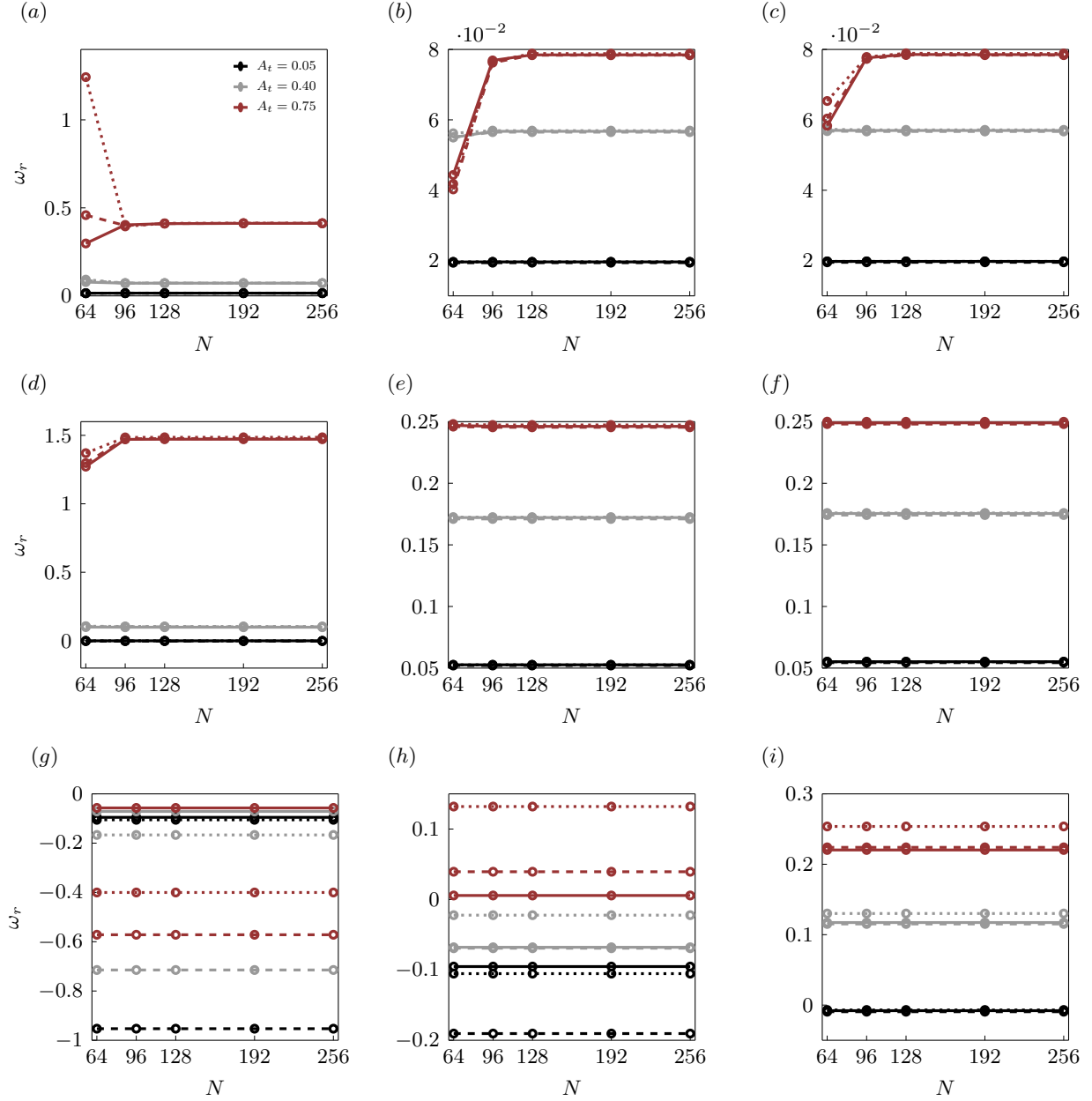

\subsection{Verification against Chandrasekhar for the variable-density solver}
\label{subsec:chandrasekhar_vd}

We first verify the variable-density solver against the non-diffusive constant-viscosity limit associated with~\citet{chandrasekhar-1961}. Starting from the general variable-density equations, this limit is obtained by setting \(\Sch^{-1}=0\) and neglecting viscosity stratification. In the present notation, the resulting linear operators are
\begin{equation}
  \begin{aligned}
    \mathscr{B}_{ww} &= -\left(\rho_0'\,\partial_z + \rho_0\,\nabla^2\right),
    &\qquad
    \mathscr{B}_{w\rho} &= 0,
    &\qquad
    \mathscr{A}_{\rho w} &= -\rho_0', \\
    \mathscr{A}_{ww} &= -\nabla^4,
    &\qquad
    \mathscr{A}_{w\rho} &= \nabla_{xy}^2,
    &\qquad
    \mathscr{A}_{\rho\rho} &= 0.
  \end{aligned}
\end{equation}
In this limit,~\citet{chandrasekhar-1961} obtained the growth rate by solving
\begin{equation}
    \label{eq:chandrasekhar_polynomial_growth_rate}
    y^{4} + \left(1-\Atw^{2}\right)y^{3} + \left(3\Atw^{2}-1\right)y^{2} - \left(1+3\Atw^{2}\right)y + \Atw^{2} - \frac{\Atw}{k^{3}} = 0,
\end{equation}
from which the temporal growth rate at wavenumber \(k\) is
\begin{equation}
    \omega_{r} = \left(y^{2}-1\right)k^{2}.
\end{equation}
The sharp-interface limit is approximated numerically by setting \(\delta=0.025\). Figure~\ref{fig:chandra_verification_vd} shows that the numerical growth rates agree very well with the analytical prediction over the full wavenumber range considered.

\begin{figure}
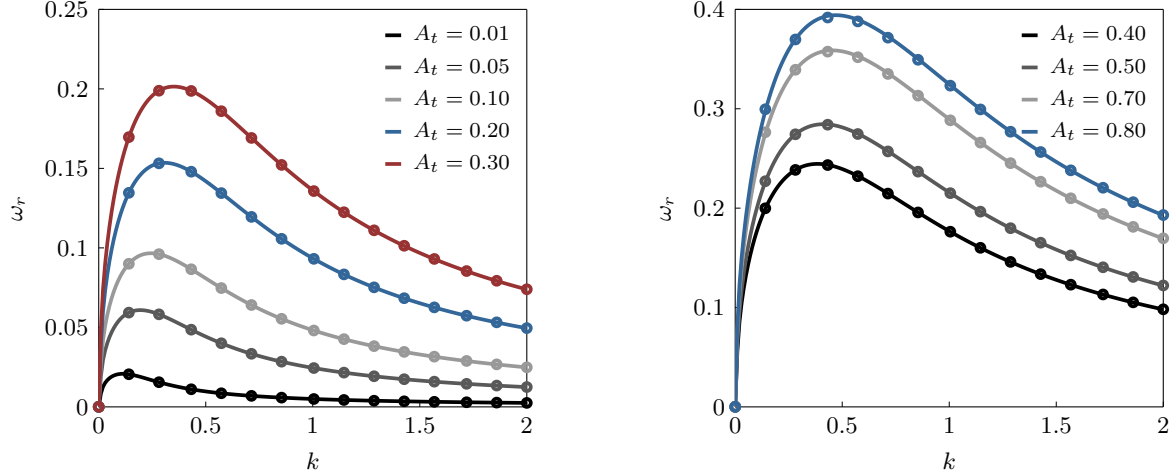

    \centering
    \begin{minipage}[t]{0.49\textwidth}
        \centering
        \ifloadfig\input{figures/verification_figures/V03_chandra_validation_VD_1.tex}\fi
    \end{minipage}
    \hfill
    \begin{minipage}[t]{0.49\textwidth}
        \centering
        \ifloadfig\input{figures/verification_figures/V03_chandra_validation_VD_2.tex}\fi
    \end{minipage}
    \caption{Verification of the variable-density solver against the analytical growth rates of~\citet{chandrasekhar-1961} in the sharp-interface limit. (\circlegend) denotes the analytical solution obtained from equation~\eqref{eq:chandrasekhar_polynomial_growth_rate}, and (\solidlegend) denotes the numerical results.}
    \label{fig:chandra_verification_vd}
\end{figure}

\subsection[Verification against Morgan et al.\ (2016)]{Verification against~\citet{morgan-likhachev-jacobs-2016}}
\label{subsec:morgan}

A further verification is provided by comparison with~\citet{morgan-likhachev-jacobs-2016}. This comparison tests whether the present formulation reproduces the growth-rate trends associated with variations in interface thickness and viscosity ratio. Figure~\ref{fig:visc_validation_VD} shows good agreement with the reported reference data.

\begin{figure}
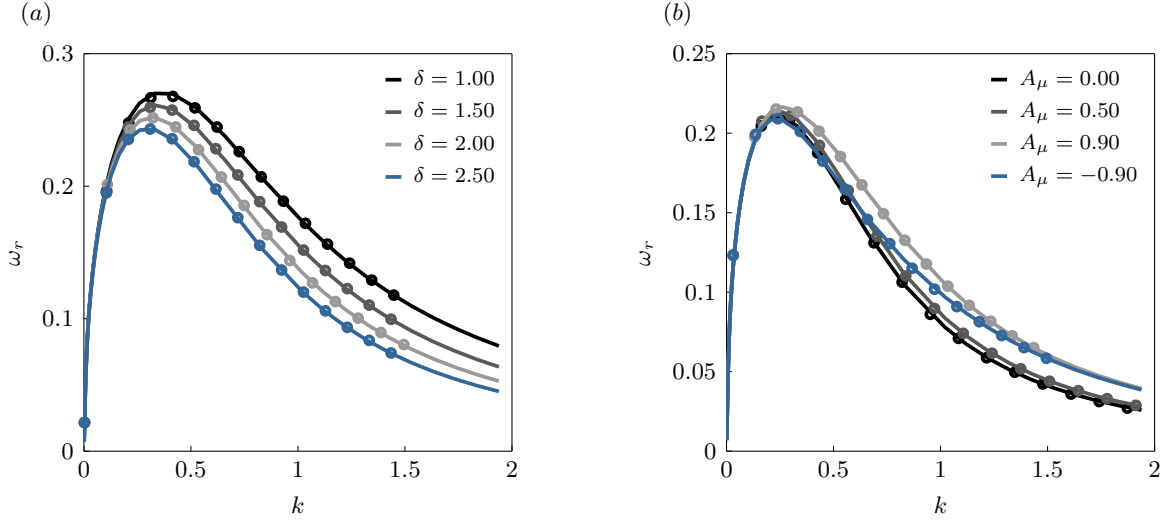

    \centering
    \begin{minipage}[t]{0.49\textwidth}
        \centering
        \ifloadfig\input{figures/verification_figures/V06_visc_validation_VD_1.tex}\fi
    \end{minipage}
    \hfill
    \begin{minipage}[t]{0.49\textwidth}
        \centering
        \ifloadfig\input{figures/verification_figures/V06_visc_validation_VD_2.tex}\fi
    \end{minipage}
    \caption{Verification against the data reported by~\citet{morgan-likhachev-jacobs-2016}, showing that the present solver reproduces the trends with interface thickness and viscosity ratio. (\solidlegend) denotes the present solver and (\circlegend) denotes the data of~\citet{morgan-likhachev-jacobs-2016}.(a) corresponds to \(\Amu=0\), \(\Atw=0.5\), with varying \(\delta\);(b) corresponds to \(\Atw=0.5\), \(\delta=5\), with varying \(\Amu\).}
    \label{fig:visc_validation_VD}
\end{figure}

\subsection[Verification against Mikaelian (1996)]{Verification against~\citet{mikaelian-1996}}
\label{subsec:mikaelian}

A final and more general verification is provided by comparison with the semi-analytical results of~\citet{mikaelian-1996}, which extend Chandrasekhar's analysis to finite layers with discontinuous density and viscosity. In that configuration,
\begin{equation}
    \rho(z) =
    \begin{cases}
        \rho_{1}, & -\,t_{1} \le z \le 0, \\
        \rho_{2}, & 0 \le z \le t_{2},
    \end{cases}
    \quad
    \mu(z) =
    \begin{cases}
        \mu_{1}, & -\,t_{1} \le z \le 0, \\
        \mu_{2}, & 0 \le z \le t_{2}.
    \end{cases}
\end{equation}
The amplitude coefficients satisfy the linear system
\begin{equation}
    \begin{pmatrix}
        1 & 1 & 1 & 1 & -1 & -1 & -1 & -1 \\
        k & q_{1} & -k & -q_{1} & k & q_{2} & -k & -q_{2} \\
        e^{-k t_{1}} & e^{-q_{1} t_{1}} & e^{k t_{1}} & e^{q_{1} t_{1}} & 0 & 0 & 0 & 0 \\
        0 & 0 & 0 & 0 & e^{-k t_{2}} & e^{-q_{2} t_{2}} & e^{k t_{2}} & e^{q_{2} t_{2}} \\
        k e^{-k t_{1}} & q_{1} e^{-q_{1} t_{1}} & -k e^{k t_{1}} & -q_{1} e^{q_{1} t_{1}} & 0 & 0 & 0 & 0 \\
        0 & 0 & 0 & 0 & -k e^{-k t_{2}} & -q_{2} e^{-q_{2} t_{2}} & k e^{k t_{2}} & q_{2} e^{q_{2} t_{2}} \\
        2k^{2}\mu_{1} & Q_{1}\mu_{1} & 2k^{2}\mu_{1} & Q_{1}\mu_{1} & -2k^{2}\mu_{2} & -Q_{2}\mu_{2} & -2k^{2}\mu_{2} & -Q_{2}\mu_{2} \\
        \rho_{1}+\alpha+\beta k & \alpha+\beta q_{1} & -\rho_{1}+\alpha-\beta k & \alpha-\beta q_{1} & \rho_{2} & 0 & -\rho_{2} & 0
    \end{pmatrix}
    \begin{pmatrix}
        A_{1}\\ B_{1}\\ C_{1}\\ D_{1}\\ A_{2}\\ B_{2}\\ C_{2}\\ D_{2}
    \end{pmatrix}
    = \boldsymbol{0},
\end{equation}
which admits nontrivial solutions when
\begin{equation}
    \label{eq:mikaelian_determinant}
    \det\bigl(\boldsymbol{M}(\gamma, k)\bigr) = 0.
\end{equation}
The definitions of the quantities entering \(\boldsymbol{M}\) are given in~\citet{mikaelian-1996}.
Figures~\ref{fig:mikaelian_verification_Amu_neg09}--\ref{fig:mikaelian_verification_Amu_pos09} compare the present results with the semi-analytical solutions for three representative Atwood numbers at fixed viscosity ratio, and show good agreement throughout the tested parameter range.

\begin{figure}
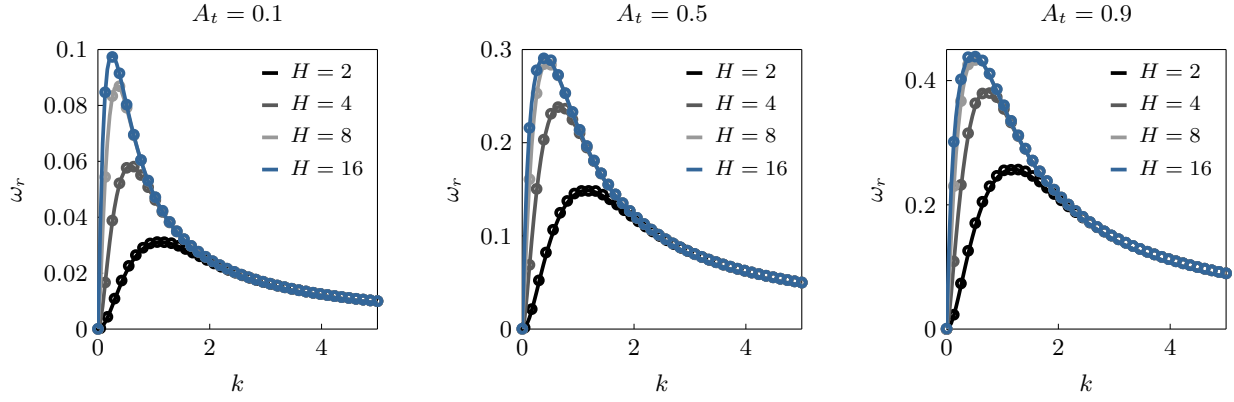
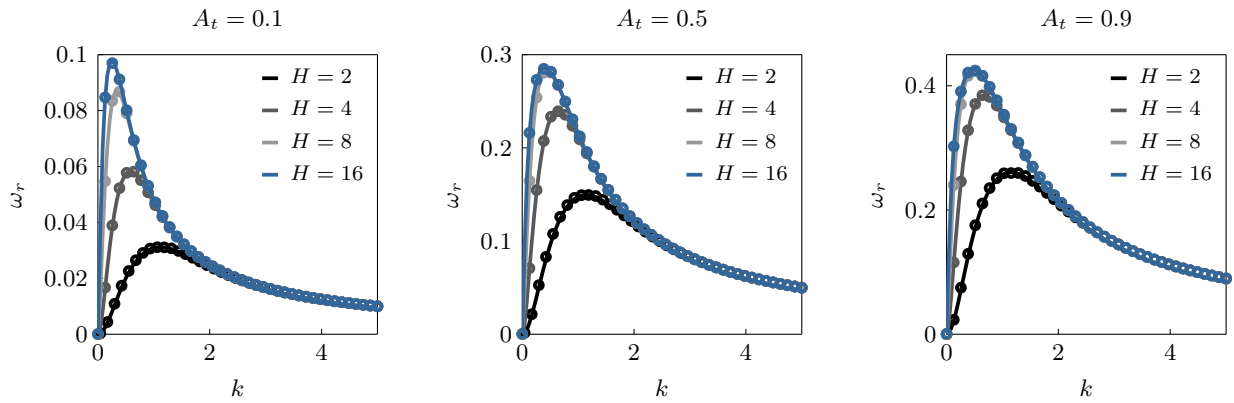

    \centering
    \begin{minipage}[t]{0.32\textwidth}
        \centering
        \ifloadfig\input{figures/verification_figures/V05_mikaelian_validation_VD_At0.1_Amu-0.9.tex}\fi
    \end{minipage}
    \hfill
    \begin{minipage}[t]{0.32\textwidth}
        \centering
        \ifloadfig\input{figures/verification_figures/V05_mikaelian_validation_VD_At0.5_Amu-0.9.tex}\fi
    \end{minipage}
    \hfill
    \begin{minipage}[t]{0.32\textwidth}
        \centering
        \ifloadfig\input{figures/verification_figures/V05_mikaelian_validation_VD_At0.9_Amu-0.9.tex}\fi
    \end{minipage}
    \caption{Verification against the semi-analytical growth rates of~\citet{mikaelian-1996} at fixed \(\Amu=-0.9\). \(H\) denotes the domain half-height. (\circlegend) denotes the semi-analytical solution obtained from equation~\eqref{eq:mikaelian_determinant}, and (\solidlegend) denotes the numerical results.}
    \label{fig:mikaelian_verification_Amu_neg09}
\end{figure}

\begin{figure}
    \centering
    \begin{minipage}[t]{0.32\textwidth}
        \centering
        \ifloadfig\input{figures/verification_figures/V05_mikaelian_validation_VD_At0.1_Amu0.0.tex}\fi
    \end{minipage}
    \hfill
    \begin{minipage}[t]{0.32\textwidth}
        \centering
        \ifloadfig\input{figures/verification_figures/V05_mikaelian_validation_VD_At0.5_Amu0.0.tex}\fi
    \end{minipage}
    \hfill
    \begin{minipage}[t]{0.32\textwidth}
        \centering
        \ifloadfig\input{figures/verification_figures/V05_mikaelian_validation_VD_At0.9_Amu0.0.tex}\fi
    \end{minipage}
    \caption{Verification against the semi-analytical growth rates of~\citet{mikaelian-1996} at fixed \(\Amu=0.0\). \(H\) denotes the domain half-height. (\circlegend) denotes the semi-analytical solution obtained from equation~\eqref{eq:mikaelian_determinant}, and (\solidlegend) denotes the numerical results.}
\end{figure}

\begin{figure}
    \centering
    \begin{minipage}[t]{0.32\textwidth}
        \centering
        \ifloadfig\input{figures/verification_figures/V05_mikaelian_validation_VD_At0.1_Amu0.9.tex}\fi
    \end{minipage}
    \hfill
    \begin{minipage}[t]{0.32\textwidth}
        \centering
        \ifloadfig\input{figures/verification_figures/V05_mikaelian_validation_VD_At0.5_Amu0.9.tex}\fi
    \end{minipage}
    \hfill
    \begin{minipage}[t]{0.32\textwidth}
        \centering
        \ifloadfig\input{figures/verification_figures/V05_mikaelian_validation_VD_At0.9_Amu0.9.tex}\fi
    \end{minipage}
    \caption{Verification against the semi-analytical growth rates of~\citet{mikaelian-1996} at fixed \(\Amu=0.9\). \(H\) denotes the domain half-height. (\circlegend) denotes the semi-analytical solution obtained from equation~\eqref{eq:mikaelian_determinant}, and (\solidlegend) denotes the numerical results.}
    \label{fig:mikaelian_verification_Amu_pos09}
\end{figure}

\end{appen}

\clearpage
\bibliographystyle{plainnat}
\bibliography{references}

\clearpage
\end{document}